\documentclass[aps,twocolumn,showkeys,showpacs,preprintnumbers,prd,superscriptaddress,nofootinbib,10pt]{revtex4-1}
\usepackage{graphicx,epsf,bm,amsmath,amsfonts,amssymb,epstopdf,natbib,hyperref,color,verbatim,multirow,bm}
\hypersetup{colorlinks=true,urlcolor=blue,citecolor=blue,linkcolor=blue,menucolor=blue,anchorcolor=blue,filecolor=blue}
\usepackage{appendix}
\usepackage{amsmath}
\date{\today}

\begin{document}

\title{Primordial regular black holes as all the dark matter. \\ II. Non-time-radial-symmetric and loop quantum gravity-inspired metrics}

\author{Marco Calz\`{a}}
\email{marco.calza@unitn.it}
\affiliation{Department of Physics, University of Trento, Via Sommarive 14, 38123 Povo (TN), Italy}
\thanks{M.C. and D.P. contributed equally to this work}
\affiliation{Trento Institute for Fundamental Physics and Applications (TIFPA)-INFN, Via Sommarive 14, 38123 Povo (TN), Italy}

\author{Davide Pedrotti}
\email{davide.pedrotti-1@unitn.it}
\affiliation{Department of Physics, University of Trento, Via Sommarive 14, 38123 Povo (TN), Italy}
\affiliation{Trento Institute for Fundamental Physics and Applications (TIFPA)-INFN, Via Sommarive 14, 38123 Povo (TN), Italy}

\author{Sunny Vagnozzi}
\email{sunny.vagnozzi@unitn.it}
\affiliation{Department of Physics, University of Trento, Via Sommarive 14, 38123 Povo (TN), Italy}
\affiliation{Trento Institute for Fundamental Physics and Applications (TIFPA)-INFN, Via Sommarive 14, 38123 Povo (TN), Italy}

\begin{abstract}
\noindent It is a common belief that a theory of quantum gravity should ultimately cure curvature singularities which are inevitable within General Relativity, and plague for instance the Schwarzschild and Kerr metrics, usually considered as prototypes for primordial black holes (PBHs) as dark matter (DM) candidates. We continue our study, initiated in a companion paper, of non-singular objects as PBHs, considering three regular non-\textit{tr} (non-time-radial)-symmetric metrics, all of which are one-parameter extensions of the Schwarzschild space-time: the Simpson-Visser, Peltola-Kunstatter, and D'Ambrosio-Rovelli space-times, with the latter two motivated by loop quantum gravity. We study evaporation constraints on PBHs described by these regular metrics, deriving upper limits on $f_{\text{pbh}}$, the fraction of DM in the form of PBHs. Compared to their Schwarzschild counterparts, these limits are weaker, and result in a larger asteroid mass window where all the DM can be in the form of PBHs, with the lower edge moving potentially more than an order of magnitude. Our work demonstrates as a proof-of-principle that quantum gravity-inspired space-times can simultaneously play an important role in the resolution of singularities and in the DM problem.
\end{abstract}

\maketitle

\section{Introduction}
\label{sec:introduction}

Once regarded as objects of pure mathematical interest, over the past decade black holes (BHs) have gone on to become some of the most fascinating objects in the Universe~\cite{Cardoso:2019rvt}. At the time of writing, observational effects associated to astrophysical BHs are detected on a regular basis allowing us to use these extreme regions of space-time as unique laboratories for testing fundamental physics in the strong-field regime~\cite{Creminelli:2017sry,Sakstein:2017xjx,Ezquiaga:2017ekz,Boran:2017rdn,Baker:2017hug,Amendola:2017orw,Visinelli:2017bny,Crisostomi:2017lbg,Dima:2017pwp,Cai:2018rzd,Casalino:2018tcd,Barack:2018yly,LIGOScientific:2018dkp,Casalino:2018wnc,Held:2019xde,Bambi:2019tjh,Vagnozzi:2019apd,Zhu:2019ura,Cunha:2019ikd,Banerjee:2019nnj,Banerjee:2019xds,Allahyari:2019jqz,Vagnozzi:2020quf,Khodadi:2020jij,Kumar:2020yem,Khodadi:2020gns,Pantig:2021zqe,Khodadi:2021gbc,Roy:2021uye,Uniyal:2022vdu,Pantig:2022ely,Ghosh:2022kit,Khodadi:2022pqh,KumarWalia:2022aop,Shaikh:2022ivr,Odintsov:2022umu,Oikonomou:2022tjm,Pantig:2023yer,Gonzalez:2023rsd,Sahoo:2023czj,Nozari:2023flq,Uniyal:2023ahv,Filho:2023ycx,Raza:2023vkn,Hoshimov:2023tlz,Chakhchi:2024tzo,Liu:2024lbi,Liu:2024lve,Khodadi:2024ubi,Liu:2024soc,Nojiri:2024txy}. On the more theoretical end of the spectrum, a widespread hope is that BHs may hold the key towards the unification of quantum mechanics and gravity, although a somewhat more humble goal could be that of using BH observations to test candidate theories of quantum gravity (QG). On the more phenomenological side, the possible role of BHs in accounting for the dark matter (DM) which makes up $\simeq 25\%$ of the energy budget of the Universe~\cite{Arbey:2021gdg,Cirelli:2024ssz} is now widely acknowledged. In our work, these two aspects -- DM and candidate theories of QG -- will naturally meet, with BHs being the common denominator, and astrophysical observations the playing ground.

The collapse of large density perturbations upon horizon re-entry in the early Universe can lead to the formation of hypothetical relics known as primordial BHs (PBHs), whose role as potential DM candidates has long been recognized~\cite{Chapline:1975ojl,Meszaros:1975ef,Khlopov:1980mg,Khlopov:1985fch,Ivanov:1994pa,Choudhury:2013woa,Belotsky:2014kca,Bird:2016dcv,Clesse:2016vqa,Poulin:2017bwe,Raccanelli:2017xee,LuisBernal:2017fmf,Clesse:2017bsw,Kohri:2018qtx,Liu:2018ess,Liu:2019rnx,Murgia:2019duy,Carr:2019kxo,Liu:2020cds,Hertzberg:2020hsz,Serpico:2020ehh,DeLuca:2020bjf,DeLuca:2020fpg,DeLuca:2020qqa,Carr:2020erq,Bhagwat:2020bzh,DeLuca:2020sae,Wong:2020yig,Carr:2020mqm,Domenech:2020ssp,DeLuca:2021wjr,Arbey:2021ysg,Franciolini:2021tla,DeLuca:2021hde,Cheek:2021odj,Cheek:2021cfe,Heydari:2021gea,Dvali:2021byy,Heydari:2021qsr,DeLuca:2021pls,Liu:2021jnw,Saha:2021pqf,Bhaumik:2022pil,Anchordoqui:2022txe,Cai:2022erk,Oguri:2022fir,Franciolini:2022tfm,Mazde:2022sdx,Cai:2022kbp,Anchordoqui:2022tgp,Liu:2022iuf,Fu:2022ypp,Choudhury:2023vuj,Papanikolaou:2023crz,deFreitasPacheco:2023hpb,Choudhury:2023jlt,Choudhury:2023rks,Musco:2023dak,Yuan:2023bvh,Choudhury:2023hvf,Ghoshal:2023sfa,Cai:2023uhc,Choudhury:2023kdb,Huang:2023chx,Choudhury:2023hfm,Bhattacharya:2023ysp,Heydari:2023xts,Heydari:2023rmq,Choudhury:2023fwk,Choudhury:2023fjs,Ghoshal:2023pcx,Hai-LongHuang:2023atg,Huang:2023mwy,Anchordoqui:2024akj,Choudhury:2024one,Thoss:2024hsr,Papanikolaou:2024kjb,Choudhury:2024ybk,Choudhury:2024jlz,Anchordoqui:2024dxu,Papanikolaou:2024fzf,Yin:2024xov,Choudhury:2024dei,Heydari:2024bxj,Dvali:2024hsb,Boccia:2024nly,Hai-LongHuang:2024kye,Choudhury:2024dzw,Anchordoqui:2024jkn,Yang:2024vij,Saha:2024ies,Anchordoqui:2024tdj,Chen:2024pge,Dai:2024guo,Hai-LongHuang:2024vvz,Zantedeschi:2024ram,Chianese:2024rsn,Barker:2024mpz,Borah:2024bcr,Hai-LongHuang:2024gtx} (for recent reviews on the topic, see Refs.~\cite{Khlopov:2008qy,Carr:2016drx,Green:2020jor,Carr:2020xqk,Villanueva-Domingo:2021spv,Carr:2021bzv,Bird:2022wvk,Carr:2023tpt,Arbey:2024ujg,Choudhury:2024aji}). A wide range of observations (mainly of astrophysical nature) severely limit the ability of PBHs to account for the entire DM component: in practice, this is potentially possible (although this possibility is not completely free of debates) only in the so-called ``\textit{asteroid mass window}'', i.e.\ $10^{17}\,{\text{g}} \lesssim M_{\text{pbh}} \lesssim 10^{23}\,{\text{g}}$, with lighter and heavier PBHs being tightly constrained by observational signatures of their evaporation and microlensing respectively~\cite{Katz:2018zrn,Bai:2018bej,Smyth:2019whb,Coogan:2020tuf,Ray:2021mxu,Auffinger:2022dic,Ghosh:2022okj,Miller:2021knj,Branco:2023frw,Bertrand:2023zkl,Tran:2023jci,Gorton:2024cdm,Dent:2024yje,Tamta:2024pow,Tinyakov:2024mcy,Loeb:2024tcc}. However, it is important to note that virtually all constraints on PBHs, including those determining the existence and extension of the asteroid mass window, are subject to the underlying assumption about these object being either Schwarzschild or Kerr BHs~\cite{Khlopov:2008qy,Carr:2016drx,Green:2020jor,Carr:2020xqk,Villanueva-Domingo:2021spv,Carr:2021bzv,Bird:2022wvk,Carr:2023tpt,Arbey:2024ujg,Choudhury:2024aji}. This assumption is perfectly reasonable from the phenomenological and observational point of view, but at the same time may be cause of some apprehensiveness on the more theoretical side. In fact, these metrics feature pathological curvature singularities, whose existence is virtually inevitable in General Relativity (GR), and is at the essence of the well-known singularity problem~\cite{Ansoldi:2008jw,Nicolini:2008aj,Sebastiani:2022wbz,Torres:2022twv,Lan:2023cvz}.

Given that significant efforts are being devoted to the study of so-called regular space-times, free of curvature singularities, a relevant question is therefore what happens if PBHs are regular. This is a question we started to systematically address in a companion paper focused on \textit{tr} (time-radius)-symmetric metrics, i.e.\ where the product of the coefficients of the $dt^2$ and $dr^2$ terms in the line element in four-dimensional Boyer–Lindquist coordinates is equal to $-1$~\cite{Calza:2024fzo}: these metrics include, for instance, the well-known Bardeen~\cite{Bardeen:1968ghw} and Hayward regular BHs~\cite{Hayward:2005gi}. As we show in our companion paper, the phenomenology of the resulting primordial regular BHs (PRBHs) can be very rich, and can result in the asteroid mass window opening by up to an extra decade in mass~\cite{Calza:2024fzo}. The choice of studying \textit{tr}-symmetric metrics was adopted to make the equations simpler to handle, but is certainly not exhaustive. Indeed, as we shall soon see, such a choice does not cover a number of well-known and well-motivated metrics, potentially including space-times rooted into candidate theories of QG. In this sense, it is worth recalling that the metrics considered in our companion paper~\cite{Calza:2024fzo} are purely phenomenological in nature. It is therefore our goal in the present work to extend our earlier study of PRBHs to non-\textit{tr}-symmetric metrics, some of which carry very strong theoretical motivation and can arise within candidate theories of QG.

To be concrete, in what follows we will consider three regular, static spherically symmetric space-times, characterized by an additional \textit{regularizing} parameter $\ell$, and recovering the Schwarzschild space-time in the $\ell \to 0$ limit. All three space-times enjoy quite different properties compared to the phenomenological ones considered in our companion paper~\cite{Calza:2024fzo}. The first metric we consider is the so-called Simpson-Visser metric: this is arguably one of the best known black-bounce space-times, and interpolates between the Schwarzschild metric, regular BHs, and traversable wormholes.~\footnote{A traversable wormhole is one for which a particle can enter through one side of the wormhole and exit through the other. In principle, these could be traversed by a human traveler without fatal effects~\cite{Morris:1988cz}, if certain conditions on the metric elements $g_{tt}$ and $g_{rr}$ are satisfied. In practice, whether this trip would actually be free from fatal effects depends on the question of how a human body would interact with the exotic matter which is required to sustain the wormhole, question which is presently open.} The other two metrics are instead deeply rooted within Loop Quantum Gravity (LQG)~\cite{Ashtekar:1986yd,Rovelli:1997yv,Ashtekar:2021kfp}: arguably one of the leading QG approaches, LQG is a fully non-perturbative and manifestly background-independent approach towards a consistent theory of QG, wherein space-time is fundamentally discrete (see e.g.\ Refs.~\cite{Bojowald:2001xe,Bojowald:2005epg,Modesto:2006qh,Engle:2007wy,Modesto:2008jz,Bianchi:2010gc,Singh:2010qa,Bojowald:2011aa,Cailleteau:2012fy,Agullo:2013ai,Haggard:2014rza,Odintsov:2014gea,Odintsov:2015uca,Haro:2015oqa,Odintsov:2016apy,DeHaro:2018hia,deHaro:2018sqw,Odintsov:2018awm,Benetti:2019kgw,Casalino:2019tho,Bouhmadi-Lopez:2020oia,Graef:2020qwe,Brahma:2020eos,Barboza:2022hng,SVicente:2022ebm,Afrin:2022ztr,Yan:2023vdg,Kumar:2023jgh,Jiang:2023img,Jiang:2024cpe} for various follow-up studies and applications). More specifically, the two regular LQG-motivated metrics we analyze as candidates for DM in the form of PRBHs are the Peltola-Kunstatter~\cite{Peltola:2008pa,Peltola:2009jm} and D'Ambrosio-Rovelli space-times~\cite{Bianchi:2018mml,DAmbrosio:2018wgv}. As a cautionary note, we remark that ours is to be intended as a pilot study in this direction, and that much more follow-up work is needed before primordial regular BHs as DM candidates are characterized to the same extent as their Schwarzschild counterparts.~\footnote{As far as we are aware, only seven earlier works considered primordial regular BHs as DM~\cite{Easson:2002tg,Dymnikova:2015yma,Pacheco:2018mvs,Arbey:2021mbl,Arbey:2022mcd,Banerjee:2024sao,Davies:2024ysj}, but mostly focused on aspects other than the ones studied here (with the exception of Refs.~\cite{Arbey:2021mbl,Arbey:2022mcd}, which however studied a different LQG-inspired metric, albeit reaching conclusions qualitatively similar to ours).}

The rest of this paper is then organized as follows. We briefly introduce the regular space-times studied in our work in Sec.~\ref{sec:regular}. Theoretical aspects of the Hawking evaporation process are presented in the next two sections, with Sec.~\ref{subsec:greybody} devoted to the derivation of the greybody factors, Sec.~\ref{subsec:spectra} to the calculation of the resulting photon spectra, and Sec.~\ref{subsec:constraints} to the derivation of constraints on the fraction of DM which may be in the form of PRBHs. The resulting limits are discussed in Sec.~\ref{sec:results}. Finally, in Sec.~\ref{sec:conclusions} we draw concluding remarks. Technical issues regarding the asymptotic solutions of the radial Teukolsky equation which may be of interest to some readers are discussed in Appendix~\ref{sec:appendix}. Unless otherwise specified, we adopt units where $G=c=\hbar=1$. We recall once again that a related study focusing on \textit{tr}-symmetric, phenomenological metrics is presented in our companion paper~\cite{Calza:2024fzo}. If time allows our recommendation is that the interested reader consult our companion paper~\cite{Calza:2024fzo} prior to reading the present work.

\section{Regular black holes}
\label{sec:regular}

As is well known, GR predicts the almost unavoidable existence of essential space-time singularities, where curvature invariants diverge. Nevertheless, it is a commonly held belief that these unwanted features are merely a reflection of our ignorance of a more fundamental theory of QG, which would ultimately cure these singularities (see, however, Refs.~\cite{Sachs:2021mcu,Ashtekar:2021dab,Ashtekar:2022oyq}). Various regular BH (RBH) metrics, free of singularities in the entire space-time, have in fact been studied in recent years, both from a more phenomenological standpoint~\cite{Borde:1996df,AyonBeato:1998ub,AyonBeato:1999rg,Bronnikov:2005gm,Berej:2006cc,Bronnikov:2012ch,Rinaldi:2012vy,Stuchlik:2014qja,Schee:2015nua,Johannsen:2015pca,Myrzakulov:2015kda,Fan:2016hvf,Sebastiani:2016ras,Toshmatov:2017zpr,Chinaglia:2017uqd,Frolov:2017dwy,Bertipagani:2020awe,Nashed:2021pah,Simpson:2021dyo,Franzin:2022iai,Chataignier:2022yic,Ghosh:2022gka,Khodadi:2022dyi,Farrah:2023opk,Fontana:2023zqz,Boshkayev:2023rhr,Luongo:2023jyz,Luongo:2023aib,Cadoni:2023lum,Giambo:2023zmy,Cadoni:2023lqe,Luongo:2023xaw,Sajadi:2023ybm,Javed:2024wbc,Ditta:2024jrv,Al-Badawi:2024lvc,Ovgun:2024zmt,Corona:2024gth,Bueno:2024dgm,Konoplya:2024hfg,Pedrotti:2024znu,Bronnikov:2024izh,Kurmanov:2024hpn,Bolokhov:2024sdy,Agrawal:2024wwt,Belfiglio:2024wel,Stashko:2024wuq,Faraoni:2024ghi,Konoplya:2024lch,Khodadi:2024efq,Calza:2024qxn} as well as from a first-principles theoretical basis~\cite{Dymnikova:1992ux,Dymnikova:2004qg,Ashtekar:2005cj,Bebronne:2009mz,Modesto:2010uh,Spallucci:2011rn,Perez:2014xca,Colleaux:2017ibe,Nicolini:2019irw,Bosma:2019aiu,Jusufi:2022cfw,Olmo:2022cui,Jusufi:2022rbt,Ashtekar:2023cod,Nicolini:2023hub}.~\footnote{We note that finiteness of curvature invariants does not necessarily imply geodesic completeness, and viceversa, and issue which plagues a number of well-known regular BHs, including the Bardeen and Hayward ones~\cite{Zhou:2022yio}. We further remark that the stability of several regular BH solutions is currently being debated~\cite{Carballo-Rubio:2021bpr,Carballo-Rubio:2022kad,Bonanno:2022jjp,Bonanno:2022rvo,Carballo-Rubio:2022twq}.} These RBHs are usually controlled by an extra \textit{regularizing parameter} (which we denote by $\ell$), and typically (but not necessarily) recover the Schwarzschild metric as $\ell \to 0$. In what follows, similarly to our companion paper~\cite{Calza:2024fzo} we will entertain the possibility that DM may be in the form of primordial RBHs.

The line element of the space-times we consider can all be written in the following general form:
\begin{eqnarray}
ds^2=-\mathfrak{f}(\tilde r) dt^2 + \frac{1}{\mathfrak{f}(\tilde r) \left [ 1-\mathfrak{g}_\ell(\tilde{r}) \right ] }d\tilde{r} ^2+\tilde{r}^2d\Omega^2\,,
\label{eq:extrinsic}
\end{eqnarray}
where $d\Omega^2=d\theta^2+\sin^2(\theta)d\phi^2$ is the metric on the 2-sphere and $\tilde{r}$ is manifestly the areal radius. On the other the function $\mathfrak{g}_\ell$, which depends on the regularizing parameter $\ell$, goes to $\mathfrak{g}_\ell(\tilde r) \rightarrow 0$ for both $\ell \rightarrow 0$ and $\tilde r \rightarrow \infty$. Such a space-time possesses horizon-like structures located at radial coordinates $\tilde{r}$ such that:
\begin{eqnarray}
\mathfrak{f}(\tilde{r}) \left [ 1-\mathfrak{g}_\ell(\tilde{r}) \right ] =0\,,
\label{eq:horizon}
\end{eqnarray}
i.e.\ at $\tilde{r}=\tilde{r}_H,\tilde{r}_0$ such that $\mathfrak{f}(\tilde{r}_H)=0$ and/or $\mathfrak{g}_\ell(\tilde{r}_0)=1$. If $\tilde{r}_0>\tilde{r}_H$, the value $\tilde{r}_0$ determines the location of a wormhole (WH) throat, whereas no event on the manifold is associated to the location of $\tilde{r}_H$. Note that, in general $\tilde{r}_0$ depends on the regularizing parameter, i.e.\ $\tilde{r}_0 = \tilde{r}_0(\ell)$. On the other hand, when $\tilde{r}_H>\tilde{r}_0$, the value $\tilde{r}_H$ characterizes the event horizon of a BH (in this case the WH throat is located within the BH event horizon and is therefore causally disconnected from the relevant BH exterior space-time). Typically, in regions where these space-times are regular, the above geometry describes a bounce into a future incarnation of the universe~\cite{Barcelo:2016hgb,Malafarina:2017csn,Barrau:2018rts}. Due to this peculiar characteristic, geometries of this type are sometimes referred to as black-bounce space-times.

For these types of metric, it generally proves advantageous to perform a change of variable for what concerns the radial coordinate, going from an extrinsic description to an intrinsic one through the coordinate transformation $\tilde{r}=\sqrt{r^2+\ell^2}$. The metric in Eq.~(\ref{eq:extrinsic}) can then be expressed in the following form:
\begin{eqnarray}
ds^2 = -f(r)dt^2+g(r)^{-1}dr^2+h(r)d\Omega^2\,,
\label{eq:intrinsic}
\end{eqnarray}
where $h(r)=r^2+\ell^2$. When expressed in the above form, the Petrov-D nature of this class of metrics is manifest. We additionally require asymptotic flatness, in other words that $\mathfrak{f}(\tilde r) \rightarrow 1$ for $\tilde r \rightarrow \infty$, from which it follows that:
\begin{eqnarray}
f(r) \xrightarrow{r \to \infty} 1\,, \quad g(r) \xrightarrow{r \to \infty} 1\,, \quad h(r) \xrightarrow{r \to \infty} r^2\,.
\label{eq:asflat}
\end{eqnarray}
Finally, we note that the metrics in question are non-\textit{tr}-symmetric, since in general $f(r) \neq g(r)$ and $h(r) \neq r^2$. The \textit{tr}-symmetric case is treated in our companion paper~\cite{Calza:2024fzo}, whereas we have chosen to deal with the non-\textit{tr}-symmetric case in a separate work both because it introduces non-trivial complications on the mathematical side, and at the same time allows us to treat metrics which are strongly motivated from first-principles theoretical considerations (unlike those considered in our companion paper, which are introduced on purely phenomenological grounds), such as LQG.

\begin{figure}
\centering
\includegraphics[width=1.0\columnwidth]{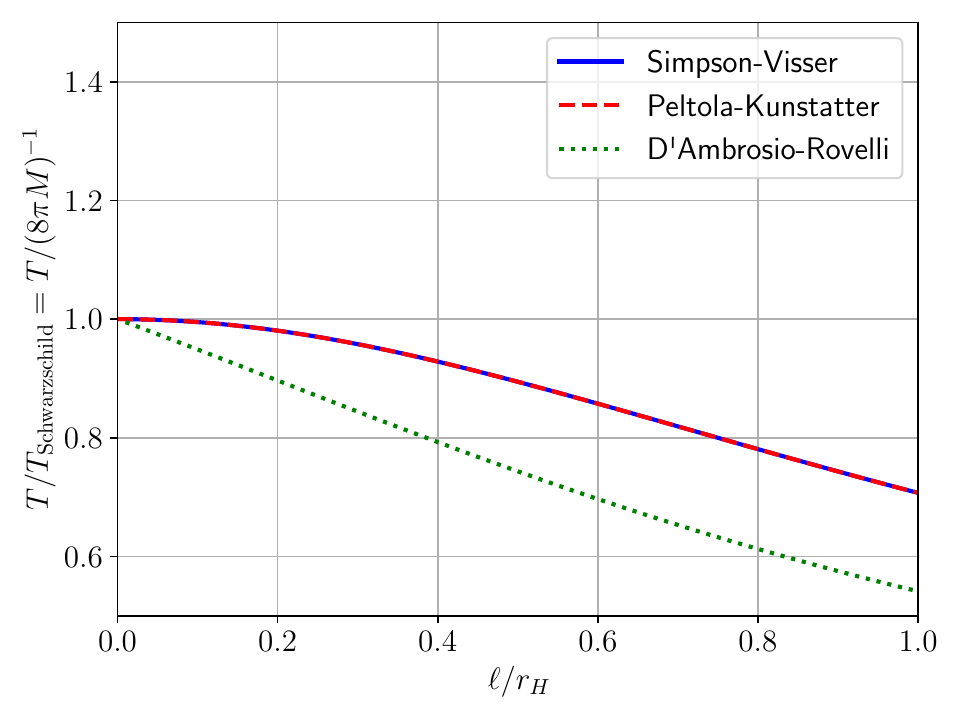}
\caption{Evolution of the temperatures (normalized to the temperature of Schwarzschild black holes, $T_{\text{Sch}}=1/8\pi M$) as a function of the regularizing parameter $\ell$ (normalized to the event horizon radius $r_H$) for the three regular space-times studied in the work: the Simpson-Visser (blue solid curve, Sec.~\ref{subsec:simpsonvisser}), Peltola-Kunstatter (red dashed curve, Sec.~\ref{subsec:peltolakunstatter}), and D'Ambrosio-Rovelli (green dotted curve, Sec.~\ref{subsec:dambrosiorovelli}) regular space-times.}
\label{fig:temperatures}
\end{figure}

A key quantity characterizing the RBHs we are considering is their temperature $T$, since this directly controls the strength of the radiation emitted from Hawking evaporation. Assuming that the temperature is the usual Gibbons-Hawking one, which in turn tacitly implies that we are assuming the standard Boltzmann-Gibbs distribution (see our companion paper for a slightly more detailed discussion on this point~\cite{Calza:2024fzo}), the temperature is given by the following:
\begin{eqnarray}
T=\sqrt{\frac{g(r)}{f(r)}}\frac{f'(r)}{4\pi}\vert_{r_H}\,,
\label{eq:temperature}
\end{eqnarray}
where the prime indicates a derivative with respect to $r$, and $r_H$ denotes the location of the event horizon. In Fig.~\ref{fig:temperatures} we show the evolution of the temperatures, normalized to the temperature of Schwarzschild BHs $T_{\text{Sch}}=1/8\pi M$, of the three RBHs we will discuss shortly, as a function of the regularizing parameter $\ell$ normalized to the event horizon radius $r_H$. As we see, for all three metrics the temperature is a monotonically decreasing function of the regularizing parameter. One may therefore qualitatively expect that the intensity of the Hawking evaporation radiation should decrease relative to that of Schwarzschild BHs of the same mass: this expectation in fact turns out to be correct, as we will explicitly show later, with important consequences for $f_{\text{pbh}}$ limits.

\subsection{Simpson-Visser space-time}
\label{subsec:simpsonvisser}

The Simpson-Visser (SV) metric is a one-parameter extension of the Schwarzschild space-time and easily one of the best known black-bounce metrics. In the words of Simpson and Visser, in some sense this metric ``represents the minimal violence to the standard Schwarzschild solution'' needed to enforce regularity~\cite{Simpson:2018tsi}. The line element analytically interpolates between black holes and traversable wormholes according to the value of the regularizing parameter. In the notation of Eq.~(\ref{eq:extrinsic}), the SV space-time is characterized by the following functions:~\footnote{For all the space-times we will consider, the parameter $M$ appearing in the function $f(r)$ always corresponds to the mass of the space-time (either the Komar, ADM, Misner-Sharp-Hernandez, or Brown-York mass).}
\begin{eqnarray}
\mathfrak{f}(\tilde{r})=1-\frac{2M}{\tilde{r}}\,, \quad \mathfrak{g}_\ell (\tilde{r}) = \frac{\ell^2}{\tilde{r}^2}\,,
\label{eq:fgssimpsonvisser}
\end{eqnarray}
whereas, in the notation of Eq.~(\ref{eq:intrinsic}), the line element of the SV space-time is given by the following:
\begin{eqnarray}
ds^2 = &&- \left ( 1-\frac{2M}{\sqrt{r^2+ \ell^2}} \right ) dt^2 + \frac{dr^2 }{1-\frac{2M}{\sqrt{r^2+ \ell^2}}} \nonumber \\
&&+(r^2+\ell^2)d\Omega^2\,.
\label{eq:metricsimpsonvisser}
\end{eqnarray}
The SV space-time encompasses a rich phenomenology, as it interpolates between the Schwarzschild BH ($\ell=0$), a regular BH with a one-way space-like throat ($0<\ell/M<2$), a one-way WH with an extremal null throat ($\ell/M=2$), and a traversable WH with a two-way time-like throat ($\ell/M>2$).~\footnote{These conditions are derived by looking at \textit{a)} the existence (or not) of horizons from the zeros of $g^{rr}$, which implies $r_H=\sqrt{4M^2-\ell^2}$, and therefore $\ell<2M$ in order to have a BH, and \textit{b)} the sign of the ``coordinate speed of light'' $\vert dr/dt \vert = 1-2M/\sqrt{r^2+\ell^2}$, which determines whether the wormhole for $\ell \geq 2M$ is traversable ($dr/dt \neq 0$ for all $r$) or not ($dr/dt \to 0$ as $r \to 0$) -- for a more detailed explanation, see the discussion above Eq.~(2.6) of Ref.~\cite{Simpson:2018tsi}.} This metric has been the subject of several follow-up studies (see e.g.\ Refs.~\cite{Tsukamoto:2020bjm,Mazza:2021rgq,Shaikh:2021yux,Islam:2021ful,Guerrero:2021ues,Bambhaniya:2021ugr,Yang:2022xxh,Riaz:2022rlx,Arora:2023ltv,Jha:2023wzo,Jha:2023nkh}) and, while originally introduced on phenomenological grounds, can potentially originate as a solution of GR coupled to non-linear electrodynamics in the presence of a minimally coupled phantom scalar field~\cite{Bronnikov:2021uta}. 

\subsection{Peltola-Kunstatter space-time}
\label{subsec:peltolakunstatter}

The Peltola-Kunstatter (PK) space-time is a LQG-motivated metric obtained upon applying effective polymerization techniques to 4D Schwarzschild BHs. Although there are indications that LQG may be capable of resolving the singularities which plague GR, the inherent difficulty in solving the complete system has led to the development of semi-classical polymer quantization techniques, which provide an unitarily inequivalent alternative to Schr\"{o}dinger quantization while maintaining the key aspect of space-time discreteness. The PK space-time is obtained polymerizing only area but not the conformal mode, and results in a space-time whose singularity is replaced by a complete and regular bounce, where the space-time reaches a minimum radius before expanding into a Kantowski-Sachs metric~\cite{Peltola:2008pa,Peltola:2009jm}. In the notation of Eq.~(\ref{eq:extrinsic}), the SV space-time is characterized by the following functions:
\begin{eqnarray}
\mathfrak{f}(\tilde{r})=\sqrt{1-\frac{\ell^2}{\tilde{r}^2}}- \frac{2M}{\tilde{r}}\,, \quad \mathfrak{g}_{\ell} (\tilde r) = \frac{\ell^2}{\tilde r^2}
\label{eq:fgpeltolakunstatter}
\end{eqnarray}
whereas, in the notation of Eq.~(\ref{eq:intrinsic}), the line element of the PK space-time is given by the following:
\begin{eqnarray}
ds^2 = - \left ( \frac{r-2M}{\sqrt{r^2+\ell^2}} \right ) dt^2 + \frac{dr^2}{\frac{r-2M}{\sqrt{r^2+\ell^2}}} + (r^2+ \ell^2)d\Omega^2\,.
\label{eq:metricpeltolakunstatter}
\end{eqnarray}
In what follows, we shall take the PK space-time as an example of regular metric motivated by first-principles quantum gravity considerations, unlike the other phenomenological metrics considered earlier.

\subsection{D'Ambrosio-Rovelli space-time}
\label{subsec:dambrosiorovelli}

The D'Ambrosio-Rovelli (DR) space-time was originally developed with motivations other than singularity avoidance, and is in fact also motivated by LQG considerations. This space-time represents a natural extension of the Schwarzschild space-time which crosses the $r=0$ singularity smoothly into the interior of a white hole, and one can see it as the $\hbar \to 0$ limit of an effective QG metric~\cite{Bianchi:2018mml,DAmbrosio:2018wgv}. It has been argued that this black hole-to-white hole tunneling mechanism can shed light on possible solutions to the information paradox. Of interest to us is the fact that the DR metric is regular, as a result of the curvature of the effective metric being bound at the Planck scale. The ansatz for the effective metric written by D'Ambrosio and Rovelli is similar to that of the SV metric, but differs in the form of the $\mathfrak{g}$ function -- specifically, the two relevant functions are given by the following:
\begin{eqnarray}
\mathfrak{f}(\tilde{r})=1-\frac{2M}{\tilde{r}}\,, \quad \mathfrak{g}_{\ell}(\tilde{r})=\frac{\ell}{\tilde{r}}\,,
\label{eq:fgdambrosiorovelli}
\end{eqnarray}
whereas, in the notation of Eq.~(\ref{eq:intrinsic}), the line element of the DR space-time is given by the following:
\begin{eqnarray}
ds^2 = &&- \left ( 1-\frac{2M}{\sqrt{r^2+ \ell^2}} \right ) dt^2 \nonumber \\
&&+ \frac{dr^2 }{1-\frac{2M}{\sqrt{r^2+ \ell^2}}} \left ( 1+\frac{\ell}{\sqrt{r^2+ \ell^2}} \right ) +(r^2+ \ell^2) d\Omega^2\,. \nonumber \\
\end{eqnarray}
Much like the PK space-time, we will take the DR space-time as another well-motivated example of QG-inspired metric. We note that the assumption of primordial DR BHs inevitably leads to the existence of long-lived primordial DR white holes from quantum transitions near the would-be singularity. This can potentially lead to an interesting phenomenology whose exploration, however, is well beyond the scope of this work.

\section{Methodology}
\label{sec:methodology}

\subsection{Greybody factors}
\label{subsec:greybody}

We now discuss the computation of the greybody factors (GBFs), functions of energy and angular momentum which characterize the shape of the emitted Hawking radiation (and in particular its deviation from a blackbody) and therefore play a key role in determining the resulting evaporation constraints~\cite{Sakalli:2022xrb,Konoplya:2024lir,Konoplya:2024vuj}. It is worth noting that, in the notation of Eq.~(\ref{eq:intrinsic}), both the SV and PK metrics share the fact that $f(r)=g(r)$, which makes the calculations somewhat easier. This is not the case, however, for the DR space-time. Therefore, in what follows, we consider the more general case where $f(r) \neq g(r)$, which differs from the much simpler \textit{tr}-symmetric case considered in our companion paper~\cite{Calza:2024fzo}.

We adopt the Newman-Penrose (NP) formalism, denoting by $\Upsilon_s$ a general perturbation of spin $s$ (defined by the appropriate NP scalars) and dropping the $l$ and $m$ indices to lighten the notation. Then, the Teukolsky equation for the evolution of massless perturbations of given spin upon a background metric characterized by functions $f(r)$, $g(r)$, $h(r)$ as in Eq.~(\ref{eq:intrinsic}) reduces to the following master equation~\cite{Teukolsky:1973ha}:
\begin{align} 
& - \frac{h}{f}\partial^2_t \Upsilon_s + s \sqrt{\frac{g}{f}}\left( \frac{hf'}{f} - h' \right ) \partial_t \Upsilon_s \nonumber \\ \nonumber
&+ g h \partial_r^2 \Upsilon_s + \left ( \frac{hg'}{2} + (s+1/2) \frac{ghf'}{f}+(s+1)gh' \right ) \partial_r \Upsilon_s \\ \nonumber
&+ \left ( \frac{1}{\sin{\theta}} \partial_\theta (\sin{\theta}\;\partial_\theta )+\frac{1}{\sin^2{\theta}} \partial_\phi^2 \right.\\ \nonumber
&  \left. + \frac{2is \cot{\theta}}{\sin{\theta}} \partial_\phi - s^2 \cot^2{\theta}-s\right ) \Upsilon_s \\ \nonumber
& \left ( s \frac{hgf''}{f} +\frac{3s-2s^2}{4}(2gh''+g'h') + \frac{s}{2} (\frac{hg'f'}{f} -\frac{ghf'^2}{f^2}) \right. \\ 
& \left. +\frac{2s^2-s}{4} \frac{gh'^2}{h} + \frac{2 s^2 +5s}{4} \frac{g f' h'}{f} \right ) \Upsilon_s = 0\,,
\label{eq:teukolsky}
\end{align}
which is separable with the following ansatz:
\begin{eqnarray}
\Upsilon_s= \sum_{l,m} e^{-i \omega t } e^{i m \phi} S^{l}_s(\theta) R_s(r)\,,
\label{eq:upsilon}
\end{eqnarray}
with $\omega$, $l$, and $m$ being the perturbation frequency, angular node number, and azimuthal node number respectively, whereas $S^l_s(\theta)$ are related to the so-called spin-weighted spherical harmonics $S^s_{l,m}(\theta, \phi)$ through $S^s_{l,m}(\theta, \phi)=\sum S^l_s(\theta) e^{im\phi}$.

We now define the functions $A_s$, $B_s$, and $C_s$ as follows:
\begin{eqnarray}
A_s= \sqrt{\frac{g}{f}} \frac{1}{(f h)^s}\,,
\label{eq:as}
\end{eqnarray}
\begin{eqnarray}
B_s=\sqrt{f g } (f h)^sh\,,
\end{eqnarray}
\begin{align}
C_s &= s \frac{g h f''}{f} + \frac{s}{2} \left ( \frac{h g' f' }{f} - \frac{g h f'^2}{f^2} \right ) \nonumber \\ 
&+ \frac{s(3-2s)}{4} \left( 2 g h'' + g' h'  \right) +\frac{s(2s-1)}{4} \frac{g h'^2}{h}\nonumber \\ 
&+\frac{s(2s+5)}{4}\frac{g f' h'}{f}-\lambda^s_l-2s\,.
\end{align}
where$\lambda^s_l = l(l+1)-s(s+1)$. With these definitions, the decoupled radial Teukolsky equation reduces to the following general form~\cite{Arbey:2021jif}:
\begin{eqnarray} 
A_s(B_s R'_s)'+ \left [ \frac{h}{f}\omega^2+i\omega s\sqrt{\frac{g}{f}} \left ( h'-\frac{h f'}{f} \right ) + C_s \right ] R_s=0\,. \nonumber \\
\label{eq:radialteukolsky}
\end{eqnarray}
We further define the tortoise coordinate $r_{\star}$ as follows:
\begin{eqnarray}
\frac{dr_{\star}}{dr}=\frac{1}{\sqrt{f(r)g(r)}}\,,
\label{eq:rstar}
\end{eqnarray}
noting that, being our space-times asymptotically flat, $r_{\star} \to r$ for large $r$. In order to compute the GBFs for the metrics in question, we set purely ingoing boundary conditions. In addition, we need to know the asymptotic behaviour of $R_s$ as at infinity and close to the horizon. These asymptotic behaviours are given as follows:
\begin{align}
\label{eq:asymptotic1}
& R_s \sim R^{\text{in}}_s \frac{e^{-i\omega r_{\star}}}{r}+ R^{out}_s \frac{e^{i\omega r_{\star}}}{r^{2s+1}} \quad (r\rightarrow \infty)  \\ 
\label{eq:asymptotic2}
& R_s \sim R^{\text{hor}}_s {A_s} e^{-i \omega r_{\star}}\quad \quad \quad \quad \quad (r \rightarrow r_H)\,,
\end{align}
as proven in more detail in Appendix~\ref{sec:appendix} (the case of the DR metric is actually far from trivial).

To compute the GBFs, we make use of the shooting method, widely used earlier in similar contexts (see e.g.\ Refs.~\cite{Rosa:2016bli,Rosa:2012uz,Calza:2021czr,Calza:2022ioe,Calza:2022ljw,Calza:2023rjt,Calza:2023gws,Calza:2023iqa}), including in our companion paper~\cite{Calza:2024fzo}. We begin by defining the rescaled coordinate $x$:
\begin{eqnarray}
x\equiv\frac{r-r_H}{r_H}\,,
\label{eq:varx}
\end{eqnarray}
where $r_H$ is the largest real root of the equation $f(r)=0$. In order to further simplify our notation, in what follows we work in units of horizon radius, setting $r_H=1$, so that $r=x+1$. The decoupled radial Teukolsky equation, Eq.~(\ref{eq:radialteukolsky}), then takes the following form:
\begin{eqnarray}
\mathcal{A}_s\ddot{R}_s+\mathcal{B}_s\dot{R}_s+\mathcal{C}_sR_s=0\,,
\label{eq:radialteukolskysimplified}
\end{eqnarray}
where the functions $\mathcal{A}$, $\mathcal{B}$, and $\mathcal{C}$ are defined as follows:
\begin{eqnarray}
\mathcal{A}_s= g^2h\,,
\end{eqnarray}
\begin{eqnarray}
\mathcal{B}_s = \left ( \left ( s + \frac{1}{2} \right) \frac{g^2 h \dot f }{f} + \frac{h g \dot g}{2} +(1+s) g^2 \dot h \right ) \,,
\end{eqnarray}
\begin{align}
    \mathcal{C}_s &= \frac{g}{4} \left ( s(2s-1)\frac{g {\dot h}^2 }{h} + 2s(3-2s)g \ddot h- 2s \frac{g h {\dot f}^2 }{f^2} \right. \nonumber  \\
    & \left. +\frac{1}{f} \left ( s(5+2s)g \dot f \dot h + 2h \left ( 2 \omega^2 + 2 s g \ddot f \right. \right. \right.  \\
    &\left. \left. \left. - s \dot f  \left ( 2 i \omega \sqrt{\frac{g}{f}+\dot g} \right) \right ) \right ) + s h \left ( (3-2s) \dot g + 4 i \omega \sqrt{\frac{g}{f}} \right ) -4 \nu^s_l\right ) \,, \nonumber
\end{align}
where the dot denotes a derivative with respect to the rescaled coordinate $x$ and $\nu^s_l=l(l+1)-s(s-1)$. For completeness, we note that $f$ and $g$ as as function of $x$ for the metrics considered here are given by the following:
\begin{equation}
\begin{split}
& h_{\text{SV}}(x)=h_{\text{PK}}(x)=h_{\text{DR}}(x)=(x+1)^2+\ell^2\,, \\
& g_{\text{SV}}(x)=f_{SV}(x)= 1 - \frac{\sqrt{1+\ell^2}}{\sqrt{\ell^2+ (x+1)^2}}\,,\\
& g_{\text{PK}}(x)=f_{\text{PK}}(x)=\frac{x}{\sqrt{\ell^2+ (x+1)^2} }\,,\\
& f_{\text{DR}}(x)=1  - \frac{\sqrt{1+\ell^2}}{\sqrt{\ell^2+ (x+1)^2}}\,,\\
& g_{\text{DR}}(x)=f_{\text{DR}}(x)\left( 1+ \frac{\ell}{\sqrt{\ell^2+ (x+1)^2}} \right)^{-1}\,.
\end{split}
\end{equation}
We express the solution to Eq.~(\ref{eq:radialteukolskysimplified}) as a Taylor expansion as follows~\cite{Leaver:1985ax,Leaver:1986vnb,Konoplya:2023ahd,Konoplya:2023ppx,Rosa:2012uz,Rosa:2016bli}:
\begin{eqnarray}
R_s(x)=x^{-s-\frac{i\omega}{\tau}}\sum_{n=0}^\infty a_n x^n\,.
\label{eq:near}
\end{eqnarray}
Here, $\tau$ is a function of the field's spin and regularizing parameter $\ell$, and varies with the metric being considered. For further details, we refer the reader to the Appendix of our companion paper~\cite{Calza:2024fzo}, where the issue is discussed in more depth. We then determine the coefficients $a_n$ iteratively, by repeatedly substituting Eq.~(\ref{eq:near}) into Eq.~(\ref{eq:radialteukolskysimplified}).

Once we have the near-horizon solution, we treat it as a boundary condition from which we numerically integrate outwards, where the solution takes the following form:
\begin{eqnarray}
R(x) \xrightarrow{r \to \infty} R^{\text{in}}_s \frac{e^{-i \omega x}}{x} +  R^{\text{out}}_s \frac{e^{i \omega x}}{x^{2s+1}}\,,
\label{eq:far}
\end{eqnarray}
with the GBFs then given by the following:
\begin{eqnarray}
\Gamma^s_{l m}(\omega)=\delta_s \vert_s R^{l m}_{in}(\omega)\vert^{-2}\,,
\label{eq:gamma}
\end{eqnarray} 
where $\delta_s$ is given by:
\begin{eqnarray}
\delta_s = \alpha\tau\frac{ie^{i\pi s}(2 \omega)^{2s-1}\Gamma(1-s-\frac{2i\omega}{\tau})}{\Gamma(s-\frac{2 i \omega}{\tau})}\,.
\label{eq:delta}
\end{eqnarray}
with $\alpha $ and $\tau$ depending on the metric considered. More specifically, for what concerns $\alpha$, we find that within the SV and PK metrics for which $f=g$ the following holds:
\begin{eqnarray}
\alpha_{\text{SV}}=\alpha_{\text{PK}}=1+\ell^2\,,
\label{eq:alphasvalphapk}
\end{eqnarray}
whereas for the DR metric we find the following:
\begin{eqnarray}
\alpha_{DR}=1+\ell(\ell+\sqrt{1+\ell^2})\,.
\label{eq:alphadr}
\end{eqnarray}
Finally, for the three metrics $\tau$ is given by the following:
\begin{align}
& \tau_{\text{SV}}=\frac{1}{1+\ell^2}\,, \nonumber \\
& \tau_{\text{PK}}=\frac{1}{\sqrt{1+\ell^2}}\,, \nonumber \\
& \tau_{\text{DR}}= \frac{\sqrt{1+ \ell(\ell-\sqrt{1+\ell^2})}}{1+\ell^2}\,.
\end{align}
As can be seen in Eqs.~(\ref{eq:alphasvalphapk},\ref{eq:alphadr}), deviations from $\alpha=1$ are associated to the non-\textit{tr}-symmetric nature of the metrics. As a general consideration, we note that the computation of GBFs for the metrics under consideration is significantly more involved compared to those considered in our companion paper~\cite{Calza:2024fzo}.

\begin{figure}
\centering
\includegraphics[width=1.0\columnwidth]{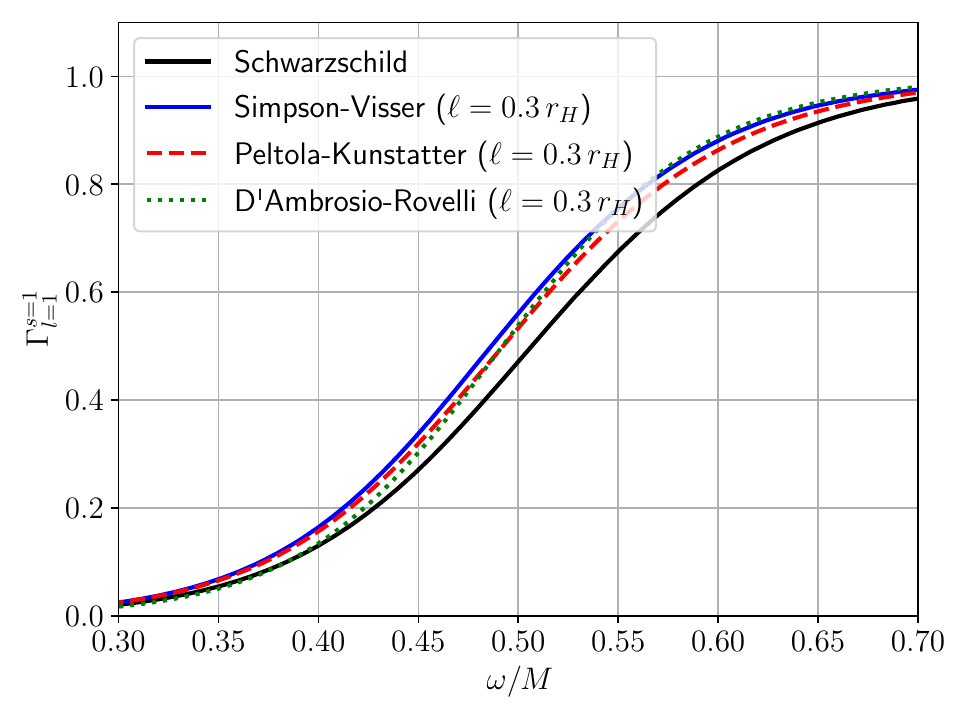}
\caption{Greybody factors $\Gamma_{l=1}^{s=1}$ as a function of $\omega/M$ for Schwarzschild BHs (black curve), as well as the Simpson-Visser (blue solid curve), Peltola-Kunstatter (red dashed curve), and D'Ambrosio-Rovelli (green dotted curve) regular space-times. For illustrative purposes we only plot $\Gamma_{l=1}^{s=1}$, since we are interested in photons ($s=1$) and the dominant emission mode is the $l=1$ one. We have fixed the regularizing parameter to $\ell=0.3r_H$ for all three regular space-times. We see that in all three cases the GBFs are consistently (slightly) higher than their Schwarzschild counterparts.}
\label{fig:gbfs}
\end{figure}

In Fig.~\ref{fig:gbfs} we plot the $\Gamma_{l=1}^{s=1}$ GBFs for the Schwarzschild BH and the three regular space-times we study, focusing for illustrative purposes on the $\Gamma_{l=1}^{s=1}$ GBF and fixing $\ell=0.3r_H$ for all three regular space-times. We observe that the GBFs are slightly higher than their Schwarzschild counterparts (by $\lesssim 10\%$ at most), and asymptote to the latter for both $\omega/M \lesssim 0.3$ and $\omega/M \gtrsim 0.7$.

\subsection{Evaporation spectra}
\label{subsec:spectra}

As in our companion paper~\cite{Calza:2024fzo}, we only consider the primary photon spectrum resulting from Hawking evaporation, while also checking that within the mass region of interest the secondary spectrum will provide a negligible contribution. For a particle species $i$ of spin $s$  (characterized by $n_i$ degrees of freedom), the rate of emission (particles per unit time per unit energy) through Hawking radiation is given by the following~\cite{Hawking:1975iha,Hawking:1975vcx,Page:1976df,Page:1976ki,Page:1977um}:
\begin{eqnarray}\label{prim}
\frac{d^2N_i}{dtdE_i}=\frac{1}{2\pi}\sum_{l,m}\frac{n_i\Gamma^s_{l,m}(\omega)}{ e^{\omega/T}\pm 1}\,,
\label{eq:d2ndtdei}
\end{eqnarray}
with $\omega=E_i$ being the mode frequency, whereas the positive (negative) sign in the denominator corresponds to fermions (bosons). To compute the (photon) GBFs $\Gamma^s_{l,m}(\omega)$ we adopt the methodology discussed earlier, going up to node number $l=4$, but verifying that including higher $l$ modes leads to negligible corrections.

\begin{figure}
\centering
\includegraphics[width=1.0\columnwidth]{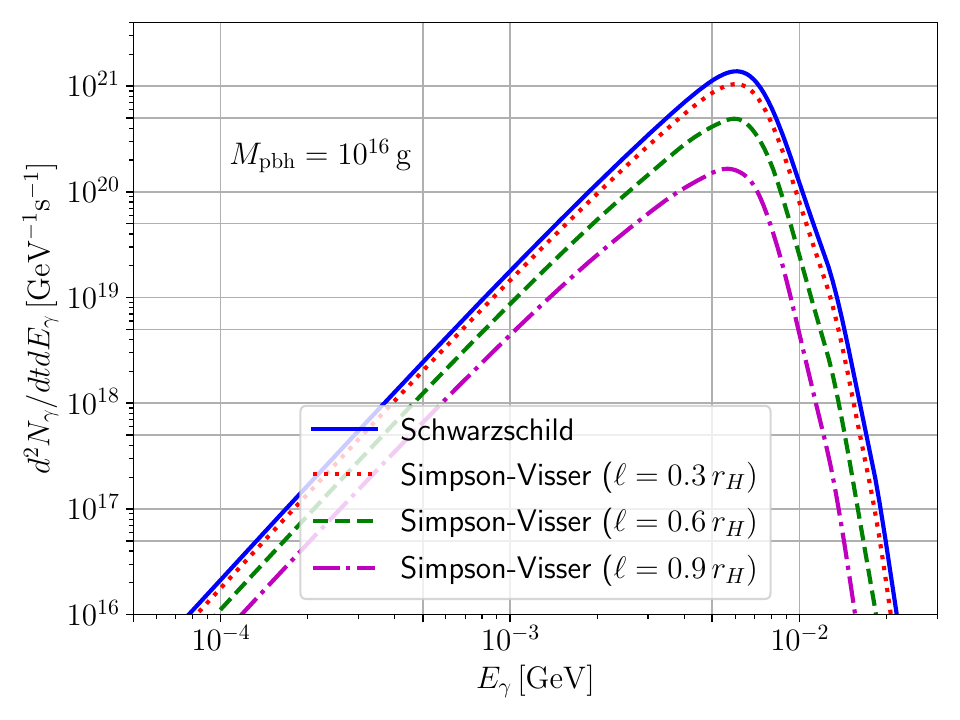}
\caption{Primary photon spectra resulting from the evaporation of Simpson-Visser black holes of mass $10^{16}\,{\text{g}}$ for different values of the regularizing parameter $\ell$ (normalized by the horizon radius $r_H$): $\ell/r_H=0.3$ (red dotted curve), $0.6$ (green dashed curve), and $0.9$ (magenta dash-dotted curve). The blue solid curve corresponds to the case $\ell/r_H=0$, which recovers the Schwarzschild black hole.}
\label{fig:spectraphotonssimpsonvisser}
\end{figure}

\begin{figure}
\centering
\includegraphics[width=1.0\columnwidth]{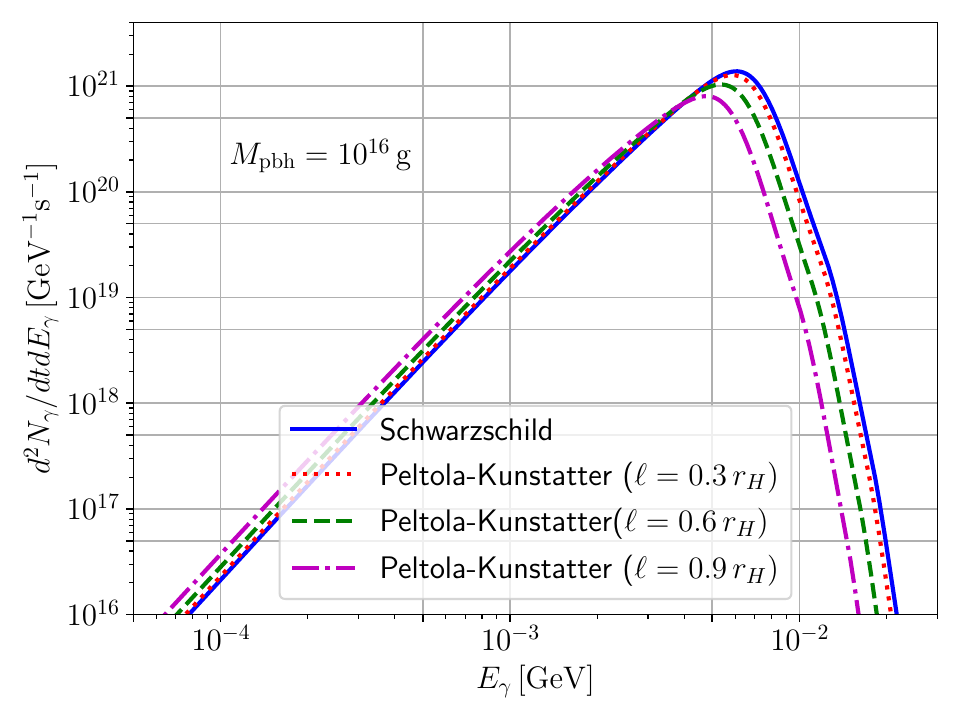}
\caption{As in Fig.~\ref{fig:spectraphotonssimpsonvisser}, but for Peltola-Kunstatter black holes, with identical values of the regularizing parameter $\ell/r_H$ and identical color coding.}
\label{fig:spectraphotonspeltolakunstatter}
\end{figure}

\begin{figure}
\centering
\includegraphics[width=1.0\columnwidth]{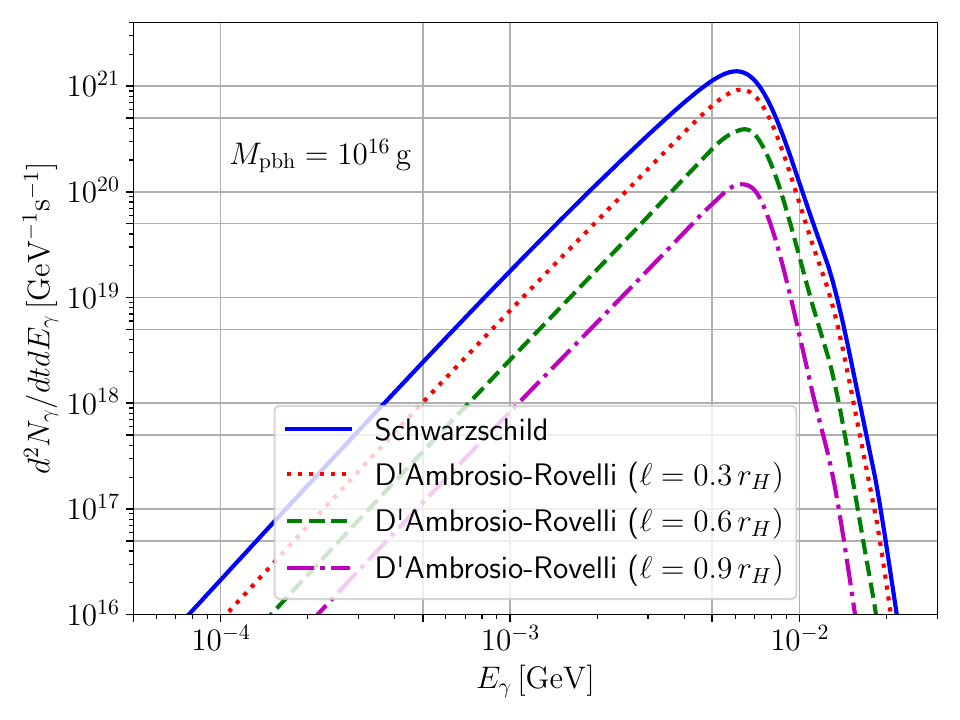}
\caption{As in Fig.~\ref{fig:spectraphotonssimpsonvisser}, but for D'Ambrosio-Rovelli black holes, with identical values of the regularizing parameter $\ell/r_H$ and identical color coding.}
\label{fig:spectraphotonsdambrosiorovelli}
\end{figure}

Examples of the resulting evaporation spectra are provided in Figs.~\ref{fig:spectraphotonssimpsonvisser},~\ref{fig:spectraphotonspeltolakunstatter}, and~\ref{fig:spectraphotonsdambrosiorovelli} for a representative PBH of mass $M_{\text{pbh}}=10^{16}\,{\text{g}}$, located somewhat halfway in the mass range of interest (although the features we discuss shortly do not depend on the chosen mass). For the SV and DR PRBHs, increasing $\ell$ leads to the intensity of the spectra decreasing at all energies. This feature confirms the intuition raised when we discussed the temperatures of these BHs (see Fig.~\ref{fig:temperatures}), which we observed to decrease with increasing $\ell$. However, we remark that a decrease in $T$ alone is not sufficient to draw this conclusion, since the GBFs also enter in Eq.~(\ref{eq:d2ndtdei}). Nevertheless, one can expect the temperature to play a more important role, as it enters exponentially into Eq.~(\ref{eq:d2ndtdei}), unlike the GBFs which enter linearly.

The role of the GBFs can be noticed in the case of the PK space-time, whose temperature evolution as a function of $\ell$ is identical to that of the SV space-time (see Fig.~\ref{fig:temperatures}). However, as $\ell$ is increased, the evaporation spectra of PK PBHs (Fig.~\ref{fig:spectraphotonspeltolakunstatter}) decreases in intensity only at energies approximately above the peak (located roughly between $5\,{\text{MeV}}$ and $10\,{\text{MeV}}$), while conversely increasing in intensity for lower energies, albeit to a lesser extent compared to the decrease at higher energies. From Fig.~\ref{fig:gbfs} we can see that the GBFs for the SV and PK space-times are close to each other for $\ell=0.3r_H$, explaining why there is little difference between the red dotted curves in Fig.~\ref{fig:spectraphotonssimpsonvisser} and Fig.~\ref{fig:spectraphotonspeltolakunstatter}. The difference between the two become more important as $\ell$ increases, and are particularly noticeable at $\ell=0.9r_H$. We have explicitly checked that, in this case (not shown in Fig.~\ref{fig:gbfs}), the PK GBFs start increasing at lower values of $\omega/M$ compared to their SV counterparts, explaining the observed trends. At any rate already at a qualitative level, inspecting the spectra just discussed, we can expect that the upper limits on $f_{\text{pbh}}$ obtained assuming Schwarzschild PBHs should loosen (thereby opening the asteroid mass window) when considering the PRBHs in question, at the very least for the SV and DR metrics -- as we shall see shortly, the expectation is in fact confirmed for all three space-times.

\subsection{Evaporation constraints}
\label{subsec:constraints}

In the mass range $10^{13}\,{\text{g}} \lesssim M_{\text{pbh}} \lesssim 10^{18}\,{\text{g}}$, the dominant constraints on the PBH abundance come from measurements of the extragalactic photon background~\cite{Auffinger:2022khh}, and more precisely of the diffuse extragalactic $\gamma$-ray background (EGRB) in the energy range $100\,{\text{keV}} \lesssim E_{\gamma} \lesssim 5\,{\text{GeV}}$, which can be directly compared against theoretical expectations for the PBH Hawking evaporation spectra. Of particular interest to us is the fact that the lower edge of the asteroid mass window, where PBHs could make up the entire DM, is precisely set by evaporation constraints. In what follows, we set evaporation constraints on the fraction of DM in the form of PBHs $f_{\text{pbh}}(M) \equiv \Omega_{\text{pbh}}/\Omega_{\text{dm}}$, assuming that PRBHs are described by the three metrics discussed so far. We work under the same set of approximations adopted in our companion paper~\cite{Calza:2024fzo}. Namely, we assume that PRBHs are isotropically distributed on sufficiently large scales and cluster in the galactic halo in the same way as other forms as DM, we only compute the primary photon spectrum, and finally we assume a monochromatic mass distribution (see e.g.\ Refs.~\cite{Kuhnel:2015vtw,Kuhnel:2017pwq,Carr:2017jsz,Raidal:2017mfl,Bellomo:2017zsr,Lehmann:2018ejc,Carr:2018poi,Gow:2019pok,DeLuca:2020ioi,Gow:2020cou,Ashoorioon:2020hln,Bagui:2021dqi,Mukhopadhyay:2022jqc,Papanikolaou:2022chm,Cai:2023ptf} for studies on the effects of an extended mass distribution). We refer the reader to Sec.~IIIC of our companion paper~\cite{Calza:2024fzo} for a more detailed discussion of why these, which are clearly all approximations, are appropriate for the scope of our work (while allowing for a more direct comparison to earlier works, including our companion paper).

We therefore focus on a population of PRBHs which all share the same mass $M_{\text{pbh}}$. Following Ref.~\cite{Carr:2009jm}, the number of emitted photons in the logarithmic energy bin $\Delta E_{\gamma} \simeq E_{\gamma}$ is approximated as $\dot{N}_{\gamma}(E_{\gamma}) \simeq E_{\gamma}(d\dot{N}_{\gamma}/dE_{\gamma})$. The rate of emitted photons with present-day energy $E_{\gamma 0}$ per unit time per unit area per unit solid angle is then obtained by integrating over the entire cosmological time, accounting for the redshift scaling of the photon energy and density, and is given by the following:
\begin{eqnarray}
I(E_{\gamma 0})= A_I\int^{z_{\star}}_{0} \frac{dz}{H(z)}\frac{d^2{N}_{\gamma}}{dt dE_{\gamma}}(M_{\text{pbh}},(1+z)E_{\gamma 0})\,,
\label{eq:flux}
\end{eqnarray}
where the normalization factor $A_I$ is given by:
\begin{eqnarray}
A_I=\frac{c}{4\pi}n_{\text{pbh}}(t_0)E_{\gamma 0}\,.
\label{eq:normalization}
\end{eqnarray}
In Eqs.~(\ref{eq:flux},\ref{eq:normalization}), $d^2{N}_{\gamma}/dtdE_{\gamma}$ is computed via Eq.~(\ref{eq:d2ndtdei}), whereas $z_{\star}$ is the redshift of recombination, $H(z)$ is the expansion rate, and $n_{\text{pbh}}(t_0)$ is the present-day PRBH number density, itself related to the parameter of interest $f_{\text{pbh}}$ via the following:
\begin{eqnarray}
f_{\text{pbh}}(M_{\text{pbh}}) \equiv \frac{\Omega_{\text{pbh}}}{\Omega_{\text{dm}}} = \frac{n_{\text{pbh}}(t_0)M_{\text{pbh}}}{\rho_{\text{crit},0}\Omega_{\text{dm}}}\,,
\label{eq:npbhtofpbh}
\end{eqnarray}
with $\rho_{\text{crit},0}=3H_0^2/8\pi G$ being the present-day critical density, $H_0$ the Hubble constant, and $\Omega_{\text{dm}}$ the present-day DM density parameter. In what follows, in order to specify $H(z)$, $\rho_{\text{crit}}$, and $\Omega_{\text{dm}}$, we adopt the same spatially flat $\Lambda$CDM cosmological model used by the seminal Ref.~\cite{Carr:2009jm}. This allows us to have a reliable reference against which we can cross-check our limits on $f_{\text{pbh}}$ in the Schwarzschild case ($\ell \to 0$), although we stress that the choice of underlying cosmology does not play a significant role in determining our constraints. Analogously to our companion paper, we focus on the mass range $M_{\text{pbh}}>10^{15}\,{\text{g}}$. A Schwarzschild PBH of this mass has a lifetime which is much longer than the age of the Universe, and is therefore far from having fully evaporated today. Importantly, it has only lost a negligible fraction of its mass from formation until today, and it is therefore safe to approximate these PBHs as being quasi-static throughout the lifetime of the Universe, while denoting by $M_{\text{pbh}}$ the values of the PBH mass both at formation and today. These considerations hold even more strongly for the regular BHs we consider, as these are colder and hence longer lived compared to their Schwarzschild counterparts at a given mass. We refer the reader to Appendix~B of our companion paper~\cite{Calza:2024fzo} for further discussions on this point.

With the cosmological model specified, the only unknown parameter in Eqs.~(\ref{eq:flux},\ref{eq:normalization}) is the present-day PRBH number density $n_{\text{pbh}}(t_0)$, or equivalently, through Eq.~(\ref{eq:npbhtofpbh}), the PRBH fraction $f_{\text{pbh}}$. For each value of $M_{\text{pbh}}$, we set upper limits on the only free parameter $f_{\text{pbh}}$ using EGRB flux measurements, and more specifically measurements from the HEAO-1 X-ray telescope in the $3$-$500\,{\text{keV}}$ range~\cite{Gruber:1999yr}, the COMPTEL imaging Compton $\gamma$-ray telescope in the $0.8$-$30\,{\text{MeV}}$ range~\cite{Schoenfelder:2000bu}, and the EGRET $\gamma$-ray telescope~\cite{Strong:2004ry}. To do so, for given values of $M_{\text{pbh}}$ and $\ell/r_H$, the maximum allowed value of $f_{\text{pbh}}$ is determined by the requirement that the theoretical prediction for the photon flux given in Eq.~(\ref{eq:flux}) does not overshoot any of the ERGB measurements by more than $1\sigma$ (see e.g.\ Fig.~6 in our companion paper~\cite{Calza:2024fzo}, and note that different datapoints are first overshot when changing $M_{\text{pbh}}$).~\footnote{This method was first discussed in the seminal Ref.~\cite{Carr:2009jm}, and later adopted by most of the works studying EGRB constraints on PBHs. While a more robust statistical analysis is of course possible, such an approach is sufficiently accurate for the purposes of our work and allows for a more direct comparison to earlier results. See Sec.~IIIC of our companion paper~\cite{Calza:2024fzo} for more detailed comments on this point, as well as on the potential use of other datasets, including local galactic measurements of the galactic $\gamma$-ray background~\cite{Carr:2016hva}, positron flux~\cite{Boudaud:2018hqb}, $0.511\,{\text{MeV}}$ annihilation radiation~\cite{DeRocco:2019fjq,Laha:2019ssq,Dasgupta:2019cae}, and more recent measurements of the EGRB from Fermi-LAT~\cite{Fermi-LAT:2014ryh}.} For each of the three metrics, we use this method to set upper limits on $f_{\text{pbh}}$ as a function of $M_{\text{pbh}}$ for fixed, representative values of $\ell/r_H=0.3$, $0.6$, and $0.9$. Finally, we note that while the origin of the EGRB is not fully understood~\cite{Dwek:2012nb}, our approach is conservative in this sense given that we remain agnostic as to the level of astrophysical (non-PBH) contribution to the EGRB.

\section{Results}
\label{sec:results}

\begin{figure}
\centering
\includegraphics[width=1.0\columnwidth]{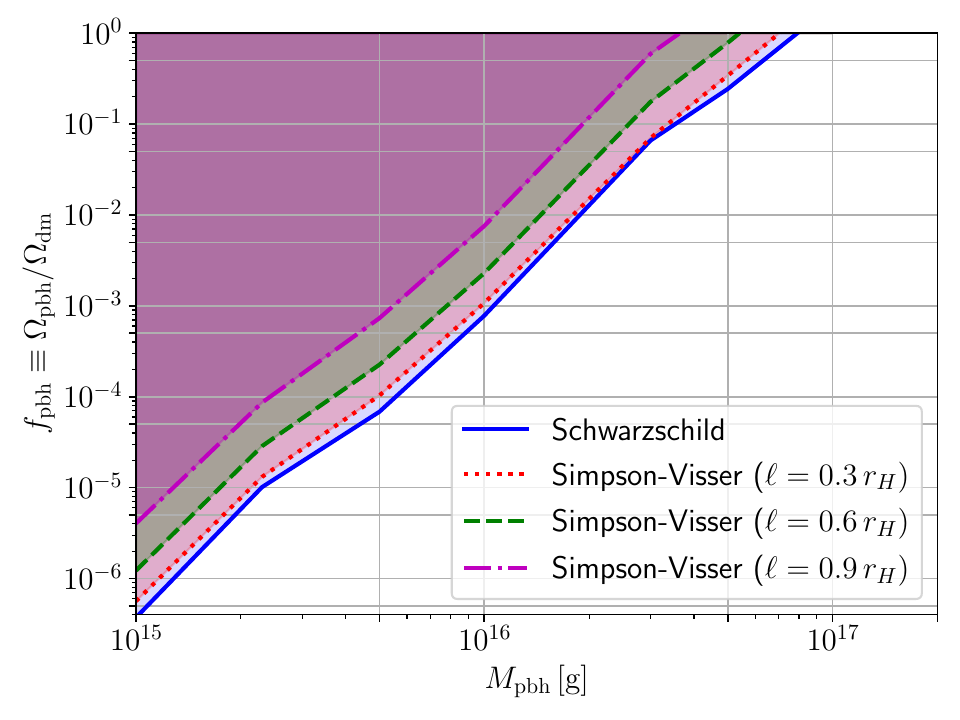}
\caption{Upper limits on $f_{\text{pbh}}$, the fraction of dark matter in the form of primordial regular Simpson-Visser black holes, as a function of the black hole mass $M_{\text{pbh}}$. The limits are derived for different values of the regularizing parameter $\ell$ (normalized to the horizon radius $r_H$), with the shaded regions excluded: $\ell/r_H=0.3$ (red dotted curve), $0.6$ (green dashed curve), and $0.9$ (magenta dash-dotted curve). Note that the blue solid curve corresponds to the case $\ell/r_H=0$, which recovers the Schwarzschild black hole, whereas the value of $M_{\text{pbh}}$ corresponding to the upper right edge of the $f_{\text{pbh}}$ constraints marks the lower edge of the asteroid mass window.}
\label{fig:fpbhlimitssimpsonvisser}
\end{figure}

\begin{figure}
\centering
\includegraphics[width=1.0\columnwidth]{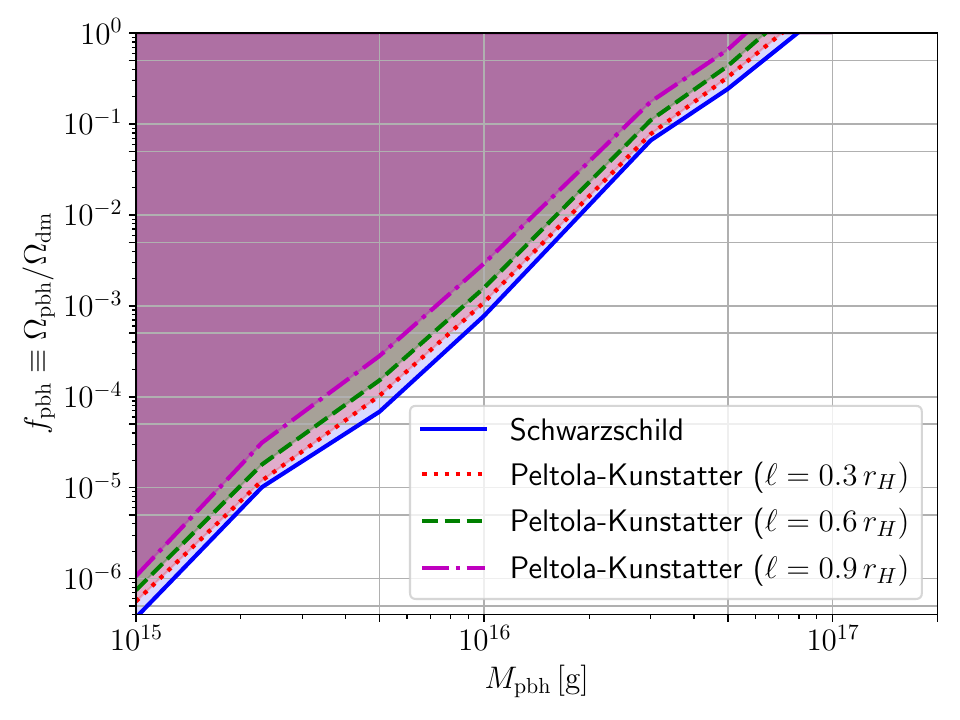}
\caption{As in Fig.~\ref{fig:fpbhlimitssimpsonvisser}, but for primordial regular Peltola-Kunstatter black holes, with identical values of the regularizing parameter $\ell/r_H$ and identical color coding.}
\label{fig:fpbhlimitspeltolakunstatter}
\end{figure}

\begin{figure}
\centering
\includegraphics[width=1.0\columnwidth]{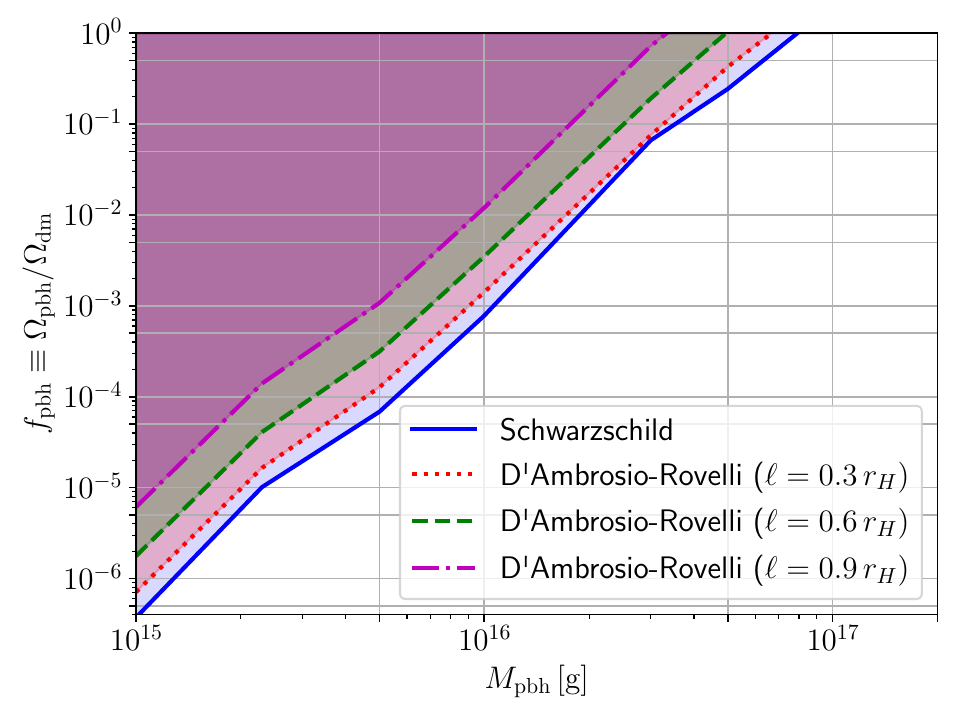}
\caption{As in Fig.~\ref{fig:fpbhlimitssimpsonvisser}, but for primordial regular D'Ambrosio-Rovelli black holes, with identical values of the regularizing parameter $\ell/r_H$ and identical color coding.}
\label{fig:fpbhlimitsdambrosiorovelli}
\end{figure}

The resulting upper limits on $f_{\text{pbh}}$ as a function of PRBH mass $M_{\text{pbh}}$, for different values of $\ell$, are shown in Figs.~\ref{fig:fpbhlimitssimpsonvisser}, ~\ref{fig:fpbhlimitspeltolakunstatter}, and~\ref{fig:fpbhlimitsdambrosiorovelli} for the SV, PK, and DR space-times respectively. In each figure, shown as blue solid curves are the corresponding constraints in the Schwarzschild PBH case ($\ell \to 0$), which we have verified to recover the results of Ref.~\cite{Carr:2009jm}. We stress that for a given value of $\ell$, the value of $M_{\text{pbh}}$ where the overclosure limit $f_{\text{pbh}}<1$ is saturated sets the lower edge of the modified asteroid mass window (potentially enlarged or contracted).

For all three cases, we see that increasing the regularizing parameter $\ell$ results (at a given value of $M_{\text{pbh}}$) in weaker limits on $f_{\text{pbh}}$. This confirms the expectation raised at the end of Sec.~\ref{subsec:spectra} upon inspection of the resulting evaporation spectra, all of which decrease in intensity relative to the Schwarzschild case (except for the slight increase in the PK PRBH case for energies below the peak, which we recall reflects the different behaviour of the GBFs). We see that, for $\ell/r_H=0.9$, the upper limits on $f_{\text{pbh}}$ weaken by up to an order of magnitude at a given $M_{\text{pbh}}$ relative to the Schwarzschild limit for the SV and DR PRBHs, whereas for PK PRBHs the extent to which the $f_{\text{pbh}}$ limits weaken is more limited -- again unsurprisingly, given that the enhanced intensity of the evaporation spectrum at low energies counteracts the decrease at higher energies in the integral of Eq.~(\ref{eq:flux}).

The aforementioned shifts result in the asteroid mass window being enlarged for all three metrics considered, because the lower edge of the window (lying roughly at $M_{\text{pbh}} \simeq 10^{17}\,{\text{g}}$ in the Schwarzschild case) moves towards lower masses. In general, we observe that the asteroid mass window further opens up by about half a decade in mass or more (an increase which is less dramatic than what we observed for the phenomenological metrics in our companion paper~\cite{Calza:2024fzo}). As a result, there is a wider available range of parameter space where PRBHs of the type we are considering could make up all the DM. We note that our constraints assume that all PRBHs in the Universe carry the same value of ``hair parameter'' $\ell/r_H$ (in the language of Ref.~\cite{Vagnozzi:2022moj}, we are treating it as an ``universal hair''). Whether or not this is a reasonable assumption requires a deeper investigation of the theoretical underpinning of the adopted metrics (in particular the LQG-inspired one) which, in the spirit of the present work being a pilot study, we defer to follow-up work.

Finally, in our companion paper~\cite{Calza:2024fzo} we extensively commented on a few caveats concerning the extension of the asteroid mass window and, more generally, on other existing constraints on $f_{\text{pbh}}$, which bear repeating here, albeit in a more condensed form (we refer the reader to Sec.~IV of our companion paper for a significantly more detailed discussion). While evaporation constraints set the lower edge of the asteroid mass window, the upper edge thereof is instead set by lensing constraints. Our claim that the asteroid mass window is enlarged because the lower edge moves towards even lower values is therefore contingent upon the upper edge remaining the standard Schwarzschild one, even within the adopted metrics. We do, in fact, expect this to be the case since, at fixed $M_{\text{pbh}}$, lensing constraints depend only on the mass of the lensing object. We can therefore assert that the asteroid mass window is indeed enlarged when considering the three PBH metrics introduced here. Finally, a variety of other constraints on $f_{\text{pbh}}$ exist, including dynamical, accretion, and CMB constraints (see e.g.\ Ref.~\cite{Carr:2021bzv} for a recent summary): however, with the exception of a few debated constraints~\cite{Capela:2012jz,Capela:2013yf,Pani:2014rca,Graham:2015apa}, we expect these to be relevant within significantly different mass ranges (unless accretion dynamics are significantly different around the RBHs under consideration), although we reserve a detailed study to follow-up work.

\section{Conclusions}
\label{sec:conclusions}

It is a commonly held belief that the singularities which plague General Relativity, and represent one of the most important open problems in theoretical physics, will eventually be solved once the long sought after theory of quantum gravity is unveiled. While a consensus theory of quantum gravity remains elusive, progress on the singularity problem can still be made by considering ans\"{a}tze for singularity-free space-times, either introduced phenomenologically or somewhat motivated from candidate quantum gravity frameworks (such as LQG). These regular black holes, if produced early on in the Universe from the collapse of large density perturbations (thus being primordial regular BHs), could also have a role to play in the dark matter problem. Our work is a pilot study which goes precisely in this direction, examining what are the consequences of PBHs being described by non-singular metrics. In fact, it bears reminding that the usual constraints on the fraction of DM in the form of PBHs, $f_{\text{pbh}}$, are derived under the assumption of PBHs being described by the Schwarzschild metric, which is well-known to be plagued by the $r=0$ singularity. In the present work, we have explored three so-called non-\textit{tr}-symmetric metrics as candidates for describing PBHs: the Simpson-Visser black-bounce, Peltola-Kunstatter, and D'Ambrosio-Rovelli black-to-white-hole-bounce space-times, with the latter two enjoying strong theoretical motivation from LQG (we note that the mathematically simpler \textit{tr}-symmetric case is covered in our companion paper~\cite{Calza:2024fzo}, and includes well-known phenomenological regular BHs such as the Bardeen and Hayward BHs).

After discussing the impact of the regularizing parameter $\ell$ (with the Schwarzschild BH corresponding to the $\ell \to 0$ limit) on the resulting evaporation spectra, we find that as $\ell$ increases, at a fixed PRBH mass $M_{\text{pbh}}$ the corresponding upper limit on $f_{\text{pbh}}$ from observations of the EGRB weakens for all three metrics considered. This results in the lower edge of the asteroid mass window shifting down by up to approximately half a decade in $M_{\text{pbh}}$ parameter space (down from $M_{\text{pbh}} \simeq 10^{17}\,{\text{g}}$ to $M_{\text{pbh}} \simeq 3 \times 10^{16}\,{\text{g}}$). As a result, there is a larger range of available parameter space where the PBHs in question could make up the entire DM in the Universe, which could be targeted by proposed probes of the asteroid mass window~\cite{Ray:2021mxu,Kainulainen:2021rbg,Ghosh:2022okj,Branco:2023frw,Amaral:2023ekd,Bertrand:2023zkl,Tran:2023jci,Dent:2024yje,Tamta:2024pow}.

Our work (alongside our companion paper~\cite{Calza:2024fzo}) demonstrates, as a proof-of-principle, that the intersection of the DM and singularity problems is a fertile terrain worthy of further studies. The most important avenue for immediate follow-up work would be to systematically revisit, within the metrics under consideration, all other non-evaporation constraints which have been extensively discussed for Schwarzschild PBHs (including lensing, accretion, and dynamical constraints). Moreover, since the Peltola-Kunstatter and D'Ambrosio-Rovelli space-times are rooted within an underlying quantum gravity theoretical framework, a first-principles study of their formation mechanism (which is otherwise not possible for metrics introduced at a phenomenological level, such as the Bardeen and Hayward BHs) and whether this leads to additional interesting complementary signatures is definitely worth pursuing. For instance, quantum transitions near the would-be singularity of the D'Ambrosio-Rovelli space-time should lead to the existence of long-lived (primordial) white holes, which in turn could potentially lead to a wide range of exotic signatures one could hope to search for. In continuing our exciting program at the interface of the DM and singularity problems, it is our intention to return to these and related issues in future follow-up work.

\begin{acknowledgments}
\noindent We acknowledge support from the Istituto Nazionale di Fisica Nucleare (INFN) through the Commissione Scientifica Nazionale 4 (CSN4) Iniziativa Specifica ``Quantum Fields in Gravity, Cosmology and Black Holes'' (FLAG). M.C. and S.V. acknowledge support from the University of Trento and the Provincia Autonoma di Trento (PAT, Autonomous Province of Trento) through the UniTrento Internal Call for Research 2023 grant ``Searching for Dark Energy off the beaten track'' (DARKTRACK, grant agreement no.\ E63C22000500003). This publication is based upon work from the COST Action CA21136 ``Addressing observational tensions in cosmology with systematics and fundamental physics'' (CosmoVerse), supported by COST (European Cooperation in Science and Technology).
\end{acknowledgments}

\appendix

\section{Asymptotic solutions to the radial Teukolsky equation}
\label{sec:appendix}

We recall that, in order to compute the GBFs for the regular BHs studied in this work, we need to know the asymptotic behaviour of the function $R_s$, introduced in the ansatz of Eq.~(\ref{eq:upsilon}) and solution to the radial Teukolsky equation given by Eq.~(\ref{eq:radialteukolsky}), both at infinity and close to the horizon. In the main text we reported these limits as being given by Eqs.~(\ref{eq:asymptotic1},\ref{eq:asymptotic2}). We now set out to prove this more formally.

The radial Teukolsky equation, Eq.~(\ref{eq:radialteukolsky}), simplifies considerably if we make the following substitution~\cite{Arbey:2021jif}:
\begin{eqnarray}
\frac{dr_{\star}}{dr}=\frac{1}{\sqrt{f(r)g(r)}}\,,\qquad Y_s=\sqrt{\frac{B_s}{\sqrt{fg}}}R_s\,.
\label{eq:substitutions}
\end{eqnarray}
This allows one to get rid of first derivatives of $Y_s$, with the radial Teukolsky equation as a function of the latter now taking the following form:
\begin{eqnarray}
&&Y_s,_{\star\star}+\nonumber \\
&& \left [ \omega^2 + i\omega s\sqrt{\frac{g}{f}} \left ( h' - \frac{hf'}{f} \right ) \frac{f}{h} + C_s \frac{f}{h} -\frac{\sqrt{\beta},_{\star\star}}{\sqrt{\beta}} \right ] Y_s=0\,, \nonumber \\
\label{eq:radialteukolskyy}
\end{eqnarray}
where $\beta \equiv (hf^2)h=B_s/\sqrt{fg}$, and $,_{\star}$ denotes differentiation with respect to the tortoise coordinate $r_{\star}$. We now consider the $r\to +\infty$ and $r\to r_H$ limits separately.

\subsection{Asymptotic behaviour at infinity}
\label{subsec:rinfty}

In this limit, we trivially see that Eq.~(\ref{eq:radialteukolskyy}) reduces to the following:
\begin{eqnarray}
Y_s,_{\star\star} + \left ( \omega^2 + \frac{2i\omega s}{r} \right ) Y_s=0\,,
\label{eq:yrinfty}
\end{eqnarray}
whose asymptotic solutions are given by:
\begin{eqnarray}
Y_s \sim r^{\pm s}e^{\mp i\omega r_{\star}}\,.
\label{eq:ysasymptoticsolutionsinfty}
\end{eqnarray}
which implies that $R_s$ scales as follows:
\begin{eqnarray}
R_s \sim \frac{e^{-i\omega r_{\star}}}{r} \quad \text{and} \quad R_s \sim \frac{e^{i\omega r_{\star}}}{r^{(2s+1)}}\,,
\label{eq:rsinfty}
\end{eqnarray}
confirming the asymptotic scaling quoted in Eq.~(\ref{eq:asymptotic1}).

\subsection{Asymptotic behaviour near the horizon}
\label{subsec:rhorizon}

In the vicinity of the event horizon, Eq.~(\ref{eq:radialteukolskyy}) reduces to the following:
\begin{eqnarray}
Y_s,_{\star\star} + \left [ \left ( \omega - \frac{i s \sqrt{\frac{g}{f}}f'}{4} \right ) ^2 + \frac{f's}{4} \left ( g' - \frac{g f'}{f} \right ) \right ] Y_s=0\,, \nonumber \\
\label{eq:yrhorizon}
\end{eqnarray}
where we kept only terms scaling as $\sim f^0, g^0$. It is easy to see that in the cases of the SV and PK space-times, for which $f=g$, the second term in square brackets vanishes. With some effort, one can show that this is true for the DR metric as well. This then leaves us with the following equation for $Y_s$:
\begin{eqnarray}
Y_s,_{\star\star} + \left ( \omega - \frac{i s \sqrt{\frac{g}{f}}f'}{4} \right ) ^2Y_s = 0\,,
\label{eq:yrhorizonsimplified}
\end{eqnarray}
whose asymptotic solutions are given by:
\begin{eqnarray}
Y_s \sim \exp \left [ \pm i \left ( \omega - \frac{isf'}{2}\sqrt{\frac{g}{f}} \right ) r_{\star} \right ] \,.
\label{eq:ysasymptoticsolutionshorizon}
\end{eqnarray}
By definition, the tortoise coordinate $r_{\star}$ is determined by the following integral:
\begin{eqnarray}
r_{\star} = \int \frac{dr}{\sqrt{f(r)g(r)}} \xrightarrow{r \to r_H} \text{K} \ln(r - r_H)\,,
\label{eq:rstarnew}
\end{eqnarray}
where $r_H$ is the radial coordinate of the event horizon and $K$ is a coefficient to be determined. Combining Eqs.~(\ref{eq:ysasymptoticsolutionshorizon},\ref{eq:rstarnew}) we reach the following expression for the asymptotic scaling of $Y_s$:
\begin{eqnarray}
Y_s \sim e^{\pm i \omega r_{\star}}\exp \left [ \pm \frac{sf'}{2}\sqrt{\frac{g}{f}}\text{K}\ln(r - r_H) \right ] \,.
\label{eq:ysasymptoticsolutionshorizonrstar}
\end{eqnarray}
It is straightforward to check that in the case of Schwarzschild BH, Eq.~(\ref{eq:rstarnew}) reduces to the following:
\begin{eqnarray}
r_{\star} = \int\frac{dr}{1 - \frac{2M}{r}} \xrightarrow{r \to r_H} 2M\ln(r - 2M)\,,
\end{eqnarray}
from which we recover the asymptotic behaviour found in Ref.~\cite{Teukolsky:1973ha}:
\begin{equation}
Y_s \sim e^{\pm i \omega r_{\star}}\Delta^{\pm s/2} \implies R_s \sim e^{i\omega r_{\star}} \; \text{or} \; R_s \sim \Delta^{-s}e^{-i\omega r_{\star}}\,,
\label{eq:ysasymptoticschwarzschild}
\end{equation}
since it is easy to show that the following holds asymptotically:
\begin{eqnarray}
\ln(r -r_H) \sim \ln(\Delta)\,,
\end{eqnarray}
where $\Delta \equiv r^2-2Mr=r^2g(r)$. Note that the $R_s \sim e^{i\omega r_{\star}}$ solution in Eq.~(\ref{eq:ysasymptoticschwarzschild}) is then discarded on the basis of the purely ingoing boundary conditions we fix at the horizon when setting up the scattering problem, thereby confirming the asymptotic scaling quoted in Eq.~(\ref{eq:asymptotic2}). The above results hinged upon the $r_{\star} \propto \ln(r-r_H)$ scaling in Eq.~(\ref{eq:rstarnew}). We now check that this scaling does indeed hold for the three space-times we consider in our work.

\subsubsection{Simpson-Visser space-time}
\label{subsubsec:simpsonvisser}

We recall that the functions $f(r)$ and $g(r)$ are given by the following:
\begin{eqnarray}
f(r)=g(r)=1-\frac{2M}{\sqrt{r^2 + \ell^2}}\,,
\end{eqnarray}
implying that the tortoise coordinate is given by the following:
\begin{eqnarray}
r_{\star} = \int \frac{dr}{1 - \frac{2M}{\sqrt{r^2 + \ell^2}}} \xrightarrow{r \to r_H^{\text{SV}}} \frac{4M^2}{\sqrt{4M^2-\ell^2}}\ln \left ( r - r_H^{\text{SV}} \right ) \,, \nonumber \\
\end{eqnarray}
where $r_H^{\text{SV}} = \sqrt{4M^2 - \ell^2}$. For this metric $g'(r)=2Mr/(r^2 + \ell^2)^{3/2}$, from which one easily finds that Eq.~(\ref{eq:ysasymptoticsolutionshorizonrstar}) can be expressed as follows:
\begin{eqnarray}
Y_s \sim e^{\pm i \omega r_{\star}}e^{\pm \frac{s}{2}\ln(r - r_H)} \sim e^{\pm i \omega r_{\star}} \left [ h(r)g(r) \right ] ^{\pm \frac{s}{2}}\,,
\label{eq:ysasymptoticsimpsonvisser}
\end{eqnarray}
where in the last step, in light of the discussion in Ref.~\cite{Sebastiani:2022wbz}, we generalized $\Delta$ to the following:
\begin{eqnarray}
\Delta \equiv r^2g(r) \longrightarrow \text{D} \equiv \left ( h(r)g(r) \right ) \,.
\end{eqnarray}
Finally, returning to $R_s$, it is easy to show that the asymptotic behaviour of the latter is given by:
\begin{eqnarray}
R_s \sim e^{i\omega r_{\star}}, \qquad \text{and} \qquad R_s \sim A_s e^{-i\omega r_{\star}}
\label{eq:rsasymptotichorizon}
\end{eqnarray}
where $A_s$ is defined according to Eq.~(\ref{eq:as}), and confirming the asymptotic scaling quoted in Eq.~(\ref{eq:asymptotic2}).

\subsubsection{Peltola-Kunstatter}
\label{subsubsec:peltolakunstatter}

For the PK space-time the functions $f(r)$ and $g(r)$ are given by the following:
\begin{eqnarray}
f(r)=g(r)=\frac{r-2M}{\sqrt{r^2 - \ell^2}}\,,
\end{eqnarray}
so the tortoise coordinate is given by the following:
\begin{eqnarray}
r_{\star} = \int \frac{dr}{1 - \frac{2M}{\sqrt{r^2+\ell^2}}} \xrightarrow{r \to r_H^{\text{PK}}} \sqrt{4M^2 + \ell^2}\ln \left ( r - r_H^{\text{PK}} \right ) \,, \nonumber \\
\label{eq:rstarpk}
\end{eqnarray}
with $r_H^{\text{PK}} = 2M$. For this metric we have $g'(r) = \ell^2+2mr/(r^2 + \ell^2)^{3/2}$, from which one can easily show that Eq.~(\ref{eq:ysasymptoticsolutionshorizonrstar}) can be expressed in the same way as Eq.~ (\ref{eq:ysasymptoticsimpsonvisser}), from which it follows that $g'\sqrt{f/g}$ cancels with the term proportional to $K$ in Eq.~(\ref{eq:ysasymptoticsolutionshorizonrstar}), and therefore that the asymptotic scaling of $R_s$ is the same as in Eq.~(\ref{eq:rsasymptotichorizon}), thereby confirming the asymptotic scaling quoted in Eq.~(\ref{eq:asymptotic2}).

\subsubsection{D'Ambrosio-Rovelli}
\label{subsubsec:dambrosiorovelli}

For the DR space-time the functions $f(r)$ and $g(r)$ are given by the following:
\begin{eqnarray}
f(r)=1-\frac{2M}{\sqrt{r^2 + \ell^2}} \quad \text{and} \quad g(r)=\frac{\sqrt{r^2 + \ell^2} - 2M}{\sqrt{r^2 + \ell^2} + \ell}\,, \nonumber \\
\end{eqnarray}
so the tortoise coordinate is given by the following:
\begin{align}
r_{\star} =& \int \frac{\sqrt{r^2 + \ell^2 + \ell\sqrt{r^2 + \ell^2}}}{\sqrt{r^2 + \ell^2} - 2M} \nonumber \\
&\xrightarrow{r \to r_H^{\text{DR}}} \frac{2\sqrt{2}M^{3/2}}{\sqrt{2M - \ell}}\ln \left ( r - r_H^{\text{DR}} \right ) \,,
\label{eq:rstardr}
\end{align}
with $r_H^{\text{DR}} = \sqrt{4M^2 - \ell^2}$. From Eq.~(\ref{eq:rstardr}) it follows that $\text{K} = 2\sqrt{2}M^{3/2}(2M-\ell)$, which cancels once again with the $f'\sqrt{g/f}$ term in Eq.~(\ref{eq:ysasymptoticsolutionshorizonrstar}). Analogously to the previous case, one can therefore conclude that the asymptotic solutions to the radial Teukolsky equation for $R_s$ in the near horizon region are indeed given by Eq.~(\ref{eq:rsasymptotichorizon}), thereby confirming the asymptotic scaling quoted in Eq.~(\ref{eq:asymptotic2}).

\bibliography{prbhii}

\begin{thebibliography}{380}%
\makeatletter
\providecommand \@ifxundefined [1]{%
 \@ifx{#1\undefined}
}%
\providecommand \@ifnum [1]{%
 \ifnum #1\expandafter \@firstoftwo
 \else \expandafter \@secondoftwo
 \fi
}%
\providecommand \@ifx [1]{%
 \ifx #1\expandafter \@firstoftwo
 \else \expandafter \@secondoftwo
 \fi
}%
\providecommand \natexlab [1]{#1}%
\providecommand \enquote  [1]{``#1''}%
\providecommand \bibnamefont  [1]{#1}%
\providecommand \bibfnamefont [1]{#1}%
\providecommand \citenamefont [1]{#1}%
\providecommand \href@noop [0]{\@secondoftwo}%
\providecommand \href [0]{\begingroup \@sanitize@url \@href}%
\providecommand \@href[1]{\@@startlink{#1}\@@href}%
\providecommand \@@href[1]{\endgroup#1\@@endlink}%
\providecommand \@sanitize@url [0]{\catcode `\\12\catcode `\$12\catcode
  `\&12\catcode `\#12\catcode `\^12\catcode `\_12\catcode `\%12\relax}%
\providecommand \@@startlink[1]{}%
\providecommand \@@endlink[0]{}%
\providecommand \url  [0]{\begingroup\@sanitize@url \@url }%
\providecommand \@url [1]{\endgroup\@href {#1}{\urlprefix }}%
\providecommand \urlprefix  [0]{URL }%
\providecommand \Eprint [0]{\href }%
\providecommand \doibase [0]{http://dx.doi.org/}%
\providecommand \selectlanguage [0]{\@gobble}%
\providecommand \bibinfo  [0]{\@secondoftwo}%
\providecommand \bibfield  [0]{\@secondoftwo}%
\providecommand \translation [1]{[#1]}%
\providecommand \BibitemOpen [0]{}%
\providecommand \bibitemStop [0]{}%
\providecommand \bibitemNoStop [0]{.\EOS\space}%
\providecommand \EOS [0]{\spacefactor3000\relax}%
\providecommand \BibitemShut  [1]{\csname bibitem#1\endcsname}%
\let\auto@bib@innerbib\@empty
\bibitem [{\citenamefont {Cardoso}\ and\ \citenamefont
  {Pani}(2019)}]{Cardoso:2019rvt}%
  \BibitemOpen
  \bibfield  {author} {\bibinfo {author} {\bibfnamefont {V.}~\bibnamefont
  {Cardoso}}\ and\ \bibinfo {author} {\bibfnamefont {P.}~\bibnamefont {Pani}},\
  }\href {\doibase 10.1007/s41114-019-0020-4} {\bibfield  {journal} {\bibinfo
  {journal} {Living Rev. Rel.}\ }\textbf {\bibinfo {volume} {22}},\ \bibinfo
  {pages} {4} (\bibinfo {year} {2019})},\ \Eprint
  {http://arxiv.org/abs/1904.05363} {arXiv:1904.05363 [gr-qc]} \BibitemShut
  {NoStop}%
\bibitem [{\citenamefont {Creminelli}\ and\ \citenamefont
  {Vernizzi}(2017)}]{Creminelli:2017sry}%
  \BibitemOpen
  \bibfield  {author} {\bibinfo {author} {\bibfnamefont {P.}~\bibnamefont
  {Creminelli}}\ and\ \bibinfo {author} {\bibfnamefont {F.}~\bibnamefont
  {Vernizzi}},\ }\href {\doibase 10.1103/PhysRevLett.119.251302} {\bibfield
  {journal} {\bibinfo  {journal} {Phys. Rev. Lett.}\ }\textbf {\bibinfo
  {volume} {119}},\ \bibinfo {pages} {251302} (\bibinfo {year} {2017})},\
  \Eprint {http://arxiv.org/abs/1710.05877} {arXiv:1710.05877 [astro-ph.CO]}
  \BibitemShut {NoStop}%
\bibitem [{\citenamefont {Sakstein}\ and\ \citenamefont
  {Jain}(2017)}]{Sakstein:2017xjx}%
  \BibitemOpen
  \bibfield  {author} {\bibinfo {author} {\bibfnamefont {J.}~\bibnamefont
  {Sakstein}}\ and\ \bibinfo {author} {\bibfnamefont {B.}~\bibnamefont
  {Jain}},\ }\href {\doibase 10.1103/PhysRevLett.119.251303} {\bibfield
  {journal} {\bibinfo  {journal} {Phys. Rev. Lett.}\ }\textbf {\bibinfo
  {volume} {119}},\ \bibinfo {pages} {251303} (\bibinfo {year} {2017})},\
  \Eprint {http://arxiv.org/abs/1710.05893} {arXiv:1710.05893 [astro-ph.CO]}
  \BibitemShut {NoStop}%
\bibitem [{\citenamefont {Ezquiaga}\ and\ \citenamefont
  {Zumalac\'arregui}(2017)}]{Ezquiaga:2017ekz}%
  \BibitemOpen
  \bibfield  {author} {\bibinfo {author} {\bibfnamefont {J.~M.}\ \bibnamefont
  {Ezquiaga}}\ and\ \bibinfo {author} {\bibfnamefont {M.}~\bibnamefont
  {Zumalac\'arregui}},\ }\href {\doibase 10.1103/PhysRevLett.119.251304}
  {\bibfield  {journal} {\bibinfo  {journal} {Phys. Rev. Lett.}\ }\textbf
  {\bibinfo {volume} {119}},\ \bibinfo {pages} {251304} (\bibinfo {year}
  {2017})},\ \Eprint {http://arxiv.org/abs/1710.05901} {arXiv:1710.05901
  [astro-ph.CO]} \BibitemShut {NoStop}%
\bibitem [{\citenamefont {Boran}\ \emph {et~al.}(2018)\citenamefont {Boran},
  \citenamefont {Desai}, \citenamefont {Kahya},\ and\ \citenamefont
  {Woodard}}]{Boran:2017rdn}%
  \BibitemOpen
  \bibfield  {author} {\bibinfo {author} {\bibfnamefont {S.}~\bibnamefont
  {Boran}}, \bibinfo {author} {\bibfnamefont {S.}~\bibnamefont {Desai}},
  \bibinfo {author} {\bibfnamefont {E.~O.}\ \bibnamefont {Kahya}}, \ and\
  \bibinfo {author} {\bibfnamefont {R.~P.}\ \bibnamefont {Woodard}},\ }\href
  {\doibase 10.1103/PhysRevD.97.041501} {\bibfield  {journal} {\bibinfo
  {journal} {Phys. Rev. D}\ }\textbf {\bibinfo {volume} {97}},\ \bibinfo
  {pages} {041501} (\bibinfo {year} {2018})},\ \Eprint
  {http://arxiv.org/abs/1710.06168} {arXiv:1710.06168 [astro-ph.HE]}
  \BibitemShut {NoStop}%
\bibitem [{\citenamefont {Baker}\ \emph {et~al.}(2017)\citenamefont {Baker},
  \citenamefont {Bellini}, \citenamefont {Ferreira}, \citenamefont {Lagos},
  \citenamefont {Noller},\ and\ \citenamefont {Sawicki}}]{Baker:2017hug}%
  \BibitemOpen
  \bibfield  {author} {\bibinfo {author} {\bibfnamefont {T.}~\bibnamefont
  {Baker}}, \bibinfo {author} {\bibfnamefont {E.}~\bibnamefont {Bellini}},
  \bibinfo {author} {\bibfnamefont {P.~G.}\ \bibnamefont {Ferreira}}, \bibinfo
  {author} {\bibfnamefont {M.}~\bibnamefont {Lagos}}, \bibinfo {author}
  {\bibfnamefont {J.}~\bibnamefont {Noller}}, \ and\ \bibinfo {author}
  {\bibfnamefont {I.}~\bibnamefont {Sawicki}},\ }\href {\doibase
  10.1103/PhysRevLett.119.251301} {\bibfield  {journal} {\bibinfo  {journal}
  {Phys. Rev. Lett.}\ }\textbf {\bibinfo {volume} {119}},\ \bibinfo {pages}
  {251301} (\bibinfo {year} {2017})},\ \Eprint
  {http://arxiv.org/abs/1710.06394} {arXiv:1710.06394 [astro-ph.CO]}
  \BibitemShut {NoStop}%
\bibitem [{\citenamefont {Amendola}\ \emph {et~al.}(2018)\citenamefont
  {Amendola}, \citenamefont {Kunz}, \citenamefont {Saltas},\ and\ \citenamefont
  {Sawicki}}]{Amendola:2017orw}%
  \BibitemOpen
  \bibfield  {author} {\bibinfo {author} {\bibfnamefont {L.}~\bibnamefont
  {Amendola}}, \bibinfo {author} {\bibfnamefont {M.}~\bibnamefont {Kunz}},
  \bibinfo {author} {\bibfnamefont {I.~D.}\ \bibnamefont {Saltas}}, \ and\
  \bibinfo {author} {\bibfnamefont {I.}~\bibnamefont {Sawicki}},\ }\href
  {\doibase 10.1103/PhysRevLett.120.131101} {\bibfield  {journal} {\bibinfo
  {journal} {Phys. Rev. Lett.}\ }\textbf {\bibinfo {volume} {120}},\ \bibinfo
  {pages} {131101} (\bibinfo {year} {2018})},\ \Eprint
  {http://arxiv.org/abs/1711.04825} {arXiv:1711.04825 [astro-ph.CO]}
  \BibitemShut {NoStop}%
\bibitem [{\citenamefont {Visinelli}\ \emph {et~al.}(2018)\citenamefont
  {Visinelli}, \citenamefont {Bolis},\ and\ \citenamefont
  {Vagnozzi}}]{Visinelli:2017bny}%
  \BibitemOpen
  \bibfield  {author} {\bibinfo {author} {\bibfnamefont {L.}~\bibnamefont
  {Visinelli}}, \bibinfo {author} {\bibfnamefont {N.}~\bibnamefont {Bolis}}, \
  and\ \bibinfo {author} {\bibfnamefont {S.}~\bibnamefont {Vagnozzi}},\ }\href
  {\doibase 10.1103/PhysRevD.97.064039} {\bibfield  {journal} {\bibinfo
  {journal} {Phys. Rev. D}\ }\textbf {\bibinfo {volume} {97}},\ \bibinfo
  {pages} {064039} (\bibinfo {year} {2018})},\ \Eprint
  {http://arxiv.org/abs/1711.06628} {arXiv:1711.06628 [gr-qc]} \BibitemShut
  {NoStop}%
\bibitem [{\citenamefont {Crisostomi}\ and\ \citenamefont
  {Koyama}(2018)}]{Crisostomi:2017lbg}%
  \BibitemOpen
  \bibfield  {author} {\bibinfo {author} {\bibfnamefont {M.}~\bibnamefont
  {Crisostomi}}\ and\ \bibinfo {author} {\bibfnamefont {K.}~\bibnamefont
  {Koyama}},\ }\href {\doibase 10.1103/PhysRevD.97.021301} {\bibfield
  {journal} {\bibinfo  {journal} {Phys. Rev. D}\ }\textbf {\bibinfo {volume}
  {97}},\ \bibinfo {pages} {021301} (\bibinfo {year} {2018})},\ \Eprint
  {http://arxiv.org/abs/1711.06661} {arXiv:1711.06661 [astro-ph.CO]}
  \BibitemShut {NoStop}%
\bibitem [{\citenamefont {Dima}\ and\ \citenamefont
  {Vernizzi}(2018)}]{Dima:2017pwp}%
  \BibitemOpen
  \bibfield  {author} {\bibinfo {author} {\bibfnamefont {A.}~\bibnamefont
  {Dima}}\ and\ \bibinfo {author} {\bibfnamefont {F.}~\bibnamefont
  {Vernizzi}},\ }\href {\doibase 10.1103/PhysRevD.97.101302} {\bibfield
  {journal} {\bibinfo  {journal} {Phys. Rev. D}\ }\textbf {\bibinfo {volume}
  {97}},\ \bibinfo {pages} {101302} (\bibinfo {year} {2018})},\ \Eprint
  {http://arxiv.org/abs/1712.04731} {arXiv:1712.04731 [gr-qc]} \BibitemShut
  {NoStop}%
\bibitem [{\citenamefont {Cai}\ \emph {et~al.}(2018)\citenamefont {Cai},
  \citenamefont {Li}, \citenamefont {Saridakis},\ and\ \citenamefont
  {Xue}}]{Cai:2018rzd}%
  \BibitemOpen
  \bibfield  {author} {\bibinfo {author} {\bibfnamefont {Y.-F.}\ \bibnamefont
  {Cai}}, \bibinfo {author} {\bibfnamefont {C.}~\bibnamefont {Li}}, \bibinfo
  {author} {\bibfnamefont {E.~N.}\ \bibnamefont {Saridakis}}, \ and\ \bibinfo
  {author} {\bibfnamefont {L.}~\bibnamefont {Xue}},\ }\href {\doibase
  10.1103/PhysRevD.97.103513} {\bibfield  {journal} {\bibinfo  {journal} {Phys.
  Rev. D}\ }\textbf {\bibinfo {volume} {97}},\ \bibinfo {pages} {103513}
  (\bibinfo {year} {2018})},\ \Eprint {http://arxiv.org/abs/1801.05827}
  {arXiv:1801.05827 [gr-qc]} \BibitemShut {NoStop}%
\bibitem [{\citenamefont {Casalino}\ \emph {et~al.}(2018)\citenamefont
  {Casalino}, \citenamefont {Rinaldi}, \citenamefont {Sebastiani},\ and\
  \citenamefont {Vagnozzi}}]{Casalino:2018tcd}%
  \BibitemOpen
  \bibfield  {author} {\bibinfo {author} {\bibfnamefont {A.}~\bibnamefont
  {Casalino}}, \bibinfo {author} {\bibfnamefont {M.}~\bibnamefont {Rinaldi}},
  \bibinfo {author} {\bibfnamefont {L.}~\bibnamefont {Sebastiani}}, \ and\
  \bibinfo {author} {\bibfnamefont {S.}~\bibnamefont {Vagnozzi}},\ }\href
  {\doibase 10.1016/j.dark.2018.10.001} {\bibfield  {journal} {\bibinfo
  {journal} {Phys. Dark Univ.}\ }\textbf {\bibinfo {volume} {22}},\ \bibinfo
  {pages} {108} (\bibinfo {year} {2018})},\ \Eprint
  {http://arxiv.org/abs/1803.02620} {arXiv:1803.02620 [gr-qc]} \BibitemShut
  {NoStop}%
\bibitem [{\citenamefont {Barack}\ \emph {et~al.}(2019)\citenamefont {Barack}
  \emph {et~al.}}]{Barack:2018yly}%
  \BibitemOpen
  \bibfield  {author} {\bibinfo {author} {\bibfnamefont {L.}~\bibnamefont
  {Barack}} \emph {et~al.},\ }\href {\doibase 10.1088/1361-6382/ab0587}
  {\bibfield  {journal} {\bibinfo  {journal} {Class. Quant. Grav.}\ }\textbf
  {\bibinfo {volume} {36}},\ \bibinfo {pages} {143001} (\bibinfo {year}
  {2019})},\ \Eprint {http://arxiv.org/abs/1806.05195} {arXiv:1806.05195
  [gr-qc]} \BibitemShut {NoStop}%
\bibitem [{\citenamefont {Abbott}\ \emph {et~al.}(2019)\citenamefont {Abbott}
  \emph {et~al.}}]{LIGOScientific:2018dkp}%
  \BibitemOpen
  \bibfield  {author} {\bibinfo {author} {\bibfnamefont {B.~P.}\ \bibnamefont
  {Abbott}} \emph {et~al.} (\bibinfo {collaboration} {LIGO Scientific,
  Virgo}),\ }\href {\doibase 10.1103/PhysRevLett.123.011102} {\bibfield
  {journal} {\bibinfo  {journal} {Phys. Rev. Lett.}\ }\textbf {\bibinfo
  {volume} {123}},\ \bibinfo {pages} {011102} (\bibinfo {year} {2019})},\
  \Eprint {http://arxiv.org/abs/1811.00364} {arXiv:1811.00364 [gr-qc]}
  \BibitemShut {NoStop}%
\bibitem [{\citenamefont {Casalino}\ \emph {et~al.}(2019)\citenamefont
  {Casalino}, \citenamefont {Rinaldi}, \citenamefont {Sebastiani},\ and\
  \citenamefont {Vagnozzi}}]{Casalino:2018wnc}%
  \BibitemOpen
  \bibfield  {author} {\bibinfo {author} {\bibfnamefont {A.}~\bibnamefont
  {Casalino}}, \bibinfo {author} {\bibfnamefont {M.}~\bibnamefont {Rinaldi}},
  \bibinfo {author} {\bibfnamefont {L.}~\bibnamefont {Sebastiani}}, \ and\
  \bibinfo {author} {\bibfnamefont {S.}~\bibnamefont {Vagnozzi}},\ }\href
  {\doibase 10.1088/1361-6382/aaf1fd} {\bibfield  {journal} {\bibinfo
  {journal} {Class. Quant. Grav.}\ }\textbf {\bibinfo {volume} {36}},\ \bibinfo
  {pages} {017001} (\bibinfo {year} {2019})},\ \Eprint
  {http://arxiv.org/abs/1811.06830} {arXiv:1811.06830 [gr-qc]} \BibitemShut
  {NoStop}%
\bibitem [{\citenamefont {Held}\ \emph {et~al.}(2019)\citenamefont {Held},
  \citenamefont {Gold},\ and\ \citenamefont {Eichhorn}}]{Held:2019xde}%
  \BibitemOpen
  \bibfield  {author} {\bibinfo {author} {\bibfnamefont {A.}~\bibnamefont
  {Held}}, \bibinfo {author} {\bibfnamefont {R.}~\bibnamefont {Gold}}, \ and\
  \bibinfo {author} {\bibfnamefont {A.}~\bibnamefont {Eichhorn}},\ }\href
  {\doibase 10.1088/1475-7516/2019/06/029} {\bibfield  {journal} {\bibinfo
  {journal} {JCAP}\ }\textbf {\bibinfo {volume} {06}},\ \bibinfo {pages} {029}
  (\bibinfo {year} {2019})},\ \Eprint {http://arxiv.org/abs/1904.07133}
  {arXiv:1904.07133 [gr-qc]} \BibitemShut {NoStop}%
\bibitem [{\citenamefont {Bambi}\ \emph {et~al.}(2019)\citenamefont {Bambi},
  \citenamefont {Freese}, \citenamefont {Vagnozzi},\ and\ \citenamefont
  {Visinelli}}]{Bambi:2019tjh}%
  \BibitemOpen
  \bibfield  {author} {\bibinfo {author} {\bibfnamefont {C.}~\bibnamefont
  {Bambi}}, \bibinfo {author} {\bibfnamefont {K.}~\bibnamefont {Freese}},
  \bibinfo {author} {\bibfnamefont {S.}~\bibnamefont {Vagnozzi}}, \ and\
  \bibinfo {author} {\bibfnamefont {L.}~\bibnamefont {Visinelli}},\ }\href
  {\doibase 10.1103/PhysRevD.100.044057} {\bibfield  {journal} {\bibinfo
  {journal} {Phys. Rev. D}\ }\textbf {\bibinfo {volume} {100}},\ \bibinfo
  {pages} {044057} (\bibinfo {year} {2019})},\ \Eprint
  {http://arxiv.org/abs/1904.12983} {arXiv:1904.12983 [gr-qc]} \BibitemShut
  {NoStop}%
\bibitem [{\citenamefont {Vagnozzi}\ and\ \citenamefont
  {Visinelli}(2019)}]{Vagnozzi:2019apd}%
  \BibitemOpen
  \bibfield  {author} {\bibinfo {author} {\bibfnamefont {S.}~\bibnamefont
  {Vagnozzi}}\ and\ \bibinfo {author} {\bibfnamefont {L.}~\bibnamefont
  {Visinelli}},\ }\href {\doibase 10.1103/PhysRevD.100.024020} {\bibfield
  {journal} {\bibinfo  {journal} {Phys. Rev. D}\ }\textbf {\bibinfo {volume}
  {100}},\ \bibinfo {pages} {024020} (\bibinfo {year} {2019})},\ \Eprint
  {http://arxiv.org/abs/1905.12421} {arXiv:1905.12421 [gr-qc]} \BibitemShut
  {NoStop}%
\bibitem [{\citenamefont {Zhu}\ \emph {et~al.}(2019)\citenamefont {Zhu},
  \citenamefont {Wu}, \citenamefont {Jamil},\ and\ \citenamefont
  {Jusufi}}]{Zhu:2019ura}%
  \BibitemOpen
  \bibfield  {author} {\bibinfo {author} {\bibfnamefont {T.}~\bibnamefont
  {Zhu}}, \bibinfo {author} {\bibfnamefont {Q.}~\bibnamefont {Wu}}, \bibinfo
  {author} {\bibfnamefont {M.}~\bibnamefont {Jamil}}, \ and\ \bibinfo {author}
  {\bibfnamefont {K.}~\bibnamefont {Jusufi}},\ }\href {\doibase
  10.1103/PhysRevD.100.044055} {\bibfield  {journal} {\bibinfo  {journal}
  {Phys. Rev. D}\ }\textbf {\bibinfo {volume} {100}},\ \bibinfo {pages}
  {044055} (\bibinfo {year} {2019})},\ \Eprint
  {http://arxiv.org/abs/1906.05673} {arXiv:1906.05673 [gr-qc]} \BibitemShut
  {NoStop}%
\bibitem [{\citenamefont {Cunha}\ \emph {et~al.}(2019)\citenamefont {Cunha},
  \citenamefont {Herdeiro},\ and\ \citenamefont {Radu}}]{Cunha:2019ikd}%
  \BibitemOpen
  \bibfield  {author} {\bibinfo {author} {\bibfnamefont {P.~V.~P.}\
  \bibnamefont {Cunha}}, \bibinfo {author} {\bibfnamefont {C.~A.~R.}\
  \bibnamefont {Herdeiro}}, \ and\ \bibinfo {author} {\bibfnamefont
  {E.}~\bibnamefont {Radu}},\ }\href {\doibase 10.3390/universe5120220}
  {\bibfield  {journal} {\bibinfo  {journal} {Universe}\ }\textbf {\bibinfo
  {volume} {5}},\ \bibinfo {pages} {220} (\bibinfo {year} {2019})},\ \Eprint
  {http://arxiv.org/abs/1909.08039} {arXiv:1909.08039 [gr-qc]} \BibitemShut
  {NoStop}%
\bibitem [{\citenamefont {Banerjee}\ \emph
  {et~al.}(2020{\natexlab{a}})\citenamefont {Banerjee}, \citenamefont
  {Chakraborty},\ and\ \citenamefont {SenGupta}}]{Banerjee:2019nnj}%
  \BibitemOpen
  \bibfield  {author} {\bibinfo {author} {\bibfnamefont {I.}~\bibnamefont
  {Banerjee}}, \bibinfo {author} {\bibfnamefont {S.}~\bibnamefont
  {Chakraborty}}, \ and\ \bibinfo {author} {\bibfnamefont {S.}~\bibnamefont
  {SenGupta}},\ }\href {\doibase 10.1103/PhysRevD.101.041301} {\bibfield
  {journal} {\bibinfo  {journal} {Phys. Rev. D}\ }\textbf {\bibinfo {volume}
  {101}},\ \bibinfo {pages} {041301} (\bibinfo {year} {2020}{\natexlab{a}})},\
  \Eprint {http://arxiv.org/abs/1909.09385} {arXiv:1909.09385 [gr-qc]}
  \BibitemShut {NoStop}%
\bibitem [{\citenamefont {Banerjee}\ \emph
  {et~al.}(2020{\natexlab{b}})\citenamefont {Banerjee}, \citenamefont {Sau},\
  and\ \citenamefont {SenGupta}}]{Banerjee:2019xds}%
  \BibitemOpen
  \bibfield  {author} {\bibinfo {author} {\bibfnamefont {I.}~\bibnamefont
  {Banerjee}}, \bibinfo {author} {\bibfnamefont {S.}~\bibnamefont {Sau}}, \
  and\ \bibinfo {author} {\bibfnamefont {S.}~\bibnamefont {SenGupta}},\ }\href
  {\doibase 10.1103/PhysRevD.101.104057} {\bibfield  {journal} {\bibinfo
  {journal} {Phys. Rev. D}\ }\textbf {\bibinfo {volume} {101}},\ \bibinfo
  {pages} {104057} (\bibinfo {year} {2020}{\natexlab{b}})},\ \Eprint
  {http://arxiv.org/abs/1911.05385} {arXiv:1911.05385 [gr-qc]} \BibitemShut
  {NoStop}%
\bibitem [{\citenamefont {Allahyari}\ \emph {et~al.}(2020)\citenamefont
  {Allahyari}, \citenamefont {Khodadi}, \citenamefont {Vagnozzi},\ and\
  \citenamefont {Mota}}]{Allahyari:2019jqz}%
  \BibitemOpen
  \bibfield  {author} {\bibinfo {author} {\bibfnamefont {A.}~\bibnamefont
  {Allahyari}}, \bibinfo {author} {\bibfnamefont {M.}~\bibnamefont {Khodadi}},
  \bibinfo {author} {\bibfnamefont {S.}~\bibnamefont {Vagnozzi}}, \ and\
  \bibinfo {author} {\bibfnamefont {D.~F.}\ \bibnamefont {Mota}},\ }\href
  {\doibase 10.1088/1475-7516/2020/02/003} {\bibfield  {journal} {\bibinfo
  {journal} {JCAP}\ }\textbf {\bibinfo {volume} {02}},\ \bibinfo {pages} {003}
  (\bibinfo {year} {2020})},\ \Eprint {http://arxiv.org/abs/1912.08231}
  {arXiv:1912.08231 [gr-qc]} \BibitemShut {NoStop}%
\bibitem [{\citenamefont {Vagnozzi}\ \emph {et~al.}(2020)\citenamefont
  {Vagnozzi}, \citenamefont {Bambi},\ and\ \citenamefont
  {Visinelli}}]{Vagnozzi:2020quf}%
  \BibitemOpen
  \bibfield  {author} {\bibinfo {author} {\bibfnamefont {S.}~\bibnamefont
  {Vagnozzi}}, \bibinfo {author} {\bibfnamefont {C.}~\bibnamefont {Bambi}}, \
  and\ \bibinfo {author} {\bibfnamefont {L.}~\bibnamefont {Visinelli}},\ }\href
  {\doibase 10.1088/1361-6382/ab7965} {\bibfield  {journal} {\bibinfo
  {journal} {Class. Quant. Grav.}\ }\textbf {\bibinfo {volume} {37}},\ \bibinfo
  {pages} {087001} (\bibinfo {year} {2020})},\ \Eprint
  {http://arxiv.org/abs/2001.02986} {arXiv:2001.02986 [gr-qc]} \BibitemShut
  {NoStop}%
\bibitem [{\citenamefont {Khodadi}\ \emph {et~al.}(2020)\citenamefont
  {Khodadi}, \citenamefont {Allahyari}, \citenamefont {Vagnozzi},\ and\
  \citenamefont {Mota}}]{Khodadi:2020jij}%
  \BibitemOpen
  \bibfield  {author} {\bibinfo {author} {\bibfnamefont {M.}~\bibnamefont
  {Khodadi}}, \bibinfo {author} {\bibfnamefont {A.}~\bibnamefont {Allahyari}},
  \bibinfo {author} {\bibfnamefont {S.}~\bibnamefont {Vagnozzi}}, \ and\
  \bibinfo {author} {\bibfnamefont {D.~F.}\ \bibnamefont {Mota}},\ }\href
  {\doibase 10.1088/1475-7516/2020/09/026} {\bibfield  {journal} {\bibinfo
  {journal} {JCAP}\ }\textbf {\bibinfo {volume} {09}},\ \bibinfo {pages} {026}
  (\bibinfo {year} {2020})},\ \Eprint {http://arxiv.org/abs/2005.05992}
  {arXiv:2005.05992 [gr-qc]} \BibitemShut {NoStop}%
\bibitem [{\citenamefont {Kumar}\ \emph {et~al.}(2020)\citenamefont {Kumar},
  \citenamefont {Kumar},\ and\ \citenamefont {Ghosh}}]{Kumar:2020yem}%
  \BibitemOpen
  \bibfield  {author} {\bibinfo {author} {\bibfnamefont {R.}~\bibnamefont
  {Kumar}}, \bibinfo {author} {\bibfnamefont {A.}~\bibnamefont {Kumar}}, \ and\
  \bibinfo {author} {\bibfnamefont {S.~G.}\ \bibnamefont {Ghosh}},\ }\href
  {\doibase 10.3847/1538-4357/ab8c4a} {\bibfield  {journal} {\bibinfo
  {journal} {Astrophys. J.}\ }\textbf {\bibinfo {volume} {896}},\ \bibinfo
  {pages} {89} (\bibinfo {year} {2020})},\ \Eprint
  {http://arxiv.org/abs/2006.09869} {arXiv:2006.09869 [gr-qc]} \BibitemShut
  {NoStop}%
\bibitem [{\citenamefont {Khodadi}\ and\ \citenamefont
  {Saridakis}(2021)}]{Khodadi:2020gns}%
  \BibitemOpen
  \bibfield  {author} {\bibinfo {author} {\bibfnamefont {M.}~\bibnamefont
  {Khodadi}}\ and\ \bibinfo {author} {\bibfnamefont {E.~N.}\ \bibnamefont
  {Saridakis}},\ }\href {\doibase 10.1016/j.dark.2021.100835} {\bibfield
  {journal} {\bibinfo  {journal} {Phys. Dark Univ.}\ }\textbf {\bibinfo
  {volume} {32}},\ \bibinfo {pages} {100835} (\bibinfo {year} {2021})},\
  \Eprint {http://arxiv.org/abs/2012.05186} {arXiv:2012.05186 [gr-qc]}
  \BibitemShut {NoStop}%
\bibitem [{\citenamefont {Pantig}\ \emph {et~al.}(2022)\citenamefont {Pantig},
  \citenamefont {Yu}, \citenamefont {Rodulfo},\ and\ \citenamefont
  {\"Ovg\"un}}]{Pantig:2021zqe}%
  \BibitemOpen
  \bibfield  {author} {\bibinfo {author} {\bibfnamefont {R.~C.}\ \bibnamefont
  {Pantig}}, \bibinfo {author} {\bibfnamefont {P.~K.}\ \bibnamefont {Yu}},
  \bibinfo {author} {\bibfnamefont {E.~T.}\ \bibnamefont {Rodulfo}}, \ and\
  \bibinfo {author} {\bibfnamefont {A.}~\bibnamefont {\"Ovg\"un}},\ }\href
  {\doibase 10.1016/j.aop.2021.168722} {\bibfield  {journal} {\bibinfo
  {journal} {Annals Phys.}\ }\textbf {\bibinfo {volume} {436}},\ \bibinfo
  {pages} {168722} (\bibinfo {year} {2022})},\ \Eprint
  {http://arxiv.org/abs/2104.04304} {arXiv:2104.04304 [gr-qc]} \BibitemShut
  {NoStop}%
\bibitem [{\citenamefont {Khodadi}\ \emph {et~al.}(2021)\citenamefont
  {Khodadi}, \citenamefont {Lambiase},\ and\ \citenamefont
  {Mota}}]{Khodadi:2021gbc}%
  \BibitemOpen
  \bibfield  {author} {\bibinfo {author} {\bibfnamefont {M.}~\bibnamefont
  {Khodadi}}, \bibinfo {author} {\bibfnamefont {G.}~\bibnamefont {Lambiase}}, \
  and\ \bibinfo {author} {\bibfnamefont {D.~F.}\ \bibnamefont {Mota}},\ }\href
  {\doibase 10.1088/1475-7516/2021/09/028} {\bibfield  {journal} {\bibinfo
  {journal} {JCAP}\ }\textbf {\bibinfo {volume} {09}},\ \bibinfo {pages} {028}
  (\bibinfo {year} {2021})},\ \Eprint {http://arxiv.org/abs/2107.00834}
  {arXiv:2107.00834 [gr-qc]} \BibitemShut {NoStop}%
\bibitem [{\citenamefont {Roy}\ \emph {et~al.}(2022)\citenamefont {Roy},
  \citenamefont {Vagnozzi},\ and\ \citenamefont {Visinelli}}]{Roy:2021uye}%
  \BibitemOpen
  \bibfield  {author} {\bibinfo {author} {\bibfnamefont {R.}~\bibnamefont
  {Roy}}, \bibinfo {author} {\bibfnamefont {S.}~\bibnamefont {Vagnozzi}}, \
  and\ \bibinfo {author} {\bibfnamefont {L.}~\bibnamefont {Visinelli}},\ }\href
  {\doibase 10.1103/PhysRevD.105.083002} {\bibfield  {journal} {\bibinfo
  {journal} {Phys. Rev. D}\ }\textbf {\bibinfo {volume} {105}},\ \bibinfo
  {pages} {083002} (\bibinfo {year} {2022})},\ \Eprint
  {http://arxiv.org/abs/2112.06932} {arXiv:2112.06932 [astro-ph.HE]}
  \BibitemShut {NoStop}%
\bibitem [{\citenamefont {Uniyal}\ \emph {et~al.}(2023)\citenamefont {Uniyal},
  \citenamefont {Pantig},\ and\ \citenamefont {\"Ovg\"un}}]{Uniyal:2022vdu}%
  \BibitemOpen
  \bibfield  {author} {\bibinfo {author} {\bibfnamefont {A.}~\bibnamefont
  {Uniyal}}, \bibinfo {author} {\bibfnamefont {R.~C.}\ \bibnamefont {Pantig}},
  \ and\ \bibinfo {author} {\bibfnamefont {A.}~\bibnamefont {\"Ovg\"un}},\
  }\href {\doibase 10.1016/j.dark.2023.101178} {\bibfield  {journal} {\bibinfo
  {journal} {Phys. Dark Univ.}\ }\textbf {\bibinfo {volume} {40}},\ \bibinfo
  {pages} {101178} (\bibinfo {year} {2023})},\ \Eprint
  {http://arxiv.org/abs/2205.11072} {arXiv:2205.11072 [gr-qc]} \BibitemShut
  {NoStop}%
\bibitem [{\citenamefont {Pantig}\ and\ \citenamefont
  {\"Ovg\"un}(2023)}]{Pantig:2022ely}%
  \BibitemOpen
  \bibfield  {author} {\bibinfo {author} {\bibfnamefont {R.~C.}\ \bibnamefont
  {Pantig}}\ and\ \bibinfo {author} {\bibfnamefont {A.}~\bibnamefont
  {\"Ovg\"un}},\ }\href {\doibase 10.1016/j.aop.2022.169197} {\bibfield
  {journal} {\bibinfo  {journal} {Annals Phys.}\ }\textbf {\bibinfo {volume}
  {448}},\ \bibinfo {pages} {169197} (\bibinfo {year} {2023})},\ \Eprint
  {http://arxiv.org/abs/2206.02161} {arXiv:2206.02161 [gr-qc]} \BibitemShut
  {NoStop}%
\bibitem [{\citenamefont {Ghosh}\ and\ \citenamefont
  {Afrin}(2023)}]{Ghosh:2022kit}%
  \BibitemOpen
  \bibfield  {author} {\bibinfo {author} {\bibfnamefont {S.~G.}\ \bibnamefont
  {Ghosh}}\ and\ \bibinfo {author} {\bibfnamefont {M.}~\bibnamefont {Afrin}},\
  }\href {\doibase 10.3847/1538-4357/acb695} {\bibfield  {journal} {\bibinfo
  {journal} {Astrophys. J.}\ }\textbf {\bibinfo {volume} {944}},\ \bibinfo
  {pages} {174} (\bibinfo {year} {2023})},\ \Eprint
  {http://arxiv.org/abs/2206.02488} {arXiv:2206.02488 [gr-qc]} \BibitemShut
  {NoStop}%
\bibitem [{\citenamefont {Khodadi}\ and\ \citenamefont
  {Lambiase}(2022)}]{Khodadi:2022pqh}%
  \BibitemOpen
  \bibfield  {author} {\bibinfo {author} {\bibfnamefont {M.}~\bibnamefont
  {Khodadi}}\ and\ \bibinfo {author} {\bibfnamefont {G.}~\bibnamefont
  {Lambiase}},\ }\href {\doibase 10.1103/PhysRevD.106.104050} {\bibfield
  {journal} {\bibinfo  {journal} {Phys. Rev. D}\ }\textbf {\bibinfo {volume}
  {106}},\ \bibinfo {pages} {104050} (\bibinfo {year} {2022})},\ \Eprint
  {http://arxiv.org/abs/2206.08601} {arXiv:2206.08601 [gr-qc]} \BibitemShut
  {NoStop}%
\bibitem [{\citenamefont {Kumar~Walia}\ \emph {et~al.}(2022)\citenamefont
  {Kumar~Walia}, \citenamefont {Ghosh},\ and\ \citenamefont
  {Maharaj}}]{KumarWalia:2022aop}%
  \BibitemOpen
  \bibfield  {author} {\bibinfo {author} {\bibfnamefont {R.}~\bibnamefont
  {Kumar~Walia}}, \bibinfo {author} {\bibfnamefont {S.~G.}\ \bibnamefont
  {Ghosh}}, \ and\ \bibinfo {author} {\bibfnamefont {S.~D.}\ \bibnamefont
  {Maharaj}},\ }\href {\doibase 10.3847/1538-4357/ac9623} {\bibfield  {journal}
  {\bibinfo  {journal} {Astrophys. J.}\ }\textbf {\bibinfo {volume} {939}},\
  \bibinfo {pages} {77} (\bibinfo {year} {2022})},\ \Eprint
  {http://arxiv.org/abs/2207.00078} {arXiv:2207.00078 [gr-qc]} \BibitemShut
  {NoStop}%
\bibitem [{\citenamefont {Shaikh}(2023)}]{Shaikh:2022ivr}%
  \BibitemOpen
  \bibfield  {author} {\bibinfo {author} {\bibfnamefont {R.}~\bibnamefont
  {Shaikh}},\ }\href {\doibase 10.1093/mnras/stad1383} {\bibfield  {journal}
  {\bibinfo  {journal} {Mon. Not. Roy. Astron. Soc.}\ }\textbf {\bibinfo
  {volume} {523}},\ \bibinfo {pages} {375} (\bibinfo {year} {2023})},\ \Eprint
  {http://arxiv.org/abs/2208.01995} {arXiv:2208.01995 [gr-qc]} \BibitemShut
  {NoStop}%
\bibitem [{\citenamefont {Odintsov}\ and\ \citenamefont
  {Oikonomou}(2022)}]{Odintsov:2022umu}%
  \BibitemOpen
  \bibfield  {author} {\bibinfo {author} {\bibfnamefont {S.~D.}\ \bibnamefont
  {Odintsov}}\ and\ \bibinfo {author} {\bibfnamefont {V.~K.}\ \bibnamefont
  {Oikonomou}},\ }\href {\doibase 10.1209/0295-5075/ac8a13} {\bibfield
  {journal} {\bibinfo  {journal} {EPL}\ }\textbf {\bibinfo {volume} {139}},\
  \bibinfo {pages} {59003} (\bibinfo {year} {2022})},\ \Eprint
  {http://arxiv.org/abs/2208.07972} {arXiv:2208.07972 [gr-qc]} \BibitemShut
  {NoStop}%
\bibitem [{\citenamefont {Oikonomou}\ \emph {et~al.}(2022)\citenamefont
  {Oikonomou}, \citenamefont {Tsyba},\ and\ \citenamefont
  {Razina}}]{Oikonomou:2022tjm}%
  \BibitemOpen
  \bibfield  {author} {\bibinfo {author} {\bibfnamefont {V.~K.}\ \bibnamefont
  {Oikonomou}}, \bibinfo {author} {\bibfnamefont {P.}~\bibnamefont {Tsyba}}, \
  and\ \bibinfo {author} {\bibfnamefont {O.}~\bibnamefont {Razina}},\ }\href
  {\doibase 10.3390/universe8090484} {\bibfield  {journal} {\bibinfo  {journal}
  {Universe}\ }\textbf {\bibinfo {volume} {8}},\ \bibinfo {pages} {484}
  (\bibinfo {year} {2022})},\ \Eprint {http://arxiv.org/abs/2209.04669}
  {arXiv:2209.04669 [gr-qc]} \BibitemShut {NoStop}%
\bibitem [{\citenamefont {Pantig}(2024)}]{Pantig:2023yer}%
  \BibitemOpen
  \bibfield  {author} {\bibinfo {author} {\bibfnamefont {R.~C.}\ \bibnamefont
  {Pantig}},\ }\href {\doibase 10.1016/j.cjph.2023.09.015} {\bibfield
  {journal} {\bibinfo  {journal} {Chin. J. Phys.}\ }\textbf {\bibinfo {volume}
  {87}},\ \bibinfo {pages} {49} (\bibinfo {year} {2024})},\ \Eprint
  {http://arxiv.org/abs/2303.01698} {arXiv:2303.01698 [gr-qc]} \BibitemShut
  {NoStop}%
\bibitem [{\citenamefont {Gonz\'alez}\ \emph {et~al.}(2023)\citenamefont
  {Gonz\'alez}, \citenamefont {Jusufi}, \citenamefont {Leon},\ and\
  \citenamefont {Saridakis}}]{Gonzalez:2023rsd}%
  \BibitemOpen
  \bibfield  {author} {\bibinfo {author} {\bibfnamefont {E.}~\bibnamefont
  {Gonz\'alez}}, \bibinfo {author} {\bibfnamefont {K.}~\bibnamefont {Jusufi}},
  \bibinfo {author} {\bibfnamefont {G.}~\bibnamefont {Leon}}, \ and\ \bibinfo
  {author} {\bibfnamefont {E.~N.}\ \bibnamefont {Saridakis}},\ }\href {\doibase
  10.1016/j.dark.2023.101304} {\bibfield  {journal} {\bibinfo  {journal} {Phys.
  Dark Univ.}\ }\textbf {\bibinfo {volume} {42}},\ \bibinfo {pages} {101304}
  (\bibinfo {year} {2023})},\ \Eprint {http://arxiv.org/abs/2305.14305}
  {arXiv:2305.14305 [astro-ph.CO]} \BibitemShut {NoStop}%
\bibitem [{\citenamefont {Sahoo}\ \emph {et~al.}(2024)\citenamefont {Sahoo},
  \citenamefont {Yadav},\ and\ \citenamefont {Banerjee}}]{Sahoo:2023czj}%
  \BibitemOpen
  \bibfield  {author} {\bibinfo {author} {\bibfnamefont {S.~K.}\ \bibnamefont
  {Sahoo}}, \bibinfo {author} {\bibfnamefont {N.}~\bibnamefont {Yadav}}, \ and\
  \bibinfo {author} {\bibfnamefont {I.}~\bibnamefont {Banerjee}},\ }\href
  {\doibase 10.1103/PhysRevD.109.044008} {\bibfield  {journal} {\bibinfo
  {journal} {Phys. Rev. D}\ }\textbf {\bibinfo {volume} {109}},\ \bibinfo
  {pages} {044008} (\bibinfo {year} {2024})},\ \Eprint
  {http://arxiv.org/abs/2305.14870} {arXiv:2305.14870 [gr-qc]} \BibitemShut
  {NoStop}%
\bibitem [{\citenamefont {Nozari}\ and\ \citenamefont
  {Saghafi}(2023)}]{Nozari:2023flq}%
  \BibitemOpen
  \bibfield  {author} {\bibinfo {author} {\bibfnamefont {K.}~\bibnamefont
  {Nozari}}\ and\ \bibinfo {author} {\bibfnamefont {S.}~\bibnamefont
  {Saghafi}},\ }\href {\doibase 10.1140/epjc/s10052-023-11755-w} {\bibfield
  {journal} {\bibinfo  {journal} {Eur. Phys. J. C}\ }\textbf {\bibinfo {volume}
  {83}},\ \bibinfo {pages} {588} (\bibinfo {year} {2023})},\ \Eprint
  {http://arxiv.org/abs/2305.17237} {arXiv:2305.17237 [gr-qc]} \BibitemShut
  {NoStop}%
\bibitem [{\citenamefont {Uniyal}\ \emph {et~al.}(2024)\citenamefont {Uniyal},
  \citenamefont {Chakrabarti}, \citenamefont {Fathi},\ and\ \citenamefont
  {\"Ovg\"un}}]{Uniyal:2023ahv}%
  \BibitemOpen
  \bibfield  {author} {\bibinfo {author} {\bibfnamefont {A.}~\bibnamefont
  {Uniyal}}, \bibinfo {author} {\bibfnamefont {S.}~\bibnamefont {Chakrabarti}},
  \bibinfo {author} {\bibfnamefont {M.}~\bibnamefont {Fathi}}, \ and\ \bibinfo
  {author} {\bibfnamefont {A.}~\bibnamefont {\"Ovg\"un}},\ }\href {\doibase
  10.1016/j.aop.2024.169614} {\bibfield  {journal} {\bibinfo  {journal} {Annals
  Phys.}\ }\textbf {\bibinfo {volume} {462}},\ \bibinfo {pages} {169614}
  (\bibinfo {year} {2024})},\ \Eprint {http://arxiv.org/abs/2309.13680}
  {arXiv:2309.13680 [gr-qc]} \BibitemShut {NoStop}%
\bibitem [{\citenamefont {Filho}\ \emph {et~al.}(2024)\citenamefont {Filho},
  \citenamefont {Reis},\ and\ \citenamefont {Hassanabadi}}]{Filho:2023ycx}%
  \BibitemOpen
  \bibfield  {author} {\bibinfo {author} {\bibfnamefont {A.~A.~A.}\
  \bibnamefont {Filho}}, \bibinfo {author} {\bibfnamefont {J.~A. A.~S.}\
  \bibnamefont {Reis}}, \ and\ \bibinfo {author} {\bibfnamefont
  {H.}~\bibnamefont {Hassanabadi}},\ }\href {\doibase
  10.1088/1475-7516/2024/05/029} {\bibfield  {journal} {\bibinfo  {journal}
  {JCAP}\ }\textbf {\bibinfo {volume} {05}},\ \bibinfo {pages} {029} (\bibinfo
  {year} {2024})},\ \Eprint {http://arxiv.org/abs/2309.15778} {arXiv:2309.15778
  [gr-qc]} \BibitemShut {NoStop}%
\bibitem [{\citenamefont {Raza}\ \emph {et~al.}(2024)\citenamefont {Raza},
  \citenamefont {Rayimbaev}, \citenamefont {Sarikulov}, \citenamefont {Zubair},
  \citenamefont {Ahmedov},\ and\ \citenamefont {Stuchlik}}]{Raza:2023vkn}%
  \BibitemOpen
  \bibfield  {author} {\bibinfo {author} {\bibfnamefont {M.~A.}\ \bibnamefont
  {Raza}}, \bibinfo {author} {\bibfnamefont {J.}~\bibnamefont {Rayimbaev}},
  \bibinfo {author} {\bibfnamefont {F.}~\bibnamefont {Sarikulov}}, \bibinfo
  {author} {\bibfnamefont {M.}~\bibnamefont {Zubair}}, \bibinfo {author}
  {\bibfnamefont {B.}~\bibnamefont {Ahmedov}}, \ and\ \bibinfo {author}
  {\bibfnamefont {Z.}~\bibnamefont {Stuchlik}},\ }\href {\doibase
  10.1016/j.dark.2024.101488} {\bibfield  {journal} {\bibinfo  {journal} {Phys.
  Dark Univ.}\ }\textbf {\bibinfo {volume} {44}},\ \bibinfo {pages} {101488}
  (\bibinfo {year} {2024})},\ \Eprint {http://arxiv.org/abs/2311.15784}
  {arXiv:2311.15784 [gr-qc]} \BibitemShut {NoStop}%
\bibitem [{\citenamefont {Hoshimov}\ \emph {et~al.}(2024)\citenamefont
  {Hoshimov}, \citenamefont {Yunusov}, \citenamefont {Atamurotov},
  \citenamefont {Jamil},\ and\ \citenamefont
  {Abdujabbarov}}]{Hoshimov:2023tlz}%
  \BibitemOpen
  \bibfield  {author} {\bibinfo {author} {\bibfnamefont {H.}~\bibnamefont
  {Hoshimov}}, \bibinfo {author} {\bibfnamefont {O.}~\bibnamefont {Yunusov}},
  \bibinfo {author} {\bibfnamefont {F.}~\bibnamefont {Atamurotov}}, \bibinfo
  {author} {\bibfnamefont {M.}~\bibnamefont {Jamil}}, \ and\ \bibinfo {author}
  {\bibfnamefont {A.}~\bibnamefont {Abdujabbarov}},\ }\href {\doibase
  10.1016/j.dark.2023.101392} {\bibfield  {journal} {\bibinfo  {journal} {Phys.
  Dark Univ.}\ }\textbf {\bibinfo {volume} {43}},\ \bibinfo {pages} {101392}
  (\bibinfo {year} {2024})},\ \Eprint {http://arxiv.org/abs/2312.10678}
  {arXiv:2312.10678 [gr-qc]} \BibitemShut {NoStop}%
\bibitem [{\citenamefont {Chakhchi}\ \emph {et~al.}(2024)\citenamefont
  {Chakhchi}, \citenamefont {El~Moumni},\ and\ \citenamefont
  {Masmar}}]{Chakhchi:2024tzo}%
  \BibitemOpen
  \bibfield  {author} {\bibinfo {author} {\bibfnamefont {L.}~\bibnamefont
  {Chakhchi}}, \bibinfo {author} {\bibfnamefont {H.}~\bibnamefont {El~Moumni}},
  \ and\ \bibinfo {author} {\bibfnamefont {K.}~\bibnamefont {Masmar}},\ }\href
  {\doibase 10.1016/j.dark.2024.101501} {\bibfield  {journal} {\bibinfo
  {journal} {Phys. Dark Univ.}\ }\textbf {\bibinfo {volume} {44}},\ \bibinfo
  {pages} {101501} (\bibinfo {year} {2024})},\ \Eprint
  {http://arxiv.org/abs/2403.09756} {arXiv:2403.09756 [gr-qc]} \BibitemShut
  {NoStop}%
\bibitem [{\citenamefont {Liu}\ \emph {et~al.}(2024{\natexlab{a}})\citenamefont
  {Liu}, \citenamefont {Wu}, \citenamefont {Fang}, \citenamefont {Jing},\ and\
  \citenamefont {Wang}}]{Liu:2024lbi}%
  \BibitemOpen
  \bibfield  {author} {\bibinfo {author} {\bibfnamefont {W.}~\bibnamefont
  {Liu}}, \bibinfo {author} {\bibfnamefont {D.}~\bibnamefont {Wu}}, \bibinfo
  {author} {\bibfnamefont {X.}~\bibnamefont {Fang}}, \bibinfo {author}
  {\bibfnamefont {J.}~\bibnamefont {Jing}}, \ and\ \bibinfo {author}
  {\bibfnamefont {J.}~\bibnamefont {Wang}},\ }\href {\doibase
  10.1088/1475-7516/2024/08/035} {\bibfield  {journal} {\bibinfo  {journal}
  {JCAP}\ }\textbf {\bibinfo {volume} {08}},\ \bibinfo {pages} {035} (\bibinfo
  {year} {2024}{\natexlab{a}})},\ \Eprint {http://arxiv.org/abs/2406.00579}
  {arXiv:2406.00579 [gr-qc]} \BibitemShut {NoStop}%
\bibitem [{\citenamefont {Liu}\ \emph {et~al.}(2024{\natexlab{b}})\citenamefont
  {Liu}, \citenamefont {Wu},\ and\ \citenamefont {Wang}}]{Liu:2024lve}%
  \BibitemOpen
  \bibfield  {author} {\bibinfo {author} {\bibfnamefont {W.}~\bibnamefont
  {Liu}}, \bibinfo {author} {\bibfnamefont {D.}~\bibnamefont {Wu}}, \ and\
  \bibinfo {author} {\bibfnamefont {J.}~\bibnamefont {Wang}},\ }\href@noop {}
  {\  (\bibinfo {year} {2024}{\natexlab{b}})},\ \Eprint
  {http://arxiv.org/abs/2407.07416} {arXiv:2407.07416 [gr-qc]} \BibitemShut
  {NoStop}%
\bibitem [{\citenamefont {Khodadi}\ \emph {et~al.}(2024)\citenamefont
  {Khodadi}, \citenamefont {Vagnozzi},\ and\ \citenamefont
  {Firouzjaee}}]{Khodadi:2024ubi}%
  \BibitemOpen
  \bibfield  {author} {\bibinfo {author} {\bibfnamefont {M.}~\bibnamefont
  {Khodadi}}, \bibinfo {author} {\bibfnamefont {S.}~\bibnamefont {Vagnozzi}}, \
  and\ \bibinfo {author} {\bibfnamefont {J.~T.}\ \bibnamefont {Firouzjaee}},\
  }\href {\doibase 10.1038/s41598-024-78264-y} {\bibfield  {journal} {\bibinfo
  {journal} {Sci. Rep.}\ }\textbf {\bibinfo {volume} {14}},\ \bibinfo {pages}
  {26932} (\bibinfo {year} {2024})},\ \Eprint {http://arxiv.org/abs/2408.03241}
  {arXiv:2408.03241 [gr-qc]} \BibitemShut {NoStop}%
\bibitem [{\citenamefont {Liu}\ \emph {et~al.}(2024{\natexlab{c}})\citenamefont
  {Liu}, \citenamefont {Wu},\ and\ \citenamefont {Wang}}]{Liu:2024soc}%
  \BibitemOpen
  \bibfield  {author} {\bibinfo {author} {\bibfnamefont {W.}~\bibnamefont
  {Liu}}, \bibinfo {author} {\bibfnamefont {D.}~\bibnamefont {Wu}}, \ and\
  \bibinfo {author} {\bibfnamefont {J.}~\bibnamefont {Wang}},\ }\href {\doibase
  10.1016/j.physletb.2024.139052} {\bibfield  {journal} {\bibinfo  {journal}
  {Phys. Lett. B}\ }\textbf {\bibinfo {volume} {858}},\ \bibinfo {pages}
  {139052} (\bibinfo {year} {2024}{\natexlab{c}})},\ \Eprint
  {http://arxiv.org/abs/2408.05569} {arXiv:2408.05569 [gr-qc]} \BibitemShut
  {NoStop}%
\bibitem [{\citenamefont {Nojiri}\ and\ \citenamefont
  {Odintsov}(2024)}]{Nojiri:2024txy}%
  \BibitemOpen
  \bibfield  {author} {\bibinfo {author} {\bibfnamefont {S.}~\bibnamefont
  {Nojiri}}\ and\ \bibinfo {author} {\bibfnamefont {S.~D.}\ \bibnamefont
  {Odintsov}},\ }\href {\doibase 10.1016/j.dark.2024.101669} {\bibfield
  {journal} {\bibinfo  {journal} {Phys. Dark Univ.}\ }\textbf {\bibinfo
  {volume} {46}},\ \bibinfo {pages} {101669} (\bibinfo {year} {2024})},\
  \Eprint {http://arxiv.org/abs/2408.05668} {arXiv:2408.05668 [gr-qc]}
  \BibitemShut {NoStop}%
\bibitem [{\citenamefont {Arbey}\ and\ \citenamefont
  {Mahmoudi}(2021)}]{Arbey:2021gdg}%
  \BibitemOpen
  \bibfield  {author} {\bibinfo {author} {\bibfnamefont {A.}~\bibnamefont
  {Arbey}}\ and\ \bibinfo {author} {\bibfnamefont {F.}~\bibnamefont
  {Mahmoudi}},\ }\href {\doibase 10.1016/j.ppnp.2021.103865} {\bibfield
  {journal} {\bibinfo  {journal} {Prog. Part. Nucl. Phys.}\ }\textbf {\bibinfo
  {volume} {119}},\ \bibinfo {pages} {103865} (\bibinfo {year} {2021})},\
  \Eprint {http://arxiv.org/abs/2104.11488} {arXiv:2104.11488 [hep-ph]}
  \BibitemShut {NoStop}%
\bibitem [{\citenamefont {Cirelli}\ \emph {et~al.}(2024)\citenamefont
  {Cirelli}, \citenamefont {Strumia},\ and\ \citenamefont
  {Zupan}}]{Cirelli:2024ssz}%
  \BibitemOpen
  \bibfield  {author} {\bibinfo {author} {\bibfnamefont {M.}~\bibnamefont
  {Cirelli}}, \bibinfo {author} {\bibfnamefont {A.}~\bibnamefont {Strumia}}, \
  and\ \bibinfo {author} {\bibfnamefont {J.}~\bibnamefont {Zupan}},\
  }\href@noop {} {\  (\bibinfo {year} {2024})},\ \Eprint
  {http://arxiv.org/abs/2406.01705} {arXiv:2406.01705 [hep-ph]} \BibitemShut
  {NoStop}%
\bibitem [{\citenamefont {Chapline}(1975)}]{Chapline:1975ojl}%
  \BibitemOpen
  \bibfield  {author} {\bibinfo {author} {\bibfnamefont {G.~F.}\ \bibnamefont
  {Chapline}},\ }\href {\doibase 10.1038/253251a0} {\bibfield  {journal}
  {\bibinfo  {journal} {Nature}\ }\textbf {\bibinfo {volume} {253}},\ \bibinfo
  {pages} {251} (\bibinfo {year} {1975})}\BibitemShut {NoStop}%
\bibitem [{\citenamefont {Meszaros}(1975)}]{Meszaros:1975ef}%
  \BibitemOpen
  \bibfield  {author} {\bibinfo {author} {\bibfnamefont {P.}~\bibnamefont
  {Meszaros}},\ }\href@noop {} {\bibfield  {journal} {\bibinfo  {journal}
  {Astron. Astrophys.}\ }\textbf {\bibinfo {volume} {38}},\ \bibinfo {pages}
  {5} (\bibinfo {year} {1975})}\BibitemShut {NoStop}%
\bibitem [{\citenamefont {Khlopov}\ and\ \citenamefont
  {Polnarev}(1980)}]{Khlopov:1980mg}%
  \BibitemOpen
  \bibfield  {author} {\bibinfo {author} {\bibfnamefont {M.~Y.}\ \bibnamefont
  {Khlopov}}\ and\ \bibinfo {author} {\bibfnamefont {A.~G.}\ \bibnamefont
  {Polnarev}},\ }\href {\doibase 10.1016/0370-2693(80)90624-3} {\bibfield
  {journal} {\bibinfo  {journal} {Phys. Lett. B}\ }\textbf {\bibinfo {volume}
  {97}},\ \bibinfo {pages} {383} (\bibinfo {year} {1980})}\BibitemShut
  {NoStop}%
\bibitem [{\citenamefont {Khlopov}\ \emph {et~al.}(1985)\citenamefont
  {Khlopov}, \citenamefont {Malomed}, \citenamefont {Zeldovich},\ and\
  \citenamefont {Zeldovich}}]{Khlopov:1985fch}%
  \BibitemOpen
  \bibfield  {author} {\bibinfo {author} {\bibfnamefont {M.~Y.}\ \bibnamefont
  {Khlopov}}, \bibinfo {author} {\bibfnamefont {B.~A.}\ \bibnamefont
  {Malomed}}, \bibinfo {author} {\bibfnamefont {I.~B.}\ \bibnamefont
  {Zeldovich}}, \ and\ \bibinfo {author} {\bibfnamefont {Y.~B.}\ \bibnamefont
  {Zeldovich}},\ }\href {\doibase 10.1093/mnras/215.4.575} {\bibfield
  {journal} {\bibinfo  {journal} {Mon. Not. Roy. Astron. Soc.}\ }\textbf
  {\bibinfo {volume} {215}},\ \bibinfo {pages} {575} (\bibinfo {year}
  {1985})}\BibitemShut {NoStop}%
\bibitem [{\citenamefont {Ivanov}\ \emph {et~al.}(1994)\citenamefont {Ivanov},
  \citenamefont {Naselsky},\ and\ \citenamefont {Novikov}}]{Ivanov:1994pa}%
  \BibitemOpen
  \bibfield  {author} {\bibinfo {author} {\bibfnamefont {P.}~\bibnamefont
  {Ivanov}}, \bibinfo {author} {\bibfnamefont {P.}~\bibnamefont {Naselsky}}, \
  and\ \bibinfo {author} {\bibfnamefont {I.}~\bibnamefont {Novikov}},\ }\href
  {\doibase 10.1103/PhysRevD.50.7173} {\bibfield  {journal} {\bibinfo
  {journal} {Phys. Rev. D}\ }\textbf {\bibinfo {volume} {50}},\ \bibinfo
  {pages} {7173} (\bibinfo {year} {1994})}\BibitemShut {NoStop}%
\bibitem [{\citenamefont {Choudhury}\ and\ \citenamefont
  {Mazumdar}(2014)}]{Choudhury:2013woa}%
  \BibitemOpen
  \bibfield  {author} {\bibinfo {author} {\bibfnamefont {S.}~\bibnamefont
  {Choudhury}}\ and\ \bibinfo {author} {\bibfnamefont {A.}~\bibnamefont
  {Mazumdar}},\ }\href {\doibase 10.1016/j.physletb.2014.04.050} {\bibfield
  {journal} {\bibinfo  {journal} {Phys. Lett. B}\ }\textbf {\bibinfo {volume}
  {733}},\ \bibinfo {pages} {270} (\bibinfo {year} {2014})},\ \Eprint
  {http://arxiv.org/abs/1307.5119} {arXiv:1307.5119 [astro-ph.CO]} \BibitemShut
  {NoStop}%
\bibitem [{\citenamefont {Belotsky}\ \emph {et~al.}(2014)\citenamefont
  {Belotsky}, \citenamefont {Dmitriev}, \citenamefont {Esipova}, \citenamefont
  {Gani}, \citenamefont {Grobov}, \citenamefont {Khlopov}, \citenamefont
  {Kirillov}, \citenamefont {Rubin},\ and\ \citenamefont
  {Svadkovsky}}]{Belotsky:2014kca}%
  \BibitemOpen
  \bibfield  {author} {\bibinfo {author} {\bibfnamefont {K.~M.}\ \bibnamefont
  {Belotsky}}, \bibinfo {author} {\bibfnamefont {A.~D.}\ \bibnamefont
  {Dmitriev}}, \bibinfo {author} {\bibfnamefont {E.~A.}\ \bibnamefont
  {Esipova}}, \bibinfo {author} {\bibfnamefont {V.~A.}\ \bibnamefont {Gani}},
  \bibinfo {author} {\bibfnamefont {A.~V.}\ \bibnamefont {Grobov}}, \bibinfo
  {author} {\bibfnamefont {M.~Y.}\ \bibnamefont {Khlopov}}, \bibinfo {author}
  {\bibfnamefont {A.~A.}\ \bibnamefont {Kirillov}}, \bibinfo {author}
  {\bibfnamefont {S.~G.}\ \bibnamefont {Rubin}}, \ and\ \bibinfo {author}
  {\bibfnamefont {I.~V.}\ \bibnamefont {Svadkovsky}},\ }\href {\doibase
  10.1142/S0217732314400057} {\bibfield  {journal} {\bibinfo  {journal} {Mod.
  Phys. Lett. A}\ }\textbf {\bibinfo {volume} {29}},\ \bibinfo {pages}
  {1440005} (\bibinfo {year} {2014})},\ \Eprint
  {http://arxiv.org/abs/1410.0203} {arXiv:1410.0203 [astro-ph.CO]} \BibitemShut
  {NoStop}%
\bibitem [{\citenamefont {Bird}\ \emph {et~al.}(2016)\citenamefont {Bird},
  \citenamefont {Cholis}, \citenamefont {Mu\~noz}, \citenamefont
  {Ali-Ha\"\i{}moud}, \citenamefont {Kamionkowski}, \citenamefont {Kovetz},
  \citenamefont {Raccanelli},\ and\ \citenamefont {Riess}}]{Bird:2016dcv}%
  \BibitemOpen
  \bibfield  {author} {\bibinfo {author} {\bibfnamefont {S.}~\bibnamefont
  {Bird}}, \bibinfo {author} {\bibfnamefont {I.}~\bibnamefont {Cholis}},
  \bibinfo {author} {\bibfnamefont {J.~B.}\ \bibnamefont {Mu\~noz}}, \bibinfo
  {author} {\bibfnamefont {Y.}~\bibnamefont {Ali-Ha\"\i{}moud}}, \bibinfo
  {author} {\bibfnamefont {M.}~\bibnamefont {Kamionkowski}}, \bibinfo {author}
  {\bibfnamefont {E.~D.}\ \bibnamefont {Kovetz}}, \bibinfo {author}
  {\bibfnamefont {A.}~\bibnamefont {Raccanelli}}, \ and\ \bibinfo {author}
  {\bibfnamefont {A.~G.}\ \bibnamefont {Riess}},\ }\href {\doibase
  10.1103/PhysRevLett.116.201301} {\bibfield  {journal} {\bibinfo  {journal}
  {Phys. Rev. Lett.}\ }\textbf {\bibinfo {volume} {116}},\ \bibinfo {pages}
  {201301} (\bibinfo {year} {2016})},\ \Eprint
  {http://arxiv.org/abs/1603.00464} {arXiv:1603.00464 [astro-ph.CO]}
  \BibitemShut {NoStop}%
\bibitem [{\citenamefont {Clesse}\ and\ \citenamefont
  {Garc\'\i{}a-Bellido}(2017)}]{Clesse:2016vqa}%
  \BibitemOpen
  \bibfield  {author} {\bibinfo {author} {\bibfnamefont {S.}~\bibnamefont
  {Clesse}}\ and\ \bibinfo {author} {\bibfnamefont {J.}~\bibnamefont
  {Garc\'\i{}a-Bellido}},\ }\href {\doibase 10.1016/j.dark.2016.10.002}
  {\bibfield  {journal} {\bibinfo  {journal} {Phys. Dark Univ.}\ }\textbf
  {\bibinfo {volume} {15}},\ \bibinfo {pages} {142} (\bibinfo {year} {2017})},\
  \Eprint {http://arxiv.org/abs/1603.05234} {arXiv:1603.05234 [astro-ph.CO]}
  \BibitemShut {NoStop}%
\bibitem [{\citenamefont {Poulin}\ \emph {et~al.}(2017)\citenamefont {Poulin},
  \citenamefont {Serpico}, \citenamefont {Calore}, \citenamefont {Clesse},\
  and\ \citenamefont {Kohri}}]{Poulin:2017bwe}%
  \BibitemOpen
  \bibfield  {author} {\bibinfo {author} {\bibfnamefont {V.}~\bibnamefont
  {Poulin}}, \bibinfo {author} {\bibfnamefont {P.~D.}\ \bibnamefont {Serpico}},
  \bibinfo {author} {\bibfnamefont {F.}~\bibnamefont {Calore}}, \bibinfo
  {author} {\bibfnamefont {S.}~\bibnamefont {Clesse}}, \ and\ \bibinfo {author}
  {\bibfnamefont {K.}~\bibnamefont {Kohri}},\ }\href {\doibase
  10.1103/PhysRevD.96.083524} {\bibfield  {journal} {\bibinfo  {journal} {Phys.
  Rev. D}\ }\textbf {\bibinfo {volume} {96}},\ \bibinfo {pages} {083524}
  (\bibinfo {year} {2017})},\ \Eprint {http://arxiv.org/abs/1707.04206}
  {arXiv:1707.04206 [astro-ph.CO]} \BibitemShut {NoStop}%
\bibitem [{\citenamefont {Raccanelli}\ \emph {et~al.}(2018)\citenamefont
  {Raccanelli}, \citenamefont {Vidotto},\ and\ \citenamefont
  {Verde}}]{Raccanelli:2017xee}%
  \BibitemOpen
  \bibfield  {author} {\bibinfo {author} {\bibfnamefont {A.}~\bibnamefont
  {Raccanelli}}, \bibinfo {author} {\bibfnamefont {F.}~\bibnamefont {Vidotto}},
  \ and\ \bibinfo {author} {\bibfnamefont {L.}~\bibnamefont {Verde}},\ }\href
  {\doibase 10.1088/1475-7516/2018/08/003} {\bibfield  {journal} {\bibinfo
  {journal} {JCAP}\ }\textbf {\bibinfo {volume} {08}},\ \bibinfo {pages} {003}
  (\bibinfo {year} {2018})},\ \Eprint {http://arxiv.org/abs/1708.02588}
  {arXiv:1708.02588 [astro-ph.CO]} \BibitemShut {NoStop}%
\bibitem [{\citenamefont {Bernal}\ \emph {et~al.}(2017)\citenamefont {Bernal},
  \citenamefont {Bellomo}, \citenamefont {Raccanelli},\ and\ \citenamefont
  {Verde}}]{LuisBernal:2017fmf}%
  \BibitemOpen
  \bibfield  {author} {\bibinfo {author} {\bibfnamefont {J.~L.}\ \bibnamefont
  {Bernal}}, \bibinfo {author} {\bibfnamefont {N.}~\bibnamefont {Bellomo}},
  \bibinfo {author} {\bibfnamefont {A.}~\bibnamefont {Raccanelli}}, \ and\
  \bibinfo {author} {\bibfnamefont {L.}~\bibnamefont {Verde}},\ }\href
  {\doibase 10.1088/1475-7516/2017/10/052} {\bibfield  {journal} {\bibinfo
  {journal} {JCAP}\ }\textbf {\bibinfo {volume} {10}},\ \bibinfo {pages} {052}
  (\bibinfo {year} {2017})},\ \Eprint {http://arxiv.org/abs/1709.07465}
  {arXiv:1709.07465 [astro-ph.CO]} \BibitemShut {NoStop}%
\bibitem [{\citenamefont {Clesse}\ and\ \citenamefont
  {Garc\'\i{}a-Bellido}(2018)}]{Clesse:2017bsw}%
  \BibitemOpen
  \bibfield  {author} {\bibinfo {author} {\bibfnamefont {S.}~\bibnamefont
  {Clesse}}\ and\ \bibinfo {author} {\bibfnamefont {J.}~\bibnamefont
  {Garc\'\i{}a-Bellido}},\ }\href {\doibase 10.1016/j.dark.2018.08.004}
  {\bibfield  {journal} {\bibinfo  {journal} {Phys. Dark Univ.}\ }\textbf
  {\bibinfo {volume} {22}},\ \bibinfo {pages} {137} (\bibinfo {year} {2018})},\
  \Eprint {http://arxiv.org/abs/1711.10458} {arXiv:1711.10458 [astro-ph.CO]}
  \BibitemShut {NoStop}%
\bibitem [{\citenamefont {Kohri}\ and\ \citenamefont
  {Terada}(2018)}]{Kohri:2018qtx}%
  \BibitemOpen
  \bibfield  {author} {\bibinfo {author} {\bibfnamefont {K.}~\bibnamefont
  {Kohri}}\ and\ \bibinfo {author} {\bibfnamefont {T.}~\bibnamefont {Terada}},\
  }\href {\doibase 10.1088/1361-6382/aaea18} {\bibfield  {journal} {\bibinfo
  {journal} {Class. Quant. Grav.}\ }\textbf {\bibinfo {volume} {35}},\ \bibinfo
  {pages} {235017} (\bibinfo {year} {2018})},\ \Eprint
  {http://arxiv.org/abs/1802.06785} {arXiv:1802.06785 [astro-ph.CO]}
  \BibitemShut {NoStop}%
\bibitem [{\citenamefont {Liu}\ \emph {et~al.}(2019{\natexlab{a}})\citenamefont
  {Liu}, \citenamefont {Guo},\ and\ \citenamefont {Cai}}]{Liu:2018ess}%
  \BibitemOpen
  \bibfield  {author} {\bibinfo {author} {\bibfnamefont {L.}~\bibnamefont
  {Liu}}, \bibinfo {author} {\bibfnamefont {Z.-K.}\ \bibnamefont {Guo}}, \ and\
  \bibinfo {author} {\bibfnamefont {R.-G.}\ \bibnamefont {Cai}},\ }\href
  {\doibase 10.1103/PhysRevD.99.063523} {\bibfield  {journal} {\bibinfo
  {journal} {Phys. Rev. D}\ }\textbf {\bibinfo {volume} {99}},\ \bibinfo
  {pages} {063523} (\bibinfo {year} {2019}{\natexlab{a}})},\ \Eprint
  {http://arxiv.org/abs/1812.05376} {arXiv:1812.05376 [astro-ph.CO]}
  \BibitemShut {NoStop}%
\bibitem [{\citenamefont {Liu}\ \emph {et~al.}(2019{\natexlab{b}})\citenamefont
  {Liu}, \citenamefont {Guo},\ and\ \citenamefont {Cai}}]{Liu:2019rnx}%
  \BibitemOpen
  \bibfield  {author} {\bibinfo {author} {\bibfnamefont {L.}~\bibnamefont
  {Liu}}, \bibinfo {author} {\bibfnamefont {Z.-K.}\ \bibnamefont {Guo}}, \ and\
  \bibinfo {author} {\bibfnamefont {R.-G.}\ \bibnamefont {Cai}},\ }\href
  {\doibase 10.1140/epjc/s10052-019-7227-0} {\bibfield  {journal} {\bibinfo
  {journal} {Eur. Phys. J. C}\ }\textbf {\bibinfo {volume} {79}},\ \bibinfo
  {pages} {717} (\bibinfo {year} {2019}{\natexlab{b}})},\ \Eprint
  {http://arxiv.org/abs/1901.07672} {arXiv:1901.07672 [astro-ph.CO]}
  \BibitemShut {NoStop}%
\bibitem [{\citenamefont {Murgia}\ \emph {et~al.}(2019)\citenamefont {Murgia},
  \citenamefont {Scelfo}, \citenamefont {Viel},\ and\ \citenamefont
  {Raccanelli}}]{Murgia:2019duy}%
  \BibitemOpen
  \bibfield  {author} {\bibinfo {author} {\bibfnamefont {R.}~\bibnamefont
  {Murgia}}, \bibinfo {author} {\bibfnamefont {G.}~\bibnamefont {Scelfo}},
  \bibinfo {author} {\bibfnamefont {M.}~\bibnamefont {Viel}}, \ and\ \bibinfo
  {author} {\bibfnamefont {A.}~\bibnamefont {Raccanelli}},\ }\href {\doibase
  10.1103/PhysRevLett.123.071102} {\bibfield  {journal} {\bibinfo  {journal}
  {Phys. Rev. Lett.}\ }\textbf {\bibinfo {volume} {123}},\ \bibinfo {pages}
  {071102} (\bibinfo {year} {2019})},\ \Eprint
  {http://arxiv.org/abs/1903.10509} {arXiv:1903.10509 [astro-ph.CO]}
  \BibitemShut {NoStop}%
\bibitem [{\citenamefont {Carr}\ \emph
  {et~al.}(2021{\natexlab{a}})\citenamefont {Carr}, \citenamefont {Clesse},
  \citenamefont {Garc\'\i{}a-Bellido},\ and\ \citenamefont
  {K\"uhnel}}]{Carr:2019kxo}%
  \BibitemOpen
  \bibfield  {author} {\bibinfo {author} {\bibfnamefont {B.}~\bibnamefont
  {Carr}}, \bibinfo {author} {\bibfnamefont {S.}~\bibnamefont {Clesse}},
  \bibinfo {author} {\bibfnamefont {J.}~\bibnamefont {Garc\'\i{}a-Bellido}}, \
  and\ \bibinfo {author} {\bibfnamefont {F.}~\bibnamefont {K\"uhnel}},\ }\href
  {\doibase 10.1016/j.dark.2020.100755} {\bibfield  {journal} {\bibinfo
  {journal} {Phys. Dark Univ.}\ }\textbf {\bibinfo {volume} {31}},\ \bibinfo
  {pages} {100755} (\bibinfo {year} {2021}{\natexlab{a}})},\ \Eprint
  {http://arxiv.org/abs/1906.08217} {arXiv:1906.08217 [astro-ph.CO]}
  \BibitemShut {NoStop}%
\bibitem [{\citenamefont {Liu}\ \emph {et~al.}(2020)\citenamefont {Liu},
  \citenamefont {Guo}, \citenamefont {Cai},\ and\ \citenamefont
  {Kim}}]{Liu:2020cds}%
  \BibitemOpen
  \bibfield  {author} {\bibinfo {author} {\bibfnamefont {L.}~\bibnamefont
  {Liu}}, \bibinfo {author} {\bibfnamefont {Z.-K.}\ \bibnamefont {Guo}},
  \bibinfo {author} {\bibfnamefont {R.-G.}\ \bibnamefont {Cai}}, \ and\
  \bibinfo {author} {\bibfnamefont {S.~P.}\ \bibnamefont {Kim}},\ }\href
  {\doibase 10.1103/PhysRevD.102.043508} {\bibfield  {journal} {\bibinfo
  {journal} {Phys. Rev. D}\ }\textbf {\bibinfo {volume} {102}},\ \bibinfo
  {pages} {043508} (\bibinfo {year} {2020})},\ \Eprint
  {http://arxiv.org/abs/2001.02984} {arXiv:2001.02984 [astro-ph.CO]}
  \BibitemShut {NoStop}%
\bibitem [{\citenamefont {Hertzberg}\ \emph {et~al.}(2020)\citenamefont
  {Hertzberg}, \citenamefont {Schiappacasse},\ and\ \citenamefont
  {Yanagida}}]{Hertzberg:2020hsz}%
  \BibitemOpen
  \bibfield  {author} {\bibinfo {author} {\bibfnamefont {M.~P.}\ \bibnamefont
  {Hertzberg}}, \bibinfo {author} {\bibfnamefont {E.~D.}\ \bibnamefont
  {Schiappacasse}}, \ and\ \bibinfo {author} {\bibfnamefont {T.~T.}\
  \bibnamefont {Yanagida}},\ }\href {\doibase 10.1103/PhysRevD.102.023013}
  {\bibfield  {journal} {\bibinfo  {journal} {Phys. Rev. D}\ }\textbf {\bibinfo
  {volume} {102}},\ \bibinfo {pages} {023013} (\bibinfo {year} {2020})},\
  \Eprint {http://arxiv.org/abs/2001.07476} {arXiv:2001.07476 [astro-ph.CO]}
  \BibitemShut {NoStop}%
\bibitem [{\citenamefont {Serpico}\ \emph {et~al.}(2020)\citenamefont
  {Serpico}, \citenamefont {Poulin}, \citenamefont {Inman},\ and\ \citenamefont
  {Kohri}}]{Serpico:2020ehh}%
  \BibitemOpen
  \bibfield  {author} {\bibinfo {author} {\bibfnamefont {P.~D.}\ \bibnamefont
  {Serpico}}, \bibinfo {author} {\bibfnamefont {V.}~\bibnamefont {Poulin}},
  \bibinfo {author} {\bibfnamefont {D.}~\bibnamefont {Inman}}, \ and\ \bibinfo
  {author} {\bibfnamefont {K.}~\bibnamefont {Kohri}},\ }\href {\doibase
  10.1103/PhysRevResearch.2.023204} {\bibfield  {journal} {\bibinfo  {journal}
  {Phys. Rev. Res.}\ }\textbf {\bibinfo {volume} {2}},\ \bibinfo {pages}
  {023204} (\bibinfo {year} {2020})},\ \Eprint
  {http://arxiv.org/abs/2002.10771} {arXiv:2002.10771 [astro-ph.CO]}
  \BibitemShut {NoStop}%
\bibitem [{\citenamefont {De~Luca}\ \emph
  {et~al.}(2020{\natexlab{a}})\citenamefont {De~Luca}, \citenamefont
  {Franciolini}, \citenamefont {Pani},\ and\ \citenamefont
  {Riotto}}]{DeLuca:2020bjf}%
  \BibitemOpen
  \bibfield  {author} {\bibinfo {author} {\bibfnamefont {V.}~\bibnamefont
  {De~Luca}}, \bibinfo {author} {\bibfnamefont {G.}~\bibnamefont
  {Franciolini}}, \bibinfo {author} {\bibfnamefont {P.}~\bibnamefont {Pani}}, \
  and\ \bibinfo {author} {\bibfnamefont {A.}~\bibnamefont {Riotto}},\ }\href
  {\doibase 10.1088/1475-7516/2020/04/052} {\bibfield  {journal} {\bibinfo
  {journal} {JCAP}\ }\textbf {\bibinfo {volume} {04}},\ \bibinfo {pages} {052}
  (\bibinfo {year} {2020}{\natexlab{a}})},\ \Eprint
  {http://arxiv.org/abs/2003.02778} {arXiv:2003.02778 [astro-ph.CO]}
  \BibitemShut {NoStop}%
\bibitem [{\citenamefont {De~Luca}\ \emph
  {et~al.}(2020{\natexlab{b}})\citenamefont {De~Luca}, \citenamefont
  {Franciolini}, \citenamefont {Pani},\ and\ \citenamefont
  {Riotto}}]{DeLuca:2020fpg}%
  \BibitemOpen
  \bibfield  {author} {\bibinfo {author} {\bibfnamefont {V.}~\bibnamefont
  {De~Luca}}, \bibinfo {author} {\bibfnamefont {G.}~\bibnamefont
  {Franciolini}}, \bibinfo {author} {\bibfnamefont {P.}~\bibnamefont {Pani}}, \
  and\ \bibinfo {author} {\bibfnamefont {A.}~\bibnamefont {Riotto}},\ }\href
  {\doibase 10.1103/PhysRevD.102.043505} {\bibfield  {journal} {\bibinfo
  {journal} {Phys. Rev. D}\ }\textbf {\bibinfo {volume} {102}},\ \bibinfo
  {pages} {043505} (\bibinfo {year} {2020}{\natexlab{b}})},\ \Eprint
  {http://arxiv.org/abs/2003.12589} {arXiv:2003.12589 [astro-ph.CO]}
  \BibitemShut {NoStop}%
\bibitem [{\citenamefont {De~Luca}\ \emph
  {et~al.}(2020{\natexlab{c}})\citenamefont {De~Luca}, \citenamefont
  {Franciolini}, \citenamefont {Pani},\ and\ \citenamefont
  {Riotto}}]{DeLuca:2020qqa}%
  \BibitemOpen
  \bibfield  {author} {\bibinfo {author} {\bibfnamefont {V.}~\bibnamefont
  {De~Luca}}, \bibinfo {author} {\bibfnamefont {G.}~\bibnamefont
  {Franciolini}}, \bibinfo {author} {\bibfnamefont {P.}~\bibnamefont {Pani}}, \
  and\ \bibinfo {author} {\bibfnamefont {A.}~\bibnamefont {Riotto}},\ }\href
  {\doibase 10.1088/1475-7516/2020/06/044} {\bibfield  {journal} {\bibinfo
  {journal} {JCAP}\ }\textbf {\bibinfo {volume} {06}},\ \bibinfo {pages} {044}
  (\bibinfo {year} {2020}{\natexlab{c}})},\ \Eprint
  {http://arxiv.org/abs/2005.05641} {arXiv:2005.05641 [astro-ph.CO]}
  \BibitemShut {NoStop}%
\bibitem [{\citenamefont {Carr}\ \emph
  {et~al.}(2021{\natexlab{b}})\citenamefont {Carr}, \citenamefont {Kuhnel},\
  and\ \citenamefont {Visinelli}}]{Carr:2020erq}%
  \BibitemOpen
  \bibfield  {author} {\bibinfo {author} {\bibfnamefont {B.}~\bibnamefont
  {Carr}}, \bibinfo {author} {\bibfnamefont {F.}~\bibnamefont {Kuhnel}}, \ and\
  \bibinfo {author} {\bibfnamefont {L.}~\bibnamefont {Visinelli}},\ }\href
  {\doibase 10.1093/mnras/staa3651} {\bibfield  {journal} {\bibinfo  {journal}
  {Mon. Not. Roy. Astron. Soc.}\ }\textbf {\bibinfo {volume} {501}},\ \bibinfo
  {pages} {2029} (\bibinfo {year} {2021}{\natexlab{b}})},\ \Eprint
  {http://arxiv.org/abs/2008.08077} {arXiv:2008.08077 [astro-ph.CO]}
  \BibitemShut {NoStop}%
\bibitem [{\citenamefont {Bhagwat}\ \emph {et~al.}(2021)\citenamefont
  {Bhagwat}, \citenamefont {De~Luca}, \citenamefont {Franciolini},
  \citenamefont {Pani},\ and\ \citenamefont {Riotto}}]{Bhagwat:2020bzh}%
  \BibitemOpen
  \bibfield  {author} {\bibinfo {author} {\bibfnamefont {S.}~\bibnamefont
  {Bhagwat}}, \bibinfo {author} {\bibfnamefont {V.}~\bibnamefont {De~Luca}},
  \bibinfo {author} {\bibfnamefont {G.}~\bibnamefont {Franciolini}}, \bibinfo
  {author} {\bibfnamefont {P.}~\bibnamefont {Pani}}, \ and\ \bibinfo {author}
  {\bibfnamefont {A.}~\bibnamefont {Riotto}},\ }\href {\doibase
  10.1088/1475-7516/2021/01/037} {\bibfield  {journal} {\bibinfo  {journal}
  {JCAP}\ }\textbf {\bibinfo {volume} {01}},\ \bibinfo {pages} {037} (\bibinfo
  {year} {2021})},\ \Eprint {http://arxiv.org/abs/2008.12320} {arXiv:2008.12320
  [astro-ph.CO]} \BibitemShut {NoStop}%
\bibitem [{\citenamefont {De~Luca}\ \emph
  {et~al.}(2021{\natexlab{a}})\citenamefont {De~Luca}, \citenamefont
  {Desjacques}, \citenamefont {Franciolini}, \citenamefont {Pani},\ and\
  \citenamefont {Riotto}}]{DeLuca:2020sae}%
  \BibitemOpen
  \bibfield  {author} {\bibinfo {author} {\bibfnamefont {V.}~\bibnamefont
  {De~Luca}}, \bibinfo {author} {\bibfnamefont {V.}~\bibnamefont {Desjacques}},
  \bibinfo {author} {\bibfnamefont {G.}~\bibnamefont {Franciolini}}, \bibinfo
  {author} {\bibfnamefont {P.}~\bibnamefont {Pani}}, \ and\ \bibinfo {author}
  {\bibfnamefont {A.}~\bibnamefont {Riotto}},\ }\href {\doibase
  10.1103/PhysRevLett.126.051101} {\bibfield  {journal} {\bibinfo  {journal}
  {Phys. Rev. Lett.}\ }\textbf {\bibinfo {volume} {126}},\ \bibinfo {pages}
  {051101} (\bibinfo {year} {2021}{\natexlab{a}})},\ \Eprint
  {http://arxiv.org/abs/2009.01728} {arXiv:2009.01728 [astro-ph.CO]}
  \BibitemShut {NoStop}%
\bibitem [{\citenamefont {Wong}\ \emph {et~al.}(2021)\citenamefont {Wong},
  \citenamefont {Franciolini}, \citenamefont {De~Luca}, \citenamefont
  {Baibhav}, \citenamefont {Berti}, \citenamefont {Pani},\ and\ \citenamefont
  {Riotto}}]{Wong:2020yig}%
  \BibitemOpen
  \bibfield  {author} {\bibinfo {author} {\bibfnamefont {K.~W.~K.}\
  \bibnamefont {Wong}}, \bibinfo {author} {\bibfnamefont {G.}~\bibnamefont
  {Franciolini}}, \bibinfo {author} {\bibfnamefont {V.}~\bibnamefont
  {De~Luca}}, \bibinfo {author} {\bibfnamefont {V.}~\bibnamefont {Baibhav}},
  \bibinfo {author} {\bibfnamefont {E.}~\bibnamefont {Berti}}, \bibinfo
  {author} {\bibfnamefont {P.}~\bibnamefont {Pani}}, \ and\ \bibinfo {author}
  {\bibfnamefont {A.}~\bibnamefont {Riotto}},\ }\href {\doibase
  10.1103/PhysRevD.103.023026} {\bibfield  {journal} {\bibinfo  {journal}
  {Phys. Rev. D}\ }\textbf {\bibinfo {volume} {103}},\ \bibinfo {pages}
  {023026} (\bibinfo {year} {2021})},\ \Eprint
  {http://arxiv.org/abs/2011.01865} {arXiv:2011.01865 [gr-qc]} \BibitemShut
  {NoStop}%
\bibitem [{\citenamefont {Carr}\ \emph
  {et~al.}(2021{\natexlab{c}})\citenamefont {Carr}, \citenamefont {Kuhnel},\
  and\ \citenamefont {Visinelli}}]{Carr:2020mqm}%
  \BibitemOpen
  \bibfield  {author} {\bibinfo {author} {\bibfnamefont {B.}~\bibnamefont
  {Carr}}, \bibinfo {author} {\bibfnamefont {F.}~\bibnamefont {Kuhnel}}, \ and\
  \bibinfo {author} {\bibfnamefont {L.}~\bibnamefont {Visinelli}},\ }\href
  {\doibase 10.1093/mnras/stab1930} {\bibfield  {journal} {\bibinfo  {journal}
  {Mon. Not. Roy. Astron. Soc.}\ }\textbf {\bibinfo {volume} {506}},\ \bibinfo
  {pages} {3648} (\bibinfo {year} {2021}{\natexlab{c}})},\ \Eprint
  {http://arxiv.org/abs/2011.01930} {arXiv:2011.01930 [astro-ph.CO]}
  \BibitemShut {NoStop}%
\bibitem [{\citenamefont {Dom\`enech}\ \emph {et~al.}(2021)\citenamefont
  {Dom\`enech}, \citenamefont {Lin},\ and\ \citenamefont
  {Sasaki}}]{Domenech:2020ssp}%
  \BibitemOpen
  \bibfield  {author} {\bibinfo {author} {\bibfnamefont {G.}~\bibnamefont
  {Dom\`enech}}, \bibinfo {author} {\bibfnamefont {C.}~\bibnamefont {Lin}}, \
  and\ \bibinfo {author} {\bibfnamefont {M.}~\bibnamefont {Sasaki}},\ }\href
  {\doibase 10.1088/1475-7516/2021/11/E01} {\bibfield  {journal} {\bibinfo
  {journal} {JCAP}\ }\textbf {\bibinfo {volume} {04}},\ \bibinfo {pages} {062}
  (\bibinfo {year} {2021})},\ \bibinfo {note} {[Erratum: JCAP 11, E01
  (2021)]},\ \Eprint {http://arxiv.org/abs/2012.08151} {arXiv:2012.08151
  [gr-qc]} \BibitemShut {NoStop}%
\bibitem [{\citenamefont {De~Luca}\ \emph
  {et~al.}(2021{\natexlab{b}})\citenamefont {De~Luca}, \citenamefont
  {Franciolini}, \citenamefont {Pani},\ and\ \citenamefont
  {Riotto}}]{DeLuca:2021wjr}%
  \BibitemOpen
  \bibfield  {author} {\bibinfo {author} {\bibfnamefont {V.}~\bibnamefont
  {De~Luca}}, \bibinfo {author} {\bibfnamefont {G.}~\bibnamefont
  {Franciolini}}, \bibinfo {author} {\bibfnamefont {P.}~\bibnamefont {Pani}}, \
  and\ \bibinfo {author} {\bibfnamefont {A.}~\bibnamefont {Riotto}},\ }\href
  {\doibase 10.1088/1475-7516/2021/05/003} {\bibfield  {journal} {\bibinfo
  {journal} {JCAP}\ }\textbf {\bibinfo {volume} {05}},\ \bibinfo {pages} {003}
  (\bibinfo {year} {2021}{\natexlab{b}})},\ \Eprint
  {http://arxiv.org/abs/2102.03809} {arXiv:2102.03809 [astro-ph.CO]}
  \BibitemShut {NoStop}%
\bibitem [{\citenamefont {Arbey}\ \emph
  {et~al.}(2021{\natexlab{a}})\citenamefont {Arbey}, \citenamefont {Auffinger},
  \citenamefont {Sandick}, \citenamefont {Shams Es~Haghi},\ and\ \citenamefont
  {Sinha}}]{Arbey:2021ysg}%
  \BibitemOpen
  \bibfield  {author} {\bibinfo {author} {\bibfnamefont {A.}~\bibnamefont
  {Arbey}}, \bibinfo {author} {\bibfnamefont {J.}~\bibnamefont {Auffinger}},
  \bibinfo {author} {\bibfnamefont {P.}~\bibnamefont {Sandick}}, \bibinfo
  {author} {\bibfnamefont {B.}~\bibnamefont {Shams Es~Haghi}}, \ and\ \bibinfo
  {author} {\bibfnamefont {K.}~\bibnamefont {Sinha}},\ }\href {\doibase
  10.1103/PhysRevD.103.123549} {\bibfield  {journal} {\bibinfo  {journal}
  {Phys. Rev. D}\ }\textbf {\bibinfo {volume} {103}},\ \bibinfo {pages}
  {123549} (\bibinfo {year} {2021}{\natexlab{a}})},\ \Eprint
  {http://arxiv.org/abs/2104.04051} {arXiv:2104.04051 [astro-ph.CO]}
  \BibitemShut {NoStop}%
\bibitem [{\citenamefont {Franciolini}\ \emph
  {et~al.}(2022{\natexlab{a}})\citenamefont {Franciolini}, \citenamefont
  {Baibhav}, \citenamefont {De~Luca}, \citenamefont {Ng}, \citenamefont {Wong},
  \citenamefont {Berti}, \citenamefont {Pani}, \citenamefont {Riotto},\ and\
  \citenamefont {Vitale}}]{Franciolini:2021tla}%
  \BibitemOpen
  \bibfield  {author} {\bibinfo {author} {\bibfnamefont {G.}~\bibnamefont
  {Franciolini}}, \bibinfo {author} {\bibfnamefont {V.}~\bibnamefont
  {Baibhav}}, \bibinfo {author} {\bibfnamefont {V.}~\bibnamefont {De~Luca}},
  \bibinfo {author} {\bibfnamefont {K.~K.~Y.}\ \bibnamefont {Ng}}, \bibinfo
  {author} {\bibfnamefont {K.~W.~K.}\ \bibnamefont {Wong}}, \bibinfo {author}
  {\bibfnamefont {E.}~\bibnamefont {Berti}}, \bibinfo {author} {\bibfnamefont
  {P.}~\bibnamefont {Pani}}, \bibinfo {author} {\bibfnamefont {A.}~\bibnamefont
  {Riotto}}, \ and\ \bibinfo {author} {\bibfnamefont {S.}~\bibnamefont
  {Vitale}},\ }\href {\doibase 10.1103/PhysRevD.105.083526} {\bibfield
  {journal} {\bibinfo  {journal} {Phys. Rev. D}\ }\textbf {\bibinfo {volume}
  {105}},\ \bibinfo {pages} {083526} (\bibinfo {year} {2022}{\natexlab{a}})},\
  \Eprint {http://arxiv.org/abs/2105.03349} {arXiv:2105.03349 [gr-qc]}
  \BibitemShut {NoStop}%
\bibitem [{\citenamefont {De~Luca}\ \emph
  {et~al.}(2021{\natexlab{c}})\citenamefont {De~Luca}, \citenamefont
  {Franciolini}, \citenamefont {Pani},\ and\ \citenamefont
  {Riotto}}]{DeLuca:2021hde}%
  \BibitemOpen
  \bibfield  {author} {\bibinfo {author} {\bibfnamefont {V.}~\bibnamefont
  {De~Luca}}, \bibinfo {author} {\bibfnamefont {G.}~\bibnamefont
  {Franciolini}}, \bibinfo {author} {\bibfnamefont {P.}~\bibnamefont {Pani}}, \
  and\ \bibinfo {author} {\bibfnamefont {A.}~\bibnamefont {Riotto}},\ }\href
  {\doibase 10.1088/1475-7516/2021/11/039} {\bibfield  {journal} {\bibinfo
  {journal} {JCAP}\ }\textbf {\bibinfo {volume} {11}},\ \bibinfo {pages} {039}
  (\bibinfo {year} {2021}{\natexlab{c}})},\ \Eprint
  {http://arxiv.org/abs/2106.13769} {arXiv:2106.13769 [astro-ph.CO]}
  \BibitemShut {NoStop}%
\bibitem [{\citenamefont {Cheek}\ \emph
  {et~al.}(2022{\natexlab{a}})\citenamefont {Cheek}, \citenamefont {Heurtier},
  \citenamefont {Perez-Gonzalez},\ and\ \citenamefont
  {Turner}}]{Cheek:2021odj}%
  \BibitemOpen
  \bibfield  {author} {\bibinfo {author} {\bibfnamefont {A.}~\bibnamefont
  {Cheek}}, \bibinfo {author} {\bibfnamefont {L.}~\bibnamefont {Heurtier}},
  \bibinfo {author} {\bibfnamefont {Y.~F.}\ \bibnamefont {Perez-Gonzalez}}, \
  and\ \bibinfo {author} {\bibfnamefont {J.}~\bibnamefont {Turner}},\ }\href
  {\doibase 10.1103/PhysRevD.105.015022} {\bibfield  {journal} {\bibinfo
  {journal} {Phys. Rev. D}\ }\textbf {\bibinfo {volume} {105}},\ \bibinfo
  {pages} {015022} (\bibinfo {year} {2022}{\natexlab{a}})},\ \Eprint
  {http://arxiv.org/abs/2107.00013} {arXiv:2107.00013 [hep-ph]} \BibitemShut
  {NoStop}%
\bibitem [{\citenamefont {Cheek}\ \emph
  {et~al.}(2022{\natexlab{b}})\citenamefont {Cheek}, \citenamefont {Heurtier},
  \citenamefont {Perez-Gonzalez},\ and\ \citenamefont
  {Turner}}]{Cheek:2021cfe}%
  \BibitemOpen
  \bibfield  {author} {\bibinfo {author} {\bibfnamefont {A.}~\bibnamefont
  {Cheek}}, \bibinfo {author} {\bibfnamefont {L.}~\bibnamefont {Heurtier}},
  \bibinfo {author} {\bibfnamefont {Y.~F.}\ \bibnamefont {Perez-Gonzalez}}, \
  and\ \bibinfo {author} {\bibfnamefont {J.}~\bibnamefont {Turner}},\ }\href
  {\doibase 10.1103/PhysRevD.105.015023} {\bibfield  {journal} {\bibinfo
  {journal} {Phys. Rev. D}\ }\textbf {\bibinfo {volume} {105}},\ \bibinfo
  {pages} {015023} (\bibinfo {year} {2022}{\natexlab{b}})},\ \Eprint
  {http://arxiv.org/abs/2107.00016} {arXiv:2107.00016 [hep-ph]} \BibitemShut
  {NoStop}%
\bibitem [{\citenamefont {Heydari}\ and\ \citenamefont
  {Karami}(2022{\natexlab{a}})}]{Heydari:2021gea}%
  \BibitemOpen
  \bibfield  {author} {\bibinfo {author} {\bibfnamefont {S.}~\bibnamefont
  {Heydari}}\ and\ \bibinfo {author} {\bibfnamefont {K.}~\bibnamefont
  {Karami}},\ }\href {\doibase 10.1140/epjc/s10052-022-10036-2} {\bibfield
  {journal} {\bibinfo  {journal} {Eur. Phys. J. C}\ }\textbf {\bibinfo {volume}
  {82}},\ \bibinfo {pages} {83} (\bibinfo {year} {2022}{\natexlab{a}})},\
  \Eprint {http://arxiv.org/abs/2107.10550} {arXiv:2107.10550 [gr-qc]}
  \BibitemShut {NoStop}%
\bibitem [{\citenamefont {Dvali}\ \emph {et~al.}(2021)\citenamefont {Dvali},
  \citenamefont {K\"uhnel},\ and\ \citenamefont {Zantedeschi}}]{Dvali:2021byy}%
  \BibitemOpen
  \bibfield  {author} {\bibinfo {author} {\bibfnamefont {G.}~\bibnamefont
  {Dvali}}, \bibinfo {author} {\bibfnamefont {F.}~\bibnamefont {K\"uhnel}}, \
  and\ \bibinfo {author} {\bibfnamefont {M.}~\bibnamefont {Zantedeschi}},\
  }\href {\doibase 10.1103/PhysRevD.104.123507} {\bibfield  {journal} {\bibinfo
   {journal} {Phys. Rev. D}\ }\textbf {\bibinfo {volume} {104}},\ \bibinfo
  {pages} {123507} (\bibinfo {year} {2021})},\ \Eprint
  {http://arxiv.org/abs/2108.09471} {arXiv:2108.09471 [hep-ph]} \BibitemShut
  {NoStop}%
\bibitem [{\citenamefont {Heydari}\ and\ \citenamefont
  {Karami}(2022{\natexlab{b}})}]{Heydari:2021qsr}%
  \BibitemOpen
  \bibfield  {author} {\bibinfo {author} {\bibfnamefont {S.}~\bibnamefont
  {Heydari}}\ and\ \bibinfo {author} {\bibfnamefont {K.}~\bibnamefont
  {Karami}},\ }\href {\doibase 10.1088/1475-7516/2022/03/033} {\bibfield
  {journal} {\bibinfo  {journal} {JCAP}\ }\textbf {\bibinfo {volume} {03}},\
  \bibinfo {pages} {033} (\bibinfo {year} {2022}{\natexlab{b}})},\ \Eprint
  {http://arxiv.org/abs/2111.00494} {arXiv:2111.00494 [gr-qc]} \BibitemShut
  {NoStop}%
\bibitem [{\citenamefont {De~Luca}\ \emph {et~al.}(2022)\citenamefont
  {De~Luca}, \citenamefont {Franciolini}, \citenamefont {Kehagias},
  \citenamefont {Pani},\ and\ \citenamefont {Riotto}}]{DeLuca:2021pls}%
  \BibitemOpen
  \bibfield  {author} {\bibinfo {author} {\bibfnamefont {V.}~\bibnamefont
  {De~Luca}}, \bibinfo {author} {\bibfnamefont {G.}~\bibnamefont
  {Franciolini}}, \bibinfo {author} {\bibfnamefont {A.}~\bibnamefont
  {Kehagias}}, \bibinfo {author} {\bibfnamefont {P.}~\bibnamefont {Pani}}, \
  and\ \bibinfo {author} {\bibfnamefont {A.}~\bibnamefont {Riotto}},\ }\href
  {\doibase 10.1016/j.physletb.2022.137265} {\bibfield  {journal} {\bibinfo
  {journal} {Phys. Lett. B}\ }\textbf {\bibinfo {volume} {832}},\ \bibinfo
  {pages} {137265} (\bibinfo {year} {2022})},\ \Eprint
  {http://arxiv.org/abs/2112.02534} {arXiv:2112.02534 [astro-ph.CO]}
  \BibitemShut {NoStop}%
\bibitem [{\citenamefont {Liu}\ \emph {et~al.}(2023{\natexlab{a}})\citenamefont
  {Liu}, \citenamefont {Yang}, \citenamefont {Guo},\ and\ \citenamefont
  {Cai}}]{Liu:2021jnw}%
  \BibitemOpen
  \bibfield  {author} {\bibinfo {author} {\bibfnamefont {L.}~\bibnamefont
  {Liu}}, \bibinfo {author} {\bibfnamefont {X.-Y.}\ \bibnamefont {Yang}},
  \bibinfo {author} {\bibfnamefont {Z.-K.}\ \bibnamefont {Guo}}, \ and\
  \bibinfo {author} {\bibfnamefont {R.-G.}\ \bibnamefont {Cai}},\ }\href
  {\doibase 10.1088/1475-7516/2023/01/006} {\bibfield  {journal} {\bibinfo
  {journal} {JCAP}\ }\textbf {\bibinfo {volume} {01}},\ \bibinfo {pages} {006}
  (\bibinfo {year} {2023}{\natexlab{a}})},\ \Eprint
  {http://arxiv.org/abs/2112.05473} {arXiv:2112.05473 [astro-ph.CO]}
  \BibitemShut {NoStop}%
\bibitem [{\citenamefont {Saha}\ and\ \citenamefont
  {Laha}(2022)}]{Saha:2021pqf}%
  \BibitemOpen
  \bibfield  {author} {\bibinfo {author} {\bibfnamefont {A.~K.}\ \bibnamefont
  {Saha}}\ and\ \bibinfo {author} {\bibfnamefont {R.}~\bibnamefont {Laha}},\
  }\href {\doibase 10.1103/PhysRevD.105.103026} {\bibfield  {journal} {\bibinfo
   {journal} {Phys. Rev. D}\ }\textbf {\bibinfo {volume} {105}},\ \bibinfo
  {pages} {103026} (\bibinfo {year} {2022})},\ \Eprint
  {http://arxiv.org/abs/2112.10794} {arXiv:2112.10794 [astro-ph.CO]}
  \BibitemShut {NoStop}%
\bibitem [{\citenamefont {Bhaumik}\ \emph {et~al.}(2022)\citenamefont
  {Bhaumik}, \citenamefont {Ghoshal},\ and\ \citenamefont
  {Lewicki}}]{Bhaumik:2022pil}%
  \BibitemOpen
  \bibfield  {author} {\bibinfo {author} {\bibfnamefont {N.}~\bibnamefont
  {Bhaumik}}, \bibinfo {author} {\bibfnamefont {A.}~\bibnamefont {Ghoshal}}, \
  and\ \bibinfo {author} {\bibfnamefont {M.}~\bibnamefont {Lewicki}},\ }\href
  {\doibase 10.1007/JHEP07(2022)130} {\bibfield  {journal} {\bibinfo  {journal}
  {JHEP}\ }\textbf {\bibinfo {volume} {07}},\ \bibinfo {pages} {130} (\bibinfo
  {year} {2022})},\ \Eprint {http://arxiv.org/abs/2205.06260} {arXiv:2205.06260
  [astro-ph.CO]} \BibitemShut {NoStop}%
\bibitem [{\citenamefont {Anchordoqui}\ \emph {et~al.}(2022)\citenamefont
  {Anchordoqui}, \citenamefont {Antoniadis},\ and\ \citenamefont
  {Lust}}]{Anchordoqui:2022txe}%
  \BibitemOpen
  \bibfield  {author} {\bibinfo {author} {\bibfnamefont {L.~A.}\ \bibnamefont
  {Anchordoqui}}, \bibinfo {author} {\bibfnamefont {I.}~\bibnamefont
  {Antoniadis}}, \ and\ \bibinfo {author} {\bibfnamefont {D.}~\bibnamefont
  {Lust}},\ }\href {\doibase 10.1103/PhysRevD.106.086001} {\bibfield  {journal}
  {\bibinfo  {journal} {Phys. Rev. D}\ }\textbf {\bibinfo {volume} {106}},\
  \bibinfo {pages} {086001} (\bibinfo {year} {2022})},\ \Eprint
  {http://arxiv.org/abs/2206.07071} {arXiv:2206.07071 [hep-th]} \BibitemShut
  {NoStop}%
\bibitem [{\citenamefont {Cai}\ \emph {et~al.}(2022)\citenamefont {Cai},
  \citenamefont {Ma}, \citenamefont {Sasaki}, \citenamefont {Wang},\ and\
  \citenamefont {Zhou}}]{Cai:2022erk}%
  \BibitemOpen
  \bibfield  {author} {\bibinfo {author} {\bibfnamefont {Y.-F.}\ \bibnamefont
  {Cai}}, \bibinfo {author} {\bibfnamefont {X.-H.}\ \bibnamefont {Ma}},
  \bibinfo {author} {\bibfnamefont {M.}~\bibnamefont {Sasaki}}, \bibinfo
  {author} {\bibfnamefont {D.-G.}\ \bibnamefont {Wang}}, \ and\ \bibinfo
  {author} {\bibfnamefont {Z.}~\bibnamefont {Zhou}},\ }\href {\doibase
  10.1088/1475-7516/2022/12/034} {\bibfield  {journal} {\bibinfo  {journal}
  {JCAP}\ }\textbf {\bibinfo {volume} {12}},\ \bibinfo {pages} {034} (\bibinfo
  {year} {2022})},\ \Eprint {http://arxiv.org/abs/2207.11910} {arXiv:2207.11910
  [astro-ph.CO]} \BibitemShut {NoStop}%
\bibitem [{\citenamefont {Oguri}\ \emph {et~al.}(2023)\citenamefont {Oguri},
  \citenamefont {Takhistov},\ and\ \citenamefont {Kohri}}]{Oguri:2022fir}%
  \BibitemOpen
  \bibfield  {author} {\bibinfo {author} {\bibfnamefont {M.}~\bibnamefont
  {Oguri}}, \bibinfo {author} {\bibfnamefont {V.}~\bibnamefont {Takhistov}}, \
  and\ \bibinfo {author} {\bibfnamefont {K.}~\bibnamefont {Kohri}},\ }\href
  {\doibase 10.1016/j.physletb.2023.138276} {\bibfield  {journal} {\bibinfo
  {journal} {Phys. Lett. B}\ }\textbf {\bibinfo {volume} {847}},\ \bibinfo
  {pages} {138276} (\bibinfo {year} {2023})},\ \Eprint
  {http://arxiv.org/abs/2208.05957} {arXiv:2208.05957 [astro-ph.CO]}
  \BibitemShut {NoStop}%
\bibitem [{\citenamefont {Franciolini}\ \emph
  {et~al.}(2022{\natexlab{b}})\citenamefont {Franciolini}, \citenamefont
  {Musco}, \citenamefont {Pani},\ and\ \citenamefont
  {Urbano}}]{Franciolini:2022tfm}%
  \BibitemOpen
  \bibfield  {author} {\bibinfo {author} {\bibfnamefont {G.}~\bibnamefont
  {Franciolini}}, \bibinfo {author} {\bibfnamefont {I.}~\bibnamefont {Musco}},
  \bibinfo {author} {\bibfnamefont {P.}~\bibnamefont {Pani}}, \ and\ \bibinfo
  {author} {\bibfnamefont {A.}~\bibnamefont {Urbano}},\ }\href {\doibase
  10.1103/PhysRevD.106.123526} {\bibfield  {journal} {\bibinfo  {journal}
  {Phys. Rev. D}\ }\textbf {\bibinfo {volume} {106}},\ \bibinfo {pages}
  {123526} (\bibinfo {year} {2022}{\natexlab{b}})},\ \Eprint
  {http://arxiv.org/abs/2209.05959} {arXiv:2209.05959 [astro-ph.CO]}
  \BibitemShut {NoStop}%
\bibitem [{\citenamefont {Mazde}\ and\ \citenamefont
  {Visinelli}(2023)}]{Mazde:2022sdx}%
  \BibitemOpen
  \bibfield  {author} {\bibinfo {author} {\bibfnamefont {K.}~\bibnamefont
  {Mazde}}\ and\ \bibinfo {author} {\bibfnamefont {L.}~\bibnamefont
  {Visinelli}},\ }\href {\doibase 10.1088/1475-7516/2023/01/021} {\bibfield
  {journal} {\bibinfo  {journal} {JCAP}\ }\textbf {\bibinfo {volume} {01}},\
  \bibinfo {pages} {021} (\bibinfo {year} {2023})},\ \Eprint
  {http://arxiv.org/abs/2209.14307} {arXiv:2209.14307 [astro-ph.CO]}
  \BibitemShut {NoStop}%
\bibitem [{\citenamefont {Cai}\ \emph {et~al.}(2023)\citenamefont {Cai},
  \citenamefont {Chen}, \citenamefont {Wang},\ and\ \citenamefont
  {Yang}}]{Cai:2022kbp}%
  \BibitemOpen
  \bibfield  {author} {\bibinfo {author} {\bibfnamefont {R.-G.}\ \bibnamefont
  {Cai}}, \bibinfo {author} {\bibfnamefont {T.}~\bibnamefont {Chen}}, \bibinfo
  {author} {\bibfnamefont {S.-J.}\ \bibnamefont {Wang}}, \ and\ \bibinfo
  {author} {\bibfnamefont {X.-Y.}\ \bibnamefont {Yang}},\ }\href {\doibase
  10.1088/1475-7516/2023/03/043} {\bibfield  {journal} {\bibinfo  {journal}
  {JCAP}\ }\textbf {\bibinfo {volume} {03}},\ \bibinfo {pages} {043} (\bibinfo
  {year} {2023})},\ \Eprint {http://arxiv.org/abs/2210.02078} {arXiv:2210.02078
  [astro-ph.CO]} \BibitemShut {NoStop}%
\bibitem [{\citenamefont {Anchordoqui}\ \emph {et~al.}(2023)\citenamefont
  {Anchordoqui}, \citenamefont {Antoniadis},\ and\ \citenamefont
  {Lust}}]{Anchordoqui:2022tgp}%
  \BibitemOpen
  \bibfield  {author} {\bibinfo {author} {\bibfnamefont {L.~A.}\ \bibnamefont
  {Anchordoqui}}, \bibinfo {author} {\bibfnamefont {I.}~\bibnamefont
  {Antoniadis}}, \ and\ \bibinfo {author} {\bibfnamefont {D.}~\bibnamefont
  {Lust}},\ }\href {\doibase 10.1016/j.physletb.2023.137844} {\bibfield
  {journal} {\bibinfo  {journal} {Phys. Lett. B}\ }\textbf {\bibinfo {volume}
  {840}},\ \bibinfo {pages} {137844} (\bibinfo {year} {2023})},\ \Eprint
  {http://arxiv.org/abs/2210.02475} {arXiv:2210.02475 [hep-th]} \BibitemShut
  {NoStop}%
\bibitem [{\citenamefont {Liu}\ \emph {et~al.}(2023{\natexlab{b}})\citenamefont
  {Liu}, \citenamefont {You}, \citenamefont {Wu},\ and\ \citenamefont
  {Chen}}]{Liu:2022iuf}%
  \BibitemOpen
  \bibfield  {author} {\bibinfo {author} {\bibfnamefont {L.}~\bibnamefont
  {Liu}}, \bibinfo {author} {\bibfnamefont {Z.-Q.}\ \bibnamefont {You}},
  \bibinfo {author} {\bibfnamefont {Y.}~\bibnamefont {Wu}}, \ and\ \bibinfo
  {author} {\bibfnamefont {Z.-C.}\ \bibnamefont {Chen}},\ }\href {\doibase
  10.1103/PhysRevD.107.063035} {\bibfield  {journal} {\bibinfo  {journal}
  {Phys. Rev. D}\ }\textbf {\bibinfo {volume} {107}},\ \bibinfo {pages}
  {063035} (\bibinfo {year} {2023}{\natexlab{b}})},\ \Eprint
  {http://arxiv.org/abs/2210.16094} {arXiv:2210.16094 [astro-ph.CO]}
  \BibitemShut {NoStop}%
\bibitem [{\citenamefont {Fu}\ and\ \citenamefont {Wang}(2023)}]{Fu:2022ypp}%
  \BibitemOpen
  \bibfield  {author} {\bibinfo {author} {\bibfnamefont {C.}~\bibnamefont
  {Fu}}\ and\ \bibinfo {author} {\bibfnamefont {S.-J.}\ \bibnamefont {Wang}},\
  }\href {\doibase 10.1088/1475-7516/2023/06/012} {\bibfield  {journal}
  {\bibinfo  {journal} {JCAP}\ }\textbf {\bibinfo {volume} {06}},\ \bibinfo
  {pages} {012} (\bibinfo {year} {2023})},\ \Eprint
  {http://arxiv.org/abs/2211.03523} {arXiv:2211.03523 [astro-ph.CO]}
  \BibitemShut {NoStop}%
\bibitem [{\citenamefont {Choudhury}\ \emph
  {et~al.}(2024{\natexlab{a}})\citenamefont {Choudhury}, \citenamefont
  {Gangopadhyay},\ and\ \citenamefont {Sami}}]{Choudhury:2023vuj}%
  \BibitemOpen
  \bibfield  {author} {\bibinfo {author} {\bibfnamefont {S.}~\bibnamefont
  {Choudhury}}, \bibinfo {author} {\bibfnamefont {M.~R.}\ \bibnamefont
  {Gangopadhyay}}, \ and\ \bibinfo {author} {\bibfnamefont {M.}~\bibnamefont
  {Sami}},\ }\href {\doibase 10.1140/epjc/s10052-024-13218-2} {\bibfield
  {journal} {\bibinfo  {journal} {Eur. Phys. J. C}\ }\textbf {\bibinfo {volume}
  {84}},\ \bibinfo {pages} {884} (\bibinfo {year} {2024}{\natexlab{a}})},\
  \Eprint {http://arxiv.org/abs/2301.10000} {arXiv:2301.10000 [astro-ph.CO]}
  \BibitemShut {NoStop}%
\bibitem [{\citenamefont {Papanikolaou}(2023)}]{Papanikolaou:2023crz}%
  \BibitemOpen
  \bibfield  {author} {\bibinfo {author} {\bibfnamefont {T.}~\bibnamefont
  {Papanikolaou}},\ }\href {\doibase 10.1088/1361-6382/acd97d} {\bibfield
  {journal} {\bibinfo  {journal} {Class. Quant. Grav.}\ }\textbf {\bibinfo
  {volume} {40}},\ \bibinfo {pages} {134001} (\bibinfo {year} {2023})},\
  \Eprint {http://arxiv.org/abs/2301.11439} {arXiv:2301.11439 [gr-qc]}
  \BibitemShut {NoStop}%
\bibitem [{\citenamefont {de~Freitas~Pacheco}\ \emph
  {et~al.}(2023)\citenamefont {de~Freitas~Pacheco}, \citenamefont {Kiritsis},
  \citenamefont {Lucca},\ and\ \citenamefont
  {Silk}}]{deFreitasPacheco:2023hpb}%
  \BibitemOpen
  \bibfield  {author} {\bibinfo {author} {\bibfnamefont {J.~A.}\ \bibnamefont
  {de~Freitas~Pacheco}}, \bibinfo {author} {\bibfnamefont {E.}~\bibnamefont
  {Kiritsis}}, \bibinfo {author} {\bibfnamefont {M.}~\bibnamefont {Lucca}}, \
  and\ \bibinfo {author} {\bibfnamefont {J.}~\bibnamefont {Silk}},\ }\href
  {\doibase 10.1103/PhysRevD.107.123525} {\bibfield  {journal} {\bibinfo
  {journal} {Phys. Rev. D}\ }\textbf {\bibinfo {volume} {107}},\ \bibinfo
  {pages} {123525} (\bibinfo {year} {2023})},\ \Eprint
  {http://arxiv.org/abs/2301.13215} {arXiv:2301.13215 [astro-ph.CO]}
  \BibitemShut {NoStop}%
\bibitem [{\citenamefont {Choudhury}\ \emph
  {et~al.}(2023{\natexlab{a}})\citenamefont {Choudhury}, \citenamefont
  {Panda},\ and\ \citenamefont {Sami}}]{Choudhury:2023jlt}%
  \BibitemOpen
  \bibfield  {author} {\bibinfo {author} {\bibfnamefont {S.}~\bibnamefont
  {Choudhury}}, \bibinfo {author} {\bibfnamefont {S.}~\bibnamefont {Panda}}, \
  and\ \bibinfo {author} {\bibfnamefont {M.}~\bibnamefont {Sami}},\ }\href
  {\doibase 10.1016/j.physletb.2023.138123} {\bibfield  {journal} {\bibinfo
  {journal} {Phys. Lett. B}\ }\textbf {\bibinfo {volume} {845}},\ \bibinfo
  {pages} {138123} (\bibinfo {year} {2023}{\natexlab{a}})},\ \Eprint
  {http://arxiv.org/abs/2302.05655} {arXiv:2302.05655 [astro-ph.CO]}
  \BibitemShut {NoStop}%
\bibitem [{\citenamefont {Choudhury}\ \emph
  {et~al.}(2023{\natexlab{b}})\citenamefont {Choudhury}, \citenamefont
  {Panda},\ and\ \citenamefont {Sami}}]{Choudhury:2023rks}%
  \BibitemOpen
  \bibfield  {author} {\bibinfo {author} {\bibfnamefont {S.}~\bibnamefont
  {Choudhury}}, \bibinfo {author} {\bibfnamefont {S.}~\bibnamefont {Panda}}, \
  and\ \bibinfo {author} {\bibfnamefont {M.}~\bibnamefont {Sami}},\ }\href
  {\doibase 10.1088/1475-7516/2023/11/066} {\bibfield  {journal} {\bibinfo
  {journal} {JCAP}\ }\textbf {\bibinfo {volume} {11}},\ \bibinfo {pages} {066}
  (\bibinfo {year} {2023}{\natexlab{b}})},\ \Eprint
  {http://arxiv.org/abs/2303.06066} {arXiv:2303.06066 [astro-ph.CO]}
  \BibitemShut {NoStop}%
\bibitem [{\citenamefont {Musco}\ \emph {et~al.}(2024)\citenamefont {Musco},
  \citenamefont {Jedamzik},\ and\ \citenamefont {Young}}]{Musco:2023dak}%
  \BibitemOpen
  \bibfield  {author} {\bibinfo {author} {\bibfnamefont {I.}~\bibnamefont
  {Musco}}, \bibinfo {author} {\bibfnamefont {K.}~\bibnamefont {Jedamzik}}, \
  and\ \bibinfo {author} {\bibfnamefont {S.}~\bibnamefont {Young}},\ }\href
  {\doibase 10.1103/PhysRevD.109.083506} {\bibfield  {journal} {\bibinfo
  {journal} {Phys. Rev. D}\ }\textbf {\bibinfo {volume} {109}},\ \bibinfo
  {pages} {083506} (\bibinfo {year} {2024})},\ \Eprint
  {http://arxiv.org/abs/2303.07980} {arXiv:2303.07980 [astro-ph.CO]}
  \BibitemShut {NoStop}%
\bibitem [{\citenamefont {Yuan}\ \emph {et~al.}(2024)\citenamefont {Yuan},
  \citenamefont {Lei}, \citenamefont {Wang}, \citenamefont {Wang},
  \citenamefont {Wang}, \citenamefont {Chen}, \citenamefont {Shen},
  \citenamefont {Cai},\ and\ \citenamefont {Fan}}]{Yuan:2023bvh}%
  \BibitemOpen
  \bibfield  {author} {\bibinfo {author} {\bibfnamefont {G.-W.}\ \bibnamefont
  {Yuan}}, \bibinfo {author} {\bibfnamefont {L.}~\bibnamefont {Lei}}, \bibinfo
  {author} {\bibfnamefont {Y.-Z.}\ \bibnamefont {Wang}}, \bibinfo {author}
  {\bibfnamefont {B.}~\bibnamefont {Wang}}, \bibinfo {author} {\bibfnamefont
  {Y.-Y.}\ \bibnamefont {Wang}}, \bibinfo {author} {\bibfnamefont
  {C.}~\bibnamefont {Chen}}, \bibinfo {author} {\bibfnamefont {Z.-Q.}\
  \bibnamefont {Shen}}, \bibinfo {author} {\bibfnamefont {Y.-F.}\ \bibnamefont
  {Cai}}, \ and\ \bibinfo {author} {\bibfnamefont {Y.-Z.}\ \bibnamefont
  {Fan}},\ }\href {\doibase 10.1007/s11433-024-2433-3} {\bibfield  {journal}
  {\bibinfo  {journal} {Sci. China Phys. Mech. Astron.}\ }\textbf {\bibinfo
  {volume} {67}},\ \bibinfo {pages} {109512} (\bibinfo {year} {2024})},\
  \Eprint {http://arxiv.org/abs/2303.09391} {arXiv:2303.09391 [astro-ph.CO]}
  \BibitemShut {NoStop}%
\bibitem [{\citenamefont {Choudhury}\ \emph
  {et~al.}(2023{\natexlab{c}})\citenamefont {Choudhury}, \citenamefont
  {Panda},\ and\ \citenamefont {Sami}}]{Choudhury:2023hvf}%
  \BibitemOpen
  \bibfield  {author} {\bibinfo {author} {\bibfnamefont {S.}~\bibnamefont
  {Choudhury}}, \bibinfo {author} {\bibfnamefont {S.}~\bibnamefont {Panda}}, \
  and\ \bibinfo {author} {\bibfnamefont {M.}~\bibnamefont {Sami}},\ }\href
  {\doibase 10.1088/1475-7516/2023/08/078} {\bibfield  {journal} {\bibinfo
  {journal} {JCAP}\ }\textbf {\bibinfo {volume} {08}},\ \bibinfo {pages} {078}
  (\bibinfo {year} {2023}{\natexlab{c}})},\ \Eprint
  {http://arxiv.org/abs/2304.04065} {arXiv:2304.04065 [astro-ph.CO]}
  \BibitemShut {NoStop}%
\bibitem [{\citenamefont {Ghoshal}\ \emph {et~al.}(2023)\citenamefont
  {Ghoshal}, \citenamefont {Gouttenoire}, \citenamefont {Heurtier},\ and\
  \citenamefont {Simakachorn}}]{Ghoshal:2023sfa}%
  \BibitemOpen
  \bibfield  {author} {\bibinfo {author} {\bibfnamefont {A.}~\bibnamefont
  {Ghoshal}}, \bibinfo {author} {\bibfnamefont {Y.}~\bibnamefont
  {Gouttenoire}}, \bibinfo {author} {\bibfnamefont {L.}~\bibnamefont
  {Heurtier}}, \ and\ \bibinfo {author} {\bibfnamefont {P.}~\bibnamefont
  {Simakachorn}},\ }\href {\doibase 10.1007/JHEP08(2023)196} {\bibfield
  {journal} {\bibinfo  {journal} {JHEP}\ }\textbf {\bibinfo {volume} {08}},\
  \bibinfo {pages} {196} (\bibinfo {year} {2023})},\ \Eprint
  {http://arxiv.org/abs/2304.04793} {arXiv:2304.04793 [hep-ph]} \BibitemShut
  {NoStop}%
\bibitem [{\citenamefont {Cai}\ \emph {et~al.}(2024{\natexlab{a}})\citenamefont
  {Cai}, \citenamefont {Zhu},\ and\ \citenamefont {Piao}}]{Cai:2023uhc}%
  \BibitemOpen
  \bibfield  {author} {\bibinfo {author} {\bibfnamefont {Y.}~\bibnamefont
  {Cai}}, \bibinfo {author} {\bibfnamefont {M.}~\bibnamefont {Zhu}}, \ and\
  \bibinfo {author} {\bibfnamefont {Y.-S.}\ \bibnamefont {Piao}},\ }\href
  {\doibase 10.1103/PhysRevLett.133.021001} {\bibfield  {journal} {\bibinfo
  {journal} {Phys. Rev. Lett.}\ }\textbf {\bibinfo {volume} {133}},\ \bibinfo
  {pages} {021001} (\bibinfo {year} {2024}{\natexlab{a}})},\ \Eprint
  {http://arxiv.org/abs/2305.10933} {arXiv:2305.10933 [gr-qc]} \BibitemShut
  {NoStop}%
\bibitem [{\citenamefont {Choudhury}\ \emph
  {et~al.}(2024{\natexlab{b}})\citenamefont {Choudhury}, \citenamefont {Karde},
  \citenamefont {Panda},\ and\ \citenamefont {Sami}}]{Choudhury:2023kdb}%
  \BibitemOpen
  \bibfield  {author} {\bibinfo {author} {\bibfnamefont {S.}~\bibnamefont
  {Choudhury}}, \bibinfo {author} {\bibfnamefont {A.}~\bibnamefont {Karde}},
  \bibinfo {author} {\bibfnamefont {S.}~\bibnamefont {Panda}}, \ and\ \bibinfo
  {author} {\bibfnamefont {M.}~\bibnamefont {Sami}},\ }\href {\doibase
  10.1088/1475-7516/2024/01/012} {\bibfield  {journal} {\bibinfo  {journal}
  {JCAP}\ }\textbf {\bibinfo {volume} {01}},\ \bibinfo {pages} {012} (\bibinfo
  {year} {2024}{\natexlab{b}})},\ \Eprint {http://arxiv.org/abs/2306.12334}
  {arXiv:2306.12334 [astro-ph.CO]} \BibitemShut {NoStop}%
\bibitem [{\citenamefont {Huang}\ \emph
  {et~al.}(2024{\natexlab{a}})\citenamefont {Huang}, \citenamefont {Cai},
  \citenamefont {Jiang}, \citenamefont {Zhang},\ and\ \citenamefont
  {Piao}}]{Huang:2023chx}%
  \BibitemOpen
  \bibfield  {author} {\bibinfo {author} {\bibfnamefont {H.-L.}\ \bibnamefont
  {Huang}}, \bibinfo {author} {\bibfnamefont {Y.}~\bibnamefont {Cai}}, \bibinfo
  {author} {\bibfnamefont {J.-Q.}\ \bibnamefont {Jiang}}, \bibinfo {author}
  {\bibfnamefont {J.}~\bibnamefont {Zhang}}, \ and\ \bibinfo {author}
  {\bibfnamefont {Y.-S.}\ \bibnamefont {Piao}},\ }\href {\doibase
  10.1088/1674-4527/ad683d} {\bibfield  {journal} {\bibinfo  {journal} {Res.
  Astron. Astrophys.}\ }\textbf {\bibinfo {volume} {24}},\ \bibinfo {pages}
  {091001} (\bibinfo {year} {2024}{\natexlab{a}})},\ \Eprint
  {http://arxiv.org/abs/2306.17577} {arXiv:2306.17577 [gr-qc]} \BibitemShut
  {NoStop}%
\bibitem [{\citenamefont {Choudhury}\ \emph
  {et~al.}(2024{\natexlab{c}})\citenamefont {Choudhury}, \citenamefont {Karde},
  \citenamefont {Panda},\ and\ \citenamefont {Sami}}]{Choudhury:2023hfm}%
  \BibitemOpen
  \bibfield  {author} {\bibinfo {author} {\bibfnamefont {S.}~\bibnamefont
  {Choudhury}}, \bibinfo {author} {\bibfnamefont {A.}~\bibnamefont {Karde}},
  \bibinfo {author} {\bibfnamefont {S.}~\bibnamefont {Panda}}, \ and\ \bibinfo
  {author} {\bibfnamefont {M.}~\bibnamefont {Sami}},\ }\href {\doibase
  10.1016/j.nuclphysb.2024.116678} {\bibfield  {journal} {\bibinfo  {journal}
  {Nucl. Phys. B}\ }\textbf {\bibinfo {volume} {1007}},\ \bibinfo {pages}
  {116678} (\bibinfo {year} {2024}{\natexlab{c}})},\ \Eprint
  {http://arxiv.org/abs/2308.09273} {arXiv:2308.09273 [astro-ph.CO]}
  \BibitemShut {NoStop}%
\bibitem [{\citenamefont {Bhattacharya}\ \emph {et~al.}(2024)\citenamefont
  {Bhattacharya}, \citenamefont {Choudhury}, \citenamefont {Dey}, \citenamefont
  {Ghosh}, \citenamefont {Karde},\ and\ \citenamefont
  {Mishra}}]{Bhattacharya:2023ysp}%
  \BibitemOpen
  \bibfield  {author} {\bibinfo {author} {\bibfnamefont {G.}~\bibnamefont
  {Bhattacharya}}, \bibinfo {author} {\bibfnamefont {S.}~\bibnamefont
  {Choudhury}}, \bibinfo {author} {\bibfnamefont {K.}~\bibnamefont {Dey}},
  \bibinfo {author} {\bibfnamefont {S.}~\bibnamefont {Ghosh}}, \bibinfo
  {author} {\bibfnamefont {A.}~\bibnamefont {Karde}}, \ and\ \bibinfo {author}
  {\bibfnamefont {N.~S.}\ \bibnamefont {Mishra}},\ }\href {\doibase
  10.1016/j.dark.2024.101602} {\bibfield  {journal} {\bibinfo  {journal} {Phys.
  Dark Univ.}\ }\textbf {\bibinfo {volume} {46}},\ \bibinfo {pages} {101602}
  (\bibinfo {year} {2024})},\ \Eprint {http://arxiv.org/abs/2309.00973}
  {arXiv:2309.00973 [astro-ph.CO]} \BibitemShut {NoStop}%
\bibitem [{\citenamefont {Heydari}\ and\ \citenamefont
  {Karami}(2024{\natexlab{a}})}]{Heydari:2023xts}%
  \BibitemOpen
  \bibfield  {author} {\bibinfo {author} {\bibfnamefont {S.}~\bibnamefont
  {Heydari}}\ and\ \bibinfo {author} {\bibfnamefont {K.}~\bibnamefont
  {Karami}},\ }\href {\doibase 10.1088/1475-7516/2024/02/047} {\bibfield
  {journal} {\bibinfo  {journal} {JCAP}\ }\textbf {\bibinfo {volume} {02}},\
  \bibinfo {pages} {047} (\bibinfo {year} {2024}{\natexlab{a}})},\ \Eprint
  {http://arxiv.org/abs/2309.01239} {arXiv:2309.01239 [astro-ph.CO]}
  \BibitemShut {NoStop}%
\bibitem [{\citenamefont {Heydari}\ and\ \citenamefont
  {Karami}(2024{\natexlab{b}})}]{Heydari:2023rmq}%
  \BibitemOpen
  \bibfield  {author} {\bibinfo {author} {\bibfnamefont {S.}~\bibnamefont
  {Heydari}}\ and\ \bibinfo {author} {\bibfnamefont {K.}~\bibnamefont
  {Karami}},\ }\href {\doibase 10.1140/epjc/s10052-024-12489-z} {\bibfield
  {journal} {\bibinfo  {journal} {Eur. Phys. J. C}\ }\textbf {\bibinfo {volume}
  {84}},\ \bibinfo {pages} {127} (\bibinfo {year} {2024}{\natexlab{b}})},\
  \Eprint {http://arxiv.org/abs/2310.11030} {arXiv:2310.11030 [gr-qc]}
  \BibitemShut {NoStop}%
\bibitem [{\citenamefont {Choudhury}\ \emph
  {et~al.}(2024{\natexlab{d}})\citenamefont {Choudhury}, \citenamefont {Dey},
  \citenamefont {Karde}, \citenamefont {Panda},\ and\ \citenamefont
  {Sami}}]{Choudhury:2023fwk}%
  \BibitemOpen
  \bibfield  {author} {\bibinfo {author} {\bibfnamefont {S.}~\bibnamefont
  {Choudhury}}, \bibinfo {author} {\bibfnamefont {K.}~\bibnamefont {Dey}},
  \bibinfo {author} {\bibfnamefont {A.}~\bibnamefont {Karde}}, \bibinfo
  {author} {\bibfnamefont {S.}~\bibnamefont {Panda}}, \ and\ \bibinfo {author}
  {\bibfnamefont {M.}~\bibnamefont {Sami}},\ }\href {\doibase
  10.1016/j.physletb.2024.138925} {\bibfield  {journal} {\bibinfo  {journal}
  {Phys. Lett. B}\ }\textbf {\bibinfo {volume} {856}},\ \bibinfo {pages}
  {138925} (\bibinfo {year} {2024}{\natexlab{d}})},\ \Eprint
  {http://arxiv.org/abs/2310.11034} {arXiv:2310.11034 [astro-ph.CO]}
  \BibitemShut {NoStop}%
\bibitem [{\citenamefont {Choudhury}\ \emph
  {et~al.}(2023{\natexlab{d}})\citenamefont {Choudhury}, \citenamefont {Dey},\
  and\ \citenamefont {Karde}}]{Choudhury:2023fjs}%
  \BibitemOpen
  \bibfield  {author} {\bibinfo {author} {\bibfnamefont {S.}~\bibnamefont
  {Choudhury}}, \bibinfo {author} {\bibfnamefont {K.}~\bibnamefont {Dey}}, \
  and\ \bibinfo {author} {\bibfnamefont {A.}~\bibnamefont {Karde}},\
  }\href@noop {} {\  (\bibinfo {year} {2023}{\natexlab{d}})},\ \Eprint
  {http://arxiv.org/abs/2311.15065} {arXiv:2311.15065 [astro-ph.CO]}
  \BibitemShut {NoStop}%
\bibitem [{\citenamefont {Ghoshal}\ and\ \citenamefont
  {Strumia}(2024)}]{Ghoshal:2023pcx}%
  \BibitemOpen
  \bibfield  {author} {\bibinfo {author} {\bibfnamefont {A.}~\bibnamefont
  {Ghoshal}}\ and\ \bibinfo {author} {\bibfnamefont {A.}~\bibnamefont
  {Strumia}},\ }\href {\doibase 10.1088/1475-7516/2024/07/011} {\bibfield
  {journal} {\bibinfo  {journal} {JCAP}\ }\textbf {\bibinfo {volume} {07}},\
  \bibinfo {pages} {011} (\bibinfo {year} {2024})},\ \Eprint
  {http://arxiv.org/abs/2311.16236} {arXiv:2311.16236 [hep-ph]} \BibitemShut
  {NoStop}%
\bibitem [{\citenamefont {Huang}\ \emph
  {et~al.}(2024{\natexlab{b}})\citenamefont {Huang}, \citenamefont {Jiang},\
  and\ \citenamefont {Piao}}]{Hai-LongHuang:2023atg}%
  \BibitemOpen
  \bibfield  {author} {\bibinfo {author} {\bibfnamefont {H.-L.}\ \bibnamefont
  {Huang}}, \bibinfo {author} {\bibfnamefont {J.-Q.}\ \bibnamefont {Jiang}}, \
  and\ \bibinfo {author} {\bibfnamefont {Y.-S.}\ \bibnamefont {Piao}},\ }\href
  {\doibase 10.1103/PhysRevD.109.063515} {\bibfield  {journal} {\bibinfo
  {journal} {Phys. Rev. D}\ }\textbf {\bibinfo {volume} {109}},\ \bibinfo
  {pages} {063515} (\bibinfo {year} {2024}{\natexlab{b}})},\ \Eprint
  {http://arxiv.org/abs/2312.00338} {arXiv:2312.00338 [astro-ph.CO]}
  \BibitemShut {NoStop}%
\bibitem [{\citenamefont {Huang}\ and\ \citenamefont
  {Piao}(2024)}]{Huang:2023mwy}%
  \BibitemOpen
  \bibfield  {author} {\bibinfo {author} {\bibfnamefont {H.-L.}\ \bibnamefont
  {Huang}}\ and\ \bibinfo {author} {\bibfnamefont {Y.-S.}\ \bibnamefont
  {Piao}},\ }\href {\doibase 10.1103/PhysRevD.110.023501} {\bibfield  {journal}
  {\bibinfo  {journal} {Phys. Rev. D}\ }\textbf {\bibinfo {volume} {110}},\
  \bibinfo {pages} {023501} (\bibinfo {year} {2024})},\ \Eprint
  {http://arxiv.org/abs/2312.11982} {arXiv:2312.11982 [astro-ph.CO]}
  \BibitemShut {NoStop}%
\bibitem [{\citenamefont {Anchordoqui}\ \emph
  {et~al.}(2024{\natexlab{a}})\citenamefont {Anchordoqui}, \citenamefont
  {Antoniadis},\ and\ \citenamefont {Lust}}]{Anchordoqui:2024akj}%
  \BibitemOpen
  \bibfield  {author} {\bibinfo {author} {\bibfnamefont {L.~A.}\ \bibnamefont
  {Anchordoqui}}, \bibinfo {author} {\bibfnamefont {I.}~\bibnamefont
  {Antoniadis}}, \ and\ \bibinfo {author} {\bibfnamefont {D.}~\bibnamefont
  {Lust}},\ }\href {\doibase 10.1103/PhysRevD.109.095008} {\bibfield  {journal}
  {\bibinfo  {journal} {Phys. Rev. D}\ }\textbf {\bibinfo {volume} {109}},\
  \bibinfo {pages} {095008} (\bibinfo {year} {2024}{\natexlab{a}})},\ \Eprint
  {http://arxiv.org/abs/2401.09087} {arXiv:2401.09087 [hep-th]} \BibitemShut
  {NoStop}%
\bibitem [{\citenamefont {Choudhury}\ \emph
  {et~al.}(2024{\natexlab{e}})\citenamefont {Choudhury}, \citenamefont {Karde},
  \citenamefont {Panda},\ and\ \citenamefont {Sami}}]{Choudhury:2024one}%
  \BibitemOpen
  \bibfield  {author} {\bibinfo {author} {\bibfnamefont {S.}~\bibnamefont
  {Choudhury}}, \bibinfo {author} {\bibfnamefont {A.}~\bibnamefont {Karde}},
  \bibinfo {author} {\bibfnamefont {S.}~\bibnamefont {Panda}}, \ and\ \bibinfo
  {author} {\bibfnamefont {M.}~\bibnamefont {Sami}},\ }\href {\doibase
  10.1088/1475-7516/2024/07/034} {\bibfield  {journal} {\bibinfo  {journal}
  {JCAP}\ }\textbf {\bibinfo {volume} {07}},\ \bibinfo {pages} {034} (\bibinfo
  {year} {2024}{\natexlab{e}})},\ \Eprint {http://arxiv.org/abs/2401.10925}
  {arXiv:2401.10925 [astro-ph.CO]} \BibitemShut {NoStop}%
\bibitem [{\citenamefont {Thoss}\ \emph {et~al.}(2024)\citenamefont {Thoss},
  \citenamefont {Burkert},\ and\ \citenamefont {Kohri}}]{Thoss:2024hsr}%
  \BibitemOpen
  \bibfield  {author} {\bibinfo {author} {\bibfnamefont {V.}~\bibnamefont
  {Thoss}}, \bibinfo {author} {\bibfnamefont {A.}~\bibnamefont {Burkert}}, \
  and\ \bibinfo {author} {\bibfnamefont {K.}~\bibnamefont {Kohri}},\ }\href
  {\doibase 10.1093/mnras/stae1098} {\bibfield  {journal} {\bibinfo  {journal}
  {Mon. Not. Roy. Astron. Soc.}\ }\textbf {\bibinfo {volume} {532}},\ \bibinfo
  {pages} {451} (\bibinfo {year} {2024})},\ \Eprint
  {http://arxiv.org/abs/2402.17823} {arXiv:2402.17823 [astro-ph.CO]}
  \BibitemShut {NoStop}%
\bibitem [{\citenamefont {Papanikolaou}\ \emph
  {et~al.}(2024{\natexlab{a}})\citenamefont {Papanikolaou}, \citenamefont {He},
  \citenamefont {Ma}, \citenamefont {Cai}, \citenamefont {Saridakis},\ and\
  \citenamefont {Sasaki}}]{Papanikolaou:2024kjb}%
  \BibitemOpen
  \bibfield  {author} {\bibinfo {author} {\bibfnamefont {T.}~\bibnamefont
  {Papanikolaou}}, \bibinfo {author} {\bibfnamefont {X.-C.}\ \bibnamefont
  {He}}, \bibinfo {author} {\bibfnamefont {X.-H.}\ \bibnamefont {Ma}}, \bibinfo
  {author} {\bibfnamefont {Y.-F.}\ \bibnamefont {Cai}}, \bibinfo {author}
  {\bibfnamefont {E.~N.}\ \bibnamefont {Saridakis}}, \ and\ \bibinfo {author}
  {\bibfnamefont {M.}~\bibnamefont {Sasaki}},\ }\href {\doibase
  10.1016/j.physletb.2024.138997} {\bibfield  {journal} {\bibinfo  {journal}
  {Phys. Lett. B}\ }\textbf {\bibinfo {volume} {857}},\ \bibinfo {pages}
  {138997} (\bibinfo {year} {2024}{\natexlab{a}})},\ \Eprint
  {http://arxiv.org/abs/2403.00660} {arXiv:2403.00660 [astro-ph.CO]}
  \BibitemShut {NoStop}%
\bibitem [{\citenamefont {Choudhury}(2024)}]{Choudhury:2024ybk}%
  \BibitemOpen
  \bibfield  {author} {\bibinfo {author} {\bibfnamefont {S.}~\bibnamefont
  {Choudhury}},\ }\href {\doibase 10.1142/S0218271824410074} {\  (\bibinfo
  {year} {2024}),\ 10.1142/S0218271824410074},\ \Eprint
  {http://arxiv.org/abs/2403.07343} {arXiv:2403.07343 [astro-ph.CO]}
  \BibitemShut {NoStop}%
\bibitem [{\citenamefont {Choudhury}\ \emph
  {et~al.}(2024{\natexlab{f}})\citenamefont {Choudhury}, \citenamefont {Karde},
  \citenamefont {Padiyar},\ and\ \citenamefont {Sami}}]{Choudhury:2024jlz}%
  \BibitemOpen
  \bibfield  {author} {\bibinfo {author} {\bibfnamefont {S.}~\bibnamefont
  {Choudhury}}, \bibinfo {author} {\bibfnamefont {A.}~\bibnamefont {Karde}},
  \bibinfo {author} {\bibfnamefont {P.}~\bibnamefont {Padiyar}}, \ and\
  \bibinfo {author} {\bibfnamefont {M.}~\bibnamefont {Sami}},\ }\href@noop {}
  {\  (\bibinfo {year} {2024}{\natexlab{f}})},\ \Eprint
  {http://arxiv.org/abs/2403.13484} {arXiv:2403.13484 [astro-ph.CO]}
  \BibitemShut {NoStop}%
\bibitem [{\citenamefont {Anchordoqui}\ \emph
  {et~al.}(2024{\natexlab{b}})\citenamefont {Anchordoqui}, \citenamefont
  {Antoniadis},\ and\ \citenamefont {Lust}}]{Anchordoqui:2024dxu}%
  \BibitemOpen
  \bibfield  {author} {\bibinfo {author} {\bibfnamefont {L.~A.}\ \bibnamefont
  {Anchordoqui}}, \bibinfo {author} {\bibfnamefont {I.}~\bibnamefont
  {Antoniadis}}, \ and\ \bibinfo {author} {\bibfnamefont {D.}~\bibnamefont
  {Lust}},\ }\href {\doibase 10.1103/PhysRevD.110.015004} {\bibfield  {journal}
  {\bibinfo  {journal} {Phys. Rev. D}\ }\textbf {\bibinfo {volume} {110}},\
  \bibinfo {pages} {015004} (\bibinfo {year} {2024}{\natexlab{b}})},\ \Eprint
  {http://arxiv.org/abs/2403.19604} {arXiv:2403.19604 [hep-th]} \BibitemShut
  {NoStop}%
\bibitem [{\citenamefont {Papanikolaou}\ \emph
  {et~al.}(2024{\natexlab{b}})\citenamefont {Papanikolaou}, \citenamefont
  {Banerjee}, \citenamefont {Cai}, \citenamefont {Capozziello},\ and\
  \citenamefont {Saridakis}}]{Papanikolaou:2024fzf}%
  \BibitemOpen
  \bibfield  {author} {\bibinfo {author} {\bibfnamefont {T.}~\bibnamefont
  {Papanikolaou}}, \bibinfo {author} {\bibfnamefont {S.}~\bibnamefont
  {Banerjee}}, \bibinfo {author} {\bibfnamefont {Y.-F.}\ \bibnamefont {Cai}},
  \bibinfo {author} {\bibfnamefont {S.}~\bibnamefont {Capozziello}}, \ and\
  \bibinfo {author} {\bibfnamefont {E.~N.}\ \bibnamefont {Saridakis}},\ }\href
  {\doibase 10.1088/1475-7516/2024/06/066} {\bibfield  {journal} {\bibinfo
  {journal} {JCAP}\ }\textbf {\bibinfo {volume} {06}},\ \bibinfo {pages} {066}
  (\bibinfo {year} {2024}{\natexlab{b}})},\ \Eprint
  {http://arxiv.org/abs/2404.03779} {arXiv:2404.03779 [gr-qc]} \BibitemShut
  {NoStop}%
\bibitem [{\citenamefont {Yin}\ and\ \citenamefont
  {Visinelli}(2024)}]{Yin:2024xov}%
  \BibitemOpen
  \bibfield  {author} {\bibinfo {author} {\bibfnamefont {Z.}~\bibnamefont
  {Yin}}\ and\ \bibinfo {author} {\bibfnamefont {L.}~\bibnamefont
  {Visinelli}},\ }\href {\doibase 10.1088/1475-7516/2024/10/013} {\bibfield
  {journal} {\bibinfo  {journal} {JCAP}\ }\textbf {\bibinfo {volume} {10}},\
  \bibinfo {pages} {013} (\bibinfo {year} {2024})},\ \Eprint
  {http://arxiv.org/abs/2404.10340} {arXiv:2404.10340 [hep-ph]} \BibitemShut
  {NoStop}%
\bibitem [{\citenamefont {Choudhury}\ \emph
  {et~al.}(2024{\natexlab{g}})\citenamefont {Choudhury}, \citenamefont {Karde},
  \citenamefont {Panda},\ and\ \citenamefont {SenGupta}}]{Choudhury:2024dei}%
  \BibitemOpen
  \bibfield  {author} {\bibinfo {author} {\bibfnamefont {S.}~\bibnamefont
  {Choudhury}}, \bibinfo {author} {\bibfnamefont {A.}~\bibnamefont {Karde}},
  \bibinfo {author} {\bibfnamefont {S.}~\bibnamefont {Panda}}, \ and\ \bibinfo
  {author} {\bibfnamefont {S.}~\bibnamefont {SenGupta}},\ }\href {\doibase
  10.1140/epjc/s10052-024-13460-8} {\bibfield  {journal} {\bibinfo  {journal}
  {Eur. Phys. J. C}\ }\textbf {\bibinfo {volume} {84}},\ \bibinfo {pages}
  {1149} (\bibinfo {year} {2024}{\natexlab{g}})},\ \Eprint
  {http://arxiv.org/abs/2405.06882} {arXiv:2405.06882 [astro-ph.CO]}
  \BibitemShut {NoStop}%
\bibitem [{\citenamefont {Heydari}\ and\ \citenamefont
  {Karami}(2024{\natexlab{c}})}]{Heydari:2024bxj}%
  \BibitemOpen
  \bibfield  {author} {\bibinfo {author} {\bibfnamefont {S.}~\bibnamefont
  {Heydari}}\ and\ \bibinfo {author} {\bibfnamefont {K.}~\bibnamefont
  {Karami}},\ }\href {\doibase 10.3847/1538-4357/ad7605} {\bibfield  {journal}
  {\bibinfo  {journal} {Astrophys. J.}\ }\textbf {\bibinfo {volume} {975}},\
  \bibinfo {pages} {148} (\bibinfo {year} {2024}{\natexlab{c}})},\ \Eprint
  {http://arxiv.org/abs/2405.08563} {arXiv:2405.08563 [gr-qc]} \BibitemShut
  {NoStop}%
\bibitem [{\citenamefont {Dvali}\ \emph {et~al.}(2024)\citenamefont {Dvali},
  \citenamefont {Valbuena-Berm\'udez},\ and\ \citenamefont
  {Zantedeschi}}]{Dvali:2024hsb}%
  \BibitemOpen
  \bibfield  {author} {\bibinfo {author} {\bibfnamefont {G.}~\bibnamefont
  {Dvali}}, \bibinfo {author} {\bibfnamefont {J.~S.}\ \bibnamefont
  {Valbuena-Berm\'udez}}, \ and\ \bibinfo {author} {\bibfnamefont
  {M.}~\bibnamefont {Zantedeschi}},\ }\href {\doibase
  10.1103/PhysRevD.110.056029} {\bibfield  {journal} {\bibinfo  {journal}
  {Phys. Rev. D}\ }\textbf {\bibinfo {volume} {110}},\ \bibinfo {pages}
  {056029} (\bibinfo {year} {2024})},\ \Eprint
  {http://arxiv.org/abs/2405.13117} {arXiv:2405.13117 [hep-th]} \BibitemShut
  {NoStop}%
\bibitem [{\citenamefont {Boccia}\ \emph {et~al.}(2024)\citenamefont {Boccia},
  \citenamefont {Iocco},\ and\ \citenamefont {Visinelli}}]{Boccia:2024nly}%
  \BibitemOpen
  \bibfield  {author} {\bibinfo {author} {\bibfnamefont {A.}~\bibnamefont
  {Boccia}}, \bibinfo {author} {\bibfnamefont {F.}~\bibnamefont {Iocco}}, \
  and\ \bibinfo {author} {\bibfnamefont {L.}~\bibnamefont {Visinelli}},\
  }\href@noop {} {\  (\bibinfo {year} {2024})},\ \Eprint
  {http://arxiv.org/abs/2405.18493} {arXiv:2405.18493 [astro-ph.CO]}
  \BibitemShut {NoStop}%
\bibitem [{\citenamefont {Huang}\ \emph
  {et~al.}(2024{\natexlab{c}})\citenamefont {Huang}, \citenamefont {Jiang},\
  and\ \citenamefont {Piao}}]{Hai-LongHuang:2024kye}%
  \BibitemOpen
  \bibfield  {author} {\bibinfo {author} {\bibfnamefont {H.-L.}\ \bibnamefont
  {Huang}}, \bibinfo {author} {\bibfnamefont {J.-Q.}\ \bibnamefont {Jiang}}, \
  and\ \bibinfo {author} {\bibfnamefont {Y.-S.}\ \bibnamefont {Piao}},\
  }\href@noop {} {\  (\bibinfo {year} {2024}{\natexlab{c}})},\ \Eprint
  {http://arxiv.org/abs/2407.15781} {arXiv:2407.15781 [astro-ph.CO]}
  \BibitemShut {NoStop}%
\bibitem [{\citenamefont {Choudhury}\ \emph
  {et~al.}(2024{\natexlab{h}})\citenamefont {Choudhury}, \citenamefont
  {Ganguly}, \citenamefont {Panda}, \citenamefont {SenGupta},\ and\
  \citenamefont {Tiwari}}]{Choudhury:2024dzw}%
  \BibitemOpen
  \bibfield  {author} {\bibinfo {author} {\bibfnamefont {S.}~\bibnamefont
  {Choudhury}}, \bibinfo {author} {\bibfnamefont {S.}~\bibnamefont {Ganguly}},
  \bibinfo {author} {\bibfnamefont {S.}~\bibnamefont {Panda}}, \bibinfo
  {author} {\bibfnamefont {S.}~\bibnamefont {SenGupta}}, \ and\ \bibinfo
  {author} {\bibfnamefont {P.}~\bibnamefont {Tiwari}},\ }\href {\doibase
  10.1088/1475-7516/2024/09/013} {\bibfield  {journal} {\bibinfo  {journal}
  {JCAP}\ }\textbf {\bibinfo {volume} {09}},\ \bibinfo {pages} {013} (\bibinfo
  {year} {2024}{\natexlab{h}})},\ \Eprint {http://arxiv.org/abs/2407.18976}
  {arXiv:2407.18976 [astro-ph.CO]} \BibitemShut {NoStop}%
\bibitem [{\citenamefont {Anchordoqui}\ \emph
  {et~al.}(2024{\natexlab{c}})\citenamefont {Anchordoqui}, \citenamefont
  {Antoniadis}, \citenamefont {Lust},\ and\ \citenamefont
  {Castillo}}]{Anchordoqui:2024jkn}%
  \BibitemOpen
  \bibfield  {author} {\bibinfo {author} {\bibfnamefont {L.~A.}\ \bibnamefont
  {Anchordoqui}}, \bibinfo {author} {\bibfnamefont {I.}~\bibnamefont
  {Antoniadis}}, \bibinfo {author} {\bibfnamefont {D.}~\bibnamefont {Lust}}, \
  and\ \bibinfo {author} {\bibfnamefont {K.~P.~n.}\ \bibnamefont {Castillo}},\
  }\href {\doibase 10.1016/j.dark.2024.101714} {\bibfield  {journal} {\bibinfo
  {journal} {Phys. Dark Univ.}\ }\textbf {\bibinfo {volume} {46}},\ \bibinfo
  {pages} {101714} (\bibinfo {year} {2024}{\natexlab{c}})},\ \Eprint
  {http://arxiv.org/abs/2407.21031} {arXiv:2407.21031 [hep-th]} \BibitemShut
  {NoStop}%
\bibitem [{\citenamefont {Yang}\ \emph {et~al.}(2024)\citenamefont {Yang},
  \citenamefont {Wang}, \citenamefont {Zhao},\ and\ \citenamefont
  {Zhang}}]{Yang:2024vij}%
  \BibitemOpen
  \bibfield  {author} {\bibinfo {author} {\bibfnamefont {C.}~\bibnamefont
  {Yang}}, \bibinfo {author} {\bibfnamefont {S.}~\bibnamefont {Wang}}, \bibinfo
  {author} {\bibfnamefont {M.-L.}\ \bibnamefont {Zhao}}, \ and\ \bibinfo
  {author} {\bibfnamefont {X.}~\bibnamefont {Zhang}},\ }\href {\doibase
  10.1088/1475-7516/2024/10/083} {\bibfield  {journal} {\bibinfo  {journal}
  {JCAP}\ }\textbf {\bibinfo {volume} {10}},\ \bibinfo {pages} {083} (\bibinfo
  {year} {2024})},\ \Eprint {http://arxiv.org/abs/2408.10897} {arXiv:2408.10897
  [astro-ph.HE]} \BibitemShut {NoStop}%
\bibitem [{\citenamefont {Saha}\ \emph {et~al.}(2024)\citenamefont {Saha},
  \citenamefont {Singh}, \citenamefont {Parashari},\ and\ \citenamefont
  {Laha}}]{Saha:2024ies}%
  \BibitemOpen
  \bibfield  {author} {\bibinfo {author} {\bibfnamefont {A.~K.}\ \bibnamefont
  {Saha}}, \bibinfo {author} {\bibfnamefont {A.}~\bibnamefont {Singh}},
  \bibinfo {author} {\bibfnamefont {P.}~\bibnamefont {Parashari}}, \ and\
  \bibinfo {author} {\bibfnamefont {R.}~\bibnamefont {Laha}},\ }\href@noop {}
  {\  (\bibinfo {year} {2024})},\ \Eprint {http://arxiv.org/abs/2409.10617}
  {arXiv:2409.10617 [astro-ph.CO]} \BibitemShut {NoStop}%
\bibitem [{\citenamefont {Anchordoqui}\ \emph
  {et~al.}(2024{\natexlab{d}})\citenamefont {Anchordoqui}, \citenamefont
  {Antoniadis}, \citenamefont {Lust},\ and\ \citenamefont
  {Castillo}}]{Anchordoqui:2024tdj}%
  \BibitemOpen
  \bibfield  {author} {\bibinfo {author} {\bibfnamefont {L.~A.}\ \bibnamefont
  {Anchordoqui}}, \bibinfo {author} {\bibfnamefont {I.}~\bibnamefont
  {Antoniadis}}, \bibinfo {author} {\bibfnamefont {D.}~\bibnamefont {Lust}}, \
  and\ \bibinfo {author} {\bibfnamefont {K.~P.~n.}\ \bibnamefont {Castillo}},\
  }\href {\doibase 10.1016/j.dark.2024.101681} {\bibfield  {journal} {\bibinfo
  {journal} {Phys. Dark Univ.}\ }\textbf {\bibinfo {volume} {46}},\ \bibinfo
  {pages} {101681} (\bibinfo {year} {2024}{\natexlab{d}})},\ \Eprint
  {http://arxiv.org/abs/2409.12904} {arXiv:2409.12904 [hep-ph]} \BibitemShut
  {NoStop}%
\bibitem [{\citenamefont {Chen}\ \emph {et~al.}(2024)\citenamefont {Chen},
  \citenamefont {Ghoshal}, \citenamefont {Tasinato},\ and\ \citenamefont
  {Tomberg}}]{Chen:2024pge}%
  \BibitemOpen
  \bibfield  {author} {\bibinfo {author} {\bibfnamefont {C.}~\bibnamefont
  {Chen}}, \bibinfo {author} {\bibfnamefont {A.}~\bibnamefont {Ghoshal}},
  \bibinfo {author} {\bibfnamefont {G.}~\bibnamefont {Tasinato}}, \ and\
  \bibinfo {author} {\bibfnamefont {E.}~\bibnamefont {Tomberg}},\ }\href@noop
  {} {\  (\bibinfo {year} {2024})},\ \Eprint {http://arxiv.org/abs/2409.12950}
  {arXiv:2409.12950 [astro-ph.CO]} \BibitemShut {NoStop}%
\bibitem [{\citenamefont {Dai}\ and\ \citenamefont
  {Stojkovic}(2024)}]{Dai:2024guo}%
  \BibitemOpen
  \bibfield  {author} {\bibinfo {author} {\bibfnamefont {D.-C.}\ \bibnamefont
  {Dai}}\ and\ \bibinfo {author} {\bibfnamefont {D.}~\bibnamefont
  {Stojkovic}},\ }\href {\doibase 10.1016/j.dark.2024.101662} {\bibfield
  {journal} {\bibinfo  {journal} {Phys. Dark Univ.}\ }\textbf {\bibinfo
  {volume} {46}},\ \bibinfo {pages} {101662} (\bibinfo {year} {2024})},\
  \Eprint {http://arxiv.org/abs/2409.14321} {arXiv:2409.14321 [gr-qc]}
  \BibitemShut {NoStop}%
\bibitem [{\citenamefont {Huang}\ \emph
  {et~al.}(2024{\natexlab{d}})\citenamefont {Huang}, \citenamefont {Wang},\
  and\ \citenamefont {Piao}}]{Hai-LongHuang:2024vvz}%
  \BibitemOpen
  \bibfield  {author} {\bibinfo {author} {\bibfnamefont {H.-L.}\ \bibnamefont
  {Huang}}, \bibinfo {author} {\bibfnamefont {Y.-T.}\ \bibnamefont {Wang}}, \
  and\ \bibinfo {author} {\bibfnamefont {Y.-S.}\ \bibnamefont {Piao}},\
  }\href@noop {} {\  (\bibinfo {year} {2024}{\natexlab{d}})},\ \Eprint
  {http://arxiv.org/abs/2410.05891} {arXiv:2410.05891 [astro-ph.GA]}
  \BibitemShut {NoStop}%
\bibitem [{\citenamefont {Zantedeschi}\ and\ \citenamefont
  {Visinelli}(2024)}]{Zantedeschi:2024ram}%
  \BibitemOpen
  \bibfield  {author} {\bibinfo {author} {\bibfnamefont {M.}~\bibnamefont
  {Zantedeschi}}\ and\ \bibinfo {author} {\bibfnamefont {L.}~\bibnamefont
  {Visinelli}},\ }\href@noop {} {\  (\bibinfo {year} {2024})},\ \Eprint
  {http://arxiv.org/abs/2410.07037} {arXiv:2410.07037 [astro-ph.HE]}
  \BibitemShut {NoStop}%
\bibitem [{\citenamefont {Chianese}\ \emph {et~al.}(2024)\citenamefont
  {Chianese}, \citenamefont {Boccia}, \citenamefont {Iocco}, \citenamefont
  {Miele},\ and\ \citenamefont {Saviano}}]{Chianese:2024rsn}%
  \BibitemOpen
  \bibfield  {author} {\bibinfo {author} {\bibfnamefont {M.}~\bibnamefont
  {Chianese}}, \bibinfo {author} {\bibfnamefont {A.}~\bibnamefont {Boccia}},
  \bibinfo {author} {\bibfnamefont {F.}~\bibnamefont {Iocco}}, \bibinfo
  {author} {\bibfnamefont {G.}~\bibnamefont {Miele}}, \ and\ \bibinfo {author}
  {\bibfnamefont {N.}~\bibnamefont {Saviano}},\ }\href@noop {} {\  (\bibinfo
  {year} {2024})},\ \Eprint {http://arxiv.org/abs/2410.07604} {arXiv:2410.07604
  [astro-ph.HE]} \BibitemShut {NoStop}%
\bibitem [{\citenamefont {Barker}\ \emph {et~al.}(2024)\citenamefont {Barker},
  \citenamefont {Gladwyn},\ and\ \citenamefont {Zell}}]{Barker:2024mpz}%
  \BibitemOpen
  \bibfield  {author} {\bibinfo {author} {\bibfnamefont {W.}~\bibnamefont
  {Barker}}, \bibinfo {author} {\bibfnamefont {B.}~\bibnamefont {Gladwyn}}, \
  and\ \bibinfo {author} {\bibfnamefont {S.}~\bibnamefont {Zell}},\ }\href@noop
  {} {\  (\bibinfo {year} {2024})},\ \Eprint {http://arxiv.org/abs/2410.11948}
  {arXiv:2410.11948 [astro-ph.CO]} \BibitemShut {NoStop}%
\bibitem [{\citenamefont {Borah}\ and\ \citenamefont
  {Das}(2024)}]{Borah:2024bcr}%
  \BibitemOpen
  \bibfield  {author} {\bibinfo {author} {\bibfnamefont {D.}~\bibnamefont
  {Borah}}\ and\ \bibinfo {author} {\bibfnamefont {N.}~\bibnamefont {Das}},\
  }\href@noop {} {\  (\bibinfo {year} {2024})},\ \Eprint
  {http://arxiv.org/abs/2410.16403} {arXiv:2410.16403 [hep-ph]} \BibitemShut
  {NoStop}%
\bibitem [{\citenamefont {Huang}\ \emph
  {et~al.}(2024{\natexlab{e}})\citenamefont {Huang}, \citenamefont {Jiang},
  \citenamefont {He}, \citenamefont {Wang},\ and\ \citenamefont
  {Piao}}]{Hai-LongHuang:2024gtx}%
  \BibitemOpen
  \bibfield  {author} {\bibinfo {author} {\bibfnamefont {H.-L.}\ \bibnamefont
  {Huang}}, \bibinfo {author} {\bibfnamefont {J.-Q.}\ \bibnamefont {Jiang}},
  \bibinfo {author} {\bibfnamefont {J.}~\bibnamefont {He}}, \bibinfo {author}
  {\bibfnamefont {Y.-T.}\ \bibnamefont {Wang}}, \ and\ \bibinfo {author}
  {\bibfnamefont {Y.-S.}\ \bibnamefont {Piao}},\ }\href@noop {} {\  (\bibinfo
  {year} {2024}{\natexlab{e}})},\ \Eprint {http://arxiv.org/abs/2410.20663}
  {arXiv:2410.20663 [astro-ph.GA]} \BibitemShut {NoStop}%
\bibitem [{\citenamefont {Khlopov}(2010)}]{Khlopov:2008qy}%
  \BibitemOpen
  \bibfield  {author} {\bibinfo {author} {\bibfnamefont {M.~Y.}\ \bibnamefont
  {Khlopov}},\ }\href {\doibase 10.1088/1674-4527/10/6/001} {\bibfield
  {journal} {\bibinfo  {journal} {Res. Astron. Astrophys.}\ }\textbf {\bibinfo
  {volume} {10}},\ \bibinfo {pages} {495} (\bibinfo {year} {2010})},\ \Eprint
  {http://arxiv.org/abs/0801.0116} {arXiv:0801.0116 [astro-ph]} \BibitemShut
  {NoStop}%
\bibitem [{\citenamefont {Carr}\ \emph
  {et~al.}(2016{\natexlab{a}})\citenamefont {Carr}, \citenamefont {Kuhnel},\
  and\ \citenamefont {Sandstad}}]{Carr:2016drx}%
  \BibitemOpen
  \bibfield  {author} {\bibinfo {author} {\bibfnamefont {B.}~\bibnamefont
  {Carr}}, \bibinfo {author} {\bibfnamefont {F.}~\bibnamefont {Kuhnel}}, \ and\
  \bibinfo {author} {\bibfnamefont {M.}~\bibnamefont {Sandstad}},\ }\href
  {\doibase 10.1103/PhysRevD.94.083504} {\bibfield  {journal} {\bibinfo
  {journal} {Phys. Rev. D}\ }\textbf {\bibinfo {volume} {94}},\ \bibinfo
  {pages} {083504} (\bibinfo {year} {2016}{\natexlab{a}})},\ \Eprint
  {http://arxiv.org/abs/1607.06077} {arXiv:1607.06077 [astro-ph.CO]}
  \BibitemShut {NoStop}%
\bibitem [{\citenamefont {Green}\ and\ \citenamefont
  {Kavanagh}(2021)}]{Green:2020jor}%
  \BibitemOpen
  \bibfield  {author} {\bibinfo {author} {\bibfnamefont {A.~M.}\ \bibnamefont
  {Green}}\ and\ \bibinfo {author} {\bibfnamefont {B.~J.}\ \bibnamefont
  {Kavanagh}},\ }\href {\doibase 10.1088/1361-6471/abc534} {\bibfield
  {journal} {\bibinfo  {journal} {J. Phys. G}\ }\textbf {\bibinfo {volume}
  {48}},\ \bibinfo {pages} {043001} (\bibinfo {year} {2021})},\ \Eprint
  {http://arxiv.org/abs/2007.10722} {arXiv:2007.10722 [astro-ph.CO]}
  \BibitemShut {NoStop}%
\bibitem [{\citenamefont {Carr}\ and\ \citenamefont
  {Kuhnel}(2020)}]{Carr:2020xqk}%
  \BibitemOpen
  \bibfield  {author} {\bibinfo {author} {\bibfnamefont {B.}~\bibnamefont
  {Carr}}\ and\ \bibinfo {author} {\bibfnamefont {F.}~\bibnamefont {Kuhnel}},\
  }\href {\doibase 10.1146/annurev-nucl-050520-125911} {\bibfield  {journal}
  {\bibinfo  {journal} {Ann. Rev. Nucl. Part. Sci.}\ }\textbf {\bibinfo
  {volume} {70}},\ \bibinfo {pages} {355} (\bibinfo {year} {2020})},\ \Eprint
  {http://arxiv.org/abs/2006.02838} {arXiv:2006.02838 [astro-ph.CO]}
  \BibitemShut {NoStop}%
\bibitem [{\citenamefont {Villanueva-Domingo}\ \emph
  {et~al.}(2021)\citenamefont {Villanueva-Domingo}, \citenamefont {Mena},\ and\
  \citenamefont {Palomares-Ruiz}}]{Villanueva-Domingo:2021spv}%
  \BibitemOpen
  \bibfield  {author} {\bibinfo {author} {\bibfnamefont {P.}~\bibnamefont
  {Villanueva-Domingo}}, \bibinfo {author} {\bibfnamefont {O.}~\bibnamefont
  {Mena}}, \ and\ \bibinfo {author} {\bibfnamefont {S.}~\bibnamefont
  {Palomares-Ruiz}},\ }\href {\doibase 10.3389/fspas.2021.681084} {\bibfield
  {journal} {\bibinfo  {journal} {Front. Astron. Space Sci.}\ }\textbf
  {\bibinfo {volume} {8}},\ \bibinfo {pages} {87} (\bibinfo {year} {2021})},\
  \Eprint {http://arxiv.org/abs/2103.12087} {arXiv:2103.12087 [astro-ph.CO]}
  \BibitemShut {NoStop}%
\bibitem [{\citenamefont {Carr}\ and\ \citenamefont
  {Kuhnel}(2022)}]{Carr:2021bzv}%
  \BibitemOpen
  \bibfield  {author} {\bibinfo {author} {\bibfnamefont {B.}~\bibnamefont
  {Carr}}\ and\ \bibinfo {author} {\bibfnamefont {F.}~\bibnamefont {Kuhnel}},\
  }\href {\doibase 10.21468/SciPostPhysLectNotes.48} {\bibfield  {journal}
  {\bibinfo  {journal} {SciPost Phys. Lect. Notes}\ }\textbf {\bibinfo {volume}
  {48}},\ \bibinfo {pages} {1} (\bibinfo {year} {2022})},\ \Eprint
  {http://arxiv.org/abs/2110.02821} {arXiv:2110.02821 [astro-ph.CO]}
  \BibitemShut {NoStop}%
\bibitem [{\citenamefont {Bird}\ \emph {et~al.}(2023)\citenamefont {Bird} \emph
  {et~al.}}]{Bird:2022wvk}%
  \BibitemOpen
  \bibfield  {author} {\bibinfo {author} {\bibfnamefont {S.}~\bibnamefont
  {Bird}} \emph {et~al.},\ }\href {\doibase 10.1016/j.dark.2023.101231}
  {\bibfield  {journal} {\bibinfo  {journal} {Phys. Dark Univ.}\ }\textbf
  {\bibinfo {volume} {41}},\ \bibinfo {pages} {101231} (\bibinfo {year}
  {2023})},\ \Eprint {http://arxiv.org/abs/2203.08967} {arXiv:2203.08967
  [hep-ph]} \BibitemShut {NoStop}%
\bibitem [{\citenamefont {Carr}\ \emph {et~al.}(2024)\citenamefont {Carr},
  \citenamefont {Clesse}, \citenamefont {Garcia-Bellido}, \citenamefont
  {Hawkins},\ and\ \citenamefont {Kuhnel}}]{Carr:2023tpt}%
  \BibitemOpen
  \bibfield  {author} {\bibinfo {author} {\bibfnamefont {B.}~\bibnamefont
  {Carr}}, \bibinfo {author} {\bibfnamefont {S.}~\bibnamefont {Clesse}},
  \bibinfo {author} {\bibfnamefont {J.}~\bibnamefont {Garcia-Bellido}},
  \bibinfo {author} {\bibfnamefont {M.}~\bibnamefont {Hawkins}}, \ and\
  \bibinfo {author} {\bibfnamefont {F.}~\bibnamefont {Kuhnel}},\ }\href
  {\doibase 10.1016/j.physrep.2023.11.005} {\bibfield  {journal} {\bibinfo
  {journal} {Phys. Rept.}\ }\textbf {\bibinfo {volume} {1054}},\ \bibinfo
  {pages} {1} (\bibinfo {year} {2024})},\ \Eprint
  {http://arxiv.org/abs/2306.03903} {arXiv:2306.03903 [astro-ph.CO]}
  \BibitemShut {NoStop}%
\bibitem [{\citenamefont {Arbey}(2024)}]{Arbey:2024ujg}%
  \BibitemOpen
  \bibfield  {author} {\bibinfo {author} {\bibfnamefont {A.}~\bibnamefont
  {Arbey}},\ }in\ \href@noop {} {\emph {\bibinfo {booktitle} {{58th Rencontres
  de Moriond on Very High Energy Phenomena in the Universe}}}}\ (\bibinfo
  {year} {2024})\ \Eprint {http://arxiv.org/abs/2405.08624} {arXiv:2405.08624
  [gr-qc]} \BibitemShut {NoStop}%
\bibitem [{\citenamefont {Choudhury}\ and\ \citenamefont
  {Sami}(2024)}]{Choudhury:2024aji}%
  \BibitemOpen
  \bibfield  {author} {\bibinfo {author} {\bibfnamefont {S.}~\bibnamefont
  {Choudhury}}\ and\ \bibinfo {author} {\bibfnamefont {M.}~\bibnamefont
  {Sami}},\ }\href@noop {} {\  (\bibinfo {year} {2024})},\ \Eprint
  {http://arxiv.org/abs/2407.17006} {arXiv:2407.17006 [gr-qc]} \BibitemShut
  {NoStop}%
\bibitem [{\citenamefont {Katz}\ \emph {et~al.}(2018)\citenamefont {Katz},
  \citenamefont {Kopp}, \citenamefont {Sibiryakov},\ and\ \citenamefont
  {Xue}}]{Katz:2018zrn}%
  \BibitemOpen
  \bibfield  {author} {\bibinfo {author} {\bibfnamefont {A.}~\bibnamefont
  {Katz}}, \bibinfo {author} {\bibfnamefont {J.}~\bibnamefont {Kopp}}, \bibinfo
  {author} {\bibfnamefont {S.}~\bibnamefont {Sibiryakov}}, \ and\ \bibinfo
  {author} {\bibfnamefont {W.}~\bibnamefont {Xue}},\ }\href {\doibase
  10.1088/1475-7516/2018/12/005} {\bibfield  {journal} {\bibinfo  {journal}
  {JCAP}\ }\textbf {\bibinfo {volume} {12}},\ \bibinfo {pages} {005} (\bibinfo
  {year} {2018})},\ \Eprint {http://arxiv.org/abs/1807.11495} {arXiv:1807.11495
  [astro-ph.CO]} \BibitemShut {NoStop}%
\bibitem [{\citenamefont {Bai}\ and\ \citenamefont
  {Orlofsky}(2019)}]{Bai:2018bej}%
  \BibitemOpen
  \bibfield  {author} {\bibinfo {author} {\bibfnamefont {Y.}~\bibnamefont
  {Bai}}\ and\ \bibinfo {author} {\bibfnamefont {N.}~\bibnamefont {Orlofsky}},\
  }\href {\doibase 10.1103/PhysRevD.99.123019} {\bibfield  {journal} {\bibinfo
  {journal} {Phys. Rev. D}\ }\textbf {\bibinfo {volume} {99}},\ \bibinfo
  {pages} {123019} (\bibinfo {year} {2019})},\ \Eprint
  {http://arxiv.org/abs/1812.01427} {arXiv:1812.01427 [astro-ph.HE]}
  \BibitemShut {NoStop}%
\bibitem [{\citenamefont {Smyth}\ \emph {et~al.}(2020)\citenamefont {Smyth},
  \citenamefont {Profumo}, \citenamefont {English}, \citenamefont {Jeltema},
  \citenamefont {McKinnon},\ and\ \citenamefont
  {Guhathakurta}}]{Smyth:2019whb}%
  \BibitemOpen
  \bibfield  {author} {\bibinfo {author} {\bibfnamefont {N.}~\bibnamefont
  {Smyth}}, \bibinfo {author} {\bibfnamefont {S.}~\bibnamefont {Profumo}},
  \bibinfo {author} {\bibfnamefont {S.}~\bibnamefont {English}}, \bibinfo
  {author} {\bibfnamefont {T.}~\bibnamefont {Jeltema}}, \bibinfo {author}
  {\bibfnamefont {K.}~\bibnamefont {McKinnon}}, \ and\ \bibinfo {author}
  {\bibfnamefont {P.}~\bibnamefont {Guhathakurta}},\ }\href {\doibase
  10.1103/PhysRevD.101.063005} {\bibfield  {journal} {\bibinfo  {journal}
  {Phys. Rev. D}\ }\textbf {\bibinfo {volume} {101}},\ \bibinfo {pages}
  {063005} (\bibinfo {year} {2020})},\ \Eprint
  {http://arxiv.org/abs/1910.01285} {arXiv:1910.01285 [astro-ph.CO]}
  \BibitemShut {NoStop}%
\bibitem [{\citenamefont {Coogan}\ \emph {et~al.}(2021)\citenamefont {Coogan},
  \citenamefont {Morrison},\ and\ \citenamefont {Profumo}}]{Coogan:2020tuf}%
  \BibitemOpen
  \bibfield  {author} {\bibinfo {author} {\bibfnamefont {A.}~\bibnamefont
  {Coogan}}, \bibinfo {author} {\bibfnamefont {L.}~\bibnamefont {Morrison}}, \
  and\ \bibinfo {author} {\bibfnamefont {S.}~\bibnamefont {Profumo}},\ }\href
  {\doibase 10.1103/PhysRevLett.126.171101} {\bibfield  {journal} {\bibinfo
  {journal} {Phys. Rev. Lett.}\ }\textbf {\bibinfo {volume} {126}},\ \bibinfo
  {pages} {171101} (\bibinfo {year} {2021})},\ \Eprint
  {http://arxiv.org/abs/2010.04797} {arXiv:2010.04797 [astro-ph.CO]}
  \BibitemShut {NoStop}%
\bibitem [{\citenamefont {Ray}\ \emph {et~al.}(2021)\citenamefont {Ray},
  \citenamefont {Laha}, \citenamefont {Mu\~noz},\ and\ \citenamefont
  {Caputo}}]{Ray:2021mxu}%
  \BibitemOpen
  \bibfield  {author} {\bibinfo {author} {\bibfnamefont {A.}~\bibnamefont
  {Ray}}, \bibinfo {author} {\bibfnamefont {R.}~\bibnamefont {Laha}}, \bibinfo
  {author} {\bibfnamefont {J.~B.}\ \bibnamefont {Mu\~noz}}, \ and\ \bibinfo
  {author} {\bibfnamefont {R.}~\bibnamefont {Caputo}},\ }\href {\doibase
  10.1103/PhysRevD.104.023516} {\bibfield  {journal} {\bibinfo  {journal}
  {Phys. Rev. D}\ }\textbf {\bibinfo {volume} {104}},\ \bibinfo {pages}
  {023516} (\bibinfo {year} {2021})},\ \Eprint
  {http://arxiv.org/abs/2102.06714} {arXiv:2102.06714 [astro-ph.CO]}
  \BibitemShut {NoStop}%
\bibitem [{\citenamefont {Auffinger}(2022)}]{Auffinger:2022dic}%
  \BibitemOpen
  \bibfield  {author} {\bibinfo {author} {\bibfnamefont {J.}~\bibnamefont
  {Auffinger}},\ }\href {\doibase 10.1140/epjc/s10052-022-10199-y} {\bibfield
  {journal} {\bibinfo  {journal} {Eur. Phys. J. C}\ }\textbf {\bibinfo {volume}
  {82}},\ \bibinfo {pages} {384} (\bibinfo {year} {2022})},\ \Eprint
  {http://arxiv.org/abs/2201.01265} {arXiv:2201.01265 [astro-ph.HE]}
  \BibitemShut {NoStop}%
\bibitem [{\citenamefont {Ghosh}\ and\ \citenamefont
  {Mishra}(2024)}]{Ghosh:2022okj}%
  \BibitemOpen
  \bibfield  {author} {\bibinfo {author} {\bibfnamefont {D.}~\bibnamefont
  {Ghosh}}\ and\ \bibinfo {author} {\bibfnamefont {A.~K.}\ \bibnamefont
  {Mishra}},\ }\href {\doibase 10.1103/PhysRevD.109.043537} {\bibfield
  {journal} {\bibinfo  {journal} {Phys. Rev. D}\ }\textbf {\bibinfo {volume}
  {109}},\ \bibinfo {pages} {043537} (\bibinfo {year} {2024})},\ \Eprint
  {http://arxiv.org/abs/2208.14279} {arXiv:2208.14279 [hep-ph]} \BibitemShut
  {NoStop}%
\bibitem [{\citenamefont {Miller}\ \emph {et~al.}(2022)\citenamefont {Miller},
  \citenamefont {Aggarwal}, \citenamefont {Clesse},\ and\ \citenamefont
  {De~Lillo}}]{Miller:2021knj}%
  \BibitemOpen
  \bibfield  {author} {\bibinfo {author} {\bibfnamefont {A.~L.}\ \bibnamefont
  {Miller}}, \bibinfo {author} {\bibfnamefont {N.}~\bibnamefont {Aggarwal}},
  \bibinfo {author} {\bibfnamefont {S.}~\bibnamefont {Clesse}}, \ and\ \bibinfo
  {author} {\bibfnamefont {F.}~\bibnamefont {De~Lillo}},\ }\href {\doibase
  10.1103/PhysRevD.105.062008} {\bibfield  {journal} {\bibinfo  {journal}
  {Phys. Rev. D}\ }\textbf {\bibinfo {volume} {105}},\ \bibinfo {pages}
  {062008} (\bibinfo {year} {2022})},\ \Eprint
  {http://arxiv.org/abs/2110.06188} {arXiv:2110.06188 [gr-qc]} \BibitemShut
  {NoStop}%
\bibitem [{\citenamefont {Branco}\ \emph {et~al.}(2023)\citenamefont {Branco},
  \citenamefont {Ferreira},\ and\ \citenamefont {Rosa}}]{Branco:2023frw}%
  \BibitemOpen
  \bibfield  {author} {\bibinfo {author} {\bibfnamefont {N.~P.}\ \bibnamefont
  {Branco}}, \bibinfo {author} {\bibfnamefont {R.~Z.}\ \bibnamefont
  {Ferreira}}, \ and\ \bibinfo {author} {\bibfnamefont {J.~a.~G.}\ \bibnamefont
  {Rosa}},\ }\href {\doibase 10.1088/1475-7516/2023/04/003} {\bibfield
  {journal} {\bibinfo  {journal} {JCAP}\ }\textbf {\bibinfo {volume} {04}},\
  \bibinfo {pages} {003} (\bibinfo {year} {2023})},\ \Eprint
  {http://arxiv.org/abs/2301.01780} {arXiv:2301.01780 [hep-ph]} \BibitemShut
  {NoStop}%
\bibitem [{\citenamefont {Bertrand}\ \emph {et~al.}(2023)\citenamefont
  {Bertrand}, \citenamefont {Cuadrat-Grzybowski}, \citenamefont {Defraigne},
  \citenamefont {Van~Camp},\ and\ \citenamefont {Clesse}}]{Bertrand:2023zkl}%
  \BibitemOpen
  \bibfield  {author} {\bibinfo {author} {\bibfnamefont {B.}~\bibnamefont
  {Bertrand}}, \bibinfo {author} {\bibfnamefont {M.}~\bibnamefont
  {Cuadrat-Grzybowski}}, \bibinfo {author} {\bibfnamefont {P.}~\bibnamefont
  {Defraigne}}, \bibinfo {author} {\bibfnamefont {M.}~\bibnamefont {Van~Camp}},
  \ and\ \bibinfo {author} {\bibfnamefont {S.}~\bibnamefont {Clesse}}\
  }(\bibinfo {year} {2023})\ \Eprint {http://arxiv.org/abs/2312.14520}
  {arXiv:2312.14520 [astro-ph.CO]} \BibitemShut {NoStop}%
\bibitem [{\citenamefont {Tran}\ \emph {et~al.}(2024)\citenamefont {Tran},
  \citenamefont {Geller}, \citenamefont {Lehmann},\ and\ \citenamefont
  {Kaiser}}]{Tran:2023jci}%
  \BibitemOpen
  \bibfield  {author} {\bibinfo {author} {\bibfnamefont {T.~X.}\ \bibnamefont
  {Tran}}, \bibinfo {author} {\bibfnamefont {S.~R.}\ \bibnamefont {Geller}},
  \bibinfo {author} {\bibfnamefont {B.~V.}\ \bibnamefont {Lehmann}}, \ and\
  \bibinfo {author} {\bibfnamefont {D.~I.}\ \bibnamefont {Kaiser}},\ }\href
  {\doibase 10.1103/PhysRevD.110.063533} {\bibfield  {journal} {\bibinfo
  {journal} {Phys. Rev. D}\ }\textbf {\bibinfo {volume} {110}},\ \bibinfo
  {pages} {063533} (\bibinfo {year} {2024})},\ \Eprint
  {http://arxiv.org/abs/2312.17217} {arXiv:2312.17217 [astro-ph.CO]}
  \BibitemShut {NoStop}%
\bibitem [{\citenamefont {Gorton}\ and\ \citenamefont
  {Green}(2024)}]{Gorton:2024cdm}%
  \BibitemOpen
  \bibfield  {author} {\bibinfo {author} {\bibfnamefont {M.}~\bibnamefont
  {Gorton}}\ and\ \bibinfo {author} {\bibfnamefont {A.~M.}\ \bibnamefont
  {Green}},\ }\href {\doibase 10.21468/SciPostPhys.17.2.032} {\bibfield
  {journal} {\bibinfo  {journal} {SciPost Phys.}\ }\textbf {\bibinfo {volume}
  {17}},\ \bibinfo {pages} {032} (\bibinfo {year} {2024})},\ \Eprint
  {http://arxiv.org/abs/2403.03839} {arXiv:2403.03839 [astro-ph.CO]}
  \BibitemShut {NoStop}%
\bibitem [{\citenamefont {Dent}\ \emph {et~al.}(2024)\citenamefont {Dent},
  \citenamefont {Dutta},\ and\ \citenamefont {Xu}}]{Dent:2024yje}%
  \BibitemOpen
  \bibfield  {author} {\bibinfo {author} {\bibfnamefont {J.~B.}\ \bibnamefont
  {Dent}}, \bibinfo {author} {\bibfnamefont {B.}~\bibnamefont {Dutta}}, \ and\
  \bibinfo {author} {\bibfnamefont {T.}~\bibnamefont {Xu}},\ }\href@noop {} {\
  (\bibinfo {year} {2024})},\ \Eprint {http://arxiv.org/abs/2404.02956}
  {arXiv:2404.02956 [hep-ph]} \BibitemShut {NoStop}%
\bibitem [{\citenamefont {Tamta}\ \emph {et~al.}(2024)\citenamefont {Tamta},
  \citenamefont {Raj},\ and\ \citenamefont {Sharma}}]{Tamta:2024pow}%
  \BibitemOpen
  \bibfield  {author} {\bibinfo {author} {\bibfnamefont {M.}~\bibnamefont
  {Tamta}}, \bibinfo {author} {\bibfnamefont {N.}~\bibnamefont {Raj}}, \ and\
  \bibinfo {author} {\bibfnamefont {P.}~\bibnamefont {Sharma}},\ }\href@noop {}
  {\  (\bibinfo {year} {2024})},\ \Eprint {http://arxiv.org/abs/2405.20365}
  {arXiv:2405.20365 [astro-ph.HE]} \BibitemShut {NoStop}%
\bibitem [{\citenamefont {Tinyakov}(2024)}]{Tinyakov:2024mcy}%
  \BibitemOpen
  \bibfield  {author} {\bibinfo {author} {\bibfnamefont {P.}~\bibnamefont
  {Tinyakov}},\ }\href@noop {} {\  (\bibinfo {year} {2024})},\ \Eprint
  {http://arxiv.org/abs/2406.03114} {arXiv:2406.03114 [astro-ph.CO]}
  \BibitemShut {NoStop}%
\bibitem [{\citenamefont {Loeb}(2024)}]{Loeb:2024tcc}%
  \BibitemOpen
  \bibfield  {author} {\bibinfo {author} {\bibfnamefont {A.}~\bibnamefont
  {Loeb}},\ }\href {\doibase 10.3847/2515-5172/ad739e} {\bibfield  {journal}
  {\bibinfo  {journal} {Res. Notes AAS}\ }\textbf {\bibinfo {volume} {8}},\
  \bibinfo {pages} {211} (\bibinfo {year} {2024})},\ \Eprint
  {http://arxiv.org/abs/2408.10799} {arXiv:2408.10799 [hep-ph]} \BibitemShut
  {NoStop}%
\bibitem [{\citenamefont {Ansoldi}(2008)}]{Ansoldi:2008jw}%
  \BibitemOpen
  \bibfield  {author} {\bibinfo {author} {\bibfnamefont {S.}~\bibnamefont
  {Ansoldi}},\ }in\ \href@noop {} {\emph {\bibinfo {booktitle} {{Conference on
  Black Holes and Naked Singularities}}}}\ (\bibinfo {year} {2008})\ \Eprint
  {http://arxiv.org/abs/0802.0330} {arXiv:0802.0330 [gr-qc]} \BibitemShut
  {NoStop}%
\bibitem [{\citenamefont {Nicolini}(2009)}]{Nicolini:2008aj}%
  \BibitemOpen
  \bibfield  {author} {\bibinfo {author} {\bibfnamefont {P.}~\bibnamefont
  {Nicolini}},\ }\href {\doibase 10.1142/S0217751X09043353} {\bibfield
  {journal} {\bibinfo  {journal} {Int. J. Mod. Phys. A}\ }\textbf {\bibinfo
  {volume} {24}},\ \bibinfo {pages} {1229} (\bibinfo {year} {2009})},\ \Eprint
  {http://arxiv.org/abs/0807.1939} {arXiv:0807.1939 [hep-th]} \BibitemShut
  {NoStop}%
\bibitem [{\citenamefont {Sebastiani}\ and\ \citenamefont
  {Zerbini}(2022)}]{Sebastiani:2022wbz}%
  \BibitemOpen
  \bibfield  {author} {\bibinfo {author} {\bibfnamefont {L.}~\bibnamefont
  {Sebastiani}}\ and\ \bibinfo {author} {\bibfnamefont {S.}~\bibnamefont
  {Zerbini}},\ }\href {\doibase 10.3390/astronomy1020010} {\bibfield  {journal}
  {\bibinfo  {journal} {Astronomy}\ }\textbf {\bibinfo {volume} {1}},\ \bibinfo
  {pages} {99} (\bibinfo {year} {2022})},\ \Eprint
  {http://arxiv.org/abs/2206.03814} {arXiv:2206.03814 [gr-qc]} \BibitemShut
  {NoStop}%
\bibitem [{\citenamefont {Torres}(2022)}]{Torres:2022twv}%
  \BibitemOpen
  \bibfield  {author} {\bibinfo {author} {\bibfnamefont {R.}~\bibnamefont
  {Torres}},\ }\href@noop {} {\  (\bibinfo {year} {2022})},\ \Eprint
  {http://arxiv.org/abs/2208.12713} {arXiv:2208.12713 [gr-qc]} \BibitemShut
  {NoStop}%
\bibitem [{\citenamefont {Lan}\ \emph {et~al.}(2023)\citenamefont {Lan},
  \citenamefont {Yang}, \citenamefont {Guo},\ and\ \citenamefont
  {Miao}}]{Lan:2023cvz}%
  \BibitemOpen
  \bibfield  {author} {\bibinfo {author} {\bibfnamefont {C.}~\bibnamefont
  {Lan}}, \bibinfo {author} {\bibfnamefont {H.}~\bibnamefont {Yang}}, \bibinfo
  {author} {\bibfnamefont {Y.}~\bibnamefont {Guo}}, \ and\ \bibinfo {author}
  {\bibfnamefont {Y.-G.}\ \bibnamefont {Miao}},\ }\href {\doibase
  10.1007/s10773-023-05454-1} {\bibfield  {journal} {\bibinfo  {journal} {Int.
  J. Theor. Phys.}\ }\textbf {\bibinfo {volume} {62}},\ \bibinfo {pages} {202}
  (\bibinfo {year} {2023})},\ \Eprint {http://arxiv.org/abs/2303.11696}
  {arXiv:2303.11696 [gr-qc]} \BibitemShut {NoStop}%
\bibitem [{\citenamefont {Calz\`a}\ \emph
  {et~al.}(2024{\natexlab{a}})\citenamefont {Calz\`a}, \citenamefont
  {Pedrotti},\ and\ \citenamefont {Vagnozzi}}]{Calza:2024fzo}%
  \BibitemOpen
  \bibfield  {author} {\bibinfo {author} {\bibfnamefont {M.}~\bibnamefont
  {Calz\`a}}, \bibinfo {author} {\bibfnamefont {D.}~\bibnamefont {Pedrotti}}, \
  and\ \bibinfo {author} {\bibfnamefont {S.}~\bibnamefont {Vagnozzi}}\
  }(\bibinfo {year} {2024})\ \Eprint {http://arxiv.org/abs/2409.02804}
  {arXiv:2409.02804 [gr-qc]} \BibitemShut {NoStop}%
\bibitem [{\citenamefont {Bardeen}(1968)}]{Bardeen:1968ghw}%
  \BibitemOpen
  \bibfield  {author} {\bibinfo {author} {\bibfnamefont {J.}~\bibnamefont
  {Bardeen}},\ }in\ \href@noop {} {\emph {\bibinfo {booktitle} {{Proceedings,
  5th International Conference on Gravitation and the theory of relativity:
  Tbilisi, USSR, December 9-13, 1968}}}}\ (\bibinfo {year} {1968})\ p.\
  \bibinfo {pages} {174}\BibitemShut {NoStop}%
\bibitem [{\citenamefont {Hayward}(2006)}]{Hayward:2005gi}%
  \BibitemOpen
  \bibfield  {author} {\bibinfo {author} {\bibfnamefont {S.~A.}\ \bibnamefont
  {Hayward}},\ }\href {\doibase 10.1103/PhysRevLett.96.031103} {\bibfield
  {journal} {\bibinfo  {journal} {Phys. Rev. Lett.}\ }\textbf {\bibinfo
  {volume} {96}},\ \bibinfo {pages} {031103} (\bibinfo {year} {2006})},\
  \Eprint {http://arxiv.org/abs/gr-qc/0506126} {arXiv:gr-qc/0506126}
  \BibitemShut {NoStop}%
\bibitem [{\citenamefont {Morris}\ and\ \citenamefont
  {Thorne}(1988)}]{Morris:1988cz}%
  \BibitemOpen
  \bibfield  {author} {\bibinfo {author} {\bibfnamefont {M.~S.}\ \bibnamefont
  {Morris}}\ and\ \bibinfo {author} {\bibfnamefont {K.~S.}\ \bibnamefont
  {Thorne}},\ }\href {\doibase 10.1119/1.15620} {\bibfield  {journal} {\bibinfo
   {journal} {Am. J. Phys.}\ }\textbf {\bibinfo {volume} {56}},\ \bibinfo
  {pages} {395} (\bibinfo {year} {1988})}\BibitemShut {NoStop}%
\bibitem [{\citenamefont {Ashtekar}(1986)}]{Ashtekar:1986yd}%
  \BibitemOpen
  \bibfield  {author} {\bibinfo {author} {\bibfnamefont {A.}~\bibnamefont
  {Ashtekar}},\ }\href {\doibase 10.1103/PhysRevLett.57.2244} {\bibfield
  {journal} {\bibinfo  {journal} {Phys. Rev. Lett.}\ }\textbf {\bibinfo
  {volume} {57}},\ \bibinfo {pages} {2244} (\bibinfo {year}
  {1986})}\BibitemShut {NoStop}%
\bibitem [{\citenamefont {Rovelli}(1998)}]{Rovelli:1997yv}%
  \BibitemOpen
  \bibfield  {author} {\bibinfo {author} {\bibfnamefont {C.}~\bibnamefont
  {Rovelli}},\ }\href {\doibase 10.12942/lrr-1998-1} {\bibfield  {journal}
  {\bibinfo  {journal} {Living Rev. Rel.}\ }\textbf {\bibinfo {volume} {1}},\
  \bibinfo {pages} {1} (\bibinfo {year} {1998})},\ \Eprint
  {http://arxiv.org/abs/gr-qc/9710008} {arXiv:gr-qc/9710008} \BibitemShut
  {NoStop}%
\bibitem [{\citenamefont {Ashtekar}\ and\ \citenamefont
  {Bianchi}(2021)}]{Ashtekar:2021kfp}%
  \BibitemOpen
  \bibfield  {author} {\bibinfo {author} {\bibfnamefont {A.}~\bibnamefont
  {Ashtekar}}\ and\ \bibinfo {author} {\bibfnamefont {E.}~\bibnamefont
  {Bianchi}},\ }\href {\doibase 10.1088/1361-6633/abed91} {\bibfield  {journal}
  {\bibinfo  {journal} {Rept. Prog. Phys.}\ }\textbf {\bibinfo {volume} {84}},\
  \bibinfo {pages} {042001} (\bibinfo {year} {2021})},\ \Eprint
  {http://arxiv.org/abs/2104.04394} {arXiv:2104.04394 [gr-qc]} \BibitemShut
  {NoStop}%
\bibitem [{\citenamefont {Bojowald}(2001)}]{Bojowald:2001xe}%
  \BibitemOpen
  \bibfield  {author} {\bibinfo {author} {\bibfnamefont {M.}~\bibnamefont
  {Bojowald}},\ }\href {\doibase 10.1103/PhysRevLett.86.5227} {\bibfield
  {journal} {\bibinfo  {journal} {Phys. Rev. Lett.}\ }\textbf {\bibinfo
  {volume} {86}},\ \bibinfo {pages} {5227} (\bibinfo {year} {2001})},\ \Eprint
  {http://arxiv.org/abs/gr-qc/0102069} {arXiv:gr-qc/0102069} \BibitemShut
  {NoStop}%
\bibitem [{\citenamefont {Bojowald}(2005)}]{Bojowald:2005epg}%
  \BibitemOpen
  \bibfield  {author} {\bibinfo {author} {\bibfnamefont {M.}~\bibnamefont
  {Bojowald}},\ }\href {\doibase 10.12942/lrr-2005-11} {\bibfield  {journal}
  {\bibinfo  {journal} {Living Rev. Rel.}\ }\textbf {\bibinfo {volume} {8}},\
  \bibinfo {pages} {11} (\bibinfo {year} {2005})},\ \Eprint
  {http://arxiv.org/abs/gr-qc/0601085} {arXiv:gr-qc/0601085} \BibitemShut
  {NoStop}%
\bibitem [{\citenamefont {Modesto}(2008)}]{Modesto:2006qh}%
  \BibitemOpen
  \bibfield  {author} {\bibinfo {author} {\bibfnamefont {L.}~\bibnamefont
  {Modesto}},\ }\href {\doibase 10.1007/s10773-007-9458-3} {\bibfield
  {journal} {\bibinfo  {journal} {Int. J. Theor. Phys.}\ }\textbf {\bibinfo
  {volume} {47}},\ \bibinfo {pages} {357} (\bibinfo {year} {2008})},\ \Eprint
  {http://arxiv.org/abs/gr-qc/0610074} {arXiv:gr-qc/0610074} \BibitemShut
  {NoStop}%
\bibitem [{\citenamefont {Engle}\ \emph {et~al.}(2008)\citenamefont {Engle},
  \citenamefont {Livine}, \citenamefont {Pereira},\ and\ \citenamefont
  {Rovelli}}]{Engle:2007wy}%
  \BibitemOpen
  \bibfield  {author} {\bibinfo {author} {\bibfnamefont {J.}~\bibnamefont
  {Engle}}, \bibinfo {author} {\bibfnamefont {E.}~\bibnamefont {Livine}},
  \bibinfo {author} {\bibfnamefont {R.}~\bibnamefont {Pereira}}, \ and\
  \bibinfo {author} {\bibfnamefont {C.}~\bibnamefont {Rovelli}},\ }\href
  {\doibase 10.1016/j.nuclphysb.2008.02.018} {\bibfield  {journal} {\bibinfo
  {journal} {Nucl. Phys. B}\ }\textbf {\bibinfo {volume} {799}},\ \bibinfo
  {pages} {136} (\bibinfo {year} {2008})},\ \Eprint
  {http://arxiv.org/abs/0711.0146} {arXiv:0711.0146 [gr-qc]} \BibitemShut
  {NoStop}%
\bibitem [{\citenamefont {Modesto}(2009)}]{Modesto:2008jz}%
  \BibitemOpen
  \bibfield  {author} {\bibinfo {author} {\bibfnamefont {L.}~\bibnamefont
  {Modesto}},\ }\href {\doibase 10.1088/0264-9381/26/24/242002} {\bibfield
  {journal} {\bibinfo  {journal} {Class. Quant. Grav.}\ }\textbf {\bibinfo
  {volume} {26}},\ \bibinfo {pages} {242002} (\bibinfo {year} {2009})},\
  \Eprint {http://arxiv.org/abs/0812.2214} {arXiv:0812.2214 [gr-qc]}
  \BibitemShut {NoStop}%
\bibitem [{\citenamefont {Bianchi}\ \emph {et~al.}(2011)\citenamefont
  {Bianchi}, \citenamefont {Dona},\ and\ \citenamefont
  {Speziale}}]{Bianchi:2010gc}%
  \BibitemOpen
  \bibfield  {author} {\bibinfo {author} {\bibfnamefont {E.}~\bibnamefont
  {Bianchi}}, \bibinfo {author} {\bibfnamefont {P.}~\bibnamefont {Dona}}, \
  and\ \bibinfo {author} {\bibfnamefont {S.}~\bibnamefont {Speziale}},\ }\href
  {\doibase 10.1103/PhysRevD.83.044035} {\bibfield  {journal} {\bibinfo
  {journal} {Phys. Rev. D}\ }\textbf {\bibinfo {volume} {83}},\ \bibinfo
  {pages} {044035} (\bibinfo {year} {2011})},\ \Eprint
  {http://arxiv.org/abs/1009.3402} {arXiv:1009.3402 [gr-qc]} \BibitemShut
  {NoStop}%
\bibitem [{\citenamefont {Singh}\ and\ \citenamefont
  {Vidotto}(2011)}]{Singh:2010qa}%
  \BibitemOpen
  \bibfield  {author} {\bibinfo {author} {\bibfnamefont {P.}~\bibnamefont
  {Singh}}\ and\ \bibinfo {author} {\bibfnamefont {F.}~\bibnamefont
  {Vidotto}},\ }\href {\doibase 10.1103/PhysRevD.83.064027} {\bibfield
  {journal} {\bibinfo  {journal} {Phys. Rev. D}\ }\textbf {\bibinfo {volume}
  {83}},\ \bibinfo {pages} {064027} (\bibinfo {year} {2011})},\ \Eprint
  {http://arxiv.org/abs/1012.1307} {arXiv:1012.1307 [gr-qc]} \BibitemShut
  {NoStop}%
\bibitem [{\citenamefont {Bojowald}\ and\ \citenamefont
  {Paily}(2012)}]{Bojowald:2011aa}%
  \BibitemOpen
  \bibfield  {author} {\bibinfo {author} {\bibfnamefont {M.}~\bibnamefont
  {Bojowald}}\ and\ \bibinfo {author} {\bibfnamefont {G.~M.}\ \bibnamefont
  {Paily}},\ }\href {\doibase 10.1103/PhysRevD.86.104018} {\bibfield  {journal}
  {\bibinfo  {journal} {Phys. Rev. D}\ }\textbf {\bibinfo {volume} {86}},\
  \bibinfo {pages} {104018} (\bibinfo {year} {2012})},\ \Eprint
  {http://arxiv.org/abs/1112.1899} {arXiv:1112.1899 [gr-qc]} \BibitemShut
  {NoStop}%
\bibitem [{\citenamefont {Cailleteau}\ \emph {et~al.}(2012)\citenamefont
  {Cailleteau}, \citenamefont {Barrau}, \citenamefont {Grain},\ and\
  \citenamefont {Vidotto}}]{Cailleteau:2012fy}%
  \BibitemOpen
  \bibfield  {author} {\bibinfo {author} {\bibfnamefont {T.}~\bibnamefont
  {Cailleteau}}, \bibinfo {author} {\bibfnamefont {A.}~\bibnamefont {Barrau}},
  \bibinfo {author} {\bibfnamefont {J.}~\bibnamefont {Grain}}, \ and\ \bibinfo
  {author} {\bibfnamefont {F.}~\bibnamefont {Vidotto}},\ }\href {\doibase
  10.1103/PhysRevD.86.087301} {\bibfield  {journal} {\bibinfo  {journal} {Phys.
  Rev. D}\ }\textbf {\bibinfo {volume} {86}},\ \bibinfo {pages} {087301}
  (\bibinfo {year} {2012})},\ \Eprint {http://arxiv.org/abs/1206.6736}
  {arXiv:1206.6736 [gr-qc]} \BibitemShut {NoStop}%
\bibitem [{\citenamefont {Agullo}\ \emph {et~al.}(2013)\citenamefont {Agullo},
  \citenamefont {Ashtekar},\ and\ \citenamefont {Nelson}}]{Agullo:2013ai}%
  \BibitemOpen
  \bibfield  {author} {\bibinfo {author} {\bibfnamefont {I.}~\bibnamefont
  {Agullo}}, \bibinfo {author} {\bibfnamefont {A.}~\bibnamefont {Ashtekar}}, \
  and\ \bibinfo {author} {\bibfnamefont {W.}~\bibnamefont {Nelson}},\ }\href
  {\doibase 10.1088/0264-9381/30/8/085014} {\bibfield  {journal} {\bibinfo
  {journal} {Class. Quant. Grav.}\ }\textbf {\bibinfo {volume} {30}},\ \bibinfo
  {pages} {085014} (\bibinfo {year} {2013})},\ \Eprint
  {http://arxiv.org/abs/1302.0254} {arXiv:1302.0254 [gr-qc]} \BibitemShut
  {NoStop}%
\bibitem [{\citenamefont {Haggard}\ and\ \citenamefont
  {Rovelli}(2015)}]{Haggard:2014rza}%
  \BibitemOpen
  \bibfield  {author} {\bibinfo {author} {\bibfnamefont {H.~M.}\ \bibnamefont
  {Haggard}}\ and\ \bibinfo {author} {\bibfnamefont {C.}~\bibnamefont
  {Rovelli}},\ }\href {\doibase 10.1103/PhysRevD.92.104020} {\bibfield
  {journal} {\bibinfo  {journal} {Phys. Rev. D}\ }\textbf {\bibinfo {volume}
  {92}},\ \bibinfo {pages} {104020} (\bibinfo {year} {2015})},\ \Eprint
  {http://arxiv.org/abs/1407.0989} {arXiv:1407.0989 [gr-qc]} \BibitemShut
  {NoStop}%
\bibitem [{\citenamefont {Odintsov}\ and\ \citenamefont
  {Oikonomou}(2014)}]{Odintsov:2014gea}%
  \BibitemOpen
  \bibfield  {author} {\bibinfo {author} {\bibfnamefont {S.~D.}\ \bibnamefont
  {Odintsov}}\ and\ \bibinfo {author} {\bibfnamefont {V.~K.}\ \bibnamefont
  {Oikonomou}},\ }\href {\doibase 10.1103/PhysRevD.90.124083} {\bibfield
  {journal} {\bibinfo  {journal} {Phys. Rev. D}\ }\textbf {\bibinfo {volume}
  {90}},\ \bibinfo {pages} {124083} (\bibinfo {year} {2014})},\ \Eprint
  {http://arxiv.org/abs/1410.8183} {arXiv:1410.8183 [gr-qc]} \BibitemShut
  {NoStop}%
\bibitem [{\citenamefont {Odintsov}\ \emph {et~al.}(2015)\citenamefont
  {Odintsov}, \citenamefont {Oikonomou},\ and\ \citenamefont
  {Saridakis}}]{Odintsov:2015uca}%
  \BibitemOpen
  \bibfield  {author} {\bibinfo {author} {\bibfnamefont {S.~D.}\ \bibnamefont
  {Odintsov}}, \bibinfo {author} {\bibfnamefont {V.~K.}\ \bibnamefont
  {Oikonomou}}, \ and\ \bibinfo {author} {\bibfnamefont {E.~N.}\ \bibnamefont
  {Saridakis}},\ }\href {\doibase 10.1016/j.aop.2015.08.021} {\bibfield
  {journal} {\bibinfo  {journal} {Annals Phys.}\ }\textbf {\bibinfo {volume}
  {363}},\ \bibinfo {pages} {141} (\bibinfo {year} {2015})},\ \Eprint
  {http://arxiv.org/abs/1501.06591} {arXiv:1501.06591 [gr-qc]} \BibitemShut
  {NoStop}%
\bibitem [{\citenamefont {Haro}\ \emph {et~al.}(2015)\citenamefont {Haro},
  \citenamefont {Makarenko}, \citenamefont {Myagky}, \citenamefont {Odintsov},\
  and\ \citenamefont {Oikonomou}}]{Haro:2015oqa}%
  \BibitemOpen
  \bibfield  {author} {\bibinfo {author} {\bibfnamefont {J.}~\bibnamefont
  {Haro}}, \bibinfo {author} {\bibfnamefont {A.~N.}\ \bibnamefont {Makarenko}},
  \bibinfo {author} {\bibfnamefont {A.~N.}\ \bibnamefont {Myagky}}, \bibinfo
  {author} {\bibfnamefont {S.~D.}\ \bibnamefont {Odintsov}}, \ and\ \bibinfo
  {author} {\bibfnamefont {V.~K.}\ \bibnamefont {Oikonomou}},\ }\href {\doibase
  10.1103/PhysRevD.92.124026} {\bibfield  {journal} {\bibinfo  {journal} {Phys.
  Rev. D}\ }\textbf {\bibinfo {volume} {92}},\ \bibinfo {pages} {124026}
  (\bibinfo {year} {2015})},\ \Eprint {http://arxiv.org/abs/1506.08273}
  {arXiv:1506.08273 [gr-qc]} \BibitemShut {NoStop}%
\bibitem [{\citenamefont {Odintsov}\ and\ \citenamefont
  {Oikonomou}(2016)}]{Odintsov:2016apy}%
  \BibitemOpen
  \bibfield  {author} {\bibinfo {author} {\bibfnamefont {S.~D.}\ \bibnamefont
  {Odintsov}}\ and\ \bibinfo {author} {\bibfnamefont {V.~K.}\ \bibnamefont
  {Oikonomou}},\ }\href {\doibase 10.1209/0295-5075/116/49001} {\bibfield
  {journal} {\bibinfo  {journal} {EPL}\ }\textbf {\bibinfo {volume} {116}},\
  \bibinfo {pages} {49001} (\bibinfo {year} {2016})},\ \Eprint
  {http://arxiv.org/abs/1610.02533} {arXiv:1610.02533 [gr-qc]} \BibitemShut
  {NoStop}%
\bibitem [{\citenamefont {De~Haro}\ \emph {et~al.}(2018)\citenamefont
  {De~Haro}, \citenamefont {Odintsov},\ and\ \citenamefont
  {Oikonomou}}]{DeHaro:2018hia}%
  \BibitemOpen
  \bibfield  {author} {\bibinfo {author} {\bibfnamefont {J.}~\bibnamefont
  {De~Haro}}, \bibinfo {author} {\bibfnamefont {S.~D.}\ \bibnamefont
  {Odintsov}}, \ and\ \bibinfo {author} {\bibfnamefont {V.~K.}\ \bibnamefont
  {Oikonomou}},\ }\href {\doibase 10.1103/PhysRevD.97.084052} {\bibfield
  {journal} {\bibinfo  {journal} {Phys. Rev. D}\ }\textbf {\bibinfo {volume}
  {97}},\ \bibinfo {pages} {084052} (\bibinfo {year} {2018})},\ \Eprint
  {http://arxiv.org/abs/1802.09024} {arXiv:1802.09024 [gr-qc]} \BibitemShut
  {NoStop}%
\bibitem [{\citenamefont {de~Haro}\ \emph {et~al.}(2019)\citenamefont
  {de~Haro}, \citenamefont {Arest\'e~Sal\'o},\ and\ \citenamefont
  {Pan}}]{deHaro:2018sqw}%
  \BibitemOpen
  \bibfield  {author} {\bibinfo {author} {\bibfnamefont {J.}~\bibnamefont
  {de~Haro}}, \bibinfo {author} {\bibfnamefont {L.}~\bibnamefont
  {Arest\'e~Sal\'o}}, \ and\ \bibinfo {author} {\bibfnamefont {S.}~\bibnamefont
  {Pan}},\ }\href {\doibase 10.1007/s10714-019-2534-1} {\bibfield  {journal}
  {\bibinfo  {journal} {Gen. Rel. Grav.}\ }\textbf {\bibinfo {volume} {51}},\
  \bibinfo {pages} {49} (\bibinfo {year} {2019})},\ \Eprint
  {http://arxiv.org/abs/1803.09653} {arXiv:1803.09653 [gr-qc]} \BibitemShut
  {NoStop}%
\bibitem [{\citenamefont {Odintsov}\ and\ \citenamefont
  {Oikonomou}(2018)}]{Odintsov:2018awm}%
  \BibitemOpen
  \bibfield  {author} {\bibinfo {author} {\bibfnamefont {S.~D.}\ \bibnamefont
  {Odintsov}}\ and\ \bibinfo {author} {\bibfnamefont {V.~K.}\ \bibnamefont
  {Oikonomou}},\ }\href {\doibase 10.1103/PhysRevD.97.124042} {\bibfield
  {journal} {\bibinfo  {journal} {Phys. Rev. D}\ }\textbf {\bibinfo {volume}
  {97}},\ \bibinfo {pages} {124042} (\bibinfo {year} {2018})},\ \Eprint
  {http://arxiv.org/abs/1806.01588} {arXiv:1806.01588 [gr-qc]} \BibitemShut
  {NoStop}%
\bibitem [{\citenamefont {Benetti}\ \emph {et~al.}(2019)\citenamefont
  {Benetti}, \citenamefont {Graef},\ and\ \citenamefont
  {Ramos}}]{Benetti:2019kgw}%
  \BibitemOpen
  \bibfield  {author} {\bibinfo {author} {\bibfnamefont {M.}~\bibnamefont
  {Benetti}}, \bibinfo {author} {\bibfnamefont {L.}~\bibnamefont {Graef}}, \
  and\ \bibinfo {author} {\bibfnamefont {R.~O.}\ \bibnamefont {Ramos}},\ }\href
  {\doibase 10.1088/1475-7516/2019/10/066} {\bibfield  {journal} {\bibinfo
  {journal} {JCAP}\ }\textbf {\bibinfo {volume} {10}},\ \bibinfo {pages} {066}
  (\bibinfo {year} {2019})},\ \Eprint {http://arxiv.org/abs/1907.03633}
  {arXiv:1907.03633 [astro-ph.CO]} \BibitemShut {NoStop}%
\bibitem [{\citenamefont {Casalino}\ \emph {et~al.}(2020)\citenamefont
  {Casalino}, \citenamefont {Sebastiani}, \citenamefont {Vanzo},\ and\
  \citenamefont {Zerbini}}]{Casalino:2019tho}%
  \BibitemOpen
  \bibfield  {author} {\bibinfo {author} {\bibfnamefont {A.}~\bibnamefont
  {Casalino}}, \bibinfo {author} {\bibfnamefont {L.}~\bibnamefont
  {Sebastiani}}, \bibinfo {author} {\bibfnamefont {L.}~\bibnamefont {Vanzo}}, \
  and\ \bibinfo {author} {\bibfnamefont {S.}~\bibnamefont {Zerbini}},\ }\href
  {\doibase 10.1016/j.dark.2020.100594} {\bibfield  {journal} {\bibinfo
  {journal} {Phys. Dark Univ.}\ }\textbf {\bibinfo {volume} {29}},\ \bibinfo
  {pages} {100594} (\bibinfo {year} {2020})},\ \Eprint
  {http://arxiv.org/abs/1912.09307} {arXiv:1912.09307 [gr-qc]} \BibitemShut
  {NoStop}%
\bibitem [{\citenamefont {Bouhmadi-L\'opez}\ \emph {et~al.}(2020)\citenamefont
  {Bouhmadi-L\'opez}, \citenamefont {Brahma}, \citenamefont {Chen},
  \citenamefont {Chen},\ and\ \citenamefont {Yeom}}]{Bouhmadi-Lopez:2020oia}%
  \BibitemOpen
  \bibfield  {author} {\bibinfo {author} {\bibfnamefont {M.}~\bibnamefont
  {Bouhmadi-L\'opez}}, \bibinfo {author} {\bibfnamefont {S.}~\bibnamefont
  {Brahma}}, \bibinfo {author} {\bibfnamefont {C.-Y.}\ \bibnamefont {Chen}},
  \bibinfo {author} {\bibfnamefont {P.}~\bibnamefont {Chen}}, \ and\ \bibinfo
  {author} {\bibfnamefont {D.-h.}\ \bibnamefont {Yeom}},\ }\href {\doibase
  10.1088/1475-7516/2020/07/066} {\bibfield  {journal} {\bibinfo  {journal}
  {JCAP}\ }\textbf {\bibinfo {volume} {07}},\ \bibinfo {pages} {066} (\bibinfo
  {year} {2020})},\ \Eprint {http://arxiv.org/abs/2004.13061} {arXiv:2004.13061
  [gr-qc]} \BibitemShut {NoStop}%
\bibitem [{\citenamefont {Graef}\ \emph {et~al.}(2020)\citenamefont {Graef},
  \citenamefont {Ramos},\ and\ \citenamefont {Vicente}}]{Graef:2020qwe}%
  \BibitemOpen
  \bibfield  {author} {\bibinfo {author} {\bibfnamefont {L.~L.}\ \bibnamefont
  {Graef}}, \bibinfo {author} {\bibfnamefont {R.~O.}\ \bibnamefont {Ramos}}, \
  and\ \bibinfo {author} {\bibfnamefont {G.~S.}\ \bibnamefont {Vicente}},\
  }\href {\doibase 10.1103/PhysRevD.102.043518} {\bibfield  {journal} {\bibinfo
   {journal} {Phys. Rev. D}\ }\textbf {\bibinfo {volume} {102}},\ \bibinfo
  {pages} {043518} (\bibinfo {year} {2020})},\ \Eprint
  {http://arxiv.org/abs/2007.02395} {arXiv:2007.02395 [gr-qc]} \BibitemShut
  {NoStop}%
\bibitem [{\citenamefont {Brahma}\ \emph {et~al.}(2021)\citenamefont {Brahma},
  \citenamefont {Chen},\ and\ \citenamefont {Yeom}}]{Brahma:2020eos}%
  \BibitemOpen
  \bibfield  {author} {\bibinfo {author} {\bibfnamefont {S.}~\bibnamefont
  {Brahma}}, \bibinfo {author} {\bibfnamefont {C.-Y.}\ \bibnamefont {Chen}}, \
  and\ \bibinfo {author} {\bibfnamefont {D.-h.}\ \bibnamefont {Yeom}},\ }\href
  {\doibase 10.1103/PhysRevLett.126.181301} {\bibfield  {journal} {\bibinfo
  {journal} {Phys. Rev. Lett.}\ }\textbf {\bibinfo {volume} {126}},\ \bibinfo
  {pages} {181301} (\bibinfo {year} {2021})},\ \Eprint
  {http://arxiv.org/abs/2012.08785} {arXiv:2012.08785 [gr-qc]} \BibitemShut
  {NoStop}%
\bibitem [{\citenamefont {Barboza}\ \emph {et~al.}(2022)\citenamefont
  {Barboza}, \citenamefont {Levy}, \citenamefont {Graef},\ and\ \citenamefont
  {O.~Ramos}}]{Barboza:2022hng}%
  \BibitemOpen
  \bibfield  {author} {\bibinfo {author} {\bibfnamefont {L.~N.}\ \bibnamefont
  {Barboza}}, \bibinfo {author} {\bibfnamefont {G.~L. L.~W.}\ \bibnamefont
  {Levy}}, \bibinfo {author} {\bibfnamefont {L.~L.}\ \bibnamefont {Graef}}, \
  and\ \bibinfo {author} {\bibfnamefont {R.}~\bibnamefont {O.~Ramos}},\ }\href
  {\doibase 10.1103/PhysRevD.106.103535} {\bibfield  {journal} {\bibinfo
  {journal} {Phys. Rev. D}\ }\textbf {\bibinfo {volume} {106}},\ \bibinfo
  {pages} {103535} (\bibinfo {year} {2022})},\ \Eprint
  {http://arxiv.org/abs/2206.14881} {arXiv:2206.14881 [gr-qc]} \BibitemShut
  {NoStop}%
\bibitem [{\citenamefont {S.~Vicente}\ \emph {et~al.}(2022)\citenamefont
  {S.~Vicente}, \citenamefont {O.~Ramos},\ and\ \citenamefont
  {Graef}}]{SVicente:2022ebm}%
  \BibitemOpen
  \bibfield  {author} {\bibinfo {author} {\bibfnamefont {G.}~\bibnamefont
  {S.~Vicente}}, \bibinfo {author} {\bibfnamefont {R.}~\bibnamefont
  {O.~Ramos}}, \ and\ \bibinfo {author} {\bibfnamefont {L.~L.}\ \bibnamefont
  {Graef}},\ }\href {\doibase 10.1103/PhysRevD.106.043518} {\bibfield
  {journal} {\bibinfo  {journal} {Phys. Rev. D}\ }\textbf {\bibinfo {volume}
  {106}},\ \bibinfo {pages} {043518} (\bibinfo {year} {2022})},\ \Eprint
  {http://arxiv.org/abs/2207.00435} {arXiv:2207.00435 [gr-qc]} \BibitemShut
  {NoStop}%
\bibitem [{\citenamefont {Afrin}\ \emph {et~al.}(2023)\citenamefont {Afrin},
  \citenamefont {Vagnozzi},\ and\ \citenamefont {Ghosh}}]{Afrin:2022ztr}%
  \BibitemOpen
  \bibfield  {author} {\bibinfo {author} {\bibfnamefont {M.}~\bibnamefont
  {Afrin}}, \bibinfo {author} {\bibfnamefont {S.}~\bibnamefont {Vagnozzi}}, \
  and\ \bibinfo {author} {\bibfnamefont {S.~G.}\ \bibnamefont {Ghosh}},\ }\href
  {\doibase 10.3847/1538-4357/acb334} {\bibfield  {journal} {\bibinfo
  {journal} {Astrophys. J.}\ }\textbf {\bibinfo {volume} {944}},\ \bibinfo
  {pages} {149} (\bibinfo {year} {2023})},\ \Eprint
  {http://arxiv.org/abs/2209.12584} {arXiv:2209.12584 [gr-qc]} \BibitemShut
  {NoStop}%
\bibitem [{\citenamefont {Yan}\ \emph {et~al.}(2023)\citenamefont {Yan},
  \citenamefont {Liu}, \citenamefont {Zhu}, \citenamefont {Wu},\ and\
  \citenamefont {Wang}}]{Yan:2023vdg}%
  \BibitemOpen
  \bibfield  {author} {\bibinfo {author} {\bibfnamefont {J.-M.}\ \bibnamefont
  {Yan}}, \bibinfo {author} {\bibfnamefont {C.}~\bibnamefont {Liu}}, \bibinfo
  {author} {\bibfnamefont {T.}~\bibnamefont {Zhu}}, \bibinfo {author}
  {\bibfnamefont {Q.}~\bibnamefont {Wu}}, \ and\ \bibinfo {author}
  {\bibfnamefont {A.}~\bibnamefont {Wang}},\ }\href {\doibase
  10.1103/PhysRevD.107.084043} {\bibfield  {journal} {\bibinfo  {journal}
  {Phys. Rev. D}\ }\textbf {\bibinfo {volume} {107}},\ \bibinfo {pages}
  {084043} (\bibinfo {year} {2023})},\ \Eprint
  {http://arxiv.org/abs/2302.10482} {arXiv:2302.10482 [gr-qc]} \BibitemShut
  {NoStop}%
\bibitem [{\citenamefont {Kumar}\ \emph {et~al.}(2023)\citenamefont {Kumar},
  \citenamefont {Islam},\ and\ \citenamefont {Ghosh}}]{Kumar:2023jgh}%
  \BibitemOpen
  \bibfield  {author} {\bibinfo {author} {\bibfnamefont {J.}~\bibnamefont
  {Kumar}}, \bibinfo {author} {\bibfnamefont {S.~U.}\ \bibnamefont {Islam}}, \
  and\ \bibinfo {author} {\bibfnamefont {S.~G.}\ \bibnamefont {Ghosh}},\ }\href
  {\doibase 10.1140/epjc/s10052-023-12205-3} {\bibfield  {journal} {\bibinfo
  {journal} {Eur. Phys. J. C}\ }\textbf {\bibinfo {volume} {83}},\ \bibinfo
  {pages} {1014} (\bibinfo {year} {2023})},\ \Eprint
  {http://arxiv.org/abs/2305.04336} {arXiv:2305.04336 [gr-qc]} \BibitemShut
  {NoStop}%
\bibitem [{\citenamefont {Jiang}\ \emph
  {et~al.}(2024{\natexlab{a}})\citenamefont {Jiang}, \citenamefont {Liu},
  \citenamefont {Dihingia}, \citenamefont {Mizuno}, \citenamefont {Xu},
  \citenamefont {Zhu},\ and\ \citenamefont {Wu}}]{Jiang:2023img}%
  \BibitemOpen
  \bibfield  {author} {\bibinfo {author} {\bibfnamefont {H.-X.}\ \bibnamefont
  {Jiang}}, \bibinfo {author} {\bibfnamefont {C.}~\bibnamefont {Liu}}, \bibinfo
  {author} {\bibfnamefont {I.~K.}\ \bibnamefont {Dihingia}}, \bibinfo {author}
  {\bibfnamefont {Y.}~\bibnamefont {Mizuno}}, \bibinfo {author} {\bibfnamefont
  {H.}~\bibnamefont {Xu}}, \bibinfo {author} {\bibfnamefont {T.}~\bibnamefont
  {Zhu}}, \ and\ \bibinfo {author} {\bibfnamefont {Q.}~\bibnamefont {Wu}},\
  }\href {\doibase 10.1088/1475-7516/2024/01/059} {\bibfield  {journal}
  {\bibinfo  {journal} {JCAP}\ }\textbf {\bibinfo {volume} {01}},\ \bibinfo
  {pages} {059} (\bibinfo {year} {2024}{\natexlab{a}})},\ \Eprint
  {http://arxiv.org/abs/2312.04288} {arXiv:2312.04288 [gr-qc]} \BibitemShut
  {NoStop}%
\bibitem [{\citenamefont {Jiang}\ \emph
  {et~al.}(2024{\natexlab{b}})\citenamefont {Jiang}, \citenamefont {Alloqulov},
  \citenamefont {Wu}, \citenamefont {Shaymatov},\ and\ \citenamefont
  {Zhu}}]{Jiang:2024cpe}%
  \BibitemOpen
  \bibfield  {author} {\bibinfo {author} {\bibfnamefont {H.}~\bibnamefont
  {Jiang}}, \bibinfo {author} {\bibfnamefont {M.}~\bibnamefont {Alloqulov}},
  \bibinfo {author} {\bibfnamefont {Q.}~\bibnamefont {Wu}}, \bibinfo {author}
  {\bibfnamefont {S.}~\bibnamefont {Shaymatov}}, \ and\ \bibinfo {author}
  {\bibfnamefont {T.}~\bibnamefont {Zhu}},\ }\href {\doibase
  10.1016/j.dark.2024.101627} {\bibfield  {journal} {\bibinfo  {journal} {Phys.
  Dark Univ.}\ }\textbf {\bibinfo {volume} {46}},\ \bibinfo {pages} {101627}
  (\bibinfo {year} {2024}{\natexlab{b}})}\BibitemShut {NoStop}%
\bibitem [{\citenamefont {Peltola}\ and\ \citenamefont
  {Kunstatter}(2009{\natexlab{a}})}]{Peltola:2008pa}%
  \BibitemOpen
  \bibfield  {author} {\bibinfo {author} {\bibfnamefont {A.}~\bibnamefont
  {Peltola}}\ and\ \bibinfo {author} {\bibfnamefont {G.}~\bibnamefont
  {Kunstatter}},\ }\href {\doibase 10.1103/PhysRevD.79.061501} {\bibfield
  {journal} {\bibinfo  {journal} {Phys. Rev. D}\ }\textbf {\bibinfo {volume}
  {79}},\ \bibinfo {pages} {061501} (\bibinfo {year} {2009}{\natexlab{a}})},\
  \Eprint {http://arxiv.org/abs/0811.3240} {arXiv:0811.3240 [gr-qc]}
  \BibitemShut {NoStop}%
\bibitem [{\citenamefont {Peltola}\ and\ \citenamefont
  {Kunstatter}(2009{\natexlab{b}})}]{Peltola:2009jm}%
  \BibitemOpen
  \bibfield  {author} {\bibinfo {author} {\bibfnamefont {A.}~\bibnamefont
  {Peltola}}\ and\ \bibinfo {author} {\bibfnamefont {G.}~\bibnamefont
  {Kunstatter}},\ }\href {\doibase 10.1103/PhysRevD.80.044031} {\bibfield
  {journal} {\bibinfo  {journal} {Phys. Rev. D}\ }\textbf {\bibinfo {volume}
  {80}},\ \bibinfo {pages} {044031} (\bibinfo {year} {2009}{\natexlab{b}})},\
  \Eprint {http://arxiv.org/abs/0902.1746} {arXiv:0902.1746 [gr-qc]}
  \BibitemShut {NoStop}%
\bibitem [{\citenamefont {Bianchi}\ \emph {et~al.}(2018)\citenamefont
  {Bianchi}, \citenamefont {Christodoulou}, \citenamefont {D'Ambrosio},
  \citenamefont {Haggard},\ and\ \citenamefont {Rovelli}}]{Bianchi:2018mml}%
  \BibitemOpen
  \bibfield  {author} {\bibinfo {author} {\bibfnamefont {E.}~\bibnamefont
  {Bianchi}}, \bibinfo {author} {\bibfnamefont {M.}~\bibnamefont
  {Christodoulou}}, \bibinfo {author} {\bibfnamefont {F.}~\bibnamefont
  {D'Ambrosio}}, \bibinfo {author} {\bibfnamefont {H.~M.}\ \bibnamefont
  {Haggard}}, \ and\ \bibinfo {author} {\bibfnamefont {C.}~\bibnamefont
  {Rovelli}},\ }\href {\doibase 10.1088/1361-6382/aae550} {\bibfield  {journal}
  {\bibinfo  {journal} {Class. Quant. Grav.}\ }\textbf {\bibinfo {volume}
  {35}},\ \bibinfo {pages} {225003} (\bibinfo {year} {2018})},\ \Eprint
  {http://arxiv.org/abs/1802.04264} {arXiv:1802.04264 [gr-qc]} \BibitemShut
  {NoStop}%
\bibitem [{\citenamefont {D'Ambrosio}\ and\ \citenamefont
  {Rovelli}(2018)}]{DAmbrosio:2018wgv}%
  \BibitemOpen
  \bibfield  {author} {\bibinfo {author} {\bibfnamefont {F.}~\bibnamefont
  {D'Ambrosio}}\ and\ \bibinfo {author} {\bibfnamefont {C.}~\bibnamefont
  {Rovelli}},\ }\href {\doibase 10.1088/1361-6382/aae499} {\bibfield  {journal}
  {\bibinfo  {journal} {Class. Quant. Grav.}\ }\textbf {\bibinfo {volume}
  {35}},\ \bibinfo {pages} {215010} (\bibinfo {year} {2018})},\ \Eprint
  {http://arxiv.org/abs/1803.05015} {arXiv:1803.05015 [gr-qc]} \BibitemShut
  {NoStop}%
\bibitem [{\citenamefont {Easson}(2003)}]{Easson:2002tg}%
  \BibitemOpen
  \bibfield  {author} {\bibinfo {author} {\bibfnamefont {D.~A.}\ \bibnamefont
  {Easson}},\ }\href {\doibase 10.1088/1126-6708/2003/02/037} {\bibfield
  {journal} {\bibinfo  {journal} {JHEP}\ }\textbf {\bibinfo {volume} {02}},\
  \bibinfo {pages} {037} (\bibinfo {year} {2003})},\ \Eprint
  {http://arxiv.org/abs/hep-th/0210016} {arXiv:hep-th/0210016} \BibitemShut
  {NoStop}%
\bibitem [{\citenamefont {Dymnikova}\ and\ \citenamefont
  {Khlopov}(2015)}]{Dymnikova:2015yma}%
  \BibitemOpen
  \bibfield  {author} {\bibinfo {author} {\bibfnamefont {I.}~\bibnamefont
  {Dymnikova}}\ and\ \bibinfo {author} {\bibfnamefont {M.}~\bibnamefont
  {Khlopov}},\ }\href {\doibase 10.1142/S0218271815450029} {\bibfield
  {journal} {\bibinfo  {journal} {Int. J. Mod. Phys. D}\ }\textbf {\bibinfo
  {volume} {24}},\ \bibinfo {pages} {1545002} (\bibinfo {year} {2015})},\
  \Eprint {http://arxiv.org/abs/1510.01351} {arXiv:1510.01351 [gr-qc]}
  \BibitemShut {NoStop}%
\bibitem [{\citenamefont {Pacheco}(2018)}]{Pacheco:2018mvs}%
  \BibitemOpen
  \bibfield  {author} {\bibinfo {author} {\bibfnamefont {J.~A. d.~F.}\
  \bibnamefont {Pacheco}},\ }\href {\doibase 10.3390/universe4050062}
  {\bibfield  {journal} {\bibinfo  {journal} {Universe}\ }\textbf {\bibinfo
  {volume} {4}},\ \bibinfo {pages} {62} (\bibinfo {year} {2018})},\ \Eprint
  {http://arxiv.org/abs/1805.03053} {arXiv:1805.03053 [gr-qc]} \BibitemShut
  {NoStop}%
\bibitem [{\citenamefont {Arbey}\ and\ \citenamefont
  {Auffinger}(2021)}]{Arbey:2021mbl}%
  \BibitemOpen
  \bibfield  {author} {\bibinfo {author} {\bibfnamefont {A.}~\bibnamefont
  {Arbey}}\ and\ \bibinfo {author} {\bibfnamefont {J.}~\bibnamefont
  {Auffinger}},\ }\href {\doibase 10.1140/epjc/s10052-021-09702-8} {\bibfield
  {journal} {\bibinfo  {journal} {Eur. Phys. J. C}\ }\textbf {\bibinfo {volume}
  {81}},\ \bibinfo {pages} {910} (\bibinfo {year} {2021})},\ \Eprint
  {http://arxiv.org/abs/2108.02737} {arXiv:2108.02737 [gr-qc]} \BibitemShut
  {NoStop}%
\bibitem [{\citenamefont {Arbey}\ \emph {et~al.}(2022)\citenamefont {Arbey},
  \citenamefont {Auffinger}, \citenamefont {Geiller}, \citenamefont {Livine},\
  and\ \citenamefont {Sartini}}]{Arbey:2022mcd}%
  \BibitemOpen
  \bibfield  {author} {\bibinfo {author} {\bibfnamefont {A.}~\bibnamefont
  {Arbey}}, \bibinfo {author} {\bibfnamefont {J.}~\bibnamefont {Auffinger}},
  \bibinfo {author} {\bibfnamefont {M.}~\bibnamefont {Geiller}}, \bibinfo
  {author} {\bibfnamefont {E.}~\bibnamefont {Livine}}, \ and\ \bibinfo {author}
  {\bibfnamefont {F.}~\bibnamefont {Sartini}},\ }\href {\doibase
  10.22323/1.398.0066} {\bibfield  {journal} {\bibinfo  {journal} {PoS}\
  }\textbf {\bibinfo {volume} {EPS-HEP2021}},\ \bibinfo {pages} {066} (\bibinfo
  {year} {2022})}\BibitemShut {NoStop}%
\bibitem [{\citenamefont {Banerjee}\ \emph {et~al.}(2024)\citenamefont
  {Banerjee}, \citenamefont {Singh~Bhandari}, \citenamefont {Jain},\ and\
  \citenamefont {Thalapillil}}]{Banerjee:2024sao}%
  \BibitemOpen
  \bibfield  {author} {\bibinfo {author} {\bibfnamefont {A.}~\bibnamefont
  {Banerjee}}, \bibinfo {author} {\bibfnamefont {L.}~\bibnamefont
  {Singh~Bhandari}}, \bibinfo {author} {\bibfnamefont {A.}~\bibnamefont
  {Jain}}, \ and\ \bibinfo {author} {\bibfnamefont {A.~M.}\ \bibnamefont
  {Thalapillil}},\ }\href@noop {} {\  (\bibinfo {year} {2024})},\ \Eprint
  {http://arxiv.org/abs/2406.08728} {arXiv:2406.08728 [astro-ph.CO]}
  \BibitemShut {NoStop}%
\bibitem [{\citenamefont {Davies}\ \emph {et~al.}(2024)\citenamefont {Davies},
  \citenamefont {Easson},\ and\ \citenamefont {Levin}}]{Davies:2024ysj}%
  \BibitemOpen
  \bibfield  {author} {\bibinfo {author} {\bibfnamefont {P.~C.~W.}\
  \bibnamefont {Davies}}, \bibinfo {author} {\bibfnamefont {D.~A.}\
  \bibnamefont {Easson}}, \ and\ \bibinfo {author} {\bibfnamefont {P.~B.}\
  \bibnamefont {Levin}},\ }\href@noop {} {\  (\bibinfo {year} {2024})},\
  \Eprint {http://arxiv.org/abs/2410.21577} {arXiv:2410.21577 [hep-th]}
  \BibitemShut {NoStop}%
\bibitem [{\citenamefont {Sachs}\ \emph {et~al.}(2021)\citenamefont {Sachs},
  \citenamefont {Schneider},\ and\ \citenamefont {Urban}}]{Sachs:2021mcu}%
  \BibitemOpen
  \bibfield  {author} {\bibinfo {author} {\bibfnamefont {I.}~\bibnamefont
  {Sachs}}, \bibinfo {author} {\bibfnamefont {M.}~\bibnamefont {Schneider}}, \
  and\ \bibinfo {author} {\bibfnamefont {M.}~\bibnamefont {Urban}},\ }\href
  {\doibase 10.1103/PhysRevD.104.125020} {\bibfield  {journal} {\bibinfo
  {journal} {Phys. Rev. D}\ }\textbf {\bibinfo {volume} {104}},\ \bibinfo
  {pages} {125020} (\bibinfo {year} {2021})},\ \Eprint
  {http://arxiv.org/abs/2105.01071} {arXiv:2105.01071 [hep-th]} \BibitemShut
  {NoStop}%
\bibitem [{\citenamefont {Ashtekar}\ \emph {et~al.}(2021)\citenamefont
  {Ashtekar}, \citenamefont {De~Lorenzo},\ and\ \citenamefont
  {Schneider}}]{Ashtekar:2021dab}%
  \BibitemOpen
  \bibfield  {author} {\bibinfo {author} {\bibfnamefont {A.}~\bibnamefont
  {Ashtekar}}, \bibinfo {author} {\bibfnamefont {T.}~\bibnamefont
  {De~Lorenzo}}, \ and\ \bibinfo {author} {\bibfnamefont {M.}~\bibnamefont
  {Schneider}},\ }\href {\doibase 10.4310/ATMP.2021.v25.n7.a1} {\bibfield
  {journal} {\bibinfo  {journal} {Adv. Theor. Math. Phys.}\ }\textbf {\bibinfo
  {volume} {25}},\ \bibinfo {pages} {1651} (\bibinfo {year} {2021})},\ \Eprint
  {http://arxiv.org/abs/2107.08506} {arXiv:2107.08506 [gr-qc]} \BibitemShut
  {NoStop}%
\bibitem [{\citenamefont {Ashtekar}\ \emph {et~al.}(2022)\citenamefont
  {Ashtekar}, \citenamefont {del R\'\i{}o},\ and\ \citenamefont
  {Schneider}}]{Ashtekar:2022oyq}%
  \BibitemOpen
  \bibfield  {author} {\bibinfo {author} {\bibfnamefont {A.}~\bibnamefont
  {Ashtekar}}, \bibinfo {author} {\bibfnamefont {A.}~\bibnamefont {del
  R\'\i{}o}}, \ and\ \bibinfo {author} {\bibfnamefont {M.}~\bibnamefont
  {Schneider}},\ }\href {\doibase 10.1007/s10714-022-02932-5} {\bibfield
  {journal} {\bibinfo  {journal} {Gen. Rel. Grav.}\ }\textbf {\bibinfo {volume}
  {54}},\ \bibinfo {pages} {45} (\bibinfo {year} {2022})},\ \Eprint
  {http://arxiv.org/abs/2205.00298} {arXiv:2205.00298 [gr-qc]} \BibitemShut
  {NoStop}%
\bibitem [{\citenamefont {Borde}(1997)}]{Borde:1996df}%
  \BibitemOpen
  \bibfield  {author} {\bibinfo {author} {\bibfnamefont {A.}~\bibnamefont
  {Borde}},\ }\href {\doibase 10.1103/PhysRevD.55.7615} {\bibfield  {journal}
  {\bibinfo  {journal} {Phys. Rev. D}\ }\textbf {\bibinfo {volume} {55}},\
  \bibinfo {pages} {7615} (\bibinfo {year} {1997})},\ \Eprint
  {http://arxiv.org/abs/gr-qc/9612057} {arXiv:gr-qc/9612057} \BibitemShut
  {NoStop}%
\bibitem [{\citenamefont {Ayon-Beato}\ and\ \citenamefont
  {Garcia}(1998)}]{AyonBeato:1998ub}%
  \BibitemOpen
  \bibfield  {author} {\bibinfo {author} {\bibfnamefont {E.}~\bibnamefont
  {Ayon-Beato}}\ and\ \bibinfo {author} {\bibfnamefont {A.}~\bibnamefont
  {Garcia}},\ }\href {\doibase 10.1103/PhysRevLett.80.5056} {\bibfield
  {journal} {\bibinfo  {journal} {Phys. Rev. Lett.}\ }\textbf {\bibinfo
  {volume} {80}},\ \bibinfo {pages} {5056} (\bibinfo {year} {1998})},\ \Eprint
  {http://arxiv.org/abs/gr-qc/9911046} {arXiv:gr-qc/9911046} \BibitemShut
  {NoStop}%
\bibitem [{\citenamefont {Ayon-Beato}\ and\ \citenamefont
  {Garcia}(1999)}]{AyonBeato:1999rg}%
  \BibitemOpen
  \bibfield  {author} {\bibinfo {author} {\bibfnamefont {E.}~\bibnamefont
  {Ayon-Beato}}\ and\ \bibinfo {author} {\bibfnamefont {A.}~\bibnamefont
  {Garcia}},\ }\href {\doibase 10.1016/S0370-2693(99)01038-2} {\bibfield
  {journal} {\bibinfo  {journal} {Phys. Lett. B}\ }\textbf {\bibinfo {volume}
  {464}},\ \bibinfo {pages} {25} (\bibinfo {year} {1999})},\ \Eprint
  {http://arxiv.org/abs/hep-th/9911174} {arXiv:hep-th/9911174} \BibitemShut
  {NoStop}%
\bibitem [{\citenamefont {Bronnikov}\ and\ \citenamefont
  {Fabris}(2006)}]{Bronnikov:2005gm}%
  \BibitemOpen
  \bibfield  {author} {\bibinfo {author} {\bibfnamefont {K.~A.}\ \bibnamefont
  {Bronnikov}}\ and\ \bibinfo {author} {\bibfnamefont {J.~C.}\ \bibnamefont
  {Fabris}},\ }\href {\doibase 10.1103/PhysRevLett.96.251101} {\bibfield
  {journal} {\bibinfo  {journal} {Phys. Rev. Lett.}\ }\textbf {\bibinfo
  {volume} {96}},\ \bibinfo {pages} {251101} (\bibinfo {year} {2006})},\
  \Eprint {http://arxiv.org/abs/gr-qc/0511109} {arXiv:gr-qc/0511109}
  \BibitemShut {NoStop}%
\bibitem [{\citenamefont {Berej}\ \emph {et~al.}(2006)\citenamefont {Berej},
  \citenamefont {Matyjasek}, \citenamefont {Tryniecki},\ and\ \citenamefont
  {Woronowicz}}]{Berej:2006cc}%
  \BibitemOpen
  \bibfield  {author} {\bibinfo {author} {\bibfnamefont {W.}~\bibnamefont
  {Berej}}, \bibinfo {author} {\bibfnamefont {J.}~\bibnamefont {Matyjasek}},
  \bibinfo {author} {\bibfnamefont {D.}~\bibnamefont {Tryniecki}}, \ and\
  \bibinfo {author} {\bibfnamefont {M.}~\bibnamefont {Woronowicz}},\ }\href
  {\doibase 10.1007/s10714-006-0270-9} {\bibfield  {journal} {\bibinfo
  {journal} {Gen. Rel. Grav.}\ }\textbf {\bibinfo {volume} {38}},\ \bibinfo
  {pages} {885} (\bibinfo {year} {2006})},\ \Eprint
  {http://arxiv.org/abs/hep-th/0606185} {arXiv:hep-th/0606185} \BibitemShut
  {NoStop}%
\bibitem [{\citenamefont {Bronnikov}\ \emph {et~al.}(2012)\citenamefont
  {Bronnikov}, \citenamefont {Konoplya},\ and\ \citenamefont
  {Zhidenko}}]{Bronnikov:2012ch}%
  \BibitemOpen
  \bibfield  {author} {\bibinfo {author} {\bibfnamefont {K.~A.}\ \bibnamefont
  {Bronnikov}}, \bibinfo {author} {\bibfnamefont {R.~A.}\ \bibnamefont
  {Konoplya}}, \ and\ \bibinfo {author} {\bibfnamefont {A.}~\bibnamefont
  {Zhidenko}},\ }\href {\doibase 10.1103/PhysRevD.86.024028} {\bibfield
  {journal} {\bibinfo  {journal} {Phys. Rev. D}\ }\textbf {\bibinfo {volume}
  {86}},\ \bibinfo {pages} {024028} (\bibinfo {year} {2012})},\ \Eprint
  {http://arxiv.org/abs/1205.2224} {arXiv:1205.2224 [gr-qc]} \BibitemShut
  {NoStop}%
\bibitem [{\citenamefont {Rinaldi}(2012)}]{Rinaldi:2012vy}%
  \BibitemOpen
  \bibfield  {author} {\bibinfo {author} {\bibfnamefont {M.}~\bibnamefont
  {Rinaldi}},\ }\href {\doibase 10.1103/PhysRevD.86.084048} {\bibfield
  {journal} {\bibinfo  {journal} {Phys. Rev. D}\ }\textbf {\bibinfo {volume}
  {86}},\ \bibinfo {pages} {084048} (\bibinfo {year} {2012})},\ \Eprint
  {http://arxiv.org/abs/1208.0103} {arXiv:1208.0103 [gr-qc]} \BibitemShut
  {NoStop}%
\bibitem [{\citenamefont {Stuchl\'\i{}k}\ and\ \citenamefont
  {Schee}(2014)}]{Stuchlik:2014qja}%
  \BibitemOpen
  \bibfield  {author} {\bibinfo {author} {\bibfnamefont {Z.}~\bibnamefont
  {Stuchl\'\i{}k}}\ and\ \bibinfo {author} {\bibfnamefont {J.}~\bibnamefont
  {Schee}},\ }\href {\doibase 10.1142/S0218271815500200} {\bibfield  {journal}
  {\bibinfo  {journal} {Int. J. Mod. Phys. D}\ }\textbf {\bibinfo {volume}
  {24}},\ \bibinfo {pages} {1550020} (\bibinfo {year} {2014})},\ \Eprint
  {http://arxiv.org/abs/1501.00015} {arXiv:1501.00015 [astro-ph.HE]}
  \BibitemShut {NoStop}%
\bibitem [{\citenamefont {Schee}\ and\ \citenamefont
  {Stuchlik}(2015)}]{Schee:2015nua}%
  \BibitemOpen
  \bibfield  {author} {\bibinfo {author} {\bibfnamefont {J.}~\bibnamefont
  {Schee}}\ and\ \bibinfo {author} {\bibfnamefont {Z.}~\bibnamefont
  {Stuchlik}},\ }\href {\doibase 10.1088/1475-7516/2015/06/048} {\bibfield
  {journal} {\bibinfo  {journal} {JCAP}\ }\textbf {\bibinfo {volume} {06}},\
  \bibinfo {pages} {048} (\bibinfo {year} {2015})},\ \Eprint
  {http://arxiv.org/abs/1501.00835} {arXiv:1501.00835 [astro-ph.HE]}
  \BibitemShut {NoStop}%
\bibitem [{\citenamefont {Johannsen}(2013)}]{Johannsen:2015pca}%
  \BibitemOpen
  \bibfield  {author} {\bibinfo {author} {\bibfnamefont {T.}~\bibnamefont
  {Johannsen}},\ }\href {\doibase 10.1103/PhysRevD.88.044002} {\bibfield
  {journal} {\bibinfo  {journal} {Phys. Rev. D}\ }\textbf {\bibinfo {volume}
  {88}},\ \bibinfo {pages} {044002} (\bibinfo {year} {2013})},\ \Eprint
  {http://arxiv.org/abs/1501.02809} {arXiv:1501.02809 [gr-qc]} \BibitemShut
  {NoStop}%
\bibitem [{\citenamefont {Myrzakulov}\ \emph {et~al.}(2016)\citenamefont
  {Myrzakulov}, \citenamefont {Sebastiani}, \citenamefont {Vagnozzi},\ and\
  \citenamefont {Zerbini}}]{Myrzakulov:2015kda}%
  \BibitemOpen
  \bibfield  {author} {\bibinfo {author} {\bibfnamefont {R.}~\bibnamefont
  {Myrzakulov}}, \bibinfo {author} {\bibfnamefont {L.}~\bibnamefont
  {Sebastiani}}, \bibinfo {author} {\bibfnamefont {S.}~\bibnamefont
  {Vagnozzi}}, \ and\ \bibinfo {author} {\bibfnamefont {S.}~\bibnamefont
  {Zerbini}},\ }\href {\doibase 10.1088/0264-9381/33/12/125005} {\bibfield
  {journal} {\bibinfo  {journal} {Class. Quant. Grav.}\ }\textbf {\bibinfo
  {volume} {33}},\ \bibinfo {pages} {125005} (\bibinfo {year} {2016})},\
  \Eprint {http://arxiv.org/abs/1510.02284} {arXiv:1510.02284 [gr-qc]}
  \BibitemShut {NoStop}%
\bibitem [{\citenamefont {Fan}\ and\ \citenamefont {Wang}(2016)}]{Fan:2016hvf}%
  \BibitemOpen
  \bibfield  {author} {\bibinfo {author} {\bibfnamefont {Z.-Y.}\ \bibnamefont
  {Fan}}\ and\ \bibinfo {author} {\bibfnamefont {X.}~\bibnamefont {Wang}},\
  }\href {\doibase 10.1103/PhysRevD.94.124027} {\bibfield  {journal} {\bibinfo
  {journal} {Phys. Rev. D}\ }\textbf {\bibinfo {volume} {94}},\ \bibinfo
  {pages} {124027} (\bibinfo {year} {2016})},\ \Eprint
  {http://arxiv.org/abs/1610.02636} {arXiv:1610.02636 [gr-qc]} \BibitemShut
  {NoStop}%
\bibitem [{\citenamefont {Sebastiani}\ \emph {et~al.}(2017)\citenamefont
  {Sebastiani}, \citenamefont {Vagnozzi},\ and\ \citenamefont
  {Myrzakulov}}]{Sebastiani:2016ras}%
  \BibitemOpen
  \bibfield  {author} {\bibinfo {author} {\bibfnamefont {L.}~\bibnamefont
  {Sebastiani}}, \bibinfo {author} {\bibfnamefont {S.}~\bibnamefont
  {Vagnozzi}}, \ and\ \bibinfo {author} {\bibfnamefont {R.}~\bibnamefont
  {Myrzakulov}},\ }\href {\doibase 10.1155/2017/3156915} {\bibfield  {journal}
  {\bibinfo  {journal} {Adv. High Energy Phys.}\ }\textbf {\bibinfo {volume}
  {2017}},\ \bibinfo {pages} {3156915} (\bibinfo {year} {2017})},\ \Eprint
  {http://arxiv.org/abs/1612.08661} {arXiv:1612.08661 [gr-qc]} \BibitemShut
  {NoStop}%
\bibitem [{\citenamefont {Toshmatov}\ \emph {et~al.}(2017)\citenamefont
  {Toshmatov}, \citenamefont {Stuchl\'\i{}k},\ and\ \citenamefont
  {Ahmedov}}]{Toshmatov:2017zpr}%
  \BibitemOpen
  \bibfield  {author} {\bibinfo {author} {\bibfnamefont {B.}~\bibnamefont
  {Toshmatov}}, \bibinfo {author} {\bibfnamefont {Z.}~\bibnamefont
  {Stuchl\'\i{}k}}, \ and\ \bibinfo {author} {\bibfnamefont {B.}~\bibnamefont
  {Ahmedov}},\ }\href {\doibase 10.1103/PhysRevD.95.084037} {\bibfield
  {journal} {\bibinfo  {journal} {Phys. Rev. D}\ }\textbf {\bibinfo {volume}
  {95}},\ \bibinfo {pages} {084037} (\bibinfo {year} {2017})},\ \Eprint
  {http://arxiv.org/abs/1704.07300} {arXiv:1704.07300 [gr-qc]} \BibitemShut
  {NoStop}%
\bibitem [{\citenamefont {Chinaglia}\ and\ \citenamefont
  {Zerbini}(2017)}]{Chinaglia:2017uqd}%
  \BibitemOpen
  \bibfield  {author} {\bibinfo {author} {\bibfnamefont {S.}~\bibnamefont
  {Chinaglia}}\ and\ \bibinfo {author} {\bibfnamefont {S.}~\bibnamefont
  {Zerbini}},\ }\href {\doibase 10.1007/s10714-017-2235-6} {\bibfield
  {journal} {\bibinfo  {journal} {Gen. Rel. Grav.}\ }\textbf {\bibinfo {volume}
  {49}},\ \bibinfo {pages} {75} (\bibinfo {year} {2017})},\ \Eprint
  {http://arxiv.org/abs/1704.08516} {arXiv:1704.08516 [gr-qc]} \BibitemShut
  {NoStop}%
\bibitem [{\citenamefont {Frolov}(2018)}]{Frolov:2017dwy}%
  \BibitemOpen
  \bibfield  {author} {\bibinfo {author} {\bibfnamefont {V.~P.}\ \bibnamefont
  {Frolov}},\ }\href {\doibase 10.1051/epjconf/201816801001} {\bibfield
  {journal} {\bibinfo  {journal} {EPJ Web Conf.}\ }\textbf {\bibinfo {volume}
  {168}},\ \bibinfo {pages} {01001} (\bibinfo {year} {2018})},\ \Eprint
  {http://arxiv.org/abs/1708.04698} {arXiv:1708.04698 [gr-qc]} \BibitemShut
  {NoStop}%
\bibitem [{\citenamefont {Bertipagani}\ \emph {et~al.}(2021)\citenamefont
  {Bertipagani}, \citenamefont {Rinaldi}, \citenamefont {Sebastiani},\ and\
  \citenamefont {Zerbini}}]{Bertipagani:2020awe}%
  \BibitemOpen
  \bibfield  {author} {\bibinfo {author} {\bibfnamefont {M.}~\bibnamefont
  {Bertipagani}}, \bibinfo {author} {\bibfnamefont {M.}~\bibnamefont
  {Rinaldi}}, \bibinfo {author} {\bibfnamefont {L.}~\bibnamefont {Sebastiani}},
  \ and\ \bibinfo {author} {\bibfnamefont {S.}~\bibnamefont {Zerbini}},\ }\href
  {\doibase 10.1016/j.dark.2021.100853} {\bibfield  {journal} {\bibinfo
  {journal} {Phys. Dark Univ.}\ }\textbf {\bibinfo {volume} {33}},\ \bibinfo
  {pages} {100853} (\bibinfo {year} {2021})},\ \Eprint
  {http://arxiv.org/abs/2012.15645} {arXiv:2012.15645 [gr-qc]} \BibitemShut
  {NoStop}%
\bibitem [{\citenamefont {Nashed}\ and\ \citenamefont
  {Saridakis}(2022)}]{Nashed:2021pah}%
  \BibitemOpen
  \bibfield  {author} {\bibinfo {author} {\bibfnamefont {G.~G.~L.}\
  \bibnamefont {Nashed}}\ and\ \bibinfo {author} {\bibfnamefont {E.~N.}\
  \bibnamefont {Saridakis}},\ }\href {\doibase 10.1088/1475-7516/2022/05/017}
  {\bibfield  {journal} {\bibinfo  {journal} {JCAP}\ }\textbf {\bibinfo
  {volume} {05}},\ \bibinfo {pages} {017} (\bibinfo {year} {2022})},\ \Eprint
  {http://arxiv.org/abs/2111.06359} {arXiv:2111.06359 [gr-qc]} \BibitemShut
  {NoStop}%
\bibitem [{\citenamefont {Simpson}\ and\ \citenamefont
  {Visser}(2022)}]{Simpson:2021dyo}%
  \BibitemOpen
  \bibfield  {author} {\bibinfo {author} {\bibfnamefont {A.}~\bibnamefont
  {Simpson}}\ and\ \bibinfo {author} {\bibfnamefont {M.}~\bibnamefont
  {Visser}},\ }\href {\doibase 10.1088/1475-7516/2022/03/011} {\bibfield
  {journal} {\bibinfo  {journal} {JCAP}\ }\textbf {\bibinfo {volume} {03}},\
  \bibinfo {pages} {011} (\bibinfo {year} {2022})},\ \Eprint
  {http://arxiv.org/abs/2111.12329} {arXiv:2111.12329 [gr-qc]} \BibitemShut
  {NoStop}%
\bibitem [{\citenamefont {Franzin}\ \emph {et~al.}(2022)\citenamefont
  {Franzin}, \citenamefont {Liberati}, \citenamefont {Mazza}, \citenamefont
  {Dey},\ and\ \citenamefont {Chakraborty}}]{Franzin:2022iai}%
  \BibitemOpen
  \bibfield  {author} {\bibinfo {author} {\bibfnamefont {E.}~\bibnamefont
  {Franzin}}, \bibinfo {author} {\bibfnamefont {S.}~\bibnamefont {Liberati}},
  \bibinfo {author} {\bibfnamefont {J.}~\bibnamefont {Mazza}}, \bibinfo
  {author} {\bibfnamefont {R.}~\bibnamefont {Dey}}, \ and\ \bibinfo {author}
  {\bibfnamefont {S.}~\bibnamefont {Chakraborty}},\ }\href {\doibase
  10.1103/PhysRevD.105.124051} {\bibfield  {journal} {\bibinfo  {journal}
  {Phys. Rev. D}\ }\textbf {\bibinfo {volume} {105}},\ \bibinfo {pages}
  {124051} (\bibinfo {year} {2022})},\ \Eprint
  {http://arxiv.org/abs/2201.01650} {arXiv:2201.01650 [gr-qc]} \BibitemShut
  {NoStop}%
\bibitem [{\citenamefont {Chataignier}\ \emph {et~al.}(2023)\citenamefont
  {Chataignier}, \citenamefont {Kamenshchik}, \citenamefont {Tronconi},\ and\
  \citenamefont {Venturi}}]{Chataignier:2022yic}%
  \BibitemOpen
  \bibfield  {author} {\bibinfo {author} {\bibfnamefont {L.}~\bibnamefont
  {Chataignier}}, \bibinfo {author} {\bibfnamefont {A.~Y.}\ \bibnamefont
  {Kamenshchik}}, \bibinfo {author} {\bibfnamefont {A.}~\bibnamefont
  {Tronconi}}, \ and\ \bibinfo {author} {\bibfnamefont {G.}~\bibnamefont
  {Venturi}},\ }\href {\doibase 10.1103/PhysRevD.107.023508} {\bibfield
  {journal} {\bibinfo  {journal} {Phys. Rev. D}\ }\textbf {\bibinfo {volume}
  {107}},\ \bibinfo {pages} {023508} (\bibinfo {year} {2023})},\ \Eprint
  {http://arxiv.org/abs/2208.02280} {arXiv:2208.02280 [gr-qc]} \BibitemShut
  {NoStop}%
\bibitem [{\citenamefont {Ghosh}\ \emph {et~al.}(2023)\citenamefont {Ghosh},
  \citenamefont {Rahman},\ and\ \citenamefont {Mishra}}]{Ghosh:2022gka}%
  \BibitemOpen
  \bibfield  {author} {\bibinfo {author} {\bibfnamefont {R.}~\bibnamefont
  {Ghosh}}, \bibinfo {author} {\bibfnamefont {M.}~\bibnamefont {Rahman}}, \
  and\ \bibinfo {author} {\bibfnamefont {A.~K.}\ \bibnamefont {Mishra}},\
  }\href {\doibase 10.1140/epjc/s10052-023-11252-0} {\bibfield  {journal}
  {\bibinfo  {journal} {Eur. Phys. J. C}\ }\textbf {\bibinfo {volume} {83}},\
  \bibinfo {pages} {91} (\bibinfo {year} {2023})},\ \Eprint
  {http://arxiv.org/abs/2209.12291} {arXiv:2209.12291 [gr-qc]} \BibitemShut
  {NoStop}%
\bibitem [{\citenamefont {Khodadi}\ and\ \citenamefont
  {Pourkhodabakhshi}(2022)}]{Khodadi:2022dyi}%
  \BibitemOpen
  \bibfield  {author} {\bibinfo {author} {\bibfnamefont {M.}~\bibnamefont
  {Khodadi}}\ and\ \bibinfo {author} {\bibfnamefont {R.}~\bibnamefont
  {Pourkhodabakhshi}},\ }\href {\doibase 10.1103/PhysRevD.106.084047}
  {\bibfield  {journal} {\bibinfo  {journal} {Phys. Rev. D}\ }\textbf {\bibinfo
  {volume} {106}},\ \bibinfo {pages} {084047} (\bibinfo {year} {2022})},\
  \Eprint {http://arxiv.org/abs/2210.06861} {arXiv:2210.06861 [gr-qc]}
  \BibitemShut {NoStop}%
\bibitem [{\citenamefont {Farrah}\ \emph {et~al.}(2023)\citenamefont {Farrah}
  \emph {et~al.}}]{Farrah:2023opk}%
  \BibitemOpen
  \bibfield  {author} {\bibinfo {author} {\bibfnamefont {D.}~\bibnamefont
  {Farrah}} \emph {et~al.},\ }\href {\doibase 10.3847/2041-8213/acb704}
  {\bibfield  {journal} {\bibinfo  {journal} {Astrophys. J. Lett.}\ }\textbf
  {\bibinfo {volume} {944}},\ \bibinfo {pages} {L31} (\bibinfo {year}
  {2023})},\ \Eprint {http://arxiv.org/abs/2302.07878} {arXiv:2302.07878
  [astro-ph.CO]} \BibitemShut {NoStop}%
\bibitem [{\citenamefont {Fontana}\ and\ \citenamefont
  {Rinaldi}(2023)}]{Fontana:2023zqz}%
  \BibitemOpen
  \bibfield  {author} {\bibinfo {author} {\bibfnamefont {M.}~\bibnamefont
  {Fontana}}\ and\ \bibinfo {author} {\bibfnamefont {M.}~\bibnamefont
  {Rinaldi}},\ }\href {\doibase 10.1103/PhysRevD.108.125003} {\bibfield
  {journal} {\bibinfo  {journal} {Phys. Rev. D}\ }\textbf {\bibinfo {volume}
  {108}},\ \bibinfo {pages} {125003} (\bibinfo {year} {2023})},\ \Eprint
  {http://arxiv.org/abs/2302.08804} {arXiv:2302.08804 [gr-qc]} \BibitemShut
  {NoStop}%
\bibitem [{\citenamefont {Boshkayev}\ \emph {et~al.}(2023)\citenamefont
  {Boshkayev}, \citenamefont {Idrissov}, \citenamefont {Luongo},\ and\
  \citenamefont {Muccino}}]{Boshkayev:2023rhr}%
  \BibitemOpen
  \bibfield  {author} {\bibinfo {author} {\bibfnamefont {K.}~\bibnamefont
  {Boshkayev}}, \bibinfo {author} {\bibfnamefont {A.}~\bibnamefont {Idrissov}},
  \bibinfo {author} {\bibfnamefont {O.}~\bibnamefont {Luongo}}, \ and\ \bibinfo
  {author} {\bibfnamefont {M.}~\bibnamefont {Muccino}},\ }\href {\doibase
  10.1103/PhysRevD.108.044063} {\bibfield  {journal} {\bibinfo  {journal}
  {Phys. Rev. D}\ }\textbf {\bibinfo {volume} {108}},\ \bibinfo {pages}
  {044063} (\bibinfo {year} {2023})},\ \Eprint
  {http://arxiv.org/abs/2303.03248} {arXiv:2303.03248 [astro-ph.HE]}
  \BibitemShut {NoStop}%
\bibitem [{\citenamefont {Luongo}\ \emph {et~al.}(2023)\citenamefont {Luongo},
  \citenamefont {Mancini},\ and\ \citenamefont {Pierosara}}]{Luongo:2023jyz}%
  \BibitemOpen
  \bibfield  {author} {\bibinfo {author} {\bibfnamefont {O.}~\bibnamefont
  {Luongo}}, \bibinfo {author} {\bibfnamefont {S.}~\bibnamefont {Mancini}}, \
  and\ \bibinfo {author} {\bibfnamefont {P.}~\bibnamefont {Pierosara}},\ }\href
  {\doibase 10.1103/PhysRevD.108.104059} {\bibfield  {journal} {\bibinfo
  {journal} {Phys. Rev. D}\ }\textbf {\bibinfo {volume} {108}},\ \bibinfo
  {pages} {104059} (\bibinfo {year} {2023})},\ \Eprint
  {http://arxiv.org/abs/2304.06593} {arXiv:2304.06593 [gr-qc]} \BibitemShut
  {NoStop}%
\bibitem [{\citenamefont {Luongo}\ and\ \citenamefont
  {Quevedo}(2024)}]{Luongo:2023aib}%
  \BibitemOpen
  \bibfield  {author} {\bibinfo {author} {\bibfnamefont {O.}~\bibnamefont
  {Luongo}}\ and\ \bibinfo {author} {\bibfnamefont {H.}~\bibnamefont
  {Quevedo}},\ }\href {\doibase 10.1088/1361-6382/ad4ae4} {\bibfield  {journal}
  {\bibinfo  {journal} {Class. Quant. Grav.}\ }\textbf {\bibinfo {volume}
  {41}},\ \bibinfo {pages} {125011} (\bibinfo {year} {2024})},\ \Eprint
  {http://arxiv.org/abs/2305.11185} {arXiv:2305.11185 [gr-qc]} \BibitemShut
  {NoStop}%
\bibitem [{\citenamefont {Cadoni}\ \emph {et~al.}(2023)\citenamefont {Cadoni},
  \citenamefont {Sanna}, \citenamefont {Pitzalis}, \citenamefont {Banerjee},
  \citenamefont {Murgia}, \citenamefont {Hazra},\ and\ \citenamefont
  {Branchesi}}]{Cadoni:2023lum}%
  \BibitemOpen
  \bibfield  {author} {\bibinfo {author} {\bibfnamefont {M.}~\bibnamefont
  {Cadoni}}, \bibinfo {author} {\bibfnamefont {A.~P.}\ \bibnamefont {Sanna}},
  \bibinfo {author} {\bibfnamefont {M.}~\bibnamefont {Pitzalis}}, \bibinfo
  {author} {\bibfnamefont {B.}~\bibnamefont {Banerjee}}, \bibinfo {author}
  {\bibfnamefont {R.}~\bibnamefont {Murgia}}, \bibinfo {author} {\bibfnamefont
  {N.}~\bibnamefont {Hazra}}, \ and\ \bibinfo {author} {\bibfnamefont
  {M.}~\bibnamefont {Branchesi}},\ }\href {\doibase
  10.1088/1475-7516/2023/11/007} {\bibfield  {journal} {\bibinfo  {journal}
  {JCAP}\ }\textbf {\bibinfo {volume} {11}},\ \bibinfo {pages} {007} (\bibinfo
  {year} {2023})},\ \Eprint {http://arxiv.org/abs/2306.11588} {arXiv:2306.11588
  [gr-qc]} \BibitemShut {NoStop}%
\bibitem [{\citenamefont {Giamb\`o}\ and\ \citenamefont
  {Luongo}(2024)}]{Giambo:2023zmy}%
  \BibitemOpen
  \bibfield  {author} {\bibinfo {author} {\bibfnamefont {R.}~\bibnamefont
  {Giamb\`o}}\ and\ \bibinfo {author} {\bibfnamefont {O.}~\bibnamefont
  {Luongo}},\ }\href {\doibase 10.1088/1361-6382/ad43a9} {\bibfield  {journal}
  {\bibinfo  {journal} {Class. Quant. Grav.}\ }\textbf {\bibinfo {volume}
  {41}},\ \bibinfo {pages} {125005} (\bibinfo {year} {2024})},\ \Eprint
  {http://arxiv.org/abs/2308.10060} {arXiv:2308.10060 [gr-qc]} \BibitemShut
  {NoStop}%
\bibitem [{\citenamefont {Cadoni}\ \emph {et~al.}(2024)\citenamefont {Cadoni},
  \citenamefont {Murgia}, \citenamefont {Pitzalis},\ and\ \citenamefont
  {Sanna}}]{Cadoni:2023lqe}%
  \BibitemOpen
  \bibfield  {author} {\bibinfo {author} {\bibfnamefont {M.}~\bibnamefont
  {Cadoni}}, \bibinfo {author} {\bibfnamefont {R.}~\bibnamefont {Murgia}},
  \bibinfo {author} {\bibfnamefont {M.}~\bibnamefont {Pitzalis}}, \ and\
  \bibinfo {author} {\bibfnamefont {A.~P.}\ \bibnamefont {Sanna}},\ }\href
  {\doibase 10.1088/1475-7516/2024/03/026} {\bibfield  {journal} {\bibinfo
  {journal} {JCAP}\ }\textbf {\bibinfo {volume} {03}},\ \bibinfo {pages} {026}
  (\bibinfo {year} {2024})},\ \Eprint {http://arxiv.org/abs/2309.16444}
  {arXiv:2309.16444 [gr-qc]} \BibitemShut {NoStop}%
\bibitem [{\citenamefont {Luongo}\ \emph {et~al.}(2024)\citenamefont {Luongo},
  \citenamefont {Quevedo},\ and\ \citenamefont {Sajadi}}]{Luongo:2023xaw}%
  \BibitemOpen
  \bibfield  {author} {\bibinfo {author} {\bibfnamefont {O.}~\bibnamefont
  {Luongo}}, \bibinfo {author} {\bibfnamefont {H.}~\bibnamefont {Quevedo}}, \
  and\ \bibinfo {author} {\bibfnamefont {S.~N.}\ \bibnamefont {Sajadi}},\
  }\href {\doibase 10.1007/s10714-024-03207-x} {\bibfield  {journal} {\bibinfo
  {journal} {Gen. Rel. Grav.}\ }\textbf {\bibinfo {volume} {56}},\ \bibinfo
  {pages} {17} (\bibinfo {year} {2024})},\ \Eprint
  {http://arxiv.org/abs/2311.13264} {arXiv:2311.13264 [gr-qc]} \BibitemShut
  {NoStop}%
\bibitem [{\citenamefont {Sajadi}\ \emph {et~al.}(2024)\citenamefont {Sajadi},
  \citenamefont {Khodadi}, \citenamefont {Luongo},\ and\ \citenamefont
  {Quevedo}}]{Sajadi:2023ybm}%
  \BibitemOpen
  \bibfield  {author} {\bibinfo {author} {\bibfnamefont {S.~N.}\ \bibnamefont
  {Sajadi}}, \bibinfo {author} {\bibfnamefont {M.}~\bibnamefont {Khodadi}},
  \bibinfo {author} {\bibfnamefont {O.}~\bibnamefont {Luongo}}, \ and\ \bibinfo
  {author} {\bibfnamefont {H.}~\bibnamefont {Quevedo}},\ }\href {\doibase
  10.1016/j.dark.2024.101525} {\bibfield  {journal} {\bibinfo  {journal} {Phys.
  Dark Univ.}\ }\textbf {\bibinfo {volume} {45}},\ \bibinfo {pages} {101525}
  (\bibinfo {year} {2024})},\ \Eprint {http://arxiv.org/abs/2312.16081}
  {arXiv:2312.16081 [gr-qc]} \BibitemShut {NoStop}%
\bibitem [{\citenamefont {Javed}\ \emph {et~al.}(2024)\citenamefont {Javed},
  \citenamefont {Mumtaz}, \citenamefont {Mustafa}, \citenamefont {Atamurotov},\
  and\ \citenamefont {Ghosh}}]{Javed:2024wbc}%
  \BibitemOpen
  \bibfield  {author} {\bibinfo {author} {\bibfnamefont {F.}~\bibnamefont
  {Javed}}, \bibinfo {author} {\bibfnamefont {S.}~\bibnamefont {Mumtaz}},
  \bibinfo {author} {\bibfnamefont {G.}~\bibnamefont {Mustafa}}, \bibinfo
  {author} {\bibfnamefont {F.}~\bibnamefont {Atamurotov}}, \ and\ \bibinfo
  {author} {\bibfnamefont {S.~G.}\ \bibnamefont {Ghosh}},\ }\href {\doibase
  10.1016/j.cjph.2023.12.029} {\bibfield  {journal} {\bibinfo  {journal} {Chin.
  J. Phys.}\ }\textbf {\bibinfo {volume} {88}},\ \bibinfo {pages} {55}
  (\bibinfo {year} {2024})}\BibitemShut {NoStop}%
\bibitem [{\citenamefont {Ditta}\ \emph {et~al.}(2024)\citenamefont {Ditta},
  \citenamefont {Xia}, \citenamefont {Ali}, \citenamefont {Mustafa},
  \citenamefont {Mustafa},\ and\ \citenamefont {Mahmood}}]{Ditta:2024jrv}%
  \BibitemOpen
  \bibfield  {author} {\bibinfo {author} {\bibfnamefont {A.}~\bibnamefont
  {Ditta}}, \bibinfo {author} {\bibfnamefont {T.}~\bibnamefont {Xia}}, \bibinfo
  {author} {\bibfnamefont {R.}~\bibnamefont {Ali}}, \bibinfo {author}
  {\bibfnamefont {G.}~\bibnamefont {Mustafa}}, \bibinfo {author} {\bibfnamefont
  {G.}~\bibnamefont {Mustafa}}, \ and\ \bibinfo {author} {\bibfnamefont
  {A.}~\bibnamefont {Mahmood}},\ }\href {\doibase 10.1016/j.dark.2023.101418}
  {\bibfield  {journal} {\bibinfo  {journal} {Phys. Dark Univ.}\ }\textbf
  {\bibinfo {volume} {43}},\ \bibinfo {pages} {101418} (\bibinfo {year}
  {2024})}\BibitemShut {NoStop}%
\bibitem [{\citenamefont {Al-Badawi}\ \emph {et~al.}(2024)\citenamefont
  {Al-Badawi}, \citenamefont {Shaymatov}, \citenamefont {Alloqulov},\ and\
  \citenamefont {Wang}}]{Al-Badawi:2024lvc}%
  \BibitemOpen
  \bibfield  {author} {\bibinfo {author} {\bibfnamefont {A.}~\bibnamefont
  {Al-Badawi}}, \bibinfo {author} {\bibfnamefont {S.}~\bibnamefont
  {Shaymatov}}, \bibinfo {author} {\bibfnamefont {M.}~\bibnamefont
  {Alloqulov}}, \ and\ \bibinfo {author} {\bibfnamefont {A.}~\bibnamefont
  {Wang}},\ }\href {\doibase 10.1088/1572-9494/ad4c55} {\bibfield  {journal}
  {\bibinfo  {journal} {Commun. Theor. Phys.}\ }\textbf {\bibinfo {volume}
  {76}},\ \bibinfo {pages} {085401} (\bibinfo {year} {2024})},\ \Eprint
  {http://arxiv.org/abs/2401.12723} {arXiv:2401.12723 [gr-qc]} \BibitemShut
  {NoStop}%
\bibitem [{\citenamefont {\"Ovg\"un}\ \emph {et~al.}(2024)\citenamefont
  {\"Ovg\"un}, \citenamefont {Pantig},\ and\ \citenamefont
  {Rinc\'on}}]{Ovgun:2024zmt}%
  \BibitemOpen
  \bibfield  {author} {\bibinfo {author} {\bibfnamefont {A.}~\bibnamefont
  {\"Ovg\"un}}, \bibinfo {author} {\bibfnamefont {R.~C.}\ \bibnamefont
  {Pantig}}, \ and\ \bibinfo {author} {\bibfnamefont {A.}~\bibnamefont
  {Rinc\'on}},\ }\href {\doibase 10.1016/j.aop.2024.169625} {\bibfield
  {journal} {\bibinfo  {journal} {Annals Phys.}\ }\textbf {\bibinfo {volume}
  {463}},\ \bibinfo {pages} {169625} (\bibinfo {year} {2024})},\ \Eprint
  {http://arxiv.org/abs/2402.14190} {arXiv:2402.14190 [gr-qc]} \BibitemShut
  {NoStop}%
\bibitem [{\citenamefont {Corona}\ \emph {et~al.}(2024)\citenamefont {Corona},
  \citenamefont {Giamb\`o},\ and\ \citenamefont {Luongo}}]{Corona:2024gth}%
  \BibitemOpen
  \bibfield  {author} {\bibinfo {author} {\bibfnamefont {D.}~\bibnamefont
  {Corona}}, \bibinfo {author} {\bibfnamefont {R.}~\bibnamefont {Giamb\`o}}, \
  and\ \bibinfo {author} {\bibfnamefont {O.}~\bibnamefont {Luongo}},\ }\href
  {\doibase 10.1142/S021988782440019X} {\bibfield  {journal} {\bibinfo
  {journal} {Int. J. Geom. Meth. Mod. Phys.}\ }\textbf {\bibinfo {volume}
  {21}},\ \bibinfo {pages} {2440019} (\bibinfo {year} {2024})},\ \Eprint
  {http://arxiv.org/abs/2402.18997} {arXiv:2402.18997 [gr-qc]} \BibitemShut
  {NoStop}%
\bibitem [{\citenamefont {Bueno}\ \emph {et~al.}(2024)\citenamefont {Bueno},
  \citenamefont {Cano},\ and\ \citenamefont {Hennigar}}]{Bueno:2024dgm}%
  \BibitemOpen
  \bibfield  {author} {\bibinfo {author} {\bibfnamefont {P.}~\bibnamefont
  {Bueno}}, \bibinfo {author} {\bibfnamefont {P.~A.}\ \bibnamefont {Cano}}, \
  and\ \bibinfo {author} {\bibfnamefont {R.~A.}\ \bibnamefont {Hennigar}},\
  }\href@noop {} {\  (\bibinfo {year} {2024})},\ \Eprint
  {http://arxiv.org/abs/2403.04827} {arXiv:2403.04827 [gr-qc]} \BibitemShut
  {NoStop}%
\bibitem [{\citenamefont {Konoplya}\ and\ \citenamefont
  {Zhidenko}(2024{\natexlab{a}})}]{Konoplya:2024hfg}%
  \BibitemOpen
  \bibfield  {author} {\bibinfo {author} {\bibfnamefont {R.~A.}\ \bibnamefont
  {Konoplya}}\ and\ \bibinfo {author} {\bibfnamefont {A.}~\bibnamefont
  {Zhidenko}},\ }\href {\doibase 10.1103/PhysRevD.109.104005} {\bibfield
  {journal} {\bibinfo  {journal} {Phys. Rev. D}\ }\textbf {\bibinfo {volume}
  {109}},\ \bibinfo {pages} {104005} (\bibinfo {year} {2024}{\natexlab{a}})},\
  \Eprint {http://arxiv.org/abs/2403.07848} {arXiv:2403.07848 [gr-qc]}
  \BibitemShut {NoStop}%
\bibitem [{\citenamefont {Pedrotti}\ and\ \citenamefont
  {Vagnozzi}(2024)}]{Pedrotti:2024znu}%
  \BibitemOpen
  \bibfield  {author} {\bibinfo {author} {\bibfnamefont {D.}~\bibnamefont
  {Pedrotti}}\ and\ \bibinfo {author} {\bibfnamefont {S.}~\bibnamefont
  {Vagnozzi}},\ }\href {\doibase 10.1103/PhysRevD.110.084075} {\bibfield
  {journal} {\bibinfo  {journal} {Phys. Rev. D}\ }\textbf {\bibinfo {volume}
  {110}},\ \bibinfo {pages} {084075} (\bibinfo {year} {2024})},\ \Eprint
  {http://arxiv.org/abs/2404.07589} {arXiv:2404.07589 [gr-qc]} \BibitemShut
  {NoStop}%
\bibitem [{\citenamefont {Bronnikov}(2024)}]{Bronnikov:2024izh}%
  \BibitemOpen
  \bibfield  {author} {\bibinfo {author} {\bibfnamefont {K.~A.}\ \bibnamefont
  {Bronnikov}},\ }\href {\doibase 10.1103/PhysRevD.110.024021} {\bibfield
  {journal} {\bibinfo  {journal} {Phys. Rev. D}\ }\textbf {\bibinfo {volume}
  {110}},\ \bibinfo {pages} {024021} (\bibinfo {year} {2024})},\ \Eprint
  {http://arxiv.org/abs/2404.14816} {arXiv:2404.14816 [gr-qc]} \BibitemShut
  {NoStop}%
\bibitem [{\citenamefont {Kurmanov}\ \emph {et~al.}(2024)\citenamefont
  {Kurmanov}, \citenamefont {Boshkayev}, \citenamefont {Konysbayev},
  \citenamefont {Luongo}, \citenamefont {Saiyp}, \citenamefont {Urazalina},
  \citenamefont {Ikhsan},\ and\ \citenamefont {Suliyeva}}]{Kurmanov:2024hpn}%
  \BibitemOpen
  \bibfield  {author} {\bibinfo {author} {\bibfnamefont {Y.}~\bibnamefont
  {Kurmanov}}, \bibinfo {author} {\bibfnamefont {K.}~\bibnamefont {Boshkayev}},
  \bibinfo {author} {\bibfnamefont {T.}~\bibnamefont {Konysbayev}}, \bibinfo
  {author} {\bibfnamefont {O.}~\bibnamefont {Luongo}}, \bibinfo {author}
  {\bibfnamefont {N.}~\bibnamefont {Saiyp}}, \bibinfo {author} {\bibfnamefont
  {A.}~\bibnamefont {Urazalina}}, \bibinfo {author} {\bibfnamefont
  {G.}~\bibnamefont {Ikhsan}}, \ and\ \bibinfo {author} {\bibfnamefont
  {G.}~\bibnamefont {Suliyeva}},\ }\href {\doibase 10.1016/j.dark.2024.101566}
  {\bibfield  {journal} {\bibinfo  {journal} {Phys. Dark Univ.}\ }\textbf
  {\bibinfo {volume} {46}},\ \bibinfo {pages} {101566} (\bibinfo {year}
  {2024})},\ \Eprint {http://arxiv.org/abs/2404.15437} {arXiv:2404.15437
  [gr-qc]} \BibitemShut {NoStop}%
\bibitem [{\citenamefont {Bolokhov}\ \emph {et~al.}(2024)\citenamefont
  {Bolokhov}, \citenamefont {Bronnikov},\ and\ \citenamefont
  {Skvortsova}}]{Bolokhov:2024sdy}%
  \BibitemOpen
  \bibfield  {author} {\bibinfo {author} {\bibfnamefont {S.~V.}\ \bibnamefont
  {Bolokhov}}, \bibinfo {author} {\bibfnamefont {K.~A.}\ \bibnamefont
  {Bronnikov}}, \ and\ \bibinfo {author} {\bibfnamefont {M.~V.}\ \bibnamefont
  {Skvortsova}},\ }\href {\doibase 10.1134/S0202289324700178} {\bibfield
  {journal} {\bibinfo  {journal} {Grav. Cosmol.}\ }\textbf {\bibinfo {volume}
  {30}},\ \bibinfo {pages} {265} (\bibinfo {year} {2024})},\ \Eprint
  {http://arxiv.org/abs/2405.09124} {arXiv:2405.09124 [gr-qc]} \BibitemShut
  {NoStop}%
\bibitem [{\citenamefont {Agrawal}\ \emph {et~al.}(2024)\citenamefont
  {Agrawal}, \citenamefont {Zerbini},\ and\ \citenamefont
  {Mishra}}]{Agrawal:2024wwt}%
  \BibitemOpen
  \bibfield  {author} {\bibinfo {author} {\bibfnamefont {A.~S.}\ \bibnamefont
  {Agrawal}}, \bibinfo {author} {\bibfnamefont {S.}~\bibnamefont {Zerbini}}, \
  and\ \bibinfo {author} {\bibfnamefont {B.}~\bibnamefont {Mishra}},\ }\href
  {\doibase 10.1016/j.dark.2024.101637} {\bibfield  {journal} {\bibinfo
  {journal} {Phys. Dark Univ.}\ }\textbf {\bibinfo {volume} {46}},\ \bibinfo
  {pages} {101637} (\bibinfo {year} {2024})},\ \Eprint
  {http://arxiv.org/abs/2406.01241} {arXiv:2406.01241 [gr-qc]} \BibitemShut
  {NoStop}%
\bibitem [{\citenamefont {Belfiglio}\ \emph {et~al.}(2024)\citenamefont
  {Belfiglio}, \citenamefont {Chandran}, \citenamefont {Luongo},\ and\
  \citenamefont {Mancini}}]{Belfiglio:2024wel}%
  \BibitemOpen
  \bibfield  {author} {\bibinfo {author} {\bibfnamefont {A.}~\bibnamefont
  {Belfiglio}}, \bibinfo {author} {\bibfnamefont {S.~M.}\ \bibnamefont
  {Chandran}}, \bibinfo {author} {\bibfnamefont {O.}~\bibnamefont {Luongo}}, \
  and\ \bibinfo {author} {\bibfnamefont {S.}~\bibnamefont {Mancini}},\
  }\href@noop {} {\  (\bibinfo {year} {2024})},\ \Eprint
  {http://arxiv.org/abs/2407.03775} {arXiv:2407.03775 [gr-qc]} \BibitemShut
  {NoStop}%
\bibitem [{\citenamefont {Stashko}(2024)}]{Stashko:2024wuq}%
  \BibitemOpen
  \bibfield  {author} {\bibinfo {author} {\bibfnamefont {O.}~\bibnamefont
  {Stashko}},\ }\href {\doibase 10.1103/PhysRevD.110.084016} {\bibfield
  {journal} {\bibinfo  {journal} {Phys. Rev. D}\ }\textbf {\bibinfo {volume}
  {110}},\ \bibinfo {pages} {084016} (\bibinfo {year} {2024})},\ \Eprint
  {http://arxiv.org/abs/2407.07892} {arXiv:2407.07892 [gr-qc]} \BibitemShut
  {NoStop}%
\bibitem [{\citenamefont {Faraoni}\ and\ \citenamefont
  {Rinaldi}(2024)}]{Faraoni:2024ghi}%
  \BibitemOpen
  \bibfield  {author} {\bibinfo {author} {\bibfnamefont {V.}~\bibnamefont
  {Faraoni}}\ and\ \bibinfo {author} {\bibfnamefont {M.}~\bibnamefont
  {Rinaldi}},\ }\href {\doibase 10.1103/PhysRevD.110.063553} {\bibfield
  {journal} {\bibinfo  {journal} {Phys. Rev. D}\ }\textbf {\bibinfo {volume}
  {110}},\ \bibinfo {pages} {063553} (\bibinfo {year} {2024})},\ \Eprint
  {http://arxiv.org/abs/2407.14549} {arXiv:2407.14549 [gr-qc]} \BibitemShut
  {NoStop}%
\bibitem [{\citenamefont {Konoplya}\ and\ \citenamefont
  {Stashko}(2024)}]{Konoplya:2024lch}%
  \BibitemOpen
  \bibfield  {author} {\bibinfo {author} {\bibfnamefont {R.~A.}\ \bibnamefont
  {Konoplya}}\ and\ \bibinfo {author} {\bibfnamefont {O.~S.}\ \bibnamefont
  {Stashko}},\ }\href@noop {} {\  (\bibinfo {year} {2024})},\ \Eprint
  {http://arxiv.org/abs/2408.02578} {arXiv:2408.02578 [gr-qc]} \BibitemShut
  {NoStop}%
\bibitem [{\citenamefont {Khodadi}\ and\ \citenamefont
  {Firouzjaee}(2024)}]{Khodadi:2024efq}%
  \BibitemOpen
  \bibfield  {author} {\bibinfo {author} {\bibfnamefont {M.}~\bibnamefont
  {Khodadi}}\ and\ \bibinfo {author} {\bibfnamefont {J.~T.}\ \bibnamefont
  {Firouzjaee}},\ }\href {\doibase 10.1016/j.physletb.2024.138986} {\bibfield
  {journal} {\bibinfo  {journal} {Phys. Lett. B}\ }\textbf {\bibinfo {volume}
  {857}},\ \bibinfo {pages} {138986} (\bibinfo {year} {2024})},\ \Eprint
  {http://arxiv.org/abs/2408.12873} {arXiv:2408.12873 [gr-qc]} \BibitemShut
  {NoStop}%
\bibitem [{\citenamefont {Calz\`a}\ \emph
  {et~al.}(2024{\natexlab{b}})\citenamefont {Calz\`a}, \citenamefont
  {Gianesello}, \citenamefont {Rinaldi},\ and\ \citenamefont
  {Vagnozzi}}]{Calza:2024qxn}%
  \BibitemOpen
  \bibfield  {author} {\bibinfo {author} {\bibfnamefont {M.}~\bibnamefont
  {Calz\`a}}, \bibinfo {author} {\bibfnamefont {F.}~\bibnamefont {Gianesello}},
  \bibinfo {author} {\bibfnamefont {M.}~\bibnamefont {Rinaldi}}, \ and\
  \bibinfo {author} {\bibfnamefont {S.}~\bibnamefont {Vagnozzi}},\ }\href@noop
  {} {\  (\bibinfo {year} {2024}{\natexlab{b}})},\ \Eprint
  {http://arxiv.org/abs/2409.01801} {arXiv:2409.01801 [gr-qc]} \BibitemShut
  {NoStop}%
\bibitem [{\citenamefont {Dymnikova}(1992)}]{Dymnikova:1992ux}%
  \BibitemOpen
  \bibfield  {author} {\bibinfo {author} {\bibfnamefont {I.}~\bibnamefont
  {Dymnikova}},\ }\href {\doibase 10.1007/BF00760226} {\bibfield  {journal}
  {\bibinfo  {journal} {Gen. Rel. Grav.}\ }\textbf {\bibinfo {volume} {24}},\
  \bibinfo {pages} {235} (\bibinfo {year} {1992})}\BibitemShut {NoStop}%
\bibitem [{\citenamefont {Dymnikova}\ and\ \citenamefont
  {Galaktionov}(2005)}]{Dymnikova:2004qg}%
  \BibitemOpen
  \bibfield  {author} {\bibinfo {author} {\bibfnamefont {I.}~\bibnamefont
  {Dymnikova}}\ and\ \bibinfo {author} {\bibfnamefont {E.}~\bibnamefont
  {Galaktionov}},\ }\href {\doibase 10.1088/0264-9381/22/12/003} {\bibfield
  {journal} {\bibinfo  {journal} {Class. Quant. Grav.}\ }\textbf {\bibinfo
  {volume} {22}},\ \bibinfo {pages} {2331} (\bibinfo {year} {2005})},\ \Eprint
  {http://arxiv.org/abs/gr-qc/0409049} {arXiv:gr-qc/0409049} \BibitemShut
  {NoStop}%
\bibitem [{\citenamefont {Ashtekar}\ and\ \citenamefont
  {Bojowald}(2005)}]{Ashtekar:2005cj}%
  \BibitemOpen
  \bibfield  {author} {\bibinfo {author} {\bibfnamefont {A.}~\bibnamefont
  {Ashtekar}}\ and\ \bibinfo {author} {\bibfnamefont {M.}~\bibnamefont
  {Bojowald}},\ }\href {\doibase 10.1088/0264-9381/22/16/014} {\bibfield
  {journal} {\bibinfo  {journal} {Class. Quant. Grav.}\ }\textbf {\bibinfo
  {volume} {22}},\ \bibinfo {pages} {3349} (\bibinfo {year} {2005})},\ \Eprint
  {http://arxiv.org/abs/gr-qc/0504029} {arXiv:gr-qc/0504029} \BibitemShut
  {NoStop}%
\bibitem [{\citenamefont {Bebronne}\ and\ \citenamefont
  {Tinyakov}(2009)}]{Bebronne:2009mz}%
  \BibitemOpen
  \bibfield  {author} {\bibinfo {author} {\bibfnamefont {M.~V.}\ \bibnamefont
  {Bebronne}}\ and\ \bibinfo {author} {\bibfnamefont {P.~G.}\ \bibnamefont
  {Tinyakov}},\ }\href {\doibase 10.1007/JHEP06(2011)018} {\bibfield  {journal}
  {\bibinfo  {journal} {JHEP}\ }\textbf {\bibinfo {volume} {04}},\ \bibinfo
  {pages} {100} (\bibinfo {year} {2009})},\ \bibinfo {note} {[Erratum: JHEP 06,
  018 (2011)]},\ \Eprint {http://arxiv.org/abs/0902.3899} {arXiv:0902.3899
  [gr-qc]} \BibitemShut {NoStop}%
\bibitem [{\citenamefont {Modesto}\ \emph {et~al.}(2011)\citenamefont
  {Modesto}, \citenamefont {Moffat},\ and\ \citenamefont
  {Nicolini}}]{Modesto:2010uh}%
  \BibitemOpen
  \bibfield  {author} {\bibinfo {author} {\bibfnamefont {L.}~\bibnamefont
  {Modesto}}, \bibinfo {author} {\bibfnamefont {J.~W.}\ \bibnamefont {Moffat}},
  \ and\ \bibinfo {author} {\bibfnamefont {P.}~\bibnamefont {Nicolini}},\
  }\href {\doibase 10.1016/j.physletb.2010.11.046} {\bibfield  {journal}
  {\bibinfo  {journal} {Phys. Lett. B}\ }\textbf {\bibinfo {volume} {695}},\
  \bibinfo {pages} {397} (\bibinfo {year} {2011})},\ \Eprint
  {http://arxiv.org/abs/1010.0680} {arXiv:1010.0680 [gr-qc]} \BibitemShut
  {NoStop}%
\bibitem [{\citenamefont {Spallucci}\ and\ \citenamefont
  {Ansoldi}(2011)}]{Spallucci:2011rn}%
  \BibitemOpen
  \bibfield  {author} {\bibinfo {author} {\bibfnamefont {E.}~\bibnamefont
  {Spallucci}}\ and\ \bibinfo {author} {\bibfnamefont {S.}~\bibnamefont
  {Ansoldi}},\ }\href {\doibase 10.1016/j.physletb.2011.06.005} {\bibfield
  {journal} {\bibinfo  {journal} {Phys. Lett. B}\ }\textbf {\bibinfo {volume}
  {701}},\ \bibinfo {pages} {471} (\bibinfo {year} {2011})},\ \Eprint
  {http://arxiv.org/abs/1101.2760} {arXiv:1101.2760 [hep-th]} \BibitemShut
  {NoStop}%
\bibitem [{\citenamefont {Perez}(2015)}]{Perez:2014xca}%
  \BibitemOpen
  \bibfield  {author} {\bibinfo {author} {\bibfnamefont {A.}~\bibnamefont
  {Perez}},\ }\href {\doibase 10.1088/0264-9381/32/8/084001} {\bibfield
  {journal} {\bibinfo  {journal} {Class. Quant. Grav.}\ }\textbf {\bibinfo
  {volume} {32}},\ \bibinfo {pages} {084001} (\bibinfo {year} {2015})},\
  \Eprint {http://arxiv.org/abs/1410.7062} {arXiv:1410.7062 [gr-qc]}
  \BibitemShut {NoStop}%
\bibitem [{\citenamefont {Coll\'eaux}\ \emph {et~al.}(2018)\citenamefont
  {Coll\'eaux}, \citenamefont {Chinaglia},\ and\ \citenamefont
  {Zerbini}}]{Colleaux:2017ibe}%
  \BibitemOpen
  \bibfield  {author} {\bibinfo {author} {\bibfnamefont {A.}~\bibnamefont
  {Coll\'eaux}}, \bibinfo {author} {\bibfnamefont {S.}~\bibnamefont
  {Chinaglia}}, \ and\ \bibinfo {author} {\bibfnamefont {S.}~\bibnamefont
  {Zerbini}},\ }\href {\doibase 10.1142/S0218271818300021} {\bibfield
  {journal} {\bibinfo  {journal} {Int. J. Mod. Phys. D}\ }\textbf {\bibinfo
  {volume} {27}},\ \bibinfo {pages} {1830002} (\bibinfo {year} {2018})},\
  \Eprint {http://arxiv.org/abs/1712.03730} {arXiv:1712.03730 [gr-qc]}
  \BibitemShut {NoStop}%
\bibitem [{\citenamefont {Nicolini}\ \emph {et~al.}(2019)\citenamefont
  {Nicolini}, \citenamefont {Spallucci},\ and\ \citenamefont
  {Wondrak}}]{Nicolini:2019irw}%
  \BibitemOpen
  \bibfield  {author} {\bibinfo {author} {\bibfnamefont {P.}~\bibnamefont
  {Nicolini}}, \bibinfo {author} {\bibfnamefont {E.}~\bibnamefont {Spallucci}},
  \ and\ \bibinfo {author} {\bibfnamefont {M.~F.}\ \bibnamefont {Wondrak}},\
  }\href {\doibase 10.1016/j.physletb.2019.134888} {\bibfield  {journal}
  {\bibinfo  {journal} {Phys. Lett. B}\ }\textbf {\bibinfo {volume} {797}},\
  \bibinfo {pages} {134888} (\bibinfo {year} {2019})},\ \Eprint
  {http://arxiv.org/abs/1902.11242} {arXiv:1902.11242 [gr-qc]} \BibitemShut
  {NoStop}%
\bibitem [{\citenamefont {Bosma}\ \emph {et~al.}(2019)\citenamefont {Bosma},
  \citenamefont {Knorr},\ and\ \citenamefont {Saueressig}}]{Bosma:2019aiu}%
  \BibitemOpen
  \bibfield  {author} {\bibinfo {author} {\bibfnamefont {L.}~\bibnamefont
  {Bosma}}, \bibinfo {author} {\bibfnamefont {B.}~\bibnamefont {Knorr}}, \ and\
  \bibinfo {author} {\bibfnamefont {F.}~\bibnamefont {Saueressig}},\ }\href
  {\doibase 10.1103/PhysRevLett.123.101301} {\bibfield  {journal} {\bibinfo
  {journal} {Phys. Rev. Lett.}\ }\textbf {\bibinfo {volume} {123}},\ \bibinfo
  {pages} {101301} (\bibinfo {year} {2019})},\ \Eprint
  {http://arxiv.org/abs/1904.04845} {arXiv:1904.04845 [hep-th]} \BibitemShut
  {NoStop}%
\bibitem [{\citenamefont {Jusufi}(2023{\natexlab{a}})}]{Jusufi:2022cfw}%
  \BibitemOpen
  \bibfield  {author} {\bibinfo {author} {\bibfnamefont {K.}~\bibnamefont
  {Jusufi}},\ }\href {\doibase 10.1016/j.aop.2022.169191} {\bibfield  {journal}
  {\bibinfo  {journal} {Annals Phys.}\ }\textbf {\bibinfo {volume} {448}},\
  \bibinfo {pages} {169191} (\bibinfo {year} {2023}{\natexlab{a}})},\ \Eprint
  {http://arxiv.org/abs/2208.12979} {arXiv:2208.12979 [gr-qc]} \BibitemShut
  {NoStop}%
\bibitem [{\citenamefont {Olmo}\ and\ \citenamefont
  {Rubiera-Garcia}(2022)}]{Olmo:2022cui}%
  \BibitemOpen
  \bibfield  {author} {\bibinfo {author} {\bibfnamefont {G.~J.}\ \bibnamefont
  {Olmo}}\ and\ \bibinfo {author} {\bibfnamefont {D.}~\bibnamefont
  {Rubiera-Garcia}},\ }\href@noop {} {\  (\bibinfo {year} {2022})},\ \Eprint
  {http://arxiv.org/abs/2209.05061} {arXiv:2209.05061 [gr-qc]} \BibitemShut
  {NoStop}%
\bibitem [{\citenamefont {Jusufi}(2023{\natexlab{b}})}]{Jusufi:2022rbt}%
  \BibitemOpen
  \bibfield  {author} {\bibinfo {author} {\bibfnamefont {K.}~\bibnamefont
  {Jusufi}},\ }\href {\doibase 10.1016/j.dark.2022.101156} {\bibfield
  {journal} {\bibinfo  {journal} {Phys. Dark Univ.}\ }\textbf {\bibinfo
  {volume} {39}},\ \bibinfo {pages} {101156} (\bibinfo {year}
  {2023}{\natexlab{b}})},\ \Eprint {http://arxiv.org/abs/2212.06760}
  {arXiv:2212.06760 [gr-qc]} \BibitemShut {NoStop}%
\bibitem [{\citenamefont {Ashtekar}\ \emph {et~al.}(2023)\citenamefont
  {Ashtekar}, \citenamefont {Olmedo},\ and\ \citenamefont
  {Singh}}]{Ashtekar:2023cod}%
  \BibitemOpen
  \bibfield  {author} {\bibinfo {author} {\bibfnamefont {A.}~\bibnamefont
  {Ashtekar}}, \bibinfo {author} {\bibfnamefont {J.}~\bibnamefont {Olmedo}}, \
  and\ \bibinfo {author} {\bibfnamefont {P.}~\bibnamefont {Singh}},\
  }\href@noop {} {\  (\bibinfo {year} {2023})},\ \Eprint
  {http://arxiv.org/abs/2301.01309} {arXiv:2301.01309 [gr-qc]} \BibitemShut
  {NoStop}%
\bibitem [{\citenamefont {Nicolini}(2023)}]{Nicolini:2023hub}%
  \BibitemOpen
  \bibfield  {author} {\bibinfo {author} {\bibfnamefont {P.}~\bibnamefont
  {Nicolini}}\ }(\bibinfo {year} {2023})\ \Eprint
  {http://arxiv.org/abs/2306.01480} {arXiv:2306.01480 [gr-qc]} \BibitemShut
  {NoStop}%
\bibitem [{\citenamefont {Zhou}\ and\ \citenamefont
  {Modesto}(2023)}]{Zhou:2022yio}%
  \BibitemOpen
  \bibfield  {author} {\bibinfo {author} {\bibfnamefont {T.}~\bibnamefont
  {Zhou}}\ and\ \bibinfo {author} {\bibfnamefont {L.}~\bibnamefont {Modesto}},\
  }\href {\doibase 10.1103/PhysRevD.107.044016} {\bibfield  {journal} {\bibinfo
   {journal} {Phys. Rev. D}\ }\textbf {\bibinfo {volume} {107}},\ \bibinfo
  {pages} {044016} (\bibinfo {year} {2023})},\ \Eprint
  {http://arxiv.org/abs/2208.02557} {arXiv:2208.02557 [gr-qc]} \BibitemShut
  {NoStop}%
\bibitem [{\citenamefont {Carballo-Rubio}\ \emph {et~al.}(2021)\citenamefont
  {Carballo-Rubio}, \citenamefont {Di~Filippo}, \citenamefont {Liberati},
  \citenamefont {Pacilio},\ and\ \citenamefont
  {Visser}}]{Carballo-Rubio:2021bpr}%
  \BibitemOpen
  \bibfield  {author} {\bibinfo {author} {\bibfnamefont {R.}~\bibnamefont
  {Carballo-Rubio}}, \bibinfo {author} {\bibfnamefont {F.}~\bibnamefont
  {Di~Filippo}}, \bibinfo {author} {\bibfnamefont {S.}~\bibnamefont
  {Liberati}}, \bibinfo {author} {\bibfnamefont {C.}~\bibnamefont {Pacilio}}, \
  and\ \bibinfo {author} {\bibfnamefont {M.}~\bibnamefont {Visser}},\ }\href
  {\doibase 10.1007/JHEP05(2021)132} {\bibfield  {journal} {\bibinfo  {journal}
  {JHEP}\ }\textbf {\bibinfo {volume} {05}},\ \bibinfo {pages} {132} (\bibinfo
  {year} {2021})},\ \Eprint {http://arxiv.org/abs/2101.05006} {arXiv:2101.05006
  [gr-qc]} \BibitemShut {NoStop}%
\bibitem [{\citenamefont {Carballo-Rubio}\ \emph {et~al.}(2022)\citenamefont
  {Carballo-Rubio}, \citenamefont {Di~Filippo}, \citenamefont {Liberati},
  \citenamefont {Pacilio},\ and\ \citenamefont
  {Visser}}]{Carballo-Rubio:2022kad}%
  \BibitemOpen
  \bibfield  {author} {\bibinfo {author} {\bibfnamefont {R.}~\bibnamefont
  {Carballo-Rubio}}, \bibinfo {author} {\bibfnamefont {F.}~\bibnamefont
  {Di~Filippo}}, \bibinfo {author} {\bibfnamefont {S.}~\bibnamefont
  {Liberati}}, \bibinfo {author} {\bibfnamefont {C.}~\bibnamefont {Pacilio}}, \
  and\ \bibinfo {author} {\bibfnamefont {M.}~\bibnamefont {Visser}},\ }\href
  {\doibase 10.1007/JHEP09(2022)118} {\bibfield  {journal} {\bibinfo  {journal}
  {JHEP}\ }\textbf {\bibinfo {volume} {09}},\ \bibinfo {pages} {118} (\bibinfo
  {year} {2022})},\ \Eprint {http://arxiv.org/abs/2205.13556} {arXiv:2205.13556
  [gr-qc]} \BibitemShut {NoStop}%
\bibitem [{\citenamefont {Bonanno}\ \emph {et~al.}(2023)\citenamefont
  {Bonanno}, \citenamefont {Khosravi},\ and\ \citenamefont
  {Saueressig}}]{Bonanno:2022jjp}%
  \BibitemOpen
  \bibfield  {author} {\bibinfo {author} {\bibfnamefont {A.}~\bibnamefont
  {Bonanno}}, \bibinfo {author} {\bibfnamefont {A.-P.}\ \bibnamefont
  {Khosravi}}, \ and\ \bibinfo {author} {\bibfnamefont {F.}~\bibnamefont
  {Saueressig}},\ }\href {\doibase 10.1103/PhysRevD.107.024005} {\bibfield
  {journal} {\bibinfo  {journal} {Phys. Rev. D}\ }\textbf {\bibinfo {volume}
  {107}},\ \bibinfo {pages} {024005} (\bibinfo {year} {2023})},\ \Eprint
  {http://arxiv.org/abs/2209.10612} {arXiv:2209.10612 [gr-qc]} \BibitemShut
  {NoStop}%
\bibitem [{\citenamefont {Bonanno}\ and\ \citenamefont
  {Saueressig}(2022)}]{Bonanno:2022rvo}%
  \BibitemOpen
  \bibfield  {author} {\bibinfo {author} {\bibfnamefont {A.}~\bibnamefont
  {Bonanno}}\ and\ \bibinfo {author} {\bibfnamefont {F.}~\bibnamefont
  {Saueressig}},\ }\href@noop {} {\  (\bibinfo {year} {2022})},\ \Eprint
  {http://arxiv.org/abs/2211.09192} {arXiv:2211.09192 [gr-qc]} \BibitemShut
  {NoStop}%
\bibitem [{\citenamefont {Carballo-Rubio}\ \emph {et~al.}(2023)\citenamefont
  {Carballo-Rubio}, \citenamefont {Di~Filippo}, \citenamefont {Liberati},
  \citenamefont {Pacilio},\ and\ \citenamefont
  {Visser}}]{Carballo-Rubio:2022twq}%
  \BibitemOpen
  \bibfield  {author} {\bibinfo {author} {\bibfnamefont {R.}~\bibnamefont
  {Carballo-Rubio}}, \bibinfo {author} {\bibfnamefont {F.}~\bibnamefont
  {Di~Filippo}}, \bibinfo {author} {\bibfnamefont {S.}~\bibnamefont
  {Liberati}}, \bibinfo {author} {\bibfnamefont {C.}~\bibnamefont {Pacilio}}, \
  and\ \bibinfo {author} {\bibfnamefont {M.}~\bibnamefont {Visser}},\ }\href
  {\doibase 10.1103/PhysRevD.108.128501} {\bibfield  {journal} {\bibinfo
  {journal} {Phys. Rev. D}\ }\textbf {\bibinfo {volume} {108}},\ \bibinfo
  {pages} {128501} (\bibinfo {year} {2023})},\ \Eprint
  {http://arxiv.org/abs/2212.07458} {arXiv:2212.07458 [gr-qc]} \BibitemShut
  {NoStop}%
\bibitem [{\citenamefont {Barcel\'o}\ \emph {et~al.}(2017)\citenamefont
  {Barcel\'o}, \citenamefont {Carballo-Rubio},\ and\ \citenamefont
  {Garay}}]{Barcelo:2016hgb}%
  \BibitemOpen
  \bibfield  {author} {\bibinfo {author} {\bibfnamefont {C.}~\bibnamefont
  {Barcel\'o}}, \bibinfo {author} {\bibfnamefont {R.}~\bibnamefont
  {Carballo-Rubio}}, \ and\ \bibinfo {author} {\bibfnamefont {L.~J.}\
  \bibnamefont {Garay}},\ }\href {\doibase 10.1088/1361-6382/aa6962} {\bibfield
   {journal} {\bibinfo  {journal} {Class. Quant. Grav.}\ }\textbf {\bibinfo
  {volume} {34}},\ \bibinfo {pages} {105007} (\bibinfo {year} {2017})},\
  \Eprint {http://arxiv.org/abs/1607.03480} {arXiv:1607.03480 [gr-qc]}
  \BibitemShut {NoStop}%
\bibitem [{\citenamefont {Malafarina}(2017)}]{Malafarina:2017csn}%
  \BibitemOpen
  \bibfield  {author} {\bibinfo {author} {\bibfnamefont {D.}~\bibnamefont
  {Malafarina}},\ }\href {\doibase 10.3390/universe3020048} {\bibfield
  {journal} {\bibinfo  {journal} {Universe}\ }\textbf {\bibinfo {volume} {3}},\
  \bibinfo {pages} {48} (\bibinfo {year} {2017})},\ \Eprint
  {http://arxiv.org/abs/1703.04138} {arXiv:1703.04138 [gr-qc]} \BibitemShut
  {NoStop}%
\bibitem [{\citenamefont {Barrau}\ \emph {et~al.}(2018)\citenamefont {Barrau},
  \citenamefont {Martineau},\ and\ \citenamefont {Moulin}}]{Barrau:2018rts}%
  \BibitemOpen
  \bibfield  {author} {\bibinfo {author} {\bibfnamefont {A.}~\bibnamefont
  {Barrau}}, \bibinfo {author} {\bibfnamefont {K.}~\bibnamefont {Martineau}}, \
  and\ \bibinfo {author} {\bibfnamefont {F.}~\bibnamefont {Moulin}},\ }\href
  {\doibase 10.3390/universe4100102} {\bibfield  {journal} {\bibinfo  {journal}
  {Universe}\ }\textbf {\bibinfo {volume} {4}},\ \bibinfo {pages} {102}
  (\bibinfo {year} {2018})},\ \Eprint {http://arxiv.org/abs/1808.08857}
  {arXiv:1808.08857 [gr-qc]} \BibitemShut {NoStop}%
\bibitem [{\citenamefont {Simpson}\ and\ \citenamefont
  {Visser}(2019)}]{Simpson:2018tsi}%
  \BibitemOpen
  \bibfield  {author} {\bibinfo {author} {\bibfnamefont {A.}~\bibnamefont
  {Simpson}}\ and\ \bibinfo {author} {\bibfnamefont {M.}~\bibnamefont
  {Visser}},\ }\href {\doibase 10.1088/1475-7516/2019/02/042} {\bibfield
  {journal} {\bibinfo  {journal} {JCAP}\ }\textbf {\bibinfo {volume} {02}},\
  \bibinfo {pages} {042} (\bibinfo {year} {2019})},\ \Eprint
  {http://arxiv.org/abs/1812.07114} {arXiv:1812.07114 [gr-qc]} \BibitemShut
  {NoStop}%
\bibitem [{\citenamefont {Tsukamoto}(2021)}]{Tsukamoto:2020bjm}%
  \BibitemOpen
  \bibfield  {author} {\bibinfo {author} {\bibfnamefont {N.}~\bibnamefont
  {Tsukamoto}},\ }\href {\doibase 10.1103/PhysRevD.103.024033} {\bibfield
  {journal} {\bibinfo  {journal} {Phys. Rev. D}\ }\textbf {\bibinfo {volume}
  {103}},\ \bibinfo {pages} {024033} (\bibinfo {year} {2021})},\ \Eprint
  {http://arxiv.org/abs/2011.03932} {arXiv:2011.03932 [gr-qc]} \BibitemShut
  {NoStop}%
\bibitem [{\citenamefont {Mazza}\ \emph {et~al.}(2021)\citenamefont {Mazza},
  \citenamefont {Franzin},\ and\ \citenamefont {Liberati}}]{Mazza:2021rgq}%
  \BibitemOpen
  \bibfield  {author} {\bibinfo {author} {\bibfnamefont {J.}~\bibnamefont
  {Mazza}}, \bibinfo {author} {\bibfnamefont {E.}~\bibnamefont {Franzin}}, \
  and\ \bibinfo {author} {\bibfnamefont {S.}~\bibnamefont {Liberati}},\ }\href
  {\doibase 10.1088/1475-7516/2021/04/082} {\bibfield  {journal} {\bibinfo
  {journal} {JCAP}\ }\textbf {\bibinfo {volume} {04}},\ \bibinfo {pages} {082}
  (\bibinfo {year} {2021})},\ \Eprint {http://arxiv.org/abs/2102.01105}
  {arXiv:2102.01105 [gr-qc]} \BibitemShut {NoStop}%
\bibitem [{\citenamefont {Shaikh}\ \emph {et~al.}(2021)\citenamefont {Shaikh},
  \citenamefont {Pal}, \citenamefont {Pal},\ and\ \citenamefont
  {Sarkar}}]{Shaikh:2021yux}%
  \BibitemOpen
  \bibfield  {author} {\bibinfo {author} {\bibfnamefont {R.}~\bibnamefont
  {Shaikh}}, \bibinfo {author} {\bibfnamefont {K.}~\bibnamefont {Pal}},
  \bibinfo {author} {\bibfnamefont {K.}~\bibnamefont {Pal}}, \ and\ \bibinfo
  {author} {\bibfnamefont {T.}~\bibnamefont {Sarkar}},\ }\href {\doibase
  10.1093/mnras/stab1779} {\bibfield  {journal} {\bibinfo  {journal} {Mon. Not.
  Roy. Astron. Soc.}\ }\textbf {\bibinfo {volume} {506}},\ \bibinfo {pages}
  {1229} (\bibinfo {year} {2021})},\ \Eprint {http://arxiv.org/abs/2102.04299}
  {arXiv:2102.04299 [gr-qc]} \BibitemShut {NoStop}%
\bibitem [{\citenamefont {Islam}\ \emph {et~al.}(2021)\citenamefont {Islam},
  \citenamefont {Kumar},\ and\ \citenamefont {Ghosh}}]{Islam:2021ful}%
  \BibitemOpen
  \bibfield  {author} {\bibinfo {author} {\bibfnamefont {S.~U.}\ \bibnamefont
  {Islam}}, \bibinfo {author} {\bibfnamefont {J.}~\bibnamefont {Kumar}}, \ and\
  \bibinfo {author} {\bibfnamefont {S.~G.}\ \bibnamefont {Ghosh}},\ }\href
  {\doibase 10.1088/1475-7516/2021/10/013} {\bibfield  {journal} {\bibinfo
  {journal} {JCAP}\ }\textbf {\bibinfo {volume} {10}},\ \bibinfo {pages} {013}
  (\bibinfo {year} {2021})},\ \Eprint {http://arxiv.org/abs/2104.00696}
  {arXiv:2104.00696 [gr-qc]} \BibitemShut {NoStop}%
\bibitem [{\citenamefont {Guerrero}\ \emph {et~al.}(2021)\citenamefont
  {Guerrero}, \citenamefont {Olmo}, \citenamefont {Rubiera-Garcia},\ and\
  \citenamefont {G\'omez}}]{Guerrero:2021ues}%
  \BibitemOpen
  \bibfield  {author} {\bibinfo {author} {\bibfnamefont {M.}~\bibnamefont
  {Guerrero}}, \bibinfo {author} {\bibfnamefont {G.~J.}\ \bibnamefont {Olmo}},
  \bibinfo {author} {\bibfnamefont {D.}~\bibnamefont {Rubiera-Garcia}}, \ and\
  \bibinfo {author} {\bibfnamefont {D.~S.-C.}\ \bibnamefont {G\'omez}},\ }\href
  {\doibase 10.1088/1475-7516/2021/08/036} {\bibfield  {journal} {\bibinfo
  {journal} {JCAP}\ }\textbf {\bibinfo {volume} {08}},\ \bibinfo {pages} {036}
  (\bibinfo {year} {2021})},\ \Eprint {http://arxiv.org/abs/2105.15073}
  {arXiv:2105.15073 [gr-qc]} \BibitemShut {NoStop}%
\bibitem [{\citenamefont {Bambhaniya}\ \emph {et~al.}(2022)\citenamefont
  {Bambhaniya}, \citenamefont {K}, \citenamefont {Jusufi},\ and\ \citenamefont
  {Joshi}}]{Bambhaniya:2021ugr}%
  \BibitemOpen
  \bibfield  {author} {\bibinfo {author} {\bibfnamefont {P.}~\bibnamefont
  {Bambhaniya}}, \bibinfo {author} {\bibfnamefont {S.}~\bibnamefont {K}},
  \bibinfo {author} {\bibfnamefont {K.}~\bibnamefont {Jusufi}}, \ and\ \bibinfo
  {author} {\bibfnamefont {P.~S.}\ \bibnamefont {Joshi}},\ }\href {\doibase
  10.1103/PhysRevD.105.023021} {\bibfield  {journal} {\bibinfo  {journal}
  {Phys. Rev. D}\ }\textbf {\bibinfo {volume} {105}},\ \bibinfo {pages}
  {023021} (\bibinfo {year} {2022})},\ \Eprint
  {http://arxiv.org/abs/2109.15054} {arXiv:2109.15054 [gr-qc]} \BibitemShut
  {NoStop}%
\bibitem [{\citenamefont {Yang}\ \emph {et~al.}(2022)\citenamefont {Yang},
  \citenamefont {Liu}, \citenamefont {\"Ovg\"un}, \citenamefont {Long},\ and\
  \citenamefont {Xu}}]{Yang:2022xxh}%
  \BibitemOpen
  \bibfield  {author} {\bibinfo {author} {\bibfnamefont {Y.}~\bibnamefont
  {Yang}}, \bibinfo {author} {\bibfnamefont {D.}~\bibnamefont {Liu}}, \bibinfo
  {author} {\bibfnamefont {A.}~\bibnamefont {\"Ovg\"un}}, \bibinfo {author}
  {\bibfnamefont {Z.-W.}\ \bibnamefont {Long}}, \ and\ \bibinfo {author}
  {\bibfnamefont {Z.}~\bibnamefont {Xu}},\ }\href@noop {} {\  (\bibinfo {year}
  {2022})},\ \Eprint {http://arxiv.org/abs/2205.07530} {arXiv:2205.07530
  [gr-qc]} \BibitemShut {NoStop}%
\bibitem [{\citenamefont {Riaz}\ \emph {et~al.}(2022)\citenamefont {Riaz},
  \citenamefont {Shashank}, \citenamefont {Roy}, \citenamefont {Abdikamalov},
  \citenamefont {Ayzenberg}, \citenamefont {Bambi}, \citenamefont {Zhang},\
  and\ \citenamefont {Zhou}}]{Riaz:2022rlx}%
  \BibitemOpen
  \bibfield  {author} {\bibinfo {author} {\bibfnamefont {S.}~\bibnamefont
  {Riaz}}, \bibinfo {author} {\bibfnamefont {S.}~\bibnamefont {Shashank}},
  \bibinfo {author} {\bibfnamefont {R.}~\bibnamefont {Roy}}, \bibinfo {author}
  {\bibfnamefont {A.~B.}\ \bibnamefont {Abdikamalov}}, \bibinfo {author}
  {\bibfnamefont {D.}~\bibnamefont {Ayzenberg}}, \bibinfo {author}
  {\bibfnamefont {C.}~\bibnamefont {Bambi}}, \bibinfo {author} {\bibfnamefont
  {Z.}~\bibnamefont {Zhang}}, \ and\ \bibinfo {author} {\bibfnamefont
  {M.}~\bibnamefont {Zhou}},\ }\href {\doibase 10.1088/1475-7516/2022/10/040}
  {\bibfield  {journal} {\bibinfo  {journal} {JCAP}\ }\textbf {\bibinfo
  {volume} {10}},\ \bibinfo {pages} {040} (\bibinfo {year} {2022})},\ \Eprint
  {http://arxiv.org/abs/2206.03729} {arXiv:2206.03729 [gr-qc]} \BibitemShut
  {NoStop}%
\bibitem [{\citenamefont {Arora}\ \emph {et~al.}(2024)\citenamefont {Arora},
  \citenamefont {Bambhaniya}, \citenamefont {Dey},\ and\ \citenamefont
  {Joshi}}]{Arora:2023ltv}%
  \BibitemOpen
  \bibfield  {author} {\bibinfo {author} {\bibfnamefont {D.}~\bibnamefont
  {Arora}}, \bibinfo {author} {\bibfnamefont {P.}~\bibnamefont {Bambhaniya}},
  \bibinfo {author} {\bibfnamefont {D.}~\bibnamefont {Dey}}, \ and\ \bibinfo
  {author} {\bibfnamefont {P.~S.}\ \bibnamefont {Joshi}},\ }\href {\doibase
  10.1016/j.dark.2024.101487} {\bibfield  {journal} {\bibinfo  {journal} {Phys.
  Dark Univ.}\ }\textbf {\bibinfo {volume} {44}},\ \bibinfo {pages} {101487}
  (\bibinfo {year} {2024})},\ \Eprint {http://arxiv.org/abs/2305.08082}
  {arXiv:2305.08082 [gr-qc]} \BibitemShut {NoStop}%
\bibitem [{\citenamefont {Jha}(2023)}]{Jha:2023wzo}%
  \BibitemOpen
  \bibfield  {author} {\bibinfo {author} {\bibfnamefont {S.~K.}\ \bibnamefont
  {Jha}},\ }\href {\doibase 10.1140/epjp/s13360-023-04384-5} {\bibfield
  {journal} {\bibinfo  {journal} {Eur. Phys. J. Plus}\ }\textbf {\bibinfo
  {volume} {138}},\ \bibinfo {pages} {757} (\bibinfo {year} {2023})},\ \Eprint
  {http://arxiv.org/abs/2309.06454} {arXiv:2309.06454 [gr-qc]} \BibitemShut
  {NoStop}%
\bibitem [{\citenamefont {Jha}\ and\ \citenamefont
  {Rahaman}(2023)}]{Jha:2023nkh}%
  \BibitemOpen
  \bibfield  {author} {\bibinfo {author} {\bibfnamefont {S.~K.}\ \bibnamefont
  {Jha}}\ and\ \bibinfo {author} {\bibfnamefont {A.}~\bibnamefont {Rahaman}},\
  }\href {\doibase 10.1016/j.dark.2023.101327} {\bibfield  {journal} {\bibinfo
  {journal} {Phys. Dark Univ.}\ }\textbf {\bibinfo {volume} {42}},\ \bibinfo
  {pages} {101327} (\bibinfo {year} {2023})}\BibitemShut {NoStop}%
\bibitem [{\citenamefont {Bronnikov}\ and\ \citenamefont
  {Walia}(2022)}]{Bronnikov:2021uta}%
  \BibitemOpen
  \bibfield  {author} {\bibinfo {author} {\bibfnamefont {K.~A.}\ \bibnamefont
  {Bronnikov}}\ and\ \bibinfo {author} {\bibfnamefont {R.~K.}\ \bibnamefont
  {Walia}},\ }\href {\doibase 10.1103/PhysRevD.105.044039} {\bibfield
  {journal} {\bibinfo  {journal} {Phys. Rev. D}\ }\textbf {\bibinfo {volume}
  {105}},\ \bibinfo {pages} {044039} (\bibinfo {year} {2022})},\ \Eprint
  {http://arxiv.org/abs/2112.13198} {arXiv:2112.13198 [gr-qc]} \BibitemShut
  {NoStop}%
\bibitem [{\citenamefont {Sakalli}\ and\ \citenamefont
  {Kanzi}(2022)}]{Sakalli:2022xrb}%
  \BibitemOpen
  \bibfield  {author} {\bibinfo {author} {\bibfnamefont {I.}~\bibnamefont
  {Sakalli}}\ and\ \bibinfo {author} {\bibfnamefont {S.}~\bibnamefont
  {Kanzi}},\ }\href {\doibase 10.55730/1300-0101.269} {\bibfield  {journal}
  {\bibinfo  {journal} {Turk. J. Phys.}\ }\textbf {\bibinfo {volume} {46}},\
  \bibinfo {pages} {51} (\bibinfo {year} {2022})},\ \Eprint
  {http://arxiv.org/abs/2205.01771} {arXiv:2205.01771 [hep-th]} \BibitemShut
  {NoStop}%
\bibitem [{\citenamefont {Konoplya}\ and\ \citenamefont
  {Zhidenko}(2024{\natexlab{b}})}]{Konoplya:2024lir}%
  \BibitemOpen
  \bibfield  {author} {\bibinfo {author} {\bibfnamefont {R.~A.}\ \bibnamefont
  {Konoplya}}\ and\ \bibinfo {author} {\bibfnamefont {A.}~\bibnamefont
  {Zhidenko}},\ }\href {\doibase 10.1088/1475-7516/2024/09/068} {\bibfield
  {journal} {\bibinfo  {journal} {JCAP}\ }\textbf {\bibinfo {volume} {09}},\
  \bibinfo {pages} {068} (\bibinfo {year} {2024}{\natexlab{b}})},\ \Eprint
  {http://arxiv.org/abs/2406.11694} {arXiv:2406.11694 [gr-qc]} \BibitemShut
  {NoStop}%
\bibitem [{\citenamefont {Konoplya}\ and\ \citenamefont
  {Zhidenko}(2024{\natexlab{c}})}]{Konoplya:2024vuj}%
  \BibitemOpen
  \bibfield  {author} {\bibinfo {author} {\bibfnamefont {R.~A.}\ \bibnamefont
  {Konoplya}}\ and\ \bibinfo {author} {\bibfnamefont {A.}~\bibnamefont
  {Zhidenko}},\ }\href@noop {} {\  (\bibinfo {year} {2024}{\natexlab{c}})},\
  \Eprint {http://arxiv.org/abs/2408.11162} {arXiv:2408.11162 [gr-qc]}
  \BibitemShut {NoStop}%
\bibitem [{\citenamefont {Teukolsky}(1973)}]{Teukolsky:1973ha}%
  \BibitemOpen
  \bibfield  {author} {\bibinfo {author} {\bibfnamefont {S.~A.}\ \bibnamefont
  {Teukolsky}},\ }\href {\doibase 10.1086/152444} {\bibfield  {journal}
  {\bibinfo  {journal} {Astrophys. J.}\ }\textbf {\bibinfo {volume} {185}},\
  \bibinfo {pages} {635} (\bibinfo {year} {1973})}\BibitemShut {NoStop}%
\bibitem [{\citenamefont {Arbey}\ \emph
  {et~al.}(2021{\natexlab{b}})\citenamefont {Arbey}, \citenamefont {Auffinger},
  \citenamefont {Geiller}, \citenamefont {Livine},\ and\ \citenamefont
  {Sartini}}]{Arbey:2021jif}%
  \BibitemOpen
  \bibfield  {author} {\bibinfo {author} {\bibfnamefont {A.}~\bibnamefont
  {Arbey}}, \bibinfo {author} {\bibfnamefont {J.}~\bibnamefont {Auffinger}},
  \bibinfo {author} {\bibfnamefont {M.}~\bibnamefont {Geiller}}, \bibinfo
  {author} {\bibfnamefont {E.~R.}\ \bibnamefont {Livine}}, \ and\ \bibinfo
  {author} {\bibfnamefont {F.}~\bibnamefont {Sartini}},\ }\href {\doibase
  10.1103/PhysRevD.103.104010} {\bibfield  {journal} {\bibinfo  {journal}
  {Phys. Rev. D}\ }\textbf {\bibinfo {volume} {103}},\ \bibinfo {pages}
  {104010} (\bibinfo {year} {2021}{\natexlab{b}})},\ \Eprint
  {http://arxiv.org/abs/2101.02951} {arXiv:2101.02951 [gr-qc]} \BibitemShut
  {NoStop}%
\bibitem [{\citenamefont {Rosa}(2017)}]{Rosa:2016bli}%
  \BibitemOpen
  \bibfield  {author} {\bibinfo {author} {\bibfnamefont {J.~G.}\ \bibnamefont
  {Rosa}},\ }\href {\doibase 10.1103/PhysRevD.95.064017} {\bibfield  {journal}
  {\bibinfo  {journal} {Phys. Rev. D}\ }\textbf {\bibinfo {volume} {95}},\
  \bibinfo {pages} {064017} (\bibinfo {year} {2017})},\ \Eprint
  {http://arxiv.org/abs/1612.01826} {arXiv:1612.01826 [gr-qc]} \BibitemShut
  {NoStop}%
\bibitem [{\citenamefont {Rosa}(2013)}]{Rosa:2012uz}%
  \BibitemOpen
  \bibfield  {author} {\bibinfo {author} {\bibfnamefont {J.~G.}\ \bibnamefont
  {Rosa}},\ }\href {\doibase 10.1007/JHEP02(2013)014} {\bibfield  {journal}
  {\bibinfo  {journal} {JHEP}\ }\textbf {\bibinfo {volume} {02}},\ \bibinfo
  {pages} {014} (\bibinfo {year} {2013})},\ \Eprint
  {http://arxiv.org/abs/1209.4211} {arXiv:1209.4211 [hep-th]} \BibitemShut
  {NoStop}%
\bibitem [{\citenamefont {Calz\`a}\ \emph {et~al.}(2021)\citenamefont
  {Calz\`a}, \citenamefont {March-Russell},\ and\ \citenamefont
  {Rosa}}]{Calza:2021czr}%
  \BibitemOpen
  \bibfield  {author} {\bibinfo {author} {\bibfnamefont {M.}~\bibnamefont
  {Calz\`a}}, \bibinfo {author} {\bibfnamefont {J.}~\bibnamefont
  {March-Russell}}, \ and\ \bibinfo {author} {\bibfnamefont {J.~a.~G.}\
  \bibnamefont {Rosa}},\ }\href@noop {} {\  (\bibinfo {year} {2021})},\ \Eprint
  {http://arxiv.org/abs/2110.13602} {arXiv:2110.13602 [astro-ph.CO]}
  \BibitemShut {NoStop}%
\bibitem [{\citenamefont {Calz\'a}(2023)}]{Calza:2022ioe}%
  \BibitemOpen
  \bibfield  {author} {\bibinfo {author} {\bibfnamefont {M.}~\bibnamefont
  {Calz\'a}},\ }\href {\doibase 10.1103/PhysRevD.107.044067} {\bibfield
  {journal} {\bibinfo  {journal} {Phys. Rev. D}\ }\textbf {\bibinfo {volume}
  {107}},\ \bibinfo {pages} {044067} (\bibinfo {year} {2023})},\ \Eprint
  {http://arxiv.org/abs/2207.10467} {arXiv:2207.10467 [gr-qc]} \BibitemShut
  {NoStop}%
\bibitem [{\citenamefont {Calz\`a}\ and\ \citenamefont
  {Rosa}(2022)}]{Calza:2022ljw}%
  \BibitemOpen
  \bibfield  {author} {\bibinfo {author} {\bibfnamefont {M.}~\bibnamefont
  {Calz\`a}}\ and\ \bibinfo {author} {\bibfnamefont {J.~a.~G.}\ \bibnamefont
  {Rosa}},\ }\href {\doibase 10.1007/JHEP12(2022)090} {\bibfield  {journal}
  {\bibinfo  {journal} {JHEP}\ }\textbf {\bibinfo {volume} {12}},\ \bibinfo
  {pages} {090} (\bibinfo {year} {2022})},\ \Eprint
  {http://arxiv.org/abs/2210.06500} {arXiv:2210.06500 [gr-qc]} \BibitemShut
  {NoStop}%
\bibitem [{\citenamefont {Calz\`a}\ \emph
  {et~al.}(2024{\natexlab{c}})\citenamefont {Calz\`a}, \citenamefont {Rosa},\
  and\ \citenamefont {Serrano}}]{Calza:2023rjt}%
  \BibitemOpen
  \bibfield  {author} {\bibinfo {author} {\bibfnamefont {M.}~\bibnamefont
  {Calz\`a}}, \bibinfo {author} {\bibfnamefont {J.~a.~G.}\ \bibnamefont
  {Rosa}}, \ and\ \bibinfo {author} {\bibfnamefont {F.}~\bibnamefont
  {Serrano}},\ }\href {\doibase 10.1007/JHEP05(2024)140} {\bibfield  {journal}
  {\bibinfo  {journal} {JHEP}\ }\textbf {\bibinfo {volume} {05}},\ \bibinfo
  {pages} {140} (\bibinfo {year} {2024}{\natexlab{c}})},\ \Eprint
  {http://arxiv.org/abs/2306.09430} {arXiv:2306.09430 [hep-ph]} \BibitemShut
  {NoStop}%
\bibitem [{\citenamefont {Calz\`a}\ and\ \citenamefont
  {Rosa}(2024)}]{Calza:2023gws}%
  \BibitemOpen
  \bibfield  {author} {\bibinfo {author} {\bibfnamefont {M.}~\bibnamefont
  {Calz\`a}}\ and\ \bibinfo {author} {\bibfnamefont {J.~a.~G.}\ \bibnamefont
  {Rosa}},\ }\href {\doibase 10.1007/JHEP08(2024)012} {\bibfield  {journal}
  {\bibinfo  {journal} {JHEP}\ }\textbf {\bibinfo {volume} {08}},\ \bibinfo
  {pages} {012} (\bibinfo {year} {2024})},\ \Eprint
  {http://arxiv.org/abs/2311.12930} {arXiv:2311.12930 [gr-qc]} \BibitemShut
  {NoStop}%
\bibitem [{\citenamefont {Calz\`a}\ and\ \citenamefont
  {Rosa}(2023)}]{Calza:2023iqa}%
  \BibitemOpen
  \bibfield  {author} {\bibinfo {author} {\bibfnamefont {M.}~\bibnamefont
  {Calz\`a}}\ and\ \bibinfo {author} {\bibfnamefont {J.~a.~G.}\ \bibnamefont
  {Rosa}},\ }\href@noop {} {\  (\bibinfo {year} {2023})},\ \Eprint
  {http://arxiv.org/abs/2312.09261} {arXiv:2312.09261 [hep-ph]} \BibitemShut
  {NoStop}%
\bibitem [{\citenamefont {Leaver}(1985)}]{Leaver:1985ax}%
  \BibitemOpen
  \bibfield  {author} {\bibinfo {author} {\bibfnamefont {E.~W.}\ \bibnamefont
  {Leaver}},\ }\href {\doibase 10.1098/rspa.1985.0119} {\bibfield  {journal}
  {\bibinfo  {journal} {Proc. Roy. Soc. Lond. A}\ }\textbf {\bibinfo {volume}
  {402}},\ \bibinfo {pages} {285} (\bibinfo {year} {1985})}\BibitemShut
  {NoStop}%
\bibitem [{\citenamefont {Leaver}(1986)}]{Leaver:1986vnb}%
  \BibitemOpen
  \bibfield  {author} {\bibinfo {author} {\bibfnamefont {E.~W.}\ \bibnamefont
  {Leaver}},\ }\href {\doibase 10.1063/1.527130} {\bibfield  {journal}
  {\bibinfo  {journal} {J. Math. Phys.}\ }\textbf {\bibinfo {volume} {27}},\
  \bibinfo {pages} {1238} (\bibinfo {year} {1986})}\BibitemShut {NoStop}%
\bibitem [{\citenamefont {Konoplya}\ \emph {et~al.}(2023)\citenamefont
  {Konoplya}, \citenamefont {Ovchinnikov},\ and\ \citenamefont
  {Ahmedov}}]{Konoplya:2023ahd}%
  \BibitemOpen
  \bibfield  {author} {\bibinfo {author} {\bibfnamefont {R.~A.}\ \bibnamefont
  {Konoplya}}, \bibinfo {author} {\bibfnamefont {D.}~\bibnamefont
  {Ovchinnikov}}, \ and\ \bibinfo {author} {\bibfnamefont {B.}~\bibnamefont
  {Ahmedov}},\ }\href {\doibase 10.1103/PhysRevD.108.104054} {\bibfield
  {journal} {\bibinfo  {journal} {Phys. Rev. D}\ }\textbf {\bibinfo {volume}
  {108}},\ \bibinfo {pages} {104054} (\bibinfo {year} {2023})},\ \Eprint
  {http://arxiv.org/abs/2307.10801} {arXiv:2307.10801 [gr-qc]} \BibitemShut
  {NoStop}%
\bibitem [{\citenamefont {Konoplya}(2023)}]{Konoplya:2023ppx}%
  \BibitemOpen
  \bibfield  {author} {\bibinfo {author} {\bibfnamefont {R.~A.}\ \bibnamefont
  {Konoplya}},\ }\href {\doibase 10.1088/1475-7516/2023/07/001} {\bibfield
  {journal} {\bibinfo  {journal} {JCAP}\ }\textbf {\bibinfo {volume} {07}},\
  \bibinfo {pages} {001} (\bibinfo {year} {2023})},\ \Eprint
  {http://arxiv.org/abs/2305.09187} {arXiv:2305.09187 [gr-qc]} \BibitemShut
  {NoStop}%
\bibitem [{\citenamefont {Hawking}(1975{\natexlab{a}})}]{Hawking:1975iha}%
  \BibitemOpen
  \bibfield  {author} {\bibinfo {author} {\bibfnamefont {S.~W.}\ \bibnamefont
  {Hawking}},\ }in\ \href@noop {} {\emph {\bibinfo {booktitle} {{1st Oxford
  Conference on Quantum Gravity}}}}\ (\bibinfo {year} {1975})\BibitemShut
  {NoStop}%
\bibitem [{\citenamefont {Hawking}(1975{\natexlab{b}})}]{Hawking:1975vcx}%
  \BibitemOpen
  \bibfield  {author} {\bibinfo {author} {\bibfnamefont {S.~W.}\ \bibnamefont
  {Hawking}},\ }\href {\doibase 10.1007/BF02345020} {\bibfield  {journal}
  {\bibinfo  {journal} {Commun. Math. Phys.}\ }\textbf {\bibinfo {volume}
  {43}},\ \bibinfo {pages} {199} (\bibinfo {year} {1975}{\natexlab{b}})},\
  \bibinfo {note} {[Erratum: Commun.Math.Phys. 46, 206 (1976)]}\BibitemShut
  {NoStop}%
\bibitem [{\citenamefont {Page}(1976{\natexlab{a}})}]{Page:1976df}%
  \BibitemOpen
  \bibfield  {author} {\bibinfo {author} {\bibfnamefont {D.~N.}\ \bibnamefont
  {Page}},\ }\href {\doibase 10.1103/PhysRevD.13.198} {\bibfield  {journal}
  {\bibinfo  {journal} {Phys. Rev. D}\ }\textbf {\bibinfo {volume} {13}},\
  \bibinfo {pages} {198} (\bibinfo {year} {1976}{\natexlab{a}})}\BibitemShut
  {NoStop}%
\bibitem [{\citenamefont {Page}(1976{\natexlab{b}})}]{Page:1976ki}%
  \BibitemOpen
  \bibfield  {author} {\bibinfo {author} {\bibfnamefont {D.~N.}\ \bibnamefont
  {Page}},\ }\href {\doibase 10.1103/PhysRevD.14.3260} {\bibfield  {journal}
  {\bibinfo  {journal} {Phys. Rev. D}\ }\textbf {\bibinfo {volume} {14}},\
  \bibinfo {pages} {3260} (\bibinfo {year} {1976}{\natexlab{b}})}\BibitemShut
  {NoStop}%
\bibitem [{\citenamefont {Page}(1977)}]{Page:1977um}%
  \BibitemOpen
  \bibfield  {author} {\bibinfo {author} {\bibfnamefont {D.~N.}\ \bibnamefont
  {Page}},\ }\href {\doibase 10.1103/PhysRevD.16.2402} {\bibfield  {journal}
  {\bibinfo  {journal} {Phys. Rev. D}\ }\textbf {\bibinfo {volume} {16}},\
  \bibinfo {pages} {2402} (\bibinfo {year} {1977})}\BibitemShut {NoStop}%
\bibitem [{\citenamefont {Auffinger}(2023)}]{Auffinger:2022khh}%
  \BibitemOpen
  \bibfield  {author} {\bibinfo {author} {\bibfnamefont {J.}~\bibnamefont
  {Auffinger}},\ }\href {\doibase 10.1016/j.ppnp.2023.104040} {\bibfield
  {journal} {\bibinfo  {journal} {Prog. Part. Nucl. Phys.}\ }\textbf {\bibinfo
  {volume} {131}},\ \bibinfo {pages} {104040} (\bibinfo {year} {2023})},\
  \Eprint {http://arxiv.org/abs/2206.02672} {arXiv:2206.02672 [astro-ph.CO]}
  \BibitemShut {NoStop}%
\bibitem [{\citenamefont {K\"uhnel}\ \emph {et~al.}(2016)\citenamefont
  {K\"uhnel}, \citenamefont {Rampf},\ and\ \citenamefont
  {Sandstad}}]{Kuhnel:2015vtw}%
  \BibitemOpen
  \bibfield  {author} {\bibinfo {author} {\bibfnamefont {F.}~\bibnamefont
  {K\"uhnel}}, \bibinfo {author} {\bibfnamefont {C.}~\bibnamefont {Rampf}}, \
  and\ \bibinfo {author} {\bibfnamefont {M.}~\bibnamefont {Sandstad}},\ }\href
  {\doibase 10.1140/epjc/s10052-016-3945-8} {\bibfield  {journal} {\bibinfo
  {journal} {Eur. Phys. J. C}\ }\textbf {\bibinfo {volume} {76}},\ \bibinfo
  {pages} {93} (\bibinfo {year} {2016})},\ \Eprint
  {http://arxiv.org/abs/1512.00488} {arXiv:1512.00488 [astro-ph.CO]}
  \BibitemShut {NoStop}%
\bibitem [{\citenamefont {K\"uhnel}\ and\ \citenamefont
  {Freese}(2017)}]{Kuhnel:2017pwq}%
  \BibitemOpen
  \bibfield  {author} {\bibinfo {author} {\bibfnamefont {F.}~\bibnamefont
  {K\"uhnel}}\ and\ \bibinfo {author} {\bibfnamefont {K.}~\bibnamefont
  {Freese}},\ }\href {\doibase 10.1103/PhysRevD.95.083508} {\bibfield
  {journal} {\bibinfo  {journal} {Phys. Rev. D}\ }\textbf {\bibinfo {volume}
  {95}},\ \bibinfo {pages} {083508} (\bibinfo {year} {2017})},\ \Eprint
  {http://arxiv.org/abs/1701.07223} {arXiv:1701.07223 [astro-ph.CO]}
  \BibitemShut {NoStop}%
\bibitem [{\citenamefont {Carr}\ \emph {et~al.}(2017)\citenamefont {Carr},
  \citenamefont {Raidal}, \citenamefont {Tenkanen}, \citenamefont {Vaskonen},\
  and\ \citenamefont {Veerm\"ae}}]{Carr:2017jsz}%
  \BibitemOpen
  \bibfield  {author} {\bibinfo {author} {\bibfnamefont {B.}~\bibnamefont
  {Carr}}, \bibinfo {author} {\bibfnamefont {M.}~\bibnamefont {Raidal}},
  \bibinfo {author} {\bibfnamefont {T.}~\bibnamefont {Tenkanen}}, \bibinfo
  {author} {\bibfnamefont {V.}~\bibnamefont {Vaskonen}}, \ and\ \bibinfo
  {author} {\bibfnamefont {H.}~\bibnamefont {Veerm\"ae}},\ }\href {\doibase
  10.1103/PhysRevD.96.023514} {\bibfield  {journal} {\bibinfo  {journal} {Phys.
  Rev. D}\ }\textbf {\bibinfo {volume} {96}},\ \bibinfo {pages} {023514}
  (\bibinfo {year} {2017})},\ \Eprint {http://arxiv.org/abs/1705.05567}
  {arXiv:1705.05567 [astro-ph.CO]} \BibitemShut {NoStop}%
\bibitem [{\citenamefont {Raidal}\ \emph {et~al.}(2017)\citenamefont {Raidal},
  \citenamefont {Vaskonen},\ and\ \citenamefont {Veerm\"ae}}]{Raidal:2017mfl}%
  \BibitemOpen
  \bibfield  {author} {\bibinfo {author} {\bibfnamefont {M.}~\bibnamefont
  {Raidal}}, \bibinfo {author} {\bibfnamefont {V.}~\bibnamefont {Vaskonen}}, \
  and\ \bibinfo {author} {\bibfnamefont {H.}~\bibnamefont {Veerm\"ae}},\ }\href
  {\doibase 10.1088/1475-7516/2017/09/037} {\bibfield  {journal} {\bibinfo
  {journal} {JCAP}\ }\textbf {\bibinfo {volume} {09}},\ \bibinfo {pages} {037}
  (\bibinfo {year} {2017})},\ \Eprint {http://arxiv.org/abs/1707.01480}
  {arXiv:1707.01480 [astro-ph.CO]} \BibitemShut {NoStop}%
\bibitem [{\citenamefont {Bellomo}\ \emph {et~al.}(2018)\citenamefont
  {Bellomo}, \citenamefont {Bernal}, \citenamefont {Raccanelli},\ and\
  \citenamefont {Verde}}]{Bellomo:2017zsr}%
  \BibitemOpen
  \bibfield  {author} {\bibinfo {author} {\bibfnamefont {N.}~\bibnamefont
  {Bellomo}}, \bibinfo {author} {\bibfnamefont {J.~L.}\ \bibnamefont {Bernal}},
  \bibinfo {author} {\bibfnamefont {A.}~\bibnamefont {Raccanelli}}, \ and\
  \bibinfo {author} {\bibfnamefont {L.}~\bibnamefont {Verde}},\ }\href
  {\doibase 10.1088/1475-7516/2018/01/004} {\bibfield  {journal} {\bibinfo
  {journal} {JCAP}\ }\textbf {\bibinfo {volume} {01}},\ \bibinfo {pages} {004}
  (\bibinfo {year} {2018})},\ \Eprint {http://arxiv.org/abs/1709.07467}
  {arXiv:1709.07467 [astro-ph.CO]} \BibitemShut {NoStop}%
\bibitem [{\citenamefont {Lehmann}\ \emph {et~al.}(2018)\citenamefont
  {Lehmann}, \citenamefont {Profumo},\ and\ \citenamefont
  {Yant}}]{Lehmann:2018ejc}%
  \BibitemOpen
  \bibfield  {author} {\bibinfo {author} {\bibfnamefont {B.~V.}\ \bibnamefont
  {Lehmann}}, \bibinfo {author} {\bibfnamefont {S.}~\bibnamefont {Profumo}}, \
  and\ \bibinfo {author} {\bibfnamefont {J.}~\bibnamefont {Yant}},\ }\href
  {\doibase 10.1088/1475-7516/2018/04/007} {\bibfield  {journal} {\bibinfo
  {journal} {JCAP}\ }\textbf {\bibinfo {volume} {04}},\ \bibinfo {pages} {007}
  (\bibinfo {year} {2018})},\ \Eprint {http://arxiv.org/abs/1801.00808}
  {arXiv:1801.00808 [astro-ph.CO]} \BibitemShut {NoStop}%
\bibitem [{\citenamefont {Carr}\ and\ \citenamefont
  {Kuhnel}(2019)}]{Carr:2018poi}%
  \BibitemOpen
  \bibfield  {author} {\bibinfo {author} {\bibfnamefont {B.}~\bibnamefont
  {Carr}}\ and\ \bibinfo {author} {\bibfnamefont {F.}~\bibnamefont {Kuhnel}},\
  }\href {\doibase 10.1103/PhysRevD.99.103535} {\bibfield  {journal} {\bibinfo
  {journal} {Phys. Rev. D}\ }\textbf {\bibinfo {volume} {99}},\ \bibinfo
  {pages} {103535} (\bibinfo {year} {2019})},\ \Eprint
  {http://arxiv.org/abs/1811.06532} {arXiv:1811.06532 [astro-ph.CO]}
  \BibitemShut {NoStop}%
\bibitem [{\citenamefont {Gow}\ \emph {et~al.}(2020)\citenamefont {Gow},
  \citenamefont {Byrnes}, \citenamefont {Hall},\ and\ \citenamefont
  {Peacock}}]{Gow:2019pok}%
  \BibitemOpen
  \bibfield  {author} {\bibinfo {author} {\bibfnamefont {A.~D.}\ \bibnamefont
  {Gow}}, \bibinfo {author} {\bibfnamefont {C.~T.}\ \bibnamefont {Byrnes}},
  \bibinfo {author} {\bibfnamefont {A.}~\bibnamefont {Hall}}, \ and\ \bibinfo
  {author} {\bibfnamefont {J.~A.}\ \bibnamefont {Peacock}},\ }\href {\doibase
  10.1088/1475-7516/2020/01/031} {\bibfield  {journal} {\bibinfo  {journal}
  {JCAP}\ }\textbf {\bibinfo {volume} {01}},\ \bibinfo {pages} {031} (\bibinfo
  {year} {2020})},\ \Eprint {http://arxiv.org/abs/1911.12685} {arXiv:1911.12685
  [astro-ph.CO]} \BibitemShut {NoStop}%
\bibitem [{\citenamefont {De~Luca}\ \emph
  {et~al.}(2020{\natexlab{d}})\citenamefont {De~Luca}, \citenamefont
  {Franciolini},\ and\ \citenamefont {Riotto}}]{DeLuca:2020ioi}%
  \BibitemOpen
  \bibfield  {author} {\bibinfo {author} {\bibfnamefont {V.}~\bibnamefont
  {De~Luca}}, \bibinfo {author} {\bibfnamefont {G.}~\bibnamefont
  {Franciolini}}, \ and\ \bibinfo {author} {\bibfnamefont {A.}~\bibnamefont
  {Riotto}},\ }\href {\doibase 10.1016/j.physletb.2020.135550} {\bibfield
  {journal} {\bibinfo  {journal} {Phys. Lett. B}\ }\textbf {\bibinfo {volume}
  {807}},\ \bibinfo {pages} {135550} (\bibinfo {year} {2020}{\natexlab{d}})},\
  \Eprint {http://arxiv.org/abs/2001.04371} {arXiv:2001.04371 [astro-ph.CO]}
  \BibitemShut {NoStop}%
\bibitem [{\citenamefont {Gow}\ \emph {et~al.}(2022)\citenamefont {Gow},
  \citenamefont {Byrnes},\ and\ \citenamefont {Hall}}]{Gow:2020cou}%
  \BibitemOpen
  \bibfield  {author} {\bibinfo {author} {\bibfnamefont {A.~D.}\ \bibnamefont
  {Gow}}, \bibinfo {author} {\bibfnamefont {C.~T.}\ \bibnamefont {Byrnes}}, \
  and\ \bibinfo {author} {\bibfnamefont {A.}~\bibnamefont {Hall}},\ }\href
  {\doibase 10.1103/PhysRevD.105.023503} {\bibfield  {journal} {\bibinfo
  {journal} {Phys. Rev. D}\ }\textbf {\bibinfo {volume} {105}},\ \bibinfo
  {pages} {023503} (\bibinfo {year} {2022})},\ \Eprint
  {http://arxiv.org/abs/2009.03204} {arXiv:2009.03204 [astro-ph.CO]}
  \BibitemShut {NoStop}%
\bibitem [{\citenamefont {Ashoorioon}\ \emph {et~al.}(2021)\citenamefont
  {Ashoorioon}, \citenamefont {Rostami},\ and\ \citenamefont
  {Firouzjaee}}]{Ashoorioon:2020hln}%
  \BibitemOpen
  \bibfield  {author} {\bibinfo {author} {\bibfnamefont {A.}~\bibnamefont
  {Ashoorioon}}, \bibinfo {author} {\bibfnamefont {A.}~\bibnamefont {Rostami}},
  \ and\ \bibinfo {author} {\bibfnamefont {J.~T.}\ \bibnamefont {Firouzjaee}},\
  }\href {\doibase 10.1103/PhysRevD.103.123512} {\bibfield  {journal} {\bibinfo
   {journal} {Phys. Rev. D}\ }\textbf {\bibinfo {volume} {103}},\ \bibinfo
  {pages} {123512} (\bibinfo {year} {2021})},\ \Eprint
  {http://arxiv.org/abs/2012.02817} {arXiv:2012.02817 [astro-ph.CO]}
  \BibitemShut {NoStop}%
\bibitem [{\citenamefont {Bagui}\ and\ \citenamefont
  {Clesse}(2022)}]{Bagui:2021dqi}%
  \BibitemOpen
  \bibfield  {author} {\bibinfo {author} {\bibfnamefont {E.}~\bibnamefont
  {Bagui}}\ and\ \bibinfo {author} {\bibfnamefont {S.}~\bibnamefont {Clesse}},\
  }\href {\doibase 10.1016/j.dark.2022.101115} {\bibfield  {journal} {\bibinfo
  {journal} {Phys. Dark Univ.}\ }\textbf {\bibinfo {volume} {38}},\ \bibinfo
  {pages} {101115} (\bibinfo {year} {2022})},\ \Eprint
  {http://arxiv.org/abs/2110.07487} {arXiv:2110.07487 [astro-ph.CO]}
  \BibitemShut {NoStop}%
\bibitem [{\citenamefont {Mukhopadhyay}\ \emph {et~al.}(2022)\citenamefont
  {Mukhopadhyay}, \citenamefont {Majumdar},\ and\ \citenamefont
  {Halder}}]{Mukhopadhyay:2022jqc}%
  \BibitemOpen
  \bibfield  {author} {\bibinfo {author} {\bibfnamefont {U.}~\bibnamefont
  {Mukhopadhyay}}, \bibinfo {author} {\bibfnamefont {D.}~\bibnamefont
  {Majumdar}}, \ and\ \bibinfo {author} {\bibfnamefont {A.}~\bibnamefont
  {Halder}},\ }\href {\doibase 10.1088/1475-7516/2022/10/099} {\bibfield
  {journal} {\bibinfo  {journal} {JCAP}\ }\textbf {\bibinfo {volume} {10}},\
  \bibinfo {pages} {099} (\bibinfo {year} {2022})},\ \Eprint
  {http://arxiv.org/abs/2203.13008} {arXiv:2203.13008 [astro-ph.CO]}
  \BibitemShut {NoStop}%
\bibitem [{\citenamefont {Papanikolaou}(2022)}]{Papanikolaou:2022chm}%
  \BibitemOpen
  \bibfield  {author} {\bibinfo {author} {\bibfnamefont {T.}~\bibnamefont
  {Papanikolaou}},\ }\href {\doibase 10.1088/1475-7516/2022/10/089} {\bibfield
  {journal} {\bibinfo  {journal} {JCAP}\ }\textbf {\bibinfo {volume} {10}},\
  \bibinfo {pages} {089} (\bibinfo {year} {2022})},\ \Eprint
  {http://arxiv.org/abs/2207.11041} {arXiv:2207.11041 [astro-ph.CO]}
  \BibitemShut {NoStop}%
\bibitem [{\citenamefont {Cai}\ \emph {et~al.}(2024{\natexlab{b}})\citenamefont
  {Cai}, \citenamefont {Tang}, \citenamefont {Mo}, \citenamefont {Yan},
  \citenamefont {Chen}, \citenamefont {Ma}, \citenamefont {Wang}, \citenamefont
  {Luo}, \citenamefont {Easson},\ and\ \citenamefont {Marciano}}]{Cai:2023ptf}%
  \BibitemOpen
  \bibfield  {author} {\bibinfo {author} {\bibfnamefont {Y.-F.}\ \bibnamefont
  {Cai}}, \bibinfo {author} {\bibfnamefont {C.}~\bibnamefont {Tang}}, \bibinfo
  {author} {\bibfnamefont {G.}~\bibnamefont {Mo}}, \bibinfo {author}
  {\bibfnamefont {S.-F.}\ \bibnamefont {Yan}}, \bibinfo {author} {\bibfnamefont
  {C.}~\bibnamefont {Chen}}, \bibinfo {author} {\bibfnamefont {X.-H.}\
  \bibnamefont {Ma}}, \bibinfo {author} {\bibfnamefont {B.}~\bibnamefont
  {Wang}}, \bibinfo {author} {\bibfnamefont {W.}~\bibnamefont {Luo}}, \bibinfo
  {author} {\bibfnamefont {D.~A.}\ \bibnamefont {Easson}}, \ and\ \bibinfo
  {author} {\bibfnamefont {A.}~\bibnamefont {Marciano}},\ }\href {\doibase
  10.1007/s11433-023-2314-1} {\bibfield  {journal} {\bibinfo  {journal} {Sci.
  China Phys. Mech. Astron.}\ }\textbf {\bibinfo {volume} {67}},\ \bibinfo
  {pages} {259512} (\bibinfo {year} {2024}{\natexlab{b}})},\ \Eprint
  {http://arxiv.org/abs/2301.09403} {arXiv:2301.09403 [astro-ph.CO]}
  \BibitemShut {NoStop}%
\bibitem [{\citenamefont {Carr}\ \emph {et~al.}(2010)\citenamefont {Carr},
  \citenamefont {Kohri}, \citenamefont {Sendouda},\ and\ \citenamefont
  {Yokoyama}}]{Carr:2009jm}%
  \BibitemOpen
  \bibfield  {author} {\bibinfo {author} {\bibfnamefont {B.~J.}\ \bibnamefont
  {Carr}}, \bibinfo {author} {\bibfnamefont {K.}~\bibnamefont {Kohri}},
  \bibinfo {author} {\bibfnamefont {Y.}~\bibnamefont {Sendouda}}, \ and\
  \bibinfo {author} {\bibfnamefont {J.}~\bibnamefont {Yokoyama}},\ }\href
  {\doibase 10.1103/PhysRevD.81.104019} {\bibfield  {journal} {\bibinfo
  {journal} {Phys. Rev. D}\ }\textbf {\bibinfo {volume} {81}},\ \bibinfo
  {pages} {104019} (\bibinfo {year} {2010})},\ \Eprint
  {http://arxiv.org/abs/0912.5297} {arXiv:0912.5297 [astro-ph.CO]} \BibitemShut
  {NoStop}%
\bibitem [{\citenamefont {Gruber}\ \emph {et~al.}(1999)\citenamefont {Gruber},
  \citenamefont {Matteson}, \citenamefont {Peterson},\ and\ \citenamefont
  {Jung}}]{Gruber:1999yr}%
  \BibitemOpen
  \bibfield  {author} {\bibinfo {author} {\bibfnamefont {D.~E.}\ \bibnamefont
  {Gruber}}, \bibinfo {author} {\bibfnamefont {J.~L.}\ \bibnamefont
  {Matteson}}, \bibinfo {author} {\bibfnamefont {L.~E.}\ \bibnamefont
  {Peterson}}, \ and\ \bibinfo {author} {\bibfnamefont {G.~V.}\ \bibnamefont
  {Jung}},\ }\href {\doibase 10.1086/307450} {\bibfield  {journal} {\bibinfo
  {journal} {Astrophys. J.}\ }\textbf {\bibinfo {volume} {520}},\ \bibinfo
  {pages} {124} (\bibinfo {year} {1999})},\ \Eprint
  {http://arxiv.org/abs/astro-ph/9903492} {arXiv:astro-ph/9903492} \BibitemShut
  {NoStop}%
\bibitem [{\citenamefont {Schoenfelder}(2000)}]{Schoenfelder:2000bu}%
  \BibitemOpen
  \bibfield  {author} {\bibinfo {author} {\bibfnamefont {V.}~\bibnamefont
  {Schoenfelder}} (\bibinfo {collaboration} {COMPTEL}),\ }\href {\doibase
  10.1051/aas:2000101} {\bibfield  {journal} {\bibinfo  {journal} {Astron.
  Astrophys. Suppl. Ser.}\ }\textbf {\bibinfo {volume} {143}},\ \bibinfo
  {pages} {145} (\bibinfo {year} {2000})},\ \Eprint
  {http://arxiv.org/abs/astro-ph/0002366} {arXiv:astro-ph/0002366} \BibitemShut
  {NoStop}%
\bibitem [{\citenamefont {Strong}\ \emph {et~al.}(2004)\citenamefont {Strong},
  \citenamefont {Moskalenko},\ and\ \citenamefont {Reimer}}]{Strong:2004ry}%
  \BibitemOpen
  \bibfield  {author} {\bibinfo {author} {\bibfnamefont {A.~W.}\ \bibnamefont
  {Strong}}, \bibinfo {author} {\bibfnamefont {I.~V.}\ \bibnamefont
  {Moskalenko}}, \ and\ \bibinfo {author} {\bibfnamefont {O.}~\bibnamefont
  {Reimer}},\ }\href {\doibase 10.1086/423196} {\bibfield  {journal} {\bibinfo
  {journal} {Astrophys. J.}\ }\textbf {\bibinfo {volume} {613}},\ \bibinfo
  {pages} {956} (\bibinfo {year} {2004})},\ \Eprint
  {http://arxiv.org/abs/astro-ph/0405441} {arXiv:astro-ph/0405441} \BibitemShut
  {NoStop}%
\bibitem [{\citenamefont {Carr}\ \emph
  {et~al.}(2016{\natexlab{b}})\citenamefont {Carr}, \citenamefont {Kohri},
  \citenamefont {Sendouda},\ and\ \citenamefont {Yokoyama}}]{Carr:2016hva}%
  \BibitemOpen
  \bibfield  {author} {\bibinfo {author} {\bibfnamefont {B.~J.}\ \bibnamefont
  {Carr}}, \bibinfo {author} {\bibfnamefont {K.}~\bibnamefont {Kohri}},
  \bibinfo {author} {\bibfnamefont {Y.}~\bibnamefont {Sendouda}}, \ and\
  \bibinfo {author} {\bibfnamefont {J.}~\bibnamefont {Yokoyama}},\ }\href
  {\doibase 10.1103/PhysRevD.94.044029} {\bibfield  {journal} {\bibinfo
  {journal} {Phys. Rev. D}\ }\textbf {\bibinfo {volume} {94}},\ \bibinfo
  {pages} {044029} (\bibinfo {year} {2016}{\natexlab{b}})},\ \Eprint
  {http://arxiv.org/abs/1604.05349} {arXiv:1604.05349 [astro-ph.CO]}
  \BibitemShut {NoStop}%
\bibitem [{\citenamefont {Boudaud}\ and\ \citenamefont
  {Cirelli}(2019)}]{Boudaud:2018hqb}%
  \BibitemOpen
  \bibfield  {author} {\bibinfo {author} {\bibfnamefont {M.}~\bibnamefont
  {Boudaud}}\ and\ \bibinfo {author} {\bibfnamefont {M.}~\bibnamefont
  {Cirelli}},\ }\href {\doibase 10.1103/PhysRevLett.122.041104} {\bibfield
  {journal} {\bibinfo  {journal} {Phys. Rev. Lett.}\ }\textbf {\bibinfo
  {volume} {122}},\ \bibinfo {pages} {041104} (\bibinfo {year} {2019})},\
  \Eprint {http://arxiv.org/abs/1807.03075} {arXiv:1807.03075 [astro-ph.HE]}
  \BibitemShut {NoStop}%
\bibitem [{\citenamefont {DeRocco}\ and\ \citenamefont
  {Graham}(2019)}]{DeRocco:2019fjq}%
  \BibitemOpen
  \bibfield  {author} {\bibinfo {author} {\bibfnamefont {W.}~\bibnamefont
  {DeRocco}}\ and\ \bibinfo {author} {\bibfnamefont {P.~W.}\ \bibnamefont
  {Graham}},\ }\href {\doibase 10.1103/PhysRevLett.123.251102} {\bibfield
  {journal} {\bibinfo  {journal} {Phys. Rev. Lett.}\ }\textbf {\bibinfo
  {volume} {123}},\ \bibinfo {pages} {251102} (\bibinfo {year} {2019})},\
  \Eprint {http://arxiv.org/abs/1906.07740} {arXiv:1906.07740 [astro-ph.CO]}
  \BibitemShut {NoStop}%
\bibitem [{\citenamefont {Laha}(2019)}]{Laha:2019ssq}%
  \BibitemOpen
  \bibfield  {author} {\bibinfo {author} {\bibfnamefont {R.}~\bibnamefont
  {Laha}},\ }\href {\doibase 10.1103/PhysRevLett.123.251101} {\bibfield
  {journal} {\bibinfo  {journal} {Phys. Rev. Lett.}\ }\textbf {\bibinfo
  {volume} {123}},\ \bibinfo {pages} {251101} (\bibinfo {year} {2019})},\
  \Eprint {http://arxiv.org/abs/1906.09994} {arXiv:1906.09994 [astro-ph.HE]}
  \BibitemShut {NoStop}%
\bibitem [{\citenamefont {Dasgupta}\ \emph {et~al.}(2020)\citenamefont
  {Dasgupta}, \citenamefont {Laha},\ and\ \citenamefont
  {Ray}}]{Dasgupta:2019cae}%
  \BibitemOpen
  \bibfield  {author} {\bibinfo {author} {\bibfnamefont {B.}~\bibnamefont
  {Dasgupta}}, \bibinfo {author} {\bibfnamefont {R.}~\bibnamefont {Laha}}, \
  and\ \bibinfo {author} {\bibfnamefont {A.}~\bibnamefont {Ray}},\ }\href
  {\doibase 10.1103/PhysRevLett.125.101101} {\bibfield  {journal} {\bibinfo
  {journal} {Phys. Rev. Lett.}\ }\textbf {\bibinfo {volume} {125}},\ \bibinfo
  {pages} {101101} (\bibinfo {year} {2020})},\ \Eprint
  {http://arxiv.org/abs/1912.01014} {arXiv:1912.01014 [hep-ph]} \BibitemShut
  {NoStop}%
\bibitem [{\citenamefont {Ackermann}\ \emph {et~al.}(2015)\citenamefont
  {Ackermann} \emph {et~al.}}]{Fermi-LAT:2014ryh}%
  \BibitemOpen
  \bibfield  {author} {\bibinfo {author} {\bibfnamefont {M.}~\bibnamefont
  {Ackermann}} \emph {et~al.} (\bibinfo {collaboration} {Fermi-LAT}),\ }\href
  {\doibase 10.1088/0004-637X/799/1/86} {\bibfield  {journal} {\bibinfo
  {journal} {Astrophys. J.}\ }\textbf {\bibinfo {volume} {799}},\ \bibinfo
  {pages} {86} (\bibinfo {year} {2015})},\ \Eprint
  {http://arxiv.org/abs/1410.3696} {arXiv:1410.3696 [astro-ph.HE]} \BibitemShut
  {NoStop}%
\bibitem [{\citenamefont {Dwek}\ and\ \citenamefont
  {Krennrich}(2013)}]{Dwek:2012nb}%
  \BibitemOpen
  \bibfield  {author} {\bibinfo {author} {\bibfnamefont {E.}~\bibnamefont
  {Dwek}}\ and\ \bibinfo {author} {\bibfnamefont {F.}~\bibnamefont
  {Krennrich}},\ }\href {\doibase 10.1016/j.astropartphys.2012.09.003}
  {\bibfield  {journal} {\bibinfo  {journal} {Astropart. Phys.}\ }\textbf
  {\bibinfo {volume} {43}},\ \bibinfo {pages} {112} (\bibinfo {year} {2013})},\
  \Eprint {http://arxiv.org/abs/1209.4661} {arXiv:1209.4661 [astro-ph.CO]}
  \BibitemShut {NoStop}%
\bibitem [{\citenamefont {Vagnozzi}\ \emph {et~al.}(2023)\citenamefont
  {Vagnozzi} \emph {et~al.}}]{Vagnozzi:2022moj}%
  \BibitemOpen
  \bibfield  {author} {\bibinfo {author} {\bibfnamefont {S.}~\bibnamefont
  {Vagnozzi}} \emph {et~al.},\ }\href {\doibase 10.1088/1361-6382/acd97b}
  {\bibfield  {journal} {\bibinfo  {journal} {Class. Quant. Grav.}\ }\textbf
  {\bibinfo {volume} {40}},\ \bibinfo {pages} {165007} (\bibinfo {year}
  {2023})},\ \Eprint {http://arxiv.org/abs/2205.07787} {arXiv:2205.07787
  [gr-qc]} \BibitemShut {NoStop}%
\bibitem [{\citenamefont {Capela}\ \emph
  {et~al.}(2013{\natexlab{a}})\citenamefont {Capela}, \citenamefont
  {Pshirkov},\ and\ \citenamefont {Tinyakov}}]{Capela:2012jz}%
  \BibitemOpen
  \bibfield  {author} {\bibinfo {author} {\bibfnamefont {F.}~\bibnamefont
  {Capela}}, \bibinfo {author} {\bibfnamefont {M.}~\bibnamefont {Pshirkov}}, \
  and\ \bibinfo {author} {\bibfnamefont {P.}~\bibnamefont {Tinyakov}},\ }\href
  {\doibase 10.1103/PhysRevD.87.023507} {\bibfield  {journal} {\bibinfo
  {journal} {Phys. Rev. D}\ }\textbf {\bibinfo {volume} {87}},\ \bibinfo
  {pages} {023507} (\bibinfo {year} {2013}{\natexlab{a}})},\ \Eprint
  {http://arxiv.org/abs/1209.6021} {arXiv:1209.6021 [astro-ph.CO]} \BibitemShut
  {NoStop}%
\bibitem [{\citenamefont {Capela}\ \emph
  {et~al.}(2013{\natexlab{b}})\citenamefont {Capela}, \citenamefont
  {Pshirkov},\ and\ \citenamefont {Tinyakov}}]{Capela:2013yf}%
  \BibitemOpen
  \bibfield  {author} {\bibinfo {author} {\bibfnamefont {F.}~\bibnamefont
  {Capela}}, \bibinfo {author} {\bibfnamefont {M.}~\bibnamefont {Pshirkov}}, \
  and\ \bibinfo {author} {\bibfnamefont {P.}~\bibnamefont {Tinyakov}},\ }\href
  {\doibase 10.1103/PhysRevD.87.123524} {\bibfield  {journal} {\bibinfo
  {journal} {Phys. Rev. D}\ }\textbf {\bibinfo {volume} {87}},\ \bibinfo
  {pages} {123524} (\bibinfo {year} {2013}{\natexlab{b}})},\ \Eprint
  {http://arxiv.org/abs/1301.4984} {arXiv:1301.4984 [astro-ph.CO]} \BibitemShut
  {NoStop}%
\bibitem [{\citenamefont {Pani}\ and\ \citenamefont
  {Loeb}(2014)}]{Pani:2014rca}%
  \BibitemOpen
  \bibfield  {author} {\bibinfo {author} {\bibfnamefont {P.}~\bibnamefont
  {Pani}}\ and\ \bibinfo {author} {\bibfnamefont {A.}~\bibnamefont {Loeb}},\
  }\href {\doibase 10.1088/1475-7516/2014/06/026} {\bibfield  {journal}
  {\bibinfo  {journal} {JCAP}\ }\textbf {\bibinfo {volume} {06}},\ \bibinfo
  {pages} {026} (\bibinfo {year} {2014})},\ \Eprint
  {http://arxiv.org/abs/1401.3025} {arXiv:1401.3025 [astro-ph.CO]} \BibitemShut
  {NoStop}%
\bibitem [{\citenamefont {Graham}\ \emph {et~al.}(2015)\citenamefont {Graham},
  \citenamefont {Rajendran},\ and\ \citenamefont {Varela}}]{Graham:2015apa}%
  \BibitemOpen
  \bibfield  {author} {\bibinfo {author} {\bibfnamefont {P.~W.}\ \bibnamefont
  {Graham}}, \bibinfo {author} {\bibfnamefont {S.}~\bibnamefont {Rajendran}}, \
  and\ \bibinfo {author} {\bibfnamefont {J.}~\bibnamefont {Varela}},\ }\href
  {\doibase 10.1103/PhysRevD.92.063007} {\bibfield  {journal} {\bibinfo
  {journal} {Phys. Rev. D}\ }\textbf {\bibinfo {volume} {92}},\ \bibinfo
  {pages} {063007} (\bibinfo {year} {2015})},\ \Eprint
  {http://arxiv.org/abs/1505.04444} {arXiv:1505.04444 [hep-ph]} \BibitemShut
  {NoStop}%
\bibitem [{\citenamefont {Kainulainen}\ \emph {et~al.}(2021)\citenamefont
  {Kainulainen}, \citenamefont {Nurmi}, \citenamefont {Schiappacasse},\ and\
  \citenamefont {Yanagida}}]{Kainulainen:2021rbg}%
  \BibitemOpen
  \bibfield  {author} {\bibinfo {author} {\bibfnamefont {K.}~\bibnamefont
  {Kainulainen}}, \bibinfo {author} {\bibfnamefont {S.}~\bibnamefont {Nurmi}},
  \bibinfo {author} {\bibfnamefont {E.~D.}\ \bibnamefont {Schiappacasse}}, \
  and\ \bibinfo {author} {\bibfnamefont {T.~T.}\ \bibnamefont {Yanagida}},\
  }\href {\doibase 10.1103/PhysRevD.104.123033} {\bibfield  {journal} {\bibinfo
   {journal} {Phys. Rev. D}\ }\textbf {\bibinfo {volume} {104}},\ \bibinfo
  {pages} {123033} (\bibinfo {year} {2021})},\ \Eprint
  {http://arxiv.org/abs/2108.08717} {arXiv:2108.08717 [astro-ph.HE]}
  \BibitemShut {NoStop}%
\bibitem [{\citenamefont {Amaral}\ and\ \citenamefont
  {Schiappacasse}(2024)}]{Amaral:2023ekd}%
  \BibitemOpen
  \bibfield  {author} {\bibinfo {author} {\bibfnamefont {D.~W.~P.}\
  \bibnamefont {Amaral}}\ and\ \bibinfo {author} {\bibfnamefont {E.~D.}\
  \bibnamefont {Schiappacasse}},\ }\href {\doibase 10.1103/PhysRevD.110.083532}
  {\bibfield  {journal} {\bibinfo  {journal} {Phys. Rev. D}\ }\textbf {\bibinfo
  {volume} {110}},\ \bibinfo {pages} {083532} (\bibinfo {year} {2024})},\
  \Eprint {http://arxiv.org/abs/2312.09285} {arXiv:2312.09285 [hep-ph]}
  \BibitemShut {NoStop}%
\end{thebibliography}%

\end{document}